%% file: sparse_FLMM_Cederbaum_etal.tex
\documentclass[submit]{smj}
\usepackage{nicefrac}
\usepackage{paralist}
\usepackage{bm}
\usepackage{dsfont}
\usepackage{xfrac} 
\usepackage{amsfonts}
\usepackage{amsthm}
\Author{Jona Cederbaum\Affil{1}, 
        Marianne Pouplier\Affil{2}, 
        Phil Hoole\Affil{2},
        and Sonja Greven\Affil{1}
}
\AuthorRunning{Jona Cederbaum \textrm{et al.}}

\Affiliations{
\item Department of Statistics,
	  Faculty of Mathematics, Computer Science and Statistics, 
	  Ludwig-Maximilians-University Munich,
	  Germany
\item Department of Phonetics and Speech Processing,
      Faculty of Languages and Literature,
      Ludwig-Maximilians-University Munich,
	  Germany
}   

\CorrAddress{Jona Cederbaum, 
             Department of Statistics, 
             Faculty of Mathematics, Computer Science and Statistics,
             Ludwig-Maximilians-University Munich,
             Ludwigstr.33, 80539 Munich,
             Germany}
\CorrEmail{Jona.Cederbaum@stat.uni-muenchen.de}
\CorrPhone{(+49)\;89\;2180\;2240}
\CorrFax{(+49)\;89\;2180\;5040}
\Title{Functional Linear Mixed Models for Irregularly or Sparsely Sampled Data}
\TitleRunning{sparse FLMM}

\Abstract{
We propose an estimation approach to analyse correlated functional data which are observed on unequal grids or even sparsely. The model we use is a functional linear mixed model, a functional analogue of the linear mixed model. Estimation is based on dimension reduction via functional principal component analysis and on mixed model methodology. Our procedure allows the decomposition of the variability in the data as well as the estimation of mean effects of interest and borrows strength across curves. Confidence bands for mean effects can be constructed conditional on estimated principal components. We provide \pkg{R}-code implementing our approach. The method is motivated by and applied to data from speech production research.\\
}


\input{header_Sparse_FLMM}

\begin{document}
\maketitle


\section{Introduction}
Advancements in technology allow today's scientists to collect an increasing amount of data consisting of functional observations rather than single data points. Most methods in functional data analysis (fda) \citep[see e.g.][]{Ramsay.2005} assume that observations are \emph{a)} independent and/or \emph{b)} observed at a typically large number of the same (equidistant) observation points across curves.\newline Linguistic research is only one of numerous fields in which the data often do not meet these strong requirements. Our motivating data come from a speech production study \citep{Pouplier.2014} on assimilation, the phenomenon that the articulation of consonants becomes more alike when they appear next to each other or across words. The data consist of the acoustic signals of nine speakers reading the same 16 target words including the consonants of interest each five times. Due to the repeated measurements for speakers and for target words, the data have a crossed design structure. Moreover, number and location of the observation points differ between the observed curves. \\
We propose a model and an estimation approach that extend existing methods by accounting for both \emph{a)} correlation between functional data and for \emph{b)} irregular spacing of -- possibly very few -- observation points. The model is a functional analogue of the linear mixed model (LMM),  
which is frequently used to analyse scalar correlated data. \\
For dimension reduction, we use functional principal component analysis \citep[FPCA, see e.g.][]{Ramsay.2005} to extract the dominant modes of variation in the data. The functional random effects are expanded in bases of eigenfunctions of their auto-covariances, which we estimate beforehand using a method of moments estimator. 
FPCA has become a key tool in fda as it yields a parsimonious representation of the data. It is attractive as the eigenfunction bases are estimated from the data and have optimal approximation properties for a fixed number of basis functions. It also allows for an explicit decomposition of the variability in the data.  \\
We propose two ways of predicting the eigenfunction weights. We either compute them directly as empirical best linear unbiased predictors (EBLUPs) of the resulting LMM or we alternatively embed our model in the more general framework of functional additive mixed models (FAMMs) introduced by \cite{Scheipl.2014}. The first approach is straight forward and computationally more efficient; it does not require additional estimation steps as a plug-in estimate is used. The latter has the advantage that all model components are estimated/predicted in one framework and it allows for approximate statistical inference conditional on the FPCA. \\
Previous work for dependent functional data 
differ in their generality and restrictions on the sampling grid, but only one approach can cope with correlated functional data observed on unequal grids and with complex correlation structures. 
\cite{Brumback.1998} consider a smoothing spline-based method for nested or crossed curves, which are modelled as fixed effect curves. 
They allow for missing observations in equal grids but do not consider any covariate effects.
A Bayesian wavelet-based functional mixed model approach is introduced by \cite{Morris.2003} and extended by \cite{Morris.2006}, \cite{Morris.2006b}, and subsequent work by this group. While this approach is quite general in the possible functional random effects structure, it assumes regular and equal grids with at most a small proportion of missings and a reasonable number of completely
 observed curves. 
\cite{Di.2009}, \cite{Greven.2010}, and \cite{Shou.2014} consider functional linear mixed models with a functional random intercept, with a functional random intercept and slope, and with nested and crossed functional random intercepts, respectively. 
While following a similar approach to estimation for these models without the options for approximate inference we provide, all three are restricted to data sampled on a fine grid. 
\cite{Di.2014} extend the random intercept model of \cite{Di.2009} to sparse functional data; the correlation structure, however, remains less general than ours. 
Also motivated by an application from linguistics, \cite{Aston.2010} perform an FPCA on all curves ignoring the correlation structure and then use the functional principal component (FPC) weights as the response variables in an LMM with random effects for speakers and words. 
Only linear effects of scalar covariates are considered, FPC bases are restricted to be the same for all latent processes, and it is assumed that the data are sampled on a common grid.
\cite{Brockhaus.2015} propose a unified class for functional regression models including group-specific functional effects, which are represented as linear array models and estimated using boosting. The array structure requires common grids. Other approaches concentrate specifically on spatially correlated functional data on equal grids, as e.g. \cite{Staicu.2010}. A very flexible class of functional response models has been developed by \cite{Scheipl.2014}, allowing for multiple partially nested and crossed functional random effects with flexible correlation structures. Both spline-based and FPC-based representations incorporated in a mixed model framework are considered. They allow for densely and sparsely sampled data, but in the case of the FPC-based representation, they assume that appropriate FPC estimates are available. We take advantage of this general framework and use our estimated FPCs to obtain improved estimates and approximate point-wise confidence bands for the mean and covariate effects. In addition to providing an interpretable variance decomposition, our FPC-based approach reduces computation time by orders of magnitude compared to the spline-based estimates from \cite{Scheipl.2014} (compare Section \ref{sec: Simulations}), allowing the analysis of realistically sized data in practice.\\ 
A number of approaches allow for irregularly or sparsely sampled functional data but assume that curves are independent. \cite{Guo.2002,Guo.2004} first introduce the term \textit{functional mixed effects models} for their model. The model does not capture between-function correlation as only curve-level random effect functions are included which are modelled using smoothing splines. The approach is not restricted to regularly sampled grid data. \cite{Chen.2011} propose a spline-based approach that is suitable for sparsely sampled data, but similar to \cite{Guo.2002,Guo.2004} they only consider curve-level random effects. 
\cite{James.2000}, \cite{Yao.2005}, and \cite{Peng.2009} among others propose FPCA approaches for sparsely observed functional data with uncorrelated curves. \\
For an extensive overview and further references for functional regression approaches, including functional response regression, see \cite{Morris.2015}.

The remainder of the paper is organized as follows. Section \ref{sec: FLMM} introduces the general functional linear mixed model and presents an important special case which is used to analyse the motivating assimilation data. Section \ref{sec: Estimation} develops our estimation framework. Our method is evaluated in an application to the assimilation data and in simulations in Sections \ref{sec: Application} and \ref{sec: Simulations}, respectively. Section \ref{sec: Discussion} closes with a discussion and outlook. Theoretical results and supplementary material including estimation details as well as additional results for application and simulations are available in the appendix.

\section{Functional linear mixed models}\label{sec: FLMM}
\subsection{The general model} 
The general functional linear mixed model (FLMM) is given by
\bea\label{eq: FLMM general}
Y_{i}(t) = \mu(t,\mx_i) + \mz_i\tr \mU(t) + E_i(t) + \eps_{i}(t), \ i=1,\ldots,n,
\eea
where $Y_{i}(t)$ is the square-integrable functional response observed at arguments $t$ in $\mathcal{T}$, a closed interval in $\mathbb{R}$, and $n$ is the number of curves.
 $\mu(t,\mx_i)$ is a fixed main effect surface dependent on a vector of known covariates $\mx_i$. To account for the functional nature of the $Y_{i}(t)$, the random effects of an LMM are replaced by a vector-valued random process $\mU(t)$. $\mz_i$ is a known covariate vector. $E_i(t)$ is a curve-specific deviation in form of a smooth residual curve. We assume that there is white noise measurement error denoted by $\eps_{i}(t)$ with variance $\sigma^2$ that captures random uncorrelated variation within each curve. Note that if needed, the error variance may also vary across $t$, $\sigma^2(t)$.
We further assume that $\mU(t)$, $E_i(t)$, and $\eps_{i}(t)$, $i=1,\ldots,n$, are zero-mean, square-integrable, mutually uncorrelated random processes.
Therefore, each of the $q$ components of $\mU(t)$ has an auto-covariance operator $K^{U_{j}}(t,t^\pr)$, $j=1,\ldots,q$, and cross-covariance operators $K^{U_{j,k}}(t,t^\pr)$, $j,k=1,\ldots,q$, some of which might be zero for uncorrelated functional random effects. $E_{i}(t)$ has an auto-covariance operator $K^E(t,t^\pr)=\Cov\left[ E_i(t),E_i(t^\pr)\right]$. In the following, mean, auto-covariances, and thus also the eigenfunctions are assumed to be smooth in $t$.\\
$\mu(t,\mx_i)$ is an additive function of $t$ and $\mx_i$. For example, it can be constant in $t$, $\mu(t,\mx_i)=\mu(\mx_i)$, or additive in $t$ and $\mx_i$, $\mu(t,\mx_i) = \mu_1(t) + \mu_2(\mx_i)$. Another special case is when all $x_{i1},\ldots,x_{ip}$ in $\mx_i$ act as index-varying coefficients, $\mu(t,\mx_i) = f_0(t) + f_1(t)x_{i1} + \ldots + f_p(t)x_{ip}$, with unknown smooth functions $f_0(\cdot),\ldots,f_p(\cdot)$.

\subsection{Special case: the FLMM for a crossed design}
For our application in speech production research (Section \ref{sec: Application}), we use an FLMM with a crossed design structure to account for correlation between measurements of the same speaker and between measurements of the same target word. 
\bea\label{eq: FLMM crossed}
Y_{ijh}(t) = \mu(t,\mx_{ijh}) + B_i(t) + C_j(t) + E_{ijh}(t) + \eps_{ijh}(t),
\eea
with $i=1,\ldots,I$ (number of speakers), $j=1,\ldots,J$ (number of target words), and $h=1,\ldots,H_{ij}$ (number of repetitions). 
Here, $Y_{ijh}(t)$ is the $h$th curve for speaker $i$ and target word $j$ observed at time $t$. 
$B_i(t)$ and $C_j(t)$ are functional random intercepts (fRIs) for the speakers and target words, respectively.
Curve-specific deviations are accommodated by the smooth residual term $E_{ijh}(t)$, which also captures interactions between speakers and words. We decided to not include an interaction effect separately based on substantive considerations and the limited sample size.  
$\eps_{ijh}(t)$ is additional white noise measurement error with variance $\sigma^2$.
We denote the auto-covariance operators by $K^B(t,t^\pr)=\Cov\left[ B_i(t),B_i(t^\pr)\right]$, $K^C(t,t^\pr)=\Cov\left[ C_j(t),C_j(t^\pr)\right]$, and $K^E(t,t^\pr)=\Cov\left[ E_{ijh}(t),E_{ijh}(t^\pr)\right]$, $i=,1,\ldots,I$, $j=1,\ldots,J$, $h=1,\ldots,H_{ij}$.
\subsection{Irregularly and sparsely sampled functional data}\label{sec: Sparse}
Let us now assume that for our general model \eqref{eq: FLMM general} we have observed $n$ curves on observation points $\lbrace t_{i1},\ldots,t_{iD_{i}} \rbrace \in \mathcal{T}, i=1,\ldots,n$. 
The number and the location of the observation points are allowed to differ from curve to curve. In the extreme, only one point may be observed for a curve. Moreover, the observation points of a curve do not have to be equally spaced. We denote realizations of the functional response $Y_i(t)$ at point $t_{ij}$ by $y_{it_{ij}}, j=1,\ldots,D_{i}$. Accordingly, we denote realizations of the response in \eqref{eq: FLMM crossed} by $y_{ijht}$, with $t \in \lbrace t_{ijh1},\ldots,t_{ijhD_{ijh}}\rbrace$.
\section{Estimation}\label{sec: Estimation}
We base our estimation on FPCA, which provides the dimension reduction so important for functional data. In addition, FPCA allows an explicit decomposition of the variability in the data. Compared to other basis approaches e.g.~using splines, the advantage of FPCA is that the eigenbases are optimal in the sense of giving the best approximation for a given number of basis functions.
An eigen decomposition of the auto-covariances of $\mU(t)$ and $E_{i}(t)$ based on Mercer's theorem \citep{Mercer.1909} is used. The estimated eigenfunctions, also known as functional principal components, describe the main modes of variation of processes $\mU(t)$ and $E_i(t)$ and the estimated eigenvalues quantify the amount of variability explained by the corresponding FPCs. To pool information across observations, which is particularly important in the case of irregularly or sparsely sampled functional data, we use smoothing of the auto-covariances, cf.~\cite{Yao.2005} for non-correlated sparse functional data. 
The four main steps of our estimation procedure are described in the following.
\begin{compactitem}
\item[\bf{Step 1}] We estimate the mean, $\mu(t,\mx_{i})$, using penalized splines based on a working independence assumption.
\item[\bf{Step 2}] For each of the functional random effects, we estimate the respective auto-covariance using a smooth method of moments estimator. Subsequently, we evaluate each of the estimates on a pre-specified equidistant, dense grid.
\item[\bf{Step 3}] We conduct an eigen decomposition of each of the estimated auto-covariance matrices. Using the Karhunen-Lo\`{e}ve (KL) expansion \citep{Karhunen.1947,Loeve.1945}, we can represent the functional random effects using truncated bases of the respective estimated eigenfunctions.
\item[\bf{Step 4}] We propose two ways of predicting the random basis weights.
\end{compactitem}
Step 1, step 3, and the first option for the prediction of the basis weights are analogous to the estimation proposed in \cite{Di.2009}, \cite{Greven.2010}, and \cite{Shou.2014} for functional data sampled on an equal, fine grid and in \cite{Di.2014} for a simpler model. 
For simplicity, we focus in the remainder of this section on model (\ref{eq: FLMM crossed}). \pkg{R}-code implementing our approach for model (\ref{eq: FLMM crossed}) can be provided upon request.

\subsection{Estimation of the mean function} \label{subsec: mean estimation}
We estimate the mean $\mu(t,\mx_{ijh})$ based on the working independence assumption
\bea \label{eq:ind assump}
Y_{ijh}(t) = \mu(t,\mx_{ijh}) + \eps_{ijh}(t),
\eea
with i.i.d.~Gaussian random variables $\eps_{ijh}(t)$.
Consider the additive mean predictor $\mu(t,\mx_{ijh}) = f_0(t) + \sum_{p=1}^P f_p(t)x_{ijhp}$ (and $\mu(t,\mx_{i}) = f_0(t) +\sum_{k=1}^P f_p(t)x_{ip}$ for model \eqref{eq: FLMM general}). We represent the unknown, smooth functions $f_p(\cdot)$ using splines and use a B-Spline basis in the following. Let $\otimes$ denote the Kronecker product and let $\mmu$ denote the stacked mean of length $\mathfrak{D}$. $\mmu$ is then approximated by 
\bea \label{eq:mean tensor}
\mmu=\sum_{p=0}^P \left(\mPsi_g^p\otimes \mathds{1}_{K^p}^\top\right)\cdot \mPsi_{t}^p\mtheta^p,
\eea where $\mPsi_g^p$ denotes an inflated vector of length $\mathfrak{D}=\sum_{i,j,h}D_{ijh}$ of covariate values, $\mPsi_t^p$ of dimension $\mathfrak{D}\times K^p$ comprises the evaluations of the $K^p$ basis functions on the $\mathfrak{D}$ time points $t_{ijh}$, and $\mathds{1}_{K^p}=\left(1,\ldots,1\right)^\top$ of length $K^p$. $\mtheta^p$ is a coefficient vector of length $K^p$. The concrete forms of $\mPsi_g^p$ and $\mPsi_t^p$ are given in Appendix A. We control the trade-off between goodness of fit and smoothness by adding a difference penalty \citep{Eilers.1996}. Using the penalized splines approximation of model \eqref{eq:ind assump} allows us to represent the model as a scalar LMM 
which has the advantage that the smoothing parameter can be estimated as a variance component using restricted maximum likelihood (REML, \cite{Patterson.1971}, cf.~\cite{Ruppert.2003}, sec.~4.9).\\
For the practical implementation, we build on existing software and either use \pkg{R}-function \citep{R} \pkg{gam} or function \pkg{bam}, both implemented in the \pkg{R}-package \pkg{mgcv} \citep{Wood.2011}. The latter is especially designed for large datasets. 
Avoiding the construction of the complete design matrix leads to a much lower memory footprint and the possibility of parallelization also gives a considerable speed-up in computation time. For further details and more general mean models see \cite{Wood.2015}.
In a next step, we center the data using the estimated mean, $\hat{\mu}(t,\mx_{ijh})$, and obtain $\tilde{y}_{ijht}:=y_{ijht}-\hat{\mu}(t,\mx_{ijh})$.
\subsection{Estimation of the auto-covariances}
The estimation of the auto-covariances is the most challenging step in the estimation procedure, also in terms of computational effort. We exploit the fact that for centred data, the expectation of the cross products corresponds to the auto-covariance $ \Cov\left[\tilde{Y}_{ijh}(t),\tilde{Y}_{i^\pr j^\pr h^\pr}(t^\pr)\right]$, which can be decomposed as follows
\bea\label{eq:cov decomp} 
\Cov\left[ \tilde{Y}_{ijh}(t) ,\tilde{Y}_{i^\pr j^\pr h^\pr }(t^\pr)\right]&=& \EV\left[ \tilde{Y}_{ijh}(t)\tilde{Y}_{i^\pr j^\pr h^\pr }(t^\pr)\right]\\ \nonumber &=& K^B(t,t^\pr)\delta_{ii^\pr } + K^C (t,t^\pr)\delta_{jj^\pr } +
\left[ K^E(t,t^\pr)+\sigma^2\delta_{tt^\pr}\right]\delta_{ii^\pr }\delta_{jj^\pr }\delta_{hh^\pr },
\eea
with $\delta_{xx^\pr }$ equal to one if $x=x^\pr $ and zero otherwise. \eqref{eq:cov decomp} can be seen as an additive, bivariate varying coefficient model, in which the auto-covariances are the unknown smooth bivariate functions to be estimated, while $\delta_{ii^\pr }, \delta_{jj^\pr }, \delta_{ii^\pr }\delta_{jj^\pr }\delta_{hh^\pr }$, and $\delta_{ii^\pr }\delta_{jj^\pr }\delta_{hh^\pr }\delta_{tt^\pr}$ represent the covariates. 
Under a working assumption of independence and homoscedastic variance of the cross products, we can use each empirical product $\tilde{y}_{ijht}\tilde{y}_{i^\pr j^\pr h^\pr t^\pr}$ for which at least $i=i^\pr $ or $j=j^\pr $
to obtain smooth estimates of $K^B(t,t^\pr)$, $K^C(t,t^\pr)$, and $K^E(t,t^\pr)$ and an estimate of the error variance $\sigma^2$. 
The total number of products $\tilde{y}_{ijht}\tilde{y}_{i^\pr j^\pr h^\pr t^\pr}$ used for the estimation of the auto-covariance then is 
$\sum_{i=1}^I\left(\sum_{j=1}^J\sum_{h=1}^{H_{ij}} D_{ijh}\right)^2 + \sum_{j=1}^J\left(\sum_{i=1}^I\sum_{h=1}^{H_{ij}} D_{ijh}\right)^2 - \sum_{i=1}^I\sum_{j=1}^J\left(\sum_{h=1}^{H_{ij}}D_{ijh}\right)^2$.
For equal $H_{ij}\equiv H$ and $D_{ijh}\equiv D$, this would reduce to $\mathfrak{D}^2\left(\nicefrac{1}{I}+\nicefrac{1}{J}-\nicefrac{1}{IJ}\right)$, with $\mathfrak{D}$ the total number of observation points.\\
For the estimation of the auto-covariances, we use bivariate tensor product P-splines \citep[see e.g.][sec.~4.1.8]{Wood.2006}. The idea is to combine low rank marginal bases for each $t, t^\pr$ in order to obtain smooth functions of the two covariates. Then, given the appropriate ordering of the parameter vector, the part of the design matrix corresponding to $K^X(t,t^\prime)$, $X\in \lbrace B,C,E\rbrace$, is given by the respective indicator matrix multiplied entry-wise by $\left(\mM_{t}^X \otimes \mathds{1}_K^\top\right) \cdot \left(\mathds{1}_K^\top\otimes \mM_{t^\pr}^X\right)$, where $\mM_{t}^X$ and $\mM_{t^\pr}^X$ denote the corresponding marginal spline design matrices of rank $K$ for covariate $t$ and $t^\pr$. 
A smoothness penalty is introduced in order to avoid over-fitting. As we estimate auto-covariances which are symmetric by definition and therefore are naturally on the same scale, we choose an isotropic penalty with the penalty matrix of the form
$S_{tt^\pr} = \mS_{t} \otimes \mS_{t^\pr}$, where $\mS_t$ and $\mS_{t^\pr}$ represent the respective marginal penalty matrices for covariate $t$ and $t^\pr$. We use marginal B-spline bases combined with marginal difference penalties. 
As for the mean estimation, we take advantage of the mixed model representation of the additive model \eqref{eq:cov decomp} using REML in order to obtain estimates for the tensor product basis coefficients and the smoothing parameter.\\
Again, we make use of existing software by applying function \pkg{bam} in \pkg{R}-package \pkg{mgcv}. Negative estimated values of $\sigma^2$ are set to zero for the final estimate. Symmetry of the auto-covariances is ensured through the model apart from numerical inaccuracies. During the auto-covariance estimation, strength is borrowed across all curves. This can be extremely advantageous for sparse functional data when some curves only have very few measurements and smoothing of curves would be infeasible.

\subsection{Eigen decompositions of estimated auto-covariances}
Based on Mercer's Theorem, the eigen decompositions of the auto-covariances are
 \bea \nonumber
K^X(t,t^\pr) =
\sum_{k=1}^{\infty} \nu^X_k \phi^X_k(t) \phi^X_k(t^\pr),\ X\in\lbrace B,C,E \rbrace,
\eea
where $\nu_1^X\geq\nu_2^X\geq\ldots\geq 0$ are the respective eigenvalues, $k\in \mathbb{N}$. The corresponding eigenfunctions $\lbrace \phi_k^X, k\in \mathbb{N} \rbrace$, $X\in\lbrace B,C,E\rbrace$, form an orthonormal basis in the Hilbert space $L^2[\mathcal{T}]$ with respect to the scalar product
\bea \label{eq: scalar product}
\langle f,g \rangle = \int f(t)g(t) \dint t.
\eea 
In practice, the smooth auto-covariances are evaluated on an equally spaced, dense grid $\lbrace t_1,\ldots,t_D\rbrace$ of pre-specified length $D$. The resulting matrices are in the following denoted as $\hat{\mK}^B= {\left[ \hat{K}^B(t_d,t_{d^\pr })\right]}_{d,d^\pr =1,\ldots,D}$, $\hat{\mK}^C = {\left[ \hat{K}^C(t_d,t_{d^\pr })\right]}_{d,d^\pr =1,\ldots,D}$, and $\hat{\mK}^E=~{\left[ \hat{K}^E(t_d,t_{d^\pr })\right]}_{d,d^\pr =1,\ldots,D}$. We conduct an eigen decomposition of each of the estimated auto-covariance matrices yielding estimated eigenvectors and estimated eigenvalues. 
Rescaling of the estimated eigenvectors and thus also of the estimated eigenvalues is necessary to ensure that the approximated eigenfunctions are orthonormal with respect to the scalar product \eqref{eq: scalar product}. More details are given in Appendix B.
Negative estimated eigenvalues are trimmed to guarantee positive semi-definiteness.

\underline{Truncation of the FPCs}. While in theory, there is an infinite number of eigenfunctions, dimension reduction and selection of the number of FPCs for each random process is necessary in practice. 
This truncation has a theoretical justification and can be seen as a form of penalization \citep[see e.g.][]{Di.2009,Peng.2009}.
Among the multiple proposals in the literature \citep[see for an overview][]{Greven.2010}, we base our choice on the proportion of variance explained. This allows us to quantify the contribution of the random processes to the variation in the observed data. It is based on the variance decomposition of the response
\bea  \nonumber
\int_{\mathcal{T}} \Var\left[ Y_{ijh}(t)\right] \dint t = \sum_{k=1}^\infty \nu^B_k + \sum_{k=1}^\infty \nu^C_k + \sum_{k=1}^\infty \nu^E_k +  \sigma^2 |\mathcal{T}|.
\eea 
The sums $\sum_{k=1}^\infty \nu^B_k$, $\sum_{k=1}^\infty \nu^C_k$, and $\sum_{k=1}^\infty \nu^E_k$ quantify the relative importance of each of the three random processes. We choose principal components of decreasing importance until a pre-specified level of explained variation is reached. See Appendix B for details and Appendix A for the derivation of the variance decomposition.

\underline{Approximation of the functional random processes}. Based on the truncation, we use KL expansions to obtain parsimonious basis representations for the random processes
\bea \nonumber
B_i(t) = \sum_{k=1}^{N^B} \xi^B_{ik} \phi^B_{k}(t),\ 
C_j(t) = \sum_{k=1}^{N^C} \xi^C_{jk} \phi^C_{k}(t),\ 
E_{ijh}(t) = \sum_{k=1}^{N^E} \xi^E_{ijhk} \phi^E_{k}(t).
\eea
Note that in the case of irregularly or sparsely sampled data, the observation points also depend on speaker $i$, target word $j$, and repetition $h$. Throughout this paper, however, we omit the additional indices for better readability. For the same reason, we do not emphasise that the truncation lags and eigenfunctions are estimated.
By construction, the basis weights or FPC weights $\xi^B_{ik}$, $\xi^C_{jk}$, and $\xi^E_{ijhk}$ are uncorrelated random variables with zero mean and variance $\nu^X_k$, $k\in \mathbb{N}$, $X\in \lbrace B,C,E \rbrace$.\\  
For prediction of the FPC weights, we first linearly interpolate the chosen eigenfunctions such that they are available on the original observation points. Due to the smoothness of all model components, this leads to a small error which could be further decreased, if desirable, by further increasing the size of the grid $D$.

\subsection{Prediction of the basis weights}
The basis weights give insight into the individual structure of each grouping level and can be used for further analyses, e.g. for classification.
For a centred random process $X_i(t)$, the basis weights of the KL expansion can be written as the scalar product of $X_i(t)$ and $\phi_k^X(t)$ and are often predicted using numerical integration. 
For irregularly or even sparsely sampled data, however, numerical integration would not work (well). 
Moreover, for correlated functional data contaminated with additional measurement error, the separation of the weights belonging to the different basis expansions remains unclear and ignoring the measurement error leads to biased predictions.\\
These considerations motivate our two proposals for the prediction of the basis weights. The first is very straight-forward and computationally efficient. It generalizes the conditional expectations introduced by \cite{Yao.2005}.
The second involves higher computational costs but has the advantage that the mean is re-estimated in the same framework and that it allows for approximate statistical inference, e.g.~for the construction of point-wise confidence bands (CBs). 

\underline{Prediction of the basis weights as EBLUPs}. 
Using the truncated KL expansions of the random processes, we can approximate model \eqref{eq: FLMM crossed} by
\bea \label{eq: rewritten FLMM crossed}
Y_{ijh}(t) = \mu(t,\mx_{ijh}) + \sum_{k=1}^{N^B} \xi_{ik}^B\phi^B_k(t) +\sum_{k=1}^{N^C}  \xi_{jk}^C\phi^C_k(t) + \sum_{k=1}^{N^E} \xi_{ijhk}^E\phi^E_k(t) + \eps_{ijh}(t)
\eea
for the discrete observation points $t \in \lbrace t_{ijh1},\ldots,t_{ijhD_{ijh}}\rbrace$.
The resulting model \eqref{eq: rewritten FLMM crossed} is a scalar LMM in which the random effects correspond to the basis weights \citep{Di.2009}. 
Once all other components are estimated, we do not need to fit the LMM \eqref{eq: rewritten FLMM crossed}, but can predict the basis weights as EBLUPs as derived in the following. Without normality assumption, they remain best linear predictors. \\
Let $\tilde{\mY}$ denote the stacked centred response vector of length $\mathfrak{D}$.
Let $L^X\in \lbrace I,J,n\rbrace$ and $N^X \in \lbrace N^B,N^C,N^E\rbrace$ denote the levels of the grouping variable and the truncation lag for process $X$, $X\in \lbrace B,C,E\rbrace$, respectively.
We define $\mxi = \left(\mxi^{B^\top},\mxi^{C^\top},\mxi^{E^\top}\right)^\top$, with $\mxi^{X}=\left({\mxi_1^X}^\top,\ldots{,\mxi_{L^X}^X}^\top\right)^\top$ the stacked vector of the basis weights of length $L^XN^X$. Thus $\mxi$ is a vector of length $\mathfrak{N}:= IN^B + JN^C + nN^E$. $\hat{\mPhi}$ is the joint design matrix of dimension $\mathfrak{D} \times \mathfrak{N}$ of the form $\hat{\mPhi} = \left[\hat{\mPhi}^B | \hat{\mPhi}^C|\hat{\mPhi}^E \right]$, where $\hat{\mPhi}^B$, $\hat{\mPhi}^C$, and $\hat{\mPhi}^E$ are the respective design matrices containing the rescaled FPC estimates evaluated on the original observation points.
The covariance matrix of $\mxi$ is a diagonal matrix with diagonal elements corresponding to the eigenvalues of the three random processes. We denote its estimate in the following by $\hat{\mG}$. A more detailed description of the form of the matrices can be found in Appendix A. Then, the EBLUP for the basis weights in model \eqref{eq: FLMM crossed} is given in the usual form by
\bea \label{eq: EBLUP} 
\hat{\mxi} = \hat{\mG}\hat{\mPhi}\tr\hat{\Cov}\left( \tilde{\mY}\right)^{-1}\tilde{\mY}
=  \hat{\mG}\hat{\mPhi}\tr\left(\hat{\sigma}^2 \mI_{\mathfrak{D}} + \hat{\mPhi} \hat{\mG}\hat{\mPhi}^\top\right)^{-1}\tilde{\mY}.
\eea
Equation \eqref{eq: EBLUP} requires the inversion of the estimated covariance matrix of $\tilde{\mY}$, which is of dimension $\mathfrak{D}\times \mathfrak{D}$. For large numbers of observation points, this can be computationally demanding. Furthermore, when $\hat{\sigma}^2$ is estimated to be close to zero, the covariance becomes singular. Transformations with the Woodbury formula yield the more favourable form 
$\hat{\mxi} = \left(\hat{\sigma}^2 \hat{\mG} + \hat{\mPhi}^\top\hat{\mPhi}\right)^{-1}\hat{\mPhi}^\top\tilde{\mY},$
for which the inversion is simplified to that of an $\mathfrak{N} \times \mathfrak{N}$ matrix which has full rank when either $\hat{\sigma}^2$ is positive or when $\hat{\mPhi}^\top\hat{\mPhi}$ has full rank. 
In practice, when neither of these requirements is met, the Moore-Penrose generalized inverse 
is used. Note that when $\sigma^2$ is estimated to be zero, the EBLUP simplifies to the least-squares estimator.\\ 
One drawback of this computationally efficient way of predicting $\mxi$ is that the mean is estimated using a working independence assumption. This may not be statistically efficient and does not provide the basis for statistical inference. This motivates our second proposal. 

\underline{Prediction of the basis weights using FAMMs}. The second option uses the fact that model \eqref{eq: rewritten FLMM crossed} together with the distribution of the basis weights implied by the KL expansion falls into the general framework of a FAMM \citep{Scheipl.2014} using suitable marginal bases and penalties. We can write model \eqref{eq: rewritten FLMM crossed} using estimated eigenfunctions and -values as 
\bea \label{eq: row tensor product bases}
\mY = \sum_{p=0}^P \left(\mPsi_g^p\otimes \mathds{1}_{K^p}^\top\right)\cdot \mPsi_{t}^p\mtheta^p + \sum_{X\in \lbrace B,C,E \rbrace}(\mPsi_{g}^X \otimes \mathds{1}_{N^X}^\top) \cdot ( \mathds{1}_{L^X}^\top \otimes \mPsi_{t}^X)\mxi^X +\meps,
\eea
with $\meps\sim\mathcal{N}(\bm{0},\sigma^2\mI_{\mathfrak{D}})$. $\mY$ is the stacked uncentred response vector of length $\mathfrak{D}$, the mean is specified as in \eqref{eq:mean tensor}, and $\mPsi_{g}^X$ denotes an inflated $\mathfrak{D}\times L^X$ matrix of grouping indicators. 
The $\mathfrak{D} \times N^X$ matrix $\mPsi_{t}^X$ comprises the evaluations of the $N^X$ 
respective estimated eigenfunctions on the original observation points. The concrete forms of the matrices $\mPsi_{g}^X$ and $\mPsi_{t}^X$ are given in Appendix A. 
Adding penalties of the form ${\mxi^X}^\top \left(\mI_{L^X} \otimes \mP_{t}^X\right) \mxi^X$, with $\mP_{t}^X =\diag \left(\hat{\nu}_1^X,\ldots,\hat{\nu}_{N^X}^X\right)^{-1}$, corresponds to the distributional assumption $\mxi_l^X\sim\mathcal{N}\left(0,\diag \left(\hat{\nu}_1^X,\ldots,\hat{\nu}_{N^X}^X\right)\right)$, $l=1,\ldots L^X$, $X\in\lbrace B,C,E\rbrace$, implied by the KL expansions.  
This set-up using linear combinations of row tensor product bases with an appropriate penalty falls naturally into the framework of a FAMM and was in fact discussed in \cite{Scheipl.2014} without, however, providing an approach to estimation of the eigenfunctions needed in $\mPsi_t^X$.
Mean and basis weights are then estimated/predicted within one framework. 
\eqref{eq: row tensor product bases} is a scalar additive LMM, which allows to take advantage of established methods for estimation and for statistical inference \citep[for more details, see][]{Scheipl.2014}. In particular, it allows us to construct point-wise CBs for the mean and for covariate effects. Note that the inference is conditional on the estimated FPCA, i.e.~it does not account for the uncertainty of the estimation of the eigenfunctions and -values and for the truncation.\\
In practice, we use the \pkg{R}-function \pkg{pffr} that \cite{Scheipl.2014} provide in the \pkg{R}-package \pkg{refundDevel} \citep{refundDevel}. Function \pkg{pffr} is a wrapper function for function \pkg{gam} and for related functions in the package \pkg{mgcv} and therefore builds on existing flexible and robust software. We use the \pkg{pffr}-constructor \pkg{pcre()} for FPC-based fRIs for the prediction of the random processes. A constraint on the functional random effects assures that they are centred. 
The resulting point-wise CBs are with a constraint correction \citep{Marra.2012}. In addition to the parsimonious basis of eigenfunctions, this approach has the advantage of not necessitating the estimation of any smoothing parameters for the random processes, as the variances of the random weights have already been estimated and the smoothing parameter can be set to one (cf.~Appendix B for details). These two features lead to a drastic decrease in computational cost compared to spline-based prediction of the random processes, as is shown in our simulations in Section \ref{sec: Simulations}.  

The estimation quality can be further improved, if desirable, by applying the four estimation steps iteratively. Several possibilities are described in Appendix B, where also further details on the estimation and implementation can be found.
\section{Application to the speech production research data}\label{sec: Application}
\subsection{Background and scientific questions}
\begin{figure}[t!]
\centering
\includegraphics[width=0.8\textwidth]{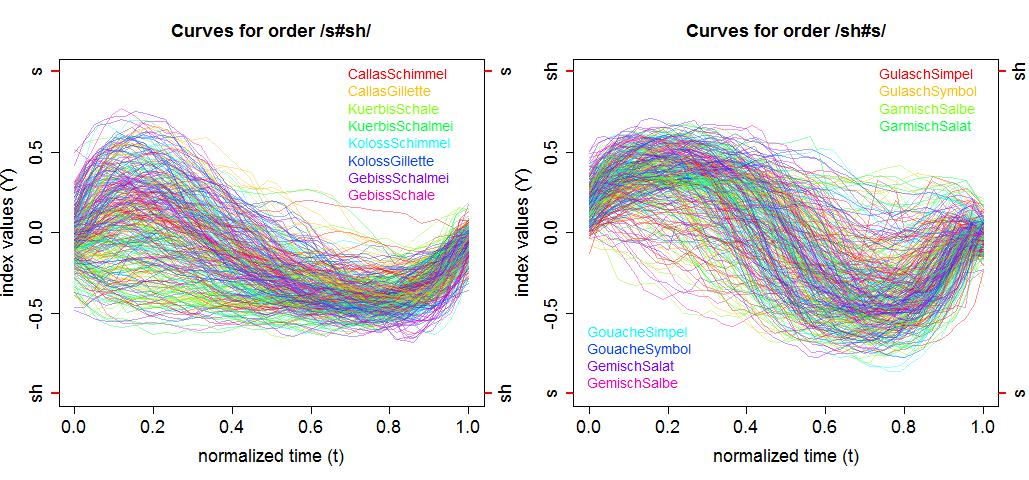}
\caption{Index curves of the consonant assimilation data over time. 
Left [right]: Curves of order \sS [\Ss]. Positive values approaching $+1$ indicate a reference /s/ [/sh/] acoustic pattern, while negative values approaching $-1$ indicate a reference /sh/ [/s/] acoustic pattern.}
\label{fig: data mirrored}
\end{figure}

In linguistics, the term assimilation refers to the common phenomenon whereby a consonant becomes phonetically more like an adjacent, usually following consonant.
For example, assimilation occurs commonly in English phrases such as ``Paris show'' in which the word-final /s/-sound is, in fluent speech, pronounced very similar to the following, word-initial /sh/-sound \citep{Pouplier.2011}. Assimilation patterns are conditioned by a complex interaction of perceptual, articulatory and language-specific factors and are therefore a central research topic in the speech sciences.
In order to investigate assimilation in German, \cite{Pouplier.2014} recorded the acoustic signals of $I=9$ speakers reading the same $J=16$ target words, each five times. Due to recording errors, for some combinations only four repetitions are included in the data, i.e.~$H_{ij}\in\lbrace 4,5\rbrace$. The authors concentrated on variation in assimilation patterns for the consonants /s/, /sh/ as a function of their order (\sS versus \Ss, where \# denotes a word boundary), lexical stress and vowel context. Target words consisted of bisyllabic noun-noun compounds. In half of the target words consonant /s/ is followed by word-initial /sh/, such as in the word ``Calla\textbf{s-Sch}immel''.
The other half contains the sequence \Ss, e.g.~``Gula\textbf{sch-S}ymbol''. 
In the following, we will refer to the syllables containing the consonants of interest as final and initial target syllables (and correspondingly to final and initial target consonants). The time interval in which the consonants of interest appear in the utterance was cut out manually from the audio recording for each repetition and the time-varying signal was summarized in a functional index over time, varying between $+1$ and $-1$. 
The index was calculated such that for both orders it ranges from $+1$ for the first consonant of the sequence to $-1$ for the second consonant of the sequence (for more details, see \cite{Pouplier.2011} and Appendix C for data pre-processing).
The resulting index curves are displayed in Figure \ref{fig: data mirrored}. \\
A special focus lies on the asymmetry arising from the order of the consonants. We investigate under which conditions (order, syllable stress, vowel context) the two consonants assimilate and whether assimilation is symmetric with respect to the orders \sS and \Ss.
A common approach to the analysis of these kind of data is to extract curve values at pre-defined points on the time axis (e.g.~25\%, 50\%, 75\%) which are subsequently used in multivariate methods \citep[e.g.][]{Pouplier.2011}. Such analyses fail to capture the continuous dynamic change characteristic of speech signals. Applying our fda-based method allows us to take into consideration the temporal dynamics and to account for the complex correlation structure in the data which arises from the repeated measurements of speakers and of target words. Moreover, we can quantify the effect of covariates and interactions and obtain a variance decomposition. \\
All utterances were recorded with the same sampling rate (32768 Hz) and then standardized to a [0,1] interval as the speaking rate and hence the target consonant duration differs across speakers and repetitions. After standardization, measurements are unequally spaced for different curves. 
In some data settings, landmark registration can be used to account for variation in time. For this application, however, registration cannot replace the standardization of the time interval as different transition speeds between the two consonants are part of the research question of interest, i.e.~of the assimilation process.
\subsection{A model for the consonant assimilation data}
In order to account for the repeated measurements of speakers and target words, we fit an FLMM with crossed fRIs, model \eqref{eq: FLMM crossed}, to the consonant assimilation data. 
The number of measurements per curve $D_{ijh}$ ranges from 22 to 57 with a median of 34. During estimation, we truncate the numbers of FPCs using a pre-specified proportion of explained variance of $0.95$ as described in Section \ref{sec: Estimation}. The equidistant grid on which the auto-covariances are evaluated is of length $D=100$. We use cubic B-splines with third order difference penalties for the estimation of the mean effects and as marginal basis functions for the estimation of the auto-covariances. We predict the FPC weights using both options. As confidence bands for the covariate and interaction effects are of interest here, the focus lies on the second approach using the FAMM framework.

\underline{Covariate effects}. We consider four dummy-coded covariates: consonant order (order), stress of the final (stress1) and of the initial (stress2) target syllable, which can be strong or weak, and vowel context (vowel), which refers to the vowels immediately adjacent to the target consonants and is either of the form ia or ai, e.g.~Call\textbf{a}s-Sch\textbf{i}mmel.
Moreover, we include the interactions of the consonant order with each of the other three covariates. All covariates enter the mean as varying coefficients yielding
\bea \label{eq: mean application}
\mu(t,\mx_{ijh}) &=& f_0(t) + f_1(t)\cdot \texttt{order$_j$} +f_2(t)\cdot \texttt{stress1$_j$} + f_3(t)\cdot \texttt{stress2$_j$} \\ \nonumber &+& f_4(t)\cdot \texttt{vowel$_j$} +f_5(t)\cdot \texttt{order$_j$}\cdot\texttt{stress1$_j$}+f_6(t)\cdot \texttt{order$_j$}\cdot\texttt{stress2$_j$}\\ \nonumber
&+&f_7(t)\cdot \texttt{order$_j$}\cdot\texttt{vowel$_j$}.
\eea
Thus, in total, eight covariates characterize the 16 target words. 

\subsection{Application results} \label{subsec: Application results}
\begin{figure}[t!]
\centering
\includegraphics[width=0.3\textwidth]{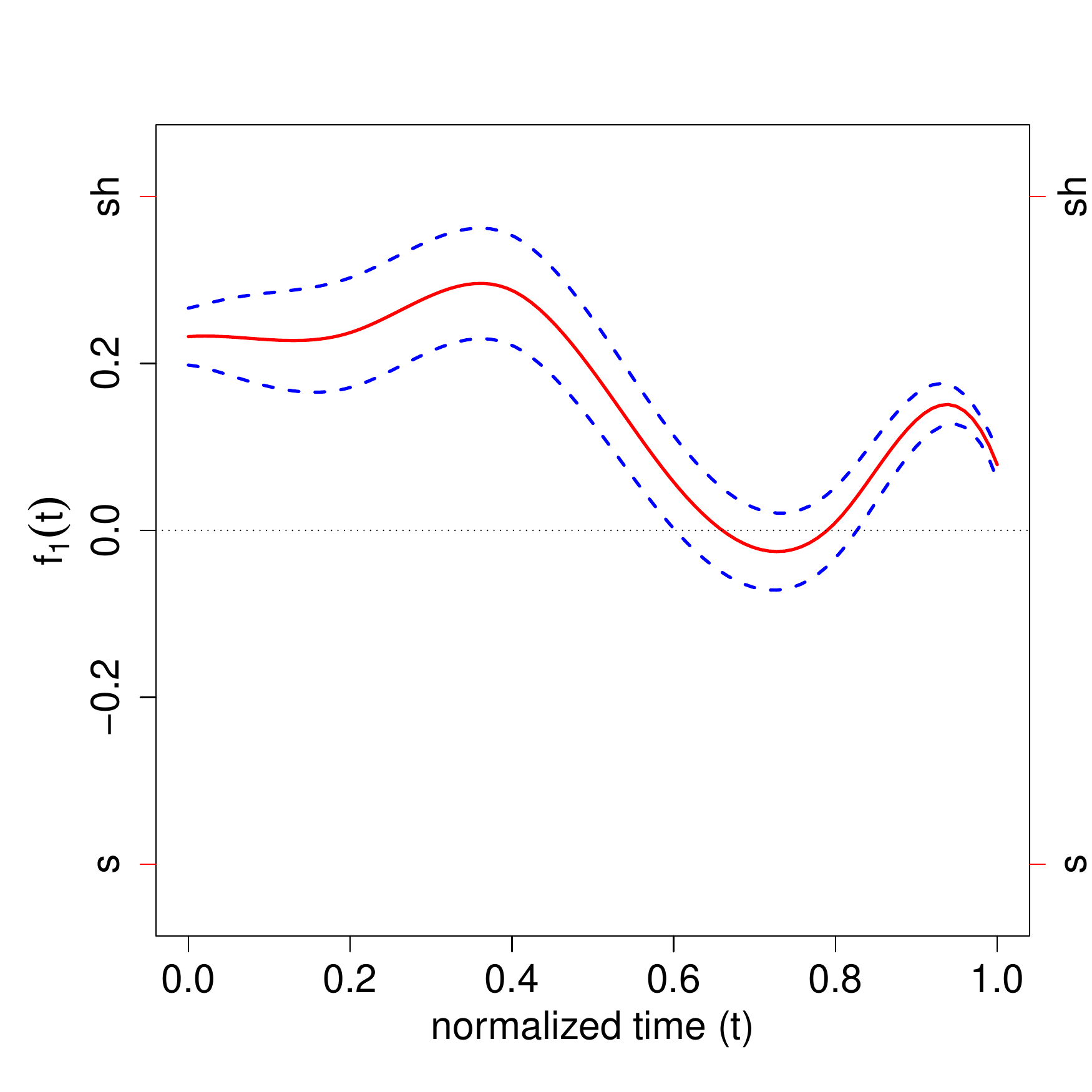}
\includegraphics[width=0.3\textwidth]{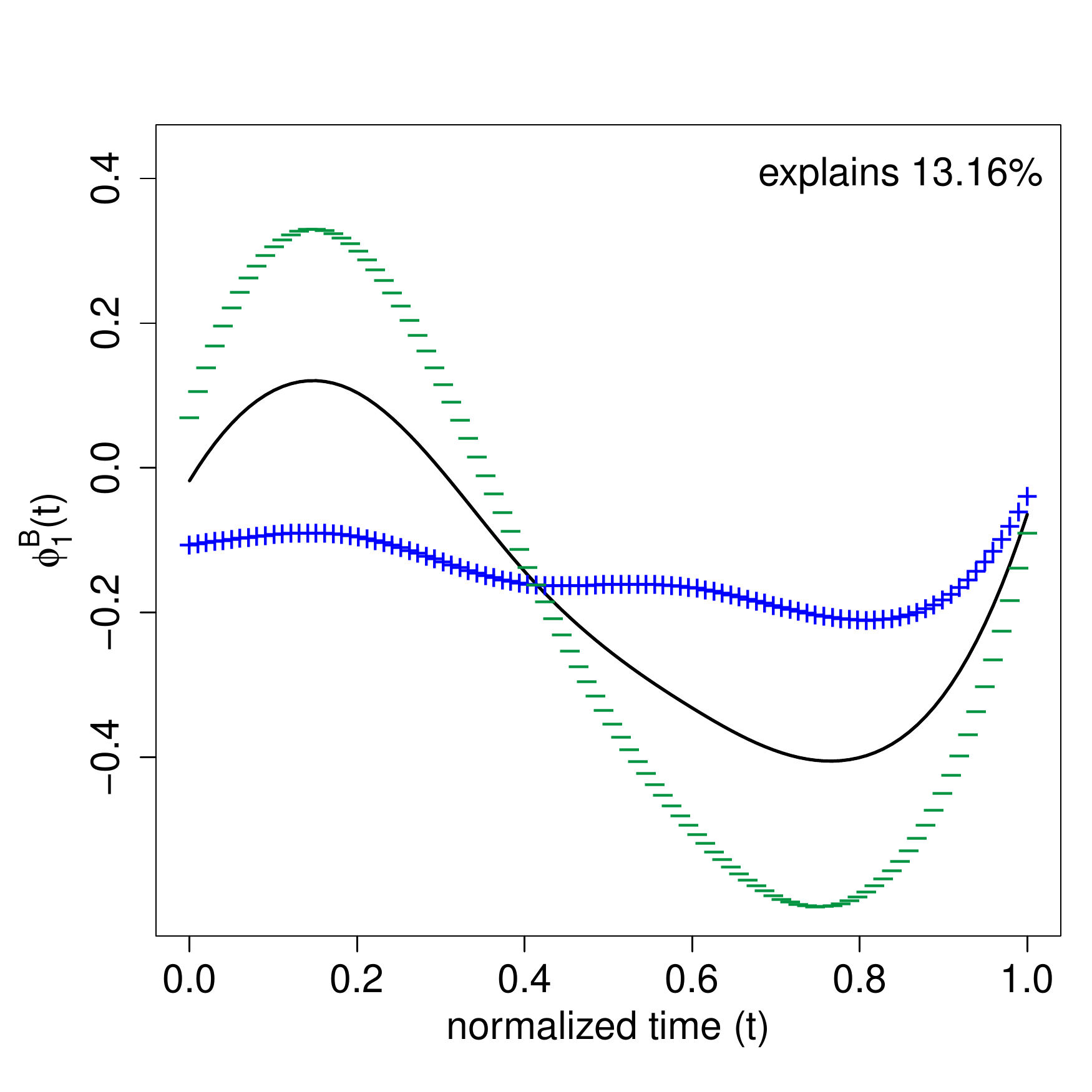}
\caption{Left: Effect of covariate order (red solid line) with point-wise confidence bands (dashed lines). Right: Mean function (solid line) and the effect of adding ($+$) and subtracting ($-$) a suitable multiple ($2\sqrt{\hat{\nu}_1^B}$) of the first FPC for speakers.}
\label{fig: effect order and FPC1 application}
\end{figure}

Our estimation yields two and three FPCs for the fRI for speakers and for the smooth error, respectively. No FPC is chosen for the fRI for target words. It is likely that the eight covariate and interaction effects describe the target words sufficiently, as confirmed by obtaining one FPC for the fRI for target words in the model without covariate effects. Most variability is explained by the curve-specific deviation which also captures interactions between speakers and target words. The variance decomposition is given in Table \ref{tab: variance decomp application}.\\
The left panel of Figure \ref{fig: effect order and FPC1 application} shows the effect of covariate order ($f_1$), which has the largest effect on the index trajectories. Covariate order is dummy-coded with reference category \sS.
Thus, the mean curves of target words with order \Ss are pulled towards the ideal reference /sh/ during the first consonant and differ slightly from the ideal /s/ during the second consonant compared to order \sS. 
We conclude that there is an asymmetry of consonant assimilation with respect to the consonant order and that the assimilation is stronger for order \sS. 
These results are consistent with results for English obtained by \cite{Pouplier.2011}.\\
Moreover, we find that assimilation is stronger for target words with unstressed final target syllables ($f_2$), for which the mean curves are pulled away from the ideal reference first consonant and slightly away from the ideal second consonant. This effect is stronger for order \sS ($f_5$).
The assimilation is not affected by the stress of the initial target syllable for order \sS ($f_3$). For order \Ss ($f_6$), however, the mean curves are pulled away from the ideal reference /s/ for unstressed initial target syllables compared to stressed initial target syllables, i.e.~assimilation of the second consonant is increased. 
For both consonant orders, the vowel context mainly affects the transition between the two consonants ($f_4$ and $f_7$). The first consonant is closer to the ideal reference value in the ai compared to the ia condition, yet the second consonant is pulled away from its reference value. These results show that the /i/ vowel perturbs an adjacent consonant away from its ideal reference pattern.\\
In the right panel of Figure \ref{fig: effect order and FPC1 application}, we show the effect of adding ($+$) and subtracting ($-$) a suitable multiple of the first FPC for speakers to the overall mean (solid line) obtained by setting all covariates to 0.5. 
\noindent The interpretation is straight forward: speakers with a negative weight for the first FPC distinguish better between the two consonants. The estimates for the basis weights can be used for further analysis. 
Further application results including plots for all mean effects can be found in Appendix C.
\begin{table}[t!]
  \setlength{\tabcolsep}{1mm}
 \small
\begin{center}
\caption{Variance explained in percent of the estimated total variance.}
\begin{tabular}{rrrrrrr}
  $\hat{\nu}^B_1$ & $\hat{\nu}^B_2$  & $\hat{\nu}^E_1$ & $\hat{\nu}^E_2$ & $\hat{\nu}^E_3$ & $\hat{\sigma}^2$\\ 
  \hline
   13.16\% & 7.29\% & 44.02\% & 17.11\% & 6.16\% & 8.88\%
\end{tabular}
\label{tab: variance decomp application}
\end{center}
\end{table}
\section{Simulations}\label{sec: Simulations}
\subsection{Simulation designs}
We conduct extensive simulation studies to investigate the performance of our method. The data generating processes can be divided into two main groups: 1) data that mimics the irregularly sampled consonant assimilation data and 2) sparsely sampled data with a higher number of observations per grouping level but fewer observations per curve. For all settings, we generate 200 data sets.

\underline{Application-based simulation scenarios}. We consider two application-based scenarios, one with an fRI for speakers and covariate mean effects (fRI scenario) and another with crossed fRIs for speakers and for target words, respectively, but no covariate mean effects (crossed-fRIs scenario). 
We generate the data based on the estimates of model \eqref{eq: FLMM crossed} for our consonant assimilation data with $\mu(t,\mx_{ijh})$ corresponding to \eqref{eq: mean application} and to a simple smooth intercept $\mu(t)$, respectively. 
The data analysis yields two FPCs for the fRI for the speakers and three FPCs for the smooth error term. For the crossed-fRIs scenario, we additionally obtain one FPC for the fRI for the target words. 
The FPC weights and the measurement errors are independently drawn from normal distributions with zero mean and with the respective estimated variances. 
Note that in similar FPC-based approaches for simpler models/non-sparse data \citep[e.g.][]{Yao.2005,Greven.2010}, simulation results were not dependent on the normal distribution. 
More details on the data generation can be found in Sections \ref{subsec: Application results}, \ref{subsec: Simulation results}, and in Appendices C and D.

\underline{Sparse simulation scenario}. In order to investigate the estimation performance in the sparse case, we additionally generate data with crossed fRIs as in \eqref{eq: FLMM crossed} consisting of observations that are sparsely sampled on [0,1]. For $B_{i}(t)$ and $C_j(t)$, we choose $I=J=40$ replications each with each combination observed $H_{ij}=3$ times. The number of observation points per curve is drawn from the discrete uniform distribution $\mathcal{U}\{3,10\}$.
For each grouping variable, we use two FPCs to generate the underlying process. Eigenvalues are generated as $\nu_k = \sfrac{2}{k}, k=1,2$ for all three random processes. We choose normalized Legendre Polynomials adapted to the interval $[0,1]$ as FPCs for $B_i(t)$ and $C_j(t)$. The first two elements of the orthonormal bases are
 \begin{table}[h!]
\small 
\begin{center}
\begin{tabular}{lll}
$\phi_1^B(t) = 1$ & $\phi_1^C(t) = \sqrt{3}(2t-1)$ & $\phi_1^E(t) = \sqrt{2}\sin(2\pi t)$\\
$\phi_2^B(t) = \sqrt{5}(6t^2-6t+1)$ & $\phi_2^C(t) = \sqrt{7} (20t^3-30t^2+12t-1)$ & 
$\phi_2^E(t) = \sqrt{2}\cos(2\pi t).$
 \end{tabular}
\end{center}
\end{table}

\vspace{0.55cm}
Note that the different bases need not be mutually orthogonal. The FPC weights and the measurement errors are independently drawn from the normal distributions $\mathcal{N}(0,\nu_k)$ and $\mathcal{N}(0,\sigma^2)$, respectively. For the smooth error term, $E_{ijh}(t)$, we choose a basis of sine and cosine functions. No covariates are included in the mean function $\mu(t)= \sin(t)+t$. We set the error variance to $\sigma^2=0.05$.\\
For all scenarios, we center the FPC weights such that the weights of each grouping variable also empirically have zero mean. Moreover, we decorrelate the basis weights belonging to one grouping variable and assure that the empirical variance corresponds to the respective eigenvalue. This is done to obtain data that meets the requirements of our model. It allows us to separate the effect of unfavourably drawn weights and of the estimation performance. This adjustment gains importance for small sample sizes $I$, $J$, and $n$ and also when the true eigenvalues are high. 
Note that in practice, we do not have centred and decorrelated FPC weights and thus estimates for small sample sizes will reflect the distribution in the sample rather than that in the population.
To assess the impact of this procedure, we also compare our results to those of simulations using the original (non-centred and non-decorrelated) FPC weights, which can be found in Appendix D.\\
 We fix the number of FPCs in order to separate the effect of the truncation from the estimation quality. We use five marginal basis functions each for the estimation of the auto-covariances and eight basis functions for the estimation of the mean. 
We predict the FPC weights as EBLUPs for all scenarios and additionally compare with the computationally more expensive FAMM prediction (FPC-FAMM) for the fRI scenario with covariates.\\ 
We compare our FPC-based approach to a spline basis representation of the functional random effects (using eight basis functions) within the FAMM framework of \cite{Scheipl.2014} (spline-FAMM). To the best of our knowledge, the work of \cite{Scheipl.2014} implemented in the \pkg{R}-function \pkg{pffr} is the only competitor to our approach as all other methods are either restricted to equal, fine grids or do not allow for a crossed structure. Due to the high computational costs of \cite{Scheipl.2014}, we restrict our comparison to the fRI scenario, in which we can compare estimation quality and CBs coverage for covariate effects. 

\subsection{Simulation results} \label{subsec: Simulation results}
We focus our discussion on the FPC-based results for the application-based scenario with crossed fRIs and compare with the other settings and estimation approaches.\\
We use root relative mean squared errors (rrMSE) as measures of goodness of fit which are of the general form 
$\sqrt{\sfrac{\left(\mbox{true-estimated}\right)^2}{\mbox{true}^2}}$.
rrMSE definitions for scalars, vectors, and bivariate functions are given in Appendix D.
For the simulations of the fRI scenario with covariate effects, we additionally evaluate the average point-wise and the simultaneous coverage of the point-wise CBs. The complete results for all simulations are given in Appendix D.

\begin{figure}[t!]
\begin{center}
\begin{minipage}{1\textwidth}
\begin{center}
\raisebox{0.15\textwidth}{\textbf{B}}
\includegraphics[width=0.2\textwidth,page=1]{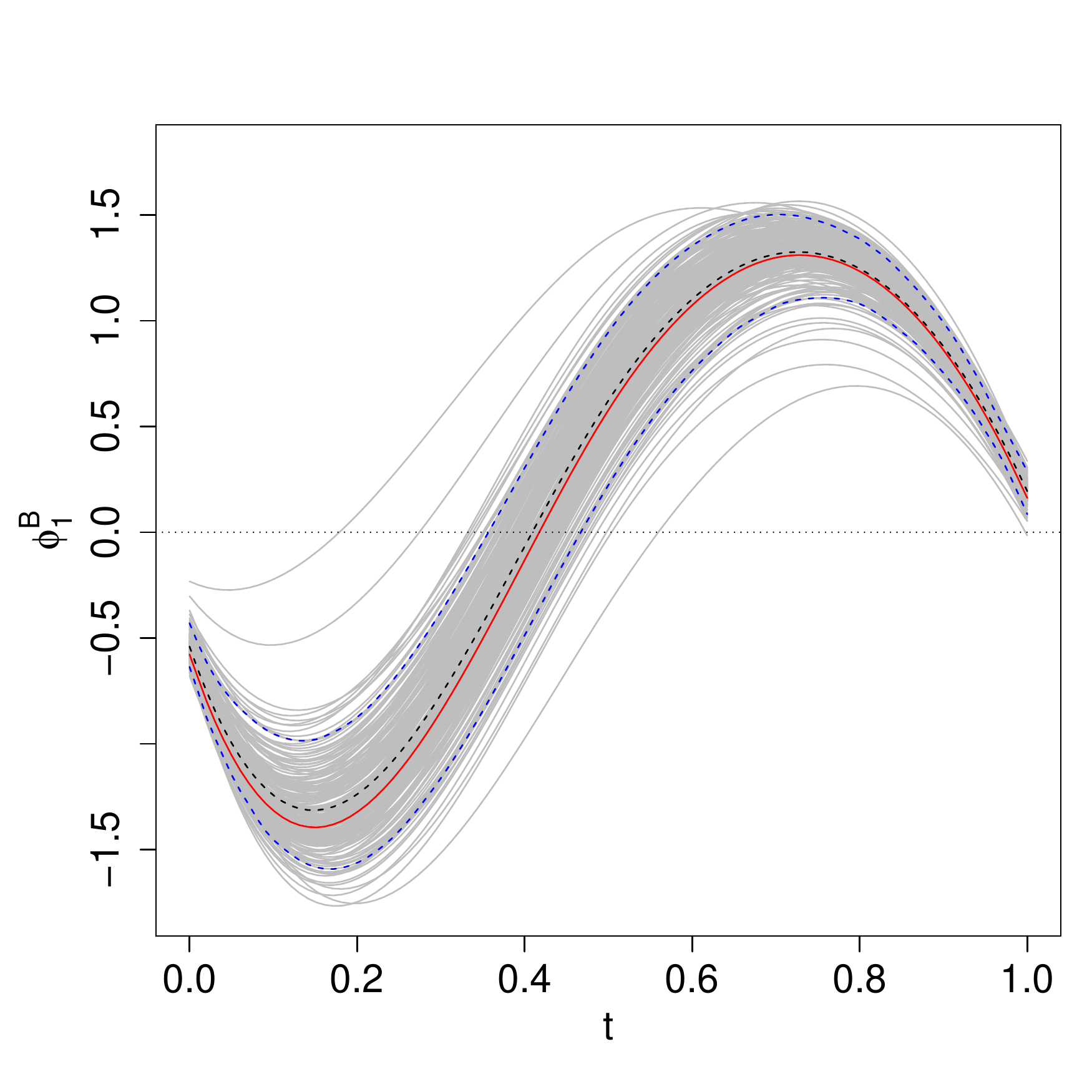}
\raisebox{0.15\textwidth}{\phantom{\textbf{B}}}
\includegraphics[width=0.2\textwidth,page=3]{figures/simulation/eigenfunctions_normal_I9_J16_crossed_as_data_26_Dec_26_Dec.pdf}
\raisebox{0.15\textwidth}{\textbf{C}}
\includegraphics[width=0.2\textwidth,page=5]{figures/simulation/eigenfunctions_normal_I9_J16_crossed_as_data_26_Dec_26_Dec.pdf}\\
\raisebox{0.15\textwidth}{\textbf{E}}
\includegraphics[width=0.2\textwidth,page=7]{figures/simulation/eigenfunctions_normal_I9_J16_crossed_as_data_26_Dec_26_Dec.pdf}
\raisebox{0.15\textwidth}{\phantom{\textbf{B}}}
\includegraphics[width=0.2\textwidth,page=9]{figures/simulation/eigenfunctions_normal_I9_J16_crossed_as_data_26_Dec_26_Dec.pdf}
\raisebox{0.15\textwidth}{\phantom{\textbf{B}}}
\includegraphics[width=0.2\textwidth,page=11]{figures/simulation/eigenfunctions_normal_I9_J16_crossed_as_data_26_Dec_26_Dec.pdf}
\end{center}
\end{minipage}
\caption{True and estimated FPCs of the crossed fRIs $B_i(t)$ and $C_j(t)$ (top row), as well as of the smooth error $E_{ijh}(t)$ (bottom row). Shown are the true functions (red solid line), the mean of the estimated functions over 200 simulation runs (black dashed line), the point-wise 5th and 95th percentiles of the estimated functions (blue dashed lines), and the estimated functions of all 200 simulation runs (grey).}
\label{fig: eigenfunctions crossed}
\end{center}
\end{figure}

\underline{Simulation results for the crossed-fRIs scenario}. 
Figure \ref{fig: eigenfunctions crossed} shows the true and estimated FPCs of the two fRIs as well as of the smooth error term. 
As expected, the FPCs are estimated better the more independent levels there are for the corresponding grouping variable which can enter the estimation of the auto-covariance. The FPCs of the smooth error term (707 levels) are estimated best, followed by the FPC of the fRI for target words (16 levels). Most variability in the estimates is found for the FPCs of the fRI for speakers due to the small number of speakers ($I=9$), but the main features of the curves are still recovered relatively well. We obtain similar results for the fRI scenario. The number and complexity of the FPCs also plays an important role for the estimation quality, as can be seen from the results for the sparse scenario, where the first FPC of $B_{i}(t)$ (40 levels) is estimated better than the first FPC of $E_{ijh}(t)$ (4800 levels). The latter has a more complex form, difficult to capture with five basis functions.\\
Table \ref{tab: mean riMSEs crossed} lists the rrMSEs averaged over 200 simulation runs for all model components. 
\begin{table}[t!]
\renewcommand{\arraystretch}{0.8} 
 \setlength{\tabcolsep}{1.5mm}
\small
\centering
\caption{rrMSEs averaged over 200 simulation runs for all model components by random process. Rows 1-3: Number of grouping levels $L^X$ and average rrMSE 
for $B_{i}(t)$, $C_{j}(t)$, and $E_{ijh}(t)$ and their covariance decompositions. Last row: Average rrMSEs for $Y_{ijh}(t)$, $\mu(t,\mx_{ijh})$, and $\sigma^2$.} 
\begin{tabular}{r|r|rrrrrrrrrrr|rr}
$X$ & $L^X$& $K^X$ & $\phi^X_1$ & $\phi^X_2$ &$\phi^X_3$&  $\nu^X_1$ & $\nu^X_2$& $\nu^X_3$ & $\xi^X_1$ & $\xi^X_2$ &$\xi^X_3$ & $X$ & $\mu$ &$\sigma^2$\\ 
  \hline
$B$ & 9 &  0.26 & 0.15 & 0.18 & & 0.15 & 0.34 & & 0.18 & 0.35 & & 0.22 & & \\ 
$C$ & 16 & 0.32 & 0.05 & & & 0.31 & & & 0.12 & & & 0.13 & & \\ 
$E$ & 707& 0.06 & 0.02 & 0.03 & 0.02 & 0.04 & 0.08 & 0.03 & 0.17 & 0.19 & 0.26 & 0.19 & & \\ 
$Y$ & & & & & & & & & & & & 0.10 & 0.02&0.09 
\end{tabular} 
\label{tab: mean riMSEs crossed}    
\end{table}
It shows that the mean function is reconstructed very well, which is also the case in the sparse scenario. 
The covariate effects for the fRI scenario are discussed below.\\ 
The auto-covariances and their eigenvalues have similar low average rrMSEs for both application-based scenarios. For the sparse scenario, the eigenvalues are estimated even better with average rrMSEs between 0.02 and 0.05. 
For the auto-covariances for the sparse scenario, we obtain average rrMSEs of 0.06 for each of the crossed fRIs and an average of 0.14 for the smooth error which is due to the complex eigenfunctions mentioned above. 
The error variance has similar low average rrMSEs for the two application-based scenarios. For the sparse scenario, the average rrMSE is higher which is due to the estimation inaccuracies in the auto-covariance of the smooth error.\\ 
The prediction quality of the basis weights clearly depends on the estimation quality of the FPCs and of the eigenvalues, as well as of the error variance, as evident from \eqref{eq: EBLUP}. 
Also important for the prediction of the basis weights is the number of curves with the given weight entering the prediction. Thus, the basis weights of $C_j(t)$ are better predicted than those of $E_{ijh}(t)$. As expected, basis weights of FPCs that explain more variability are estimated better. Similar results can be found for the fRI and for the sparse scenario. \\
For all scenarios, we obtain good results for the functional random effects as well as for the functional response. The rrMSEs for the functional response are lowest, which is due to the fact that even if the FPC bases are not perfectly estimated, they can still serve as a good empirical basis. Thus, the data can be reconstructed very well.\\ 
We found considerably more outliers of the relative errors for the sparse scenario than for the other two scenarios, which is most probably due to an unfavourable distribution of the few observation points across the curves in a few data sets.
Overall, we can conclude that all components are estimated well and especially for the functional response we obtain very small rrMSEs across all simulations.

\underline{Comparison of the different estimation results for the fRI scenario}.
We find that the functional random processes as well as the functional response are estimated equally well for the two options of the basis weights prediction. 
The functional response is again estimated very well with an average rrMSE of 0.09 for both EBLUP and FPC-FAMM estimation. The spline-FAMM results are considerably worse for the random processes (almost three (smooth error) and almost seven (fRI) times higher average rrMSEs), which results from the fact that the constraint $\sum_{l=1}^{L^X}X_l(t)\equiv 0$, $X\in\lbrace B,E\rbrace$, is not fulfilled and parts are shifted between terms. The functional response is recovered reasonably well, but has a more than 1.5 times higher average rrMSE than the EBLUP and FPC-FAMM estimates. 
Note that due to high computation times (see below), we only consider 100 simulation runs for the spline-FAMM simulation.\\
For the covariate effects, the FPC-FAMM estimation gives better results than the estimation under an independence assumption and considerably better results than the spline-FAMM estimation (between 2.8 and six times higher average rrMSEs). 
The average point-wise coverage of the point-wise CBs is very good for most effects for FPC-FAMM and the simultaneous coverage is reasonable. Both are considerably better than for the spline-FAMM alternative. The coverage for the latter would most probably improve by increasing the number of spline basis functions which is, however, limited by the high computation time. 

\underline{Computation times}.
Our simulations show that the FPC-based approach has clear advantages in terms of computational complexity, despite the computational cost of auto-covariance estimation. 
We compare times for one simulation run of the fRI scenario for each estimation option obtained under the same conditions (without parallelization in function \pkg{bam} that would speed up the estimation). The study was run on a 64 Bit Linux platform with 660 Gb of RAM memory. The FPC-based approach with the basis weights predicted as EBLUPs took five hours and predicting the basis weights using FPC-FAMM took slightly less than ten hours longer. The spline-FAMM took by far the longest with a duration of ten days which is due to the two extra smoothing parameters each for the fRI and the smooth error which have to be estimated. Moreover, using FPCs reduces the number of necessary basis functions.

\section{Discussion and outlook}\label{sec: Discussion}
We propose an FPC-based estimation approach for functional linear mixed models that is particularly suited to irregularly or sparsely sampled observations. To pool information, we smooth both the mean and auto-covariance functions.
We propose and compare two options for the prediction of the FPC weights and obtain conditional point-wise confidence bands for the functional covariate effects. Our simulations show that our method reliably recovers the features of interest. The parsimonious representation of the functional random effects in bases of eigenfunctions outperforms the spline-based alternative of \cite{Scheipl.2014} with which we compare, both in terms of error rates and coverage as well as in terms of computation time. To the best of our knowledge, there is no other competitor to our approach as all other methods are either restricted to regular grid data or simpler correlation structures. In our application to speech production data, we show that our method allows to draw conclusions about the asymmetry of consonant assimilation to an extent which is not achievable using conventional methods with data reduction.\\
Building on existing methods for our estimation approach allows us to take advantage of robust, flexible algorithms with a high functionality. The computational efficiency, however, could potentially be improved by exploiting the special structure of our model. In future work,
we plan to improve the estimation of the auto-covariances in order to better account for their symmetry and positive semi-definiteness and for the fact that the cross products in \eqref{eq:cov decomp} are not homoscedastic. Moreover, it would be interesting to compare the different options for iterative estimation in detail. \\
The construction of point-wise and simultaneous confidence bands that account for the variability of the estimated FPC decomposition is beyond the scope of this work, but would be of interest. 
For uncorrelated functions, \cite{Goldsmith.2013} propose bootstrap-based corrected confidence bands for densely and sparsely sampled functional data. However, it remains an open question how to extend their non-parametric bootstrap to our correlated curves, and computational cost is another issue.\\ 

\subsection*{Acknowledgements}
We thank Fabian Scheipl and Simon Wood for making available and further improving their extensive software at our requests, for technical support, and for fruitful discussions. 
Sonja Greven and Jona Cederbaum were funded by the Emmy Noether grant GR 3793/1-1 from the
German Research Foundation. Marianne Pouplier was supported by the ERC under the EU's 7th Framework Programme (FP/2007-2013)/Grant Agreement n.~283349-SCSPL.

\vspace{5mm}
\bibliography{references}


\appendix
\section{Derivations} \label{appendix: proofs}
\renewcommand{\thesubsection}{\thesection.\arabic{subsection}}
\renewcommand{\thesubsection}{\thesection.\arabic{subsection}}

Derivation of the variance decomposition of model (\ref{eq: FLMM crossed}).
Using iterated expectations allows us to decompose the variance of the response as follows
\bea \nonumber
\int_{\mathcal{T}}{\Var\left[ Y_{ijh}(t)\right]}\dint t &=& 
\int_{\mathcal{T}}{\Var\left[B_i(t)\right]} \dint t + \int_{\mathcal{T}}{\Var\left[C_j(t)\right]} \dint t + \int_{\mathcal{T}}{\Var\left[ E_{ijh}(t)\right]} \dint t\\ \nonumber &+& \int_{\mathcal{T}}{\varepsilon_{ijh}(t)}\dint t \\ \nonumber
&=& \sum_{k=1}^{\infty}\nu_k^B \underbrace{\int_{\mathcal{T}}{\phi_{k}^B(t)\phi_{k}^B(t)}\dint t}_{=1} + \sum_{k=1}^{\infty}\nu_k^C \underbrace{\int_{\mathcal{T}}{\phi_{k}^C(t)\phi_{k}^C(t)}\dint t}_{=1}  \\ \nonumber &+&\sum_{k=1}^{\infty}\nu_k^E \underbrace{\int_{\mathcal{T}}{\phi_{k}^E(t)\phi_{k}^E(t)}\dint t }_{=1} + \int_{\mathcal{T}}\sigma^2 \dint t\\ \nonumber
&=& \sum_{k=1}^{\infty}\nu_k^B + \sum_{k=1}^{\infty}\nu_k^C +\sum_{k=1}^{\infty}\nu_k^E + \sigma^2 |\mathcal{T}|.
\eea

\subsection{Marginal bases for the estimation of the mean}
$\mPsi_g^p$, $p=1,\ldots,P$, is an inflated vector of length $\mathfrak{D}$ containing the values of covariate $\mx_{p}=\left(x_{111p},\ldots,x_{IJH_{IJ}p}\right)^\top$. For each curve, the covariate values are replicated for each observation point yielding
{\footnotesize\bea \nonumber
\mPsi_g^p = \left[
\begin{array}{*{7}{c}}
x_{111p}\\
\vdots\\
x_{111p}\\
\vdots\\
x_{IJH_{IJ}p}\\
\vdots\\
x_{IJH_{IJ}p}
\end{array}\right].
\eea
}
For the functional intercept $f_0(t)$, $\mPsi_g^p= \mathds{1}_{\mathfrak{D}}$. 

\bigskip
$\mPsi_t^p$, $p=0,\ldots,P$, is a $\mathfrak{D}\times K^p$ matrix comprising the evaluations of the basis functions on the original observations points of the form
{\footnotesize
\bea  \nonumber
\mPsi_{t}^p &=&\left[\begin{array}{*{3}{c}}
\psi_1^p(t_{1111}) & \cdots & \psi_{K^p}^p(t_{1111})\\
\vdots & & \vdots \\
\psi_1^p(t_{111T_{111}}) & \cdots & \psi_{K^p}^p(t_{111T_{111}})\\
\vdots & & \vdots \\
\psi_1^p(t_{IJH_{IJ}1}) & \cdots & \psi_{K^p}^p(t_{IJH_{IJ}1})\\
\vdots & & \vdots \\
\psi_1^p(t_{IJH_{IJ}T_{IJH_{IJ}}}) & \cdots & \psi_{K^p}^p(t_{IJH_{IJ}T_{IJH_{IJ}}})\\
\end{array}\right].
\eea
}

\subsection{Matrices in the prediction of the basis weights as EBLUPs}
 $\hat{\mPhi}^B$ is a block-diagonal matrix of dimension $\mathfrak{D} \times IN^B$. 
 {\footnotesize
\bea \nonumber
\hat{\mPhi}^B = \left[\begin{array}{*{7}{c}}
\hat{\phi}^B_1(t_{1111}) & \cdots & \hat{\phi}^B_{N^B}(t_{1111}) & \cdots & 0 & \cdots & 0\\
\vdots &  & \vdots  & & \vdots  &  &\vdots \\
\hat{\phi}^B_1(t_{1JH_{1J}T_{1JH_{1J}}}) & \cdots & \hat{\phi}^B_{N^B}(t_{t_{1JH_{1J}T_{1JH_{1J}}}}) & \cdots & 0 & \cdots & 0\\
\vdots &  & \vdots  & & \vdots  &  &\vdots \\
0 & \cdots &  \cdots 0 &  &  \hat{\phi^B_1}(t_{I111}) & \cdots & \hat{\phi^B_{N^B}}(t_{I111})\\
\vdots &  & \vdots  & & \vdots  &  &\vdots \\
0 & \cdots &  \cdots 0 &  &  \hat{\phi^B_1}(t_{IJH_{IJ}T_{IJH_{IJ}}}) & \cdots & \hat{\phi^B_{N^B}}(t_{IJH_{IJ}T_{IJH_{IJ}}})\\
\end{array}\right]
\eea
}
$\hat{\mPhi}^C$ consists of $I$ block-diagonal matrices -- one for each speaker -- and is of dimension $\mathfrak{D} \times JN^C$. 
 {\footnotesize
\bea \nonumber
\hat{\mPhi}^C = \left[\begin{array}{*{7}{c}}
\hat{\phi}^C_1(t_{1111}) & \cdots & \hat{\phi}^C_{N^C}(t_{1111}) & & & &\\
\vdots &  & \vdots & & & &\\
\hat{\phi}^C_1(t_{11H_{11}T_{11H_{11}}}) & \cdots & \hat{\phi}^C_{N^C}(t_{11H_{11}T_{11H_{11}}}) & & & &\\
&&& \ddots &&&\\
&&&& \hat{\phi}^C_1(t_{1J11}) & \cdots & \hat{\phi}^C_{N^C}(t_{1J11})\\
&&&& \vdots & & \vdots\\
&&&& \hat{\phi}^C_1(t_{1JH_{1JT_{1JH_{1J}}}}) & \cdots & \hat{\phi}^C_{N^C}(t_{1JH_{1JT_{1JH_{1J}}}})\\
&&& \vdots &&&\\
\hat{\phi}^C_1(t_{I111}) & \cdots & \hat{\phi}^C_{N^C}(t_{I111}) & & & &\\
\vdots &  & \vdots & & & &\\
\hat{\phi}^C_1(t_{I1H_{11}T_{I1H_{I1}}}) & \cdots & \hat{\phi}^C_{N^C}(t_{I1H_{I1}T_{11H_{I1}}}) & & & &\\
&&& \ddots &&&\\
&&&& \hat{\phi}^C_1(t_{IJ11}) & \cdots & \hat{\phi}^C_{N^C}(t_{IJ11})\\
&&&& \vdots & & \vdots\\
&&&& \hat{\phi}^C_1(t_{IJH_{IJT_{IJH_{IJ}}}}) & \cdots & \hat{\phi}^C_{N^C}(t_{IJH_{IJT_{IJH_{IJ}}}})
\end{array}\right]
\eea
} 
Obviously, the role of speakers and target words is exchangeable. For present purposes, however, we assume that all vectors and matrices are first ordered by speakers and within each speaker ordered by target words.
$\hat{\mPhi}^E$ is a block-diagonal matrix of dimension $\mathfrak{D} \times nN^E$ with blocks
{\footnotesize
\bea \nonumber
\left[\begin{array}{*{4}{c}}
\hat{\phi}^E_1(t_{1111}) & \cdots & \hat{\phi}^E_{N^E}(t_{1111})\\
\vdots & &  \vdots\\
\hat{\phi}^E_1(t_{111T_{111}}) & \cdots & \hat{\phi}^E_{N^E}(t_{111T_{111}})\\
\end{array}\right], \ldots, 
\left[\begin{array}{*{3}{c}}
\hat{\phi}^E_1(t_{IJH_{IJ}1}) & \cdots & \hat{\phi}^E_{N^E}(t_{IJH_{IJ}1})\\
\vdots & &  \vdots\\
\hat{\phi}^E_1(t_{IJH_{IJ}T_{IJH_{IJ}}}) & \cdots & \hat{\phi}^E_{N^E}(t_{IJH_{IJ}T_{IJH_{IJ}}})
\end{array}\right].
\eea
}
Note that for irregularly spaced functional data, the FPCs evaluated on the original observation points are curve-specific. \\
The estimated covariance matrix of the $\mxi$, $\hat{\mG}$, is of the form
{\footnotesize
\bea \nonumber
\hat{\mG} = \left[\begin{array}{*{3}{c}}
\hat{\Cov}(\mxi^B) & & \\
& \hat{\Cov}(\mxi^C) &\\
&&\hat{\Cov}(\mxi^E)
\end{array}\right],
\eea
}
where each covariance matrix $\hat{\Cov}(\mxi^X), X\in \lbrace B,C,E\rbrace$ is a diagonal matrix with elements $\hat{\nu}^X_1, \ldots, \hat{\nu}^X_{N^X},\ldots,\hat{\nu}^X_1,\ldots,\hat{\nu}^X_{N^X}$.

\subsection{Marginal bases for FPC-FAMM}
$\mPsi_{g}^B$ is a $\mathfrak{D} \times I$ incidence matrix of which the entries in the $d$th column are one wherever the row belongs to the $d$th speaker and zero otherwise. $\mPsi_{g}^C$ and $\mPsi_{g}^E$ are analogously $\mathfrak{D} \times J$ and $\mathfrak{D} \times n$ matrices with entries in the $d$th column equal to one wherever the row belongs to the $d$th target word or to the $d$th curve, respectively.
For lack of space, we exemplarily show in the following the part of the matrices $\mPsi_{g}^X$, $X\in\lbrace B,C,E\rbrace$, that belongs to the $i$th speaker (denoted by $\mPsi_{g}^{X,i}$), assuming that the data are ordered by speakers, within each speaker ordered by target words, within each target word ordered by repetition, then by curves, and finally by observation points.
{\footnotesize
\bea \nonumber
\mPsi_{g}^{B,i}=
\overset{i}{\left[\ 
\begin{array}{*{13}{c}}
\cline{3-3}
0& \cdots & \multicolumn{1}{|c|}{1} &\cdots & 0  \\
\vdots & & \multicolumn{1}{|c|}{\vdots}&& \vdots    \\
0& \cdots & \multicolumn{1}{|c|}{1}&\cdots  &0 \\
\vdots & & \multicolumn{1}{|c|}{\vdots}& & \vdots \\
0& \cdots & \multicolumn{1}{|c|}{1} &\cdots & 0 \\
\vdots & & \multicolumn{1}{|c|}{\vdots}& &\vdots \\
0& \cdots & \multicolumn{1}{|c|}{1} &\cdots  &0 \\
\vdots & & \multicolumn{1}{|c|}{\vdots}& & \vdots \\
0& \cdots & \multicolumn{1}{|c|}{1}& \cdots &0\\
\vdots & & \multicolumn{1}{|c|}{\vdots}&  & \vdots\\
0& \cdots & \multicolumn{1}{|c|}{1} &\cdots  &0\\
\vdots & & \multicolumn{1}{|c|}{\vdots}&  & \vdots \\
0& \cdots & \multicolumn{1}{|c|}{1}  &\cdots & 0 \\
\vdots & & \multicolumn{1}{|c|}{\vdots} && \vdots\\
0& \cdots & \multicolumn{1}{|c|}{1} &\cdots  &0 \\
\cline{3-3}
\end{array}\right]}
\mPsi_{g}^{C,i}=
\left[
\begin{array}{*{13}{c}}
\cline{1-1}
\multicolumn{1}{|c|}{1}  & 0 & \cdots & 0\\
\multicolumn{1}{|c|}{\vdots} &  \vdots & & \vdots  \\
\multicolumn{1}{|c|}{1}  & 0 & \cdots  & 0\\
\multicolumn{1}{|c|}{\vdots} & \vdots &   & \vdots\\
\multicolumn{1}{|c|}{1} &  0 & \cdots & 0 \\
\multicolumn{1}{|c|}{\vdots}&  \vdots & & \vdots\\
\multicolumn{1}{|c|}{1}  & 0 & \cdots & 0\\
\cline{1-1}
\vdots &  \vdots & & \vdots\\
\cline{4-4}
0 & 0 & \cdots & \multicolumn{1}{|c|}{1} \\
\vdots &  \vdots & &\multicolumn{1}{|c|}{\vdots}\\
0& 0 & \cdots & \multicolumn{1}{|c|}{1} \\
\vdots & \vdots & & \multicolumn{1}{|c|}{\vdots}\\
0  & 0 &\cdots& \multicolumn{1}{|c|}{1} \\
\vdots &  \vdots & &\multicolumn{1}{|c|}{\vdots}\\
0  & 0 & \cdots &\multicolumn{1}{|c|}{1} \\
\cline{4-4}
\end{array}\right],
\mPsi_{g}^{E,i}=
\left[
\begin{array}{*{13}{c}}
\cline{1-1}
\multicolumn{1}{|c|}{1} & 0 & \cdots & 0\\
\multicolumn{1}{|c|}{\vdots} &   \vdots & & \vdots  \\
\multicolumn{1}{|c|}{1}  & 0 & \cdots  & 0\\
\cline{1-1}
\vdots &  \vdots &   & \vdots\\
0  & 0 & \cdots & 0 \\
\vdots & \vdots & & \vdots\\
0& 0 & \cdots & 0\\
\vdots &  \vdots & & \vdots\\
0 & 0 & \cdots & 0\\
\vdots &  \vdots & & \vdots\\
0  & 0 & \cdots & 0\\
\vdots &  \vdots & & \vdots\\
\cline{4-4}
0  & 0 &\cdots& \multicolumn{1}{|c|}{1}\\
\vdots &  \vdots & & \multicolumn{1}{|c|}{\vdots}\\
0  & 0 & \cdots & \multicolumn{1}{|c|}{1}\\
\cline{4-4}
\end{array}\right]
\eea
}
 $\mPsi_{g}^X$ then contains the stacked partial matrices for all speakers.
 
\bigskip
$\mPsi_{t}^X$ is a $\mathfrak{D} \times N^X$ matrix containing the evaluations of the respective eigenfunctions on the original observation points of the form
{\footnotesize
\bea  \nonumber
\mPsi_{t}^X &=&\left[\begin{array}{*{3}{c}}
\phi_1^X(t_{1111}) & \cdots & \phi_{N^X}^X(t_{1111})\\
\vdots & & \vdots \\
\phi_1^X(t_{111T_{111}}) & \cdots & \phi_{N^X}^X(t_{111T_{111}})\\
\vdots & & \vdots \\
\phi_1^X(t_{IJH_{IJ}1}) & \cdots & \phi_{N^X}^X(t_{IJH_{IJ}1})\\
\vdots & & \vdots \\
\phi_1^X(t_{IJH_{IJ}T_{IJH_{IJ}}}) & \cdots & \phi_{N^X}^X(t_{IJH_{IJ}T_{IJH_{IJ}}})\\
\end{array}\right], \ X\in \lbrace B,C,E\rbrace.
\eea
}
\section{Supplementary details on the estimation and implementation} \label{appendix: supplementary details on estimation}

\subsection{Implementation of the auto-covariance estimation}
Model (\ref{eq:cov decomp}) does not contain an intercept. The implementation for the fRI design, however, requires that an intercept is included and added to the auto-covariance of the fRI due to the centring constraint in function \pkg{bam}. This is not the case for the crossed-fRI design, where each smooth is varied by an indicator variable and thus no constraint is applied. For more details see the description of the function \pkg{gam} and of \pkg{gam.models}.

\subsection{Rescaling of the eigenvectors and eigenvalues}
Denote the estimated eigenvectors by $\hat{\tilde{\mphi}}^X_k = \left(\hat{\tilde{\phi}}^X_k(t_1),\ldots \hat{\tilde{\phi}}^X_k(t_D)\right)\tr$,
and the estimated eigenvalues by $\hat{\tilde{\nu}}_k^X$, $k\in\mathbb{N}$, $X \in \lbrace B,C,E \rbrace$. In order to ensure that the approximated functions are orthonormal with respect to the scalar product $\langle f,g \rangle = \int f(t)g(t) \dint t$, we rescale the eigenvectors by
\bea \nonumber
\hat{\mphi}^X_k = \frac{1}{\sqrt{a}} \hat{\tilde{\mphi}}^{X}_k,
\eea
with $a$ denoting the constant interval width of the equidistant grid $\lbrace t_1,\ldots,t_D\rbrace$. As a consequence, the eigenvalues to the rescaled eigenvectors need to be adjusted as $\hat{\nu}^X_k = a \hat{\tilde{\nu}}^X_k$. 

\subsection{Truncation of the FPCs}
Due to the additive structure in the variance decomposition, we can choose the truncation lags $N^B$, $N^C$, and $N^E$ in the following way:
\begin{enumerate}
\itemsep0pt
\item specify the proportion of explained variance $L$, e.g.~$L=0.95$ as used in the application in Section 4.
\item select the FPCs corresponding to their eigenvalues in decreasing order, until 
\bea \label{eq: truncation}
\frac{\sum_{k=1}^{N^B}\nu^B_k + \sum_{k=1}^{N^C}\nu^C_k +\sum_{k=1}^{N^E}\nu^E_k +\sigma^2|T|}{\sum_{k=1}^{\infty}\nu^B_k + \sum_{k=1}^{\infty}\nu^C_k +\sum_{k=1}^{\infty}\nu^E_k +\sigma^2|T|} \geq L.
\eea
\end{enumerate}
Note that the three random processes are treated equally, i.e.~in each step, the FPC with the highest corresponding eigenvalue is chosen regardless of the associated process. 
In practice, all infinite sums in \eqref{eq: truncation} are approximated by the finite sums of all obtained eigenvalues. The eigenvalues and the error variance are replaced by their estimates.

\subsection{Fixing the smoothing parameter in FPC-FAMM}
Technically, in order that a fixed smoothing parameter $\lambda_{\star}$ (here $\lambda_{\star}=1$) is used in the prediction of the FPC-based fRIs, we need to specify the smoothing parameter in function \pkg{pffr} as 
\bea \label{eq: sp_fix}
\lambda_{fix} = \lambda_{\star}\cdot S.scale \cdot \hat{\sigma}^2,
\eea
where $S.scale$ is a scaling factor used in the set-up of the design matrices in order to numerically stabilize the prediction. It can be obtained by setting \texttt{fit=FALSE} in the call of function \pkg{pffr}. $\hat{\sigma}^2$ is an  estimate of the error variance. We can use the estimate obtained in step 2 of our estimation procedure. Note that equation \eqref{eq: sp_fix} makes clear that the point-wise confidence bands for the mean function are not only conditional on the estimated FPCs and the truncation level, but also on the estimated error variance.

\subsection{Iterative estimation}
If desired, the estimation accuracy can be improved by applying the estimation steps iteratively. At least two possibilities exist: We can either perform steps 1-4 and then re-start with step 1 with the adjusted observations $Y_{ijh}^*(t) := Y_{ijh}(t)-\hat{B}_i(t)-\hat{C}_j(t)-\hat{E}_{ijh}(t)$ until a pre-defined criterion is reached. Alternatively, we can replace the mean with that obtained in the FAMM framework in step 4 and restart with step 2. 

\newpage
\section{Supplementary application details and results}

\subsection{Pre-processing}
The index calculation is based on the calculation of the power spectrum over a time window of approximately 20 ms, shifted in 5 ms steps over the time interval of the consonants.
In order to be able to compare the index curves for the two consonant orders (\sS, \Ss) directly, the index curves of \Ss were mirrored along the time axis (mapping +1 to -1 and vice versa) such that for both orders the index dynamic ranges from +1 for the first consonant to -1 for the second consonant rather than from +1 for /s/ to -1 for /sh/. To achieve this, we first estimated smooth mean curves of available reference curves of orders /sh\#sh/ and /s\#s/ per speaker and for each combination of the covariates vowel, stress1, stress2 using penalized splines and evaluated them on the measurement points. We then mirrored the index curves of order \Ss at the speaker-condition specific mean values, averaging over the mean curves for \Ss and \sS.

\subsection{Supplementary application results}
\begin{figure}[h!]
\centering
\includegraphics[width=0.25\textwidth]{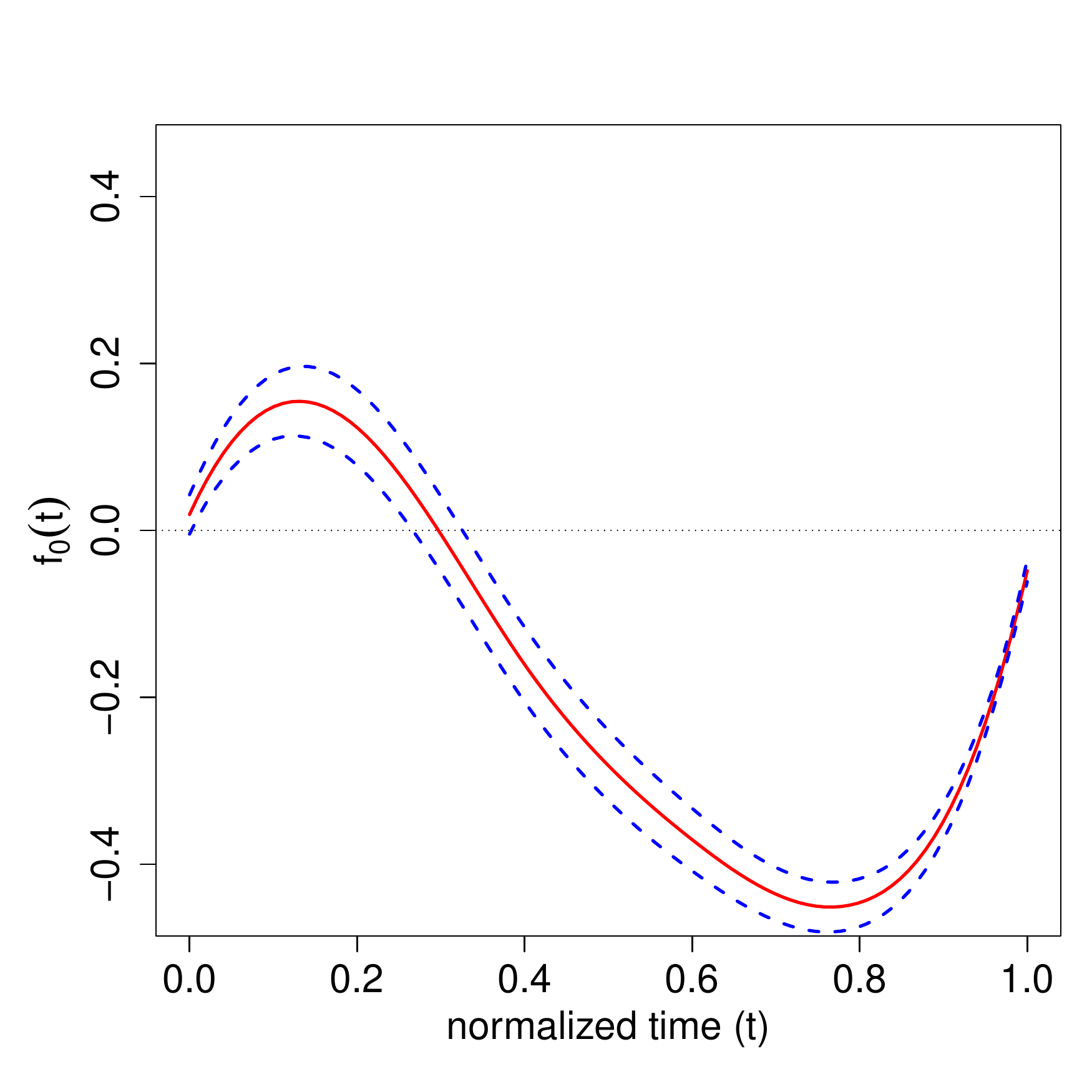}\\
\includegraphics[width=0.25\textwidth]{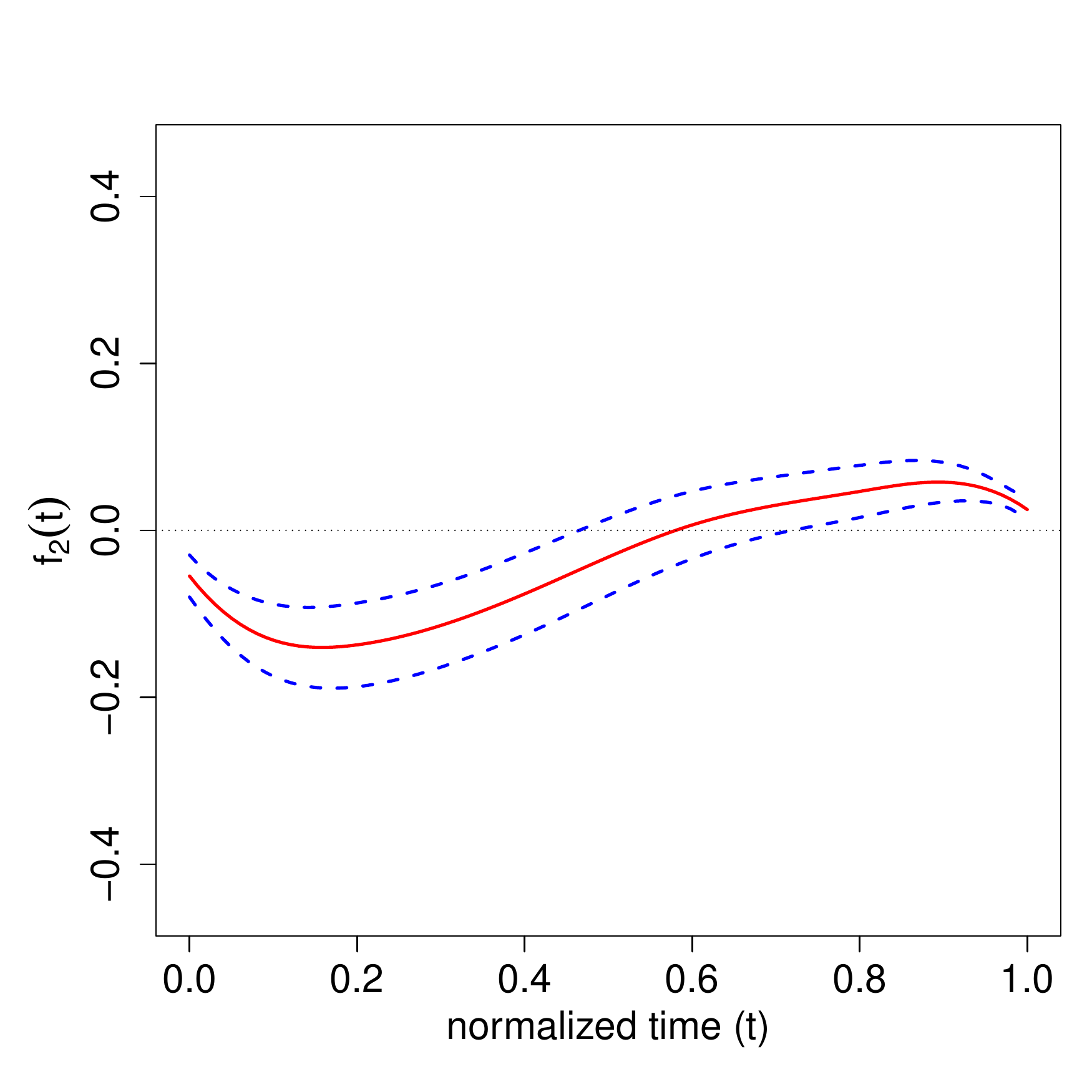}
\includegraphics[width=0.25\textwidth]{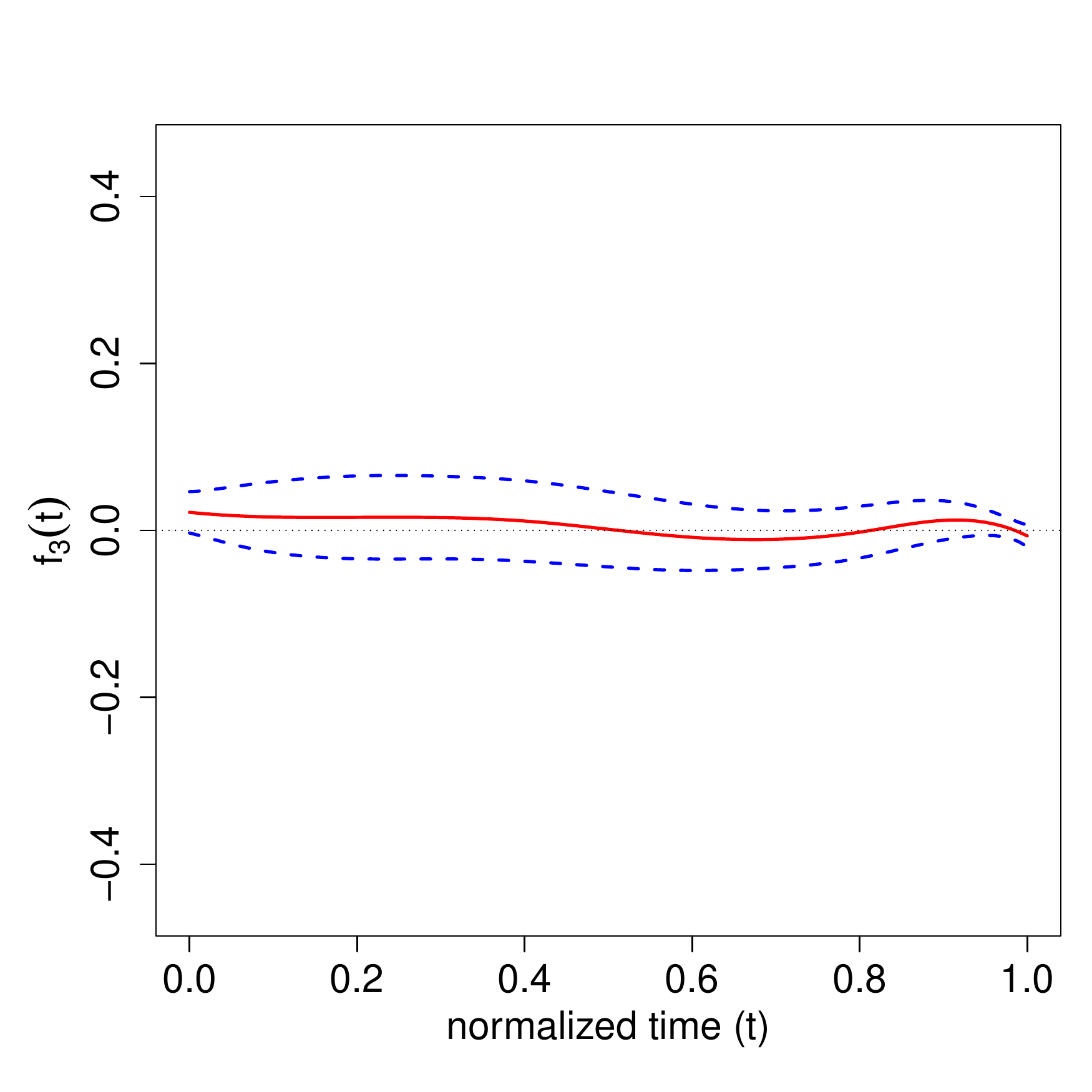}
\includegraphics[width=0.25\textwidth]{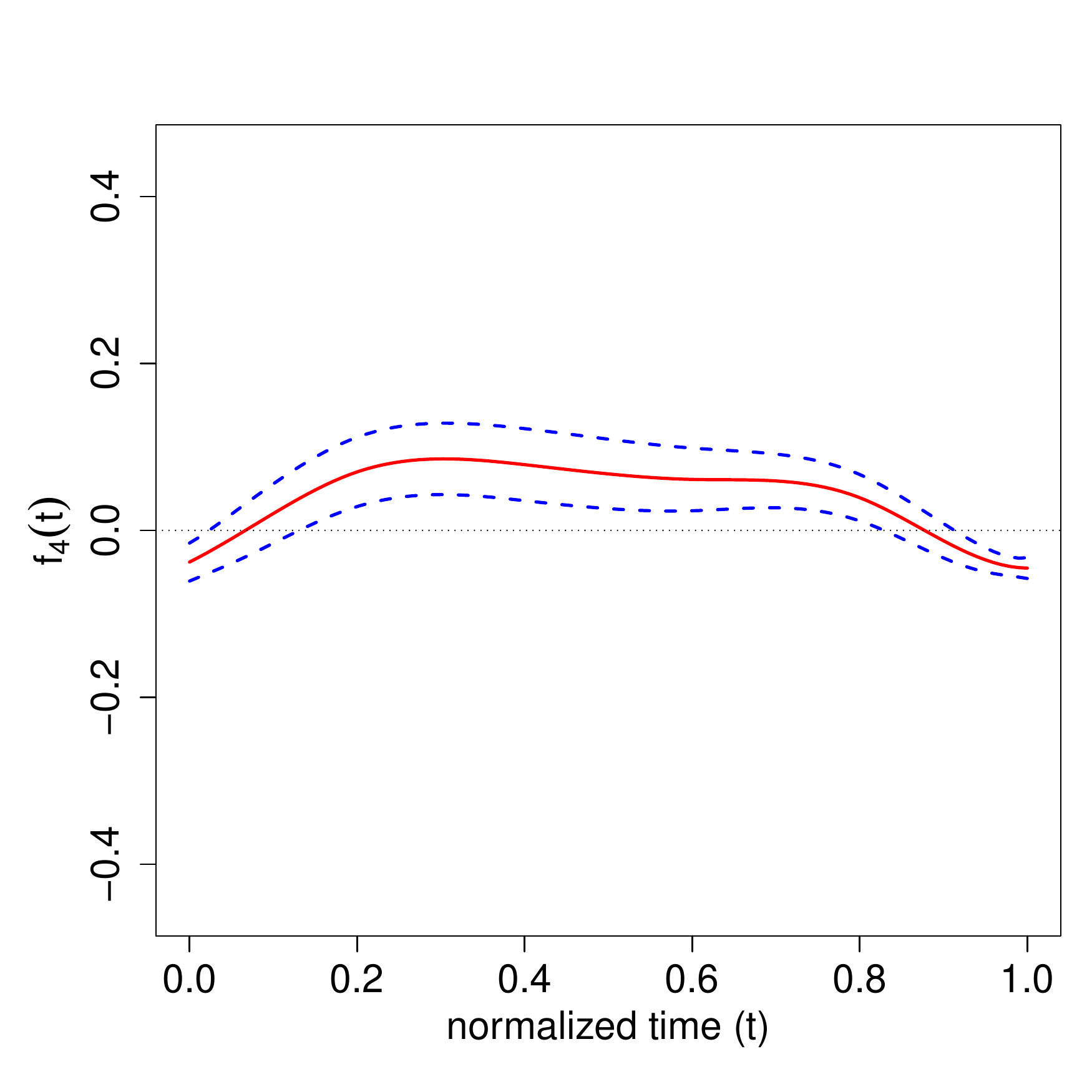}\\
\includegraphics[width=0.25\textwidth]{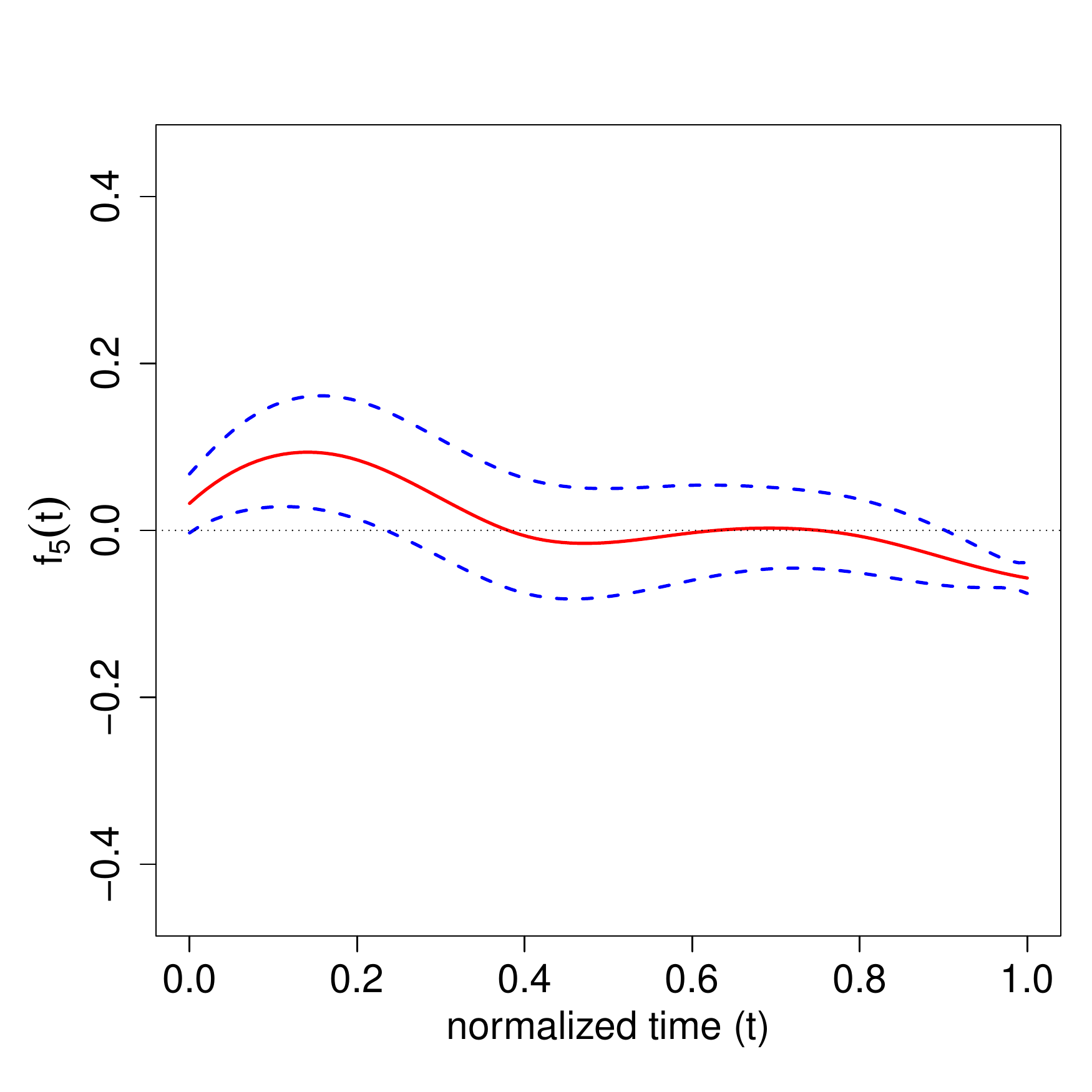}
\includegraphics[width=0.25\textwidth]{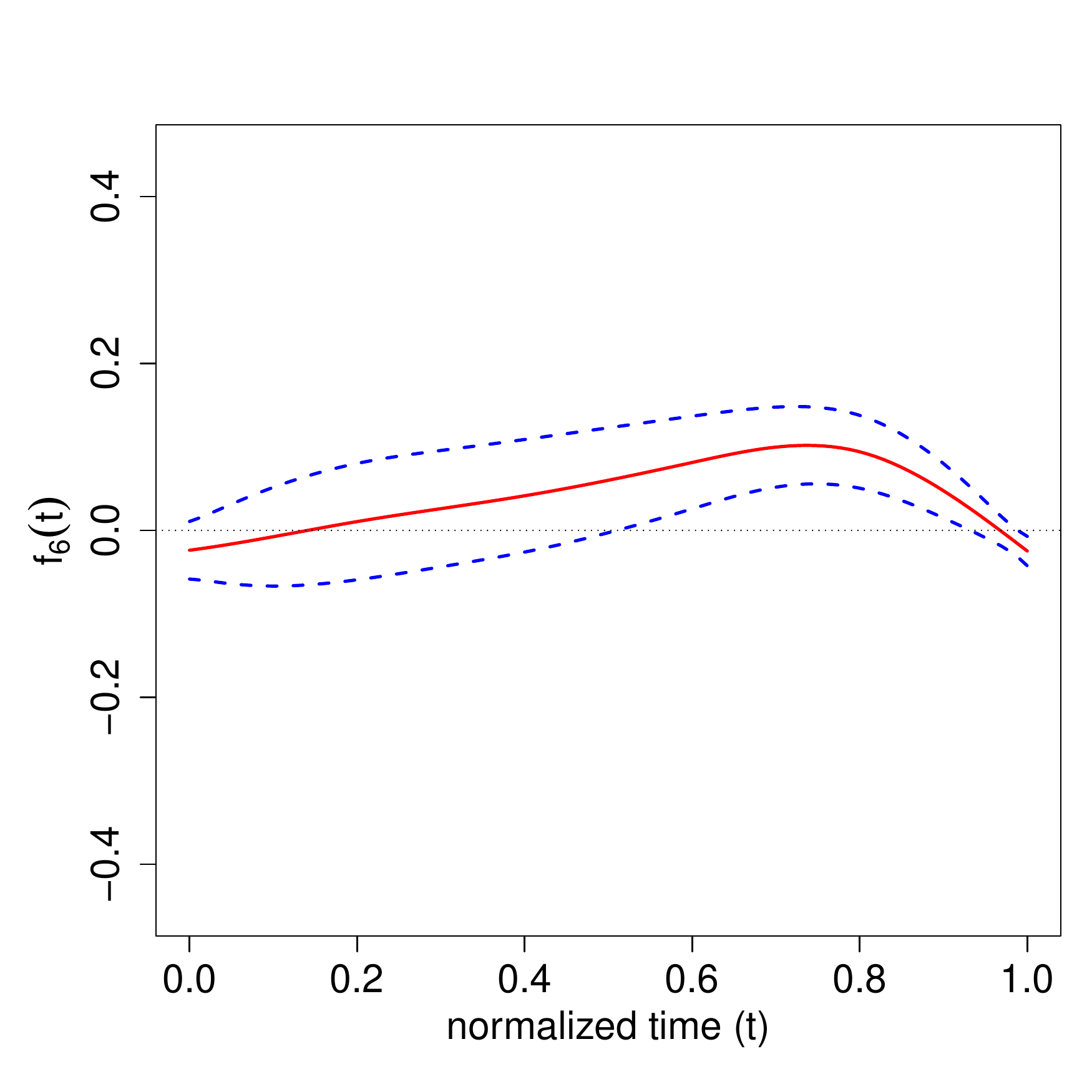}
\includegraphics[width=0.25\textwidth]{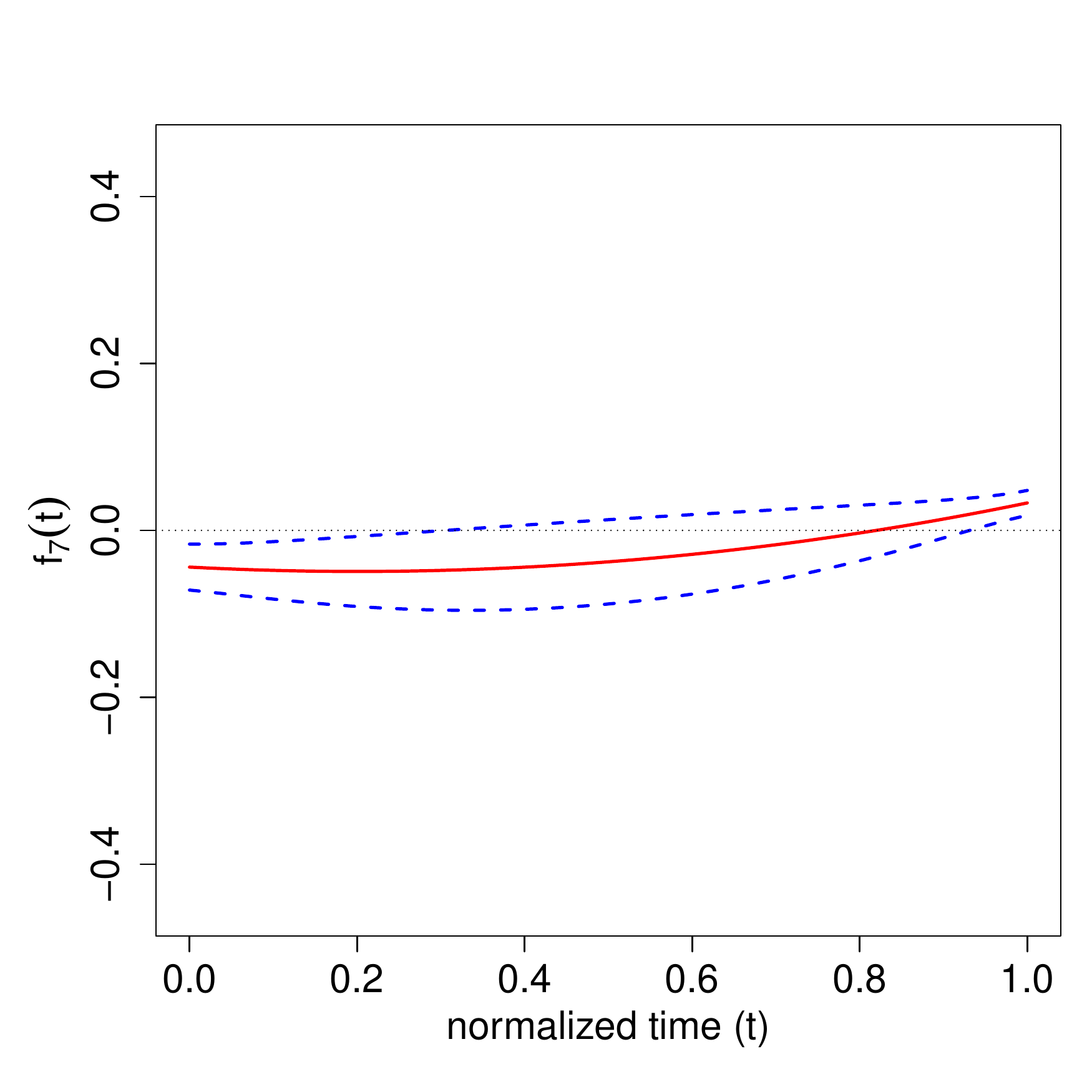}
\caption{Covariate mean effects (red solid lines) with conditional point-wise confidence bands (dashed lines). Upper: Reference mean $f_0(t)$. Middle (from left to right): covariate effects of covariates stress1, stress2, vowel. Lower (from left to right): interaction effects of order and stress1, order and stress2, order and vowel.}
\label{fig: covariate effects application}
\end{figure}
\begin{figure}[t!]
\centering
\includegraphics[width=0.25\textwidth]{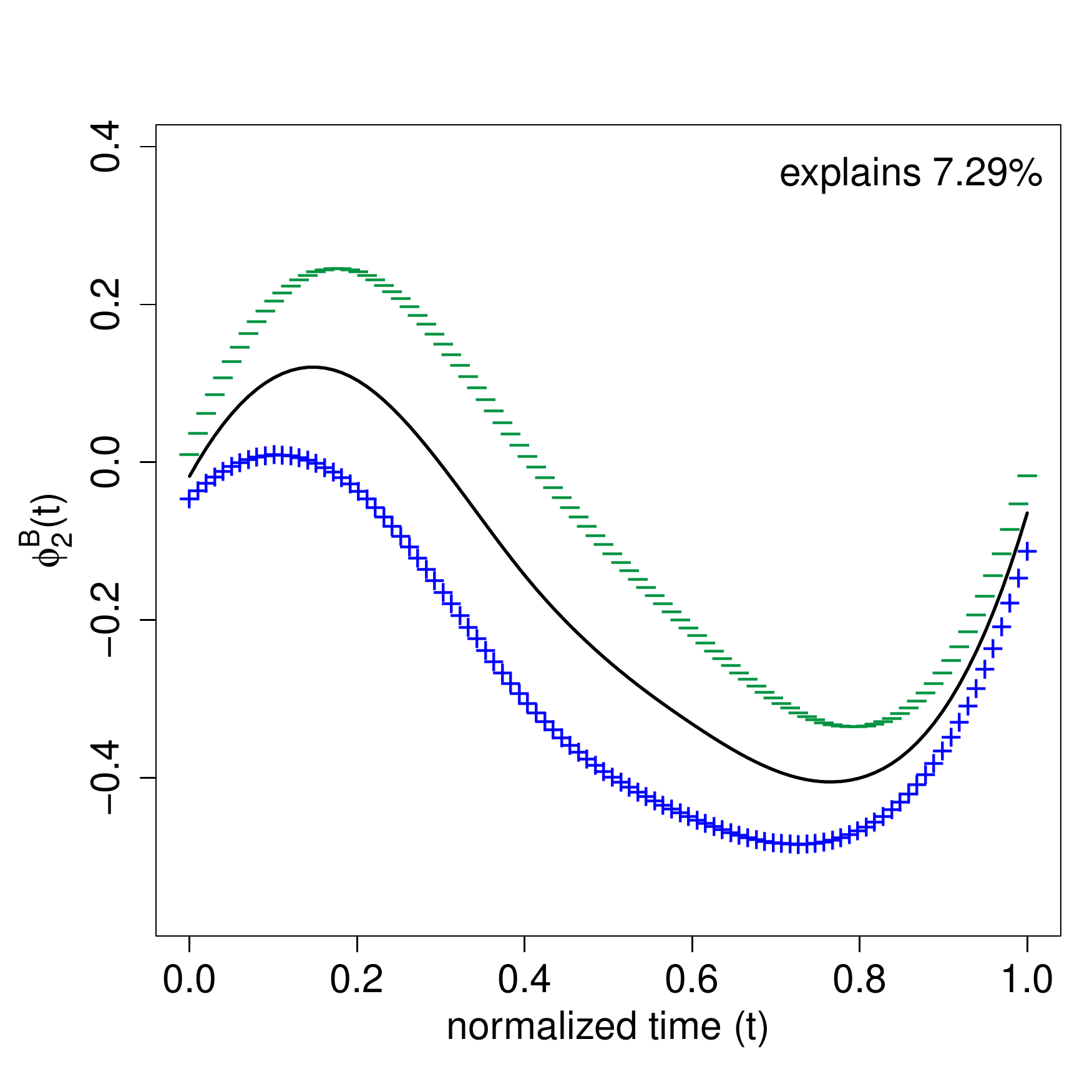}
\caption{Second FPC for speakers. Shown is the mean function (solid line) and the effect of adding (+) and subtracting (-) a suitable multiple ($2\sqrt{\hat{\nu}_2^B}$) of the first FPC for speakers.}
\label{fig: FPC2 application}
\end{figure}

In the following, we show additional application results, including the eigenvalues, the effects of the covariates and interactions, the second FPC for speakers, and the variance decomposition of the model without covariates on which we base our crossed-fRI simulation setting.\\
Table \ref{tab: eigenvalues application} gives the estimated eigenvalues for the model with covariates included. 
Figure \ref{fig: covariate effects application} shows the estimated effects and point-wise confidence bands of the covariates stress1 (0:Strong, 1:Weak), stress2 (0:Strong, 1:Weak), and vowel (0: ia, 1:ai) as well as of the interactions between covariate order and the other three covariates.\\
The second FPC for speakers is depicted in Figure \ref{fig: FPC2 application}. The interpretation is as follows: The index curves of speakers with positive FPC weights for the second FPC are pulled towards the ideal reference value of the second consonant, whereas the index curves of speakers with negative FPC weights are pulled towards the ideal reference value of the first consonant.\\
Table \ref{tab: variance decomp no covariates application} gives the variance decomposition of the model with no covariate included. 
 \begin{table}[t!]
  \setlength{\tabcolsep}{1mm}
 \small
\begin{center}
\caption{Estimated eigenvalues for the model with covariates.}
\begin{tabular}{rrrrrrr}
  $\hat{\nu}^B_1$ & $\hat{\nu}^B_2$  & $\hat{\nu}^E_1$ & $\hat{\nu}^E_2$ & $\hat{\nu}^E_3$ & $\hat{\sigma}^2$\\ 
  \hline
 5.84 $\cdot 10^{-3}$ & 3.23$\cdot 10^{-3}$ & 19.53$\cdot 10^{-3}$ & 7.59$\cdot 10^{-3}$ & 2.73$\cdot 10^{-3}$ & 3.94 $\cdot 10^{-3}$
\end{tabular}
\label{tab: eigenvalues application}
\end{center}
\end{table}
 \setlength{\tabcolsep}{1mm}
\begin{table}[h!]
\small
\begin{center}
\caption{First row: Estimated eigenvalues of $\hat{\mK}^B$ and $\hat{\mK}^E$ and estimated error variance for the crossed model without covariates. Second row: Variance explained in percent of the estimated total variance.}
\begin{tabular}{rrrrrrr}
  $\hat{\nu}^B_1$ & $\hat{\nu}^B_2$  & $\hat{\nu}^C_1$ &  $\hat{\nu}^E_1$ & $\hat{\nu}^E_2$ & $\hat{\nu}^E_3$ & $\hat{\sigma}^2$\\ 
  \hline
$5.86\cdot 10^{-3}$ & $2.71 \cdot10^{-3}$ & $8.89 \cdot 10^{-3}$ & $19.1\cdot 10^{-3}$ & $7.53\cdot 10^{-3}$ & $2.66\cdot 10^{-3}$ & $ 5.62 \cdot10^{-3}$\\
    \hline
 10.83\% & 5.00\% & 16.44\% & 35.23\% & 13.93\% & 4.92\% & 10.39\%
\end{tabular}
\label{tab: variance decomp no covariates application}
\end{center}
\end{table}
\clearpage
\section{Supplementary simulation details and results}
\subsection{Measures of goodness of fit}
We use the root relative mean squared error
\bea  \label{eq: rrMSE}
\rrMSE(\mtheta,\hat{\mtheta}) = \sqrt{\frac{\dfrac{1}{L}\sum_{l=1}^L \left(\theta_{l} - \hat{\theta}_{l} \right)^2}{\dfrac{1}{L}\sum_{l=1}^L {\theta_{l}}^2}},
\eea 
as a measure of goodness of fit for vector-valued estimates $\hat{\mtheta}$ of $\mtheta=\left(\theta_1,\ldots,\theta_L\right)^\top$ (FPC weights $\mxi_{k}^X = \left(\xi_{1k}^X,\ldots,\xi_{L^Xk}^X\right)$, $k=1,\ldots,N^X$, $X\in\lbrace B,C,E\rbrace$) and for scalar estimates (eigenvalues $\nu_k^X$, $k=1,\ldots,N^X$, $X\in\lbrace B,C,E\rbrace$ and the error variance $\sigma^2$) as a special case with $L=1$. For the FPC weights, \eqref{eq: rrMSE} is approximately $\sqrt{\nicefrac{\nicefrac{1}{L^X}\sum_{l=1}^{L^X} \left(\xi_{lk} - \hat{\xi}_{lk} \right)^2}{\nu_k^X}}$.\\
For all functions $\theta(t)$ (mean function with and without covariates $\mu(t)$, $f_0(t)$,\ldots, $f_7(t)$, eigenfunctions $\phi_k^X(t)$, $k=1,\ldots,N^X$, $X \in \lbrace B,C,E\rbrace$), we approximate the integrals by sums and obtain 
\bea \label{eq: rrISE}
\rrMSE(\theta(\cdot),\hat{\theta}(\cdot)) =  \sqrt{\frac{\dfrac{1}{D}\sum_{d=1}^D\left(\theta(t_d)-\hat{\theta}(t_d)\right)^2}{\dfrac{1}{D}\sum_{d=1}^D \theta(t_d)^2}}.
\eea

Note that for the eigenfunctions, the denominator simplifies to approximately one as $\int \phi_k(t)^2 \dint t= 1$. As the eigenfunctions are only unique up to sign, we mirror the estimated eigenfunctions around the x-axis, also compute the rrMSE for the mirrored estimates, and choose the smaller rrMSE. For the fRIs and for the response function, we additionally average over the respective levels. As the fRIs are centred, the denominator simplifies to the average variance. 
The rrMSE for the bivariate functions (auto-covariances) are defined analogously.
\subsection{Results for simulations with centered and decorrelated basis weights}
\begin{figure}[h!]
\begin{center}
\includegraphics[width=0.25\textwidth]{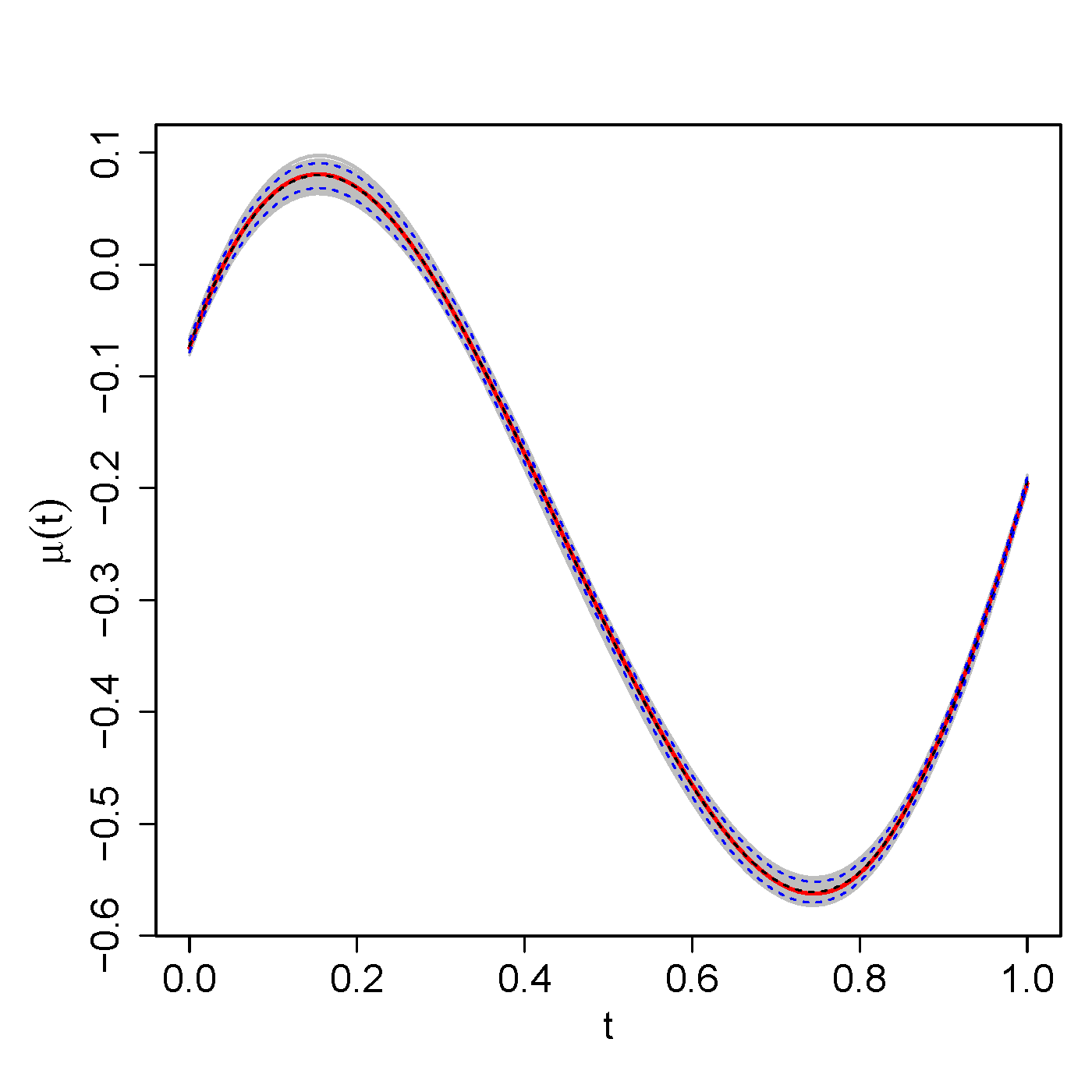}
\caption{True and estimated mean functions. Shown are the true function (red), the mean of the estimated functions over 200 simulation runs (black dashed line), the point-wise 5th and 95th percentiles of the estimated functions (blue dashed lines), and the estimated functions of all 200 simulation runs (grey).}
\label{fig: mean crossed}
\end{center}
\end{figure}

\begin{figure}[h!]
\begin{center}
\begin{minipage}{1\textwidth}
\begin{center}
\raisebox{0.15\textwidth}{\textbf{B}}
\includegraphics[width=0.25\textwidth,page=1]{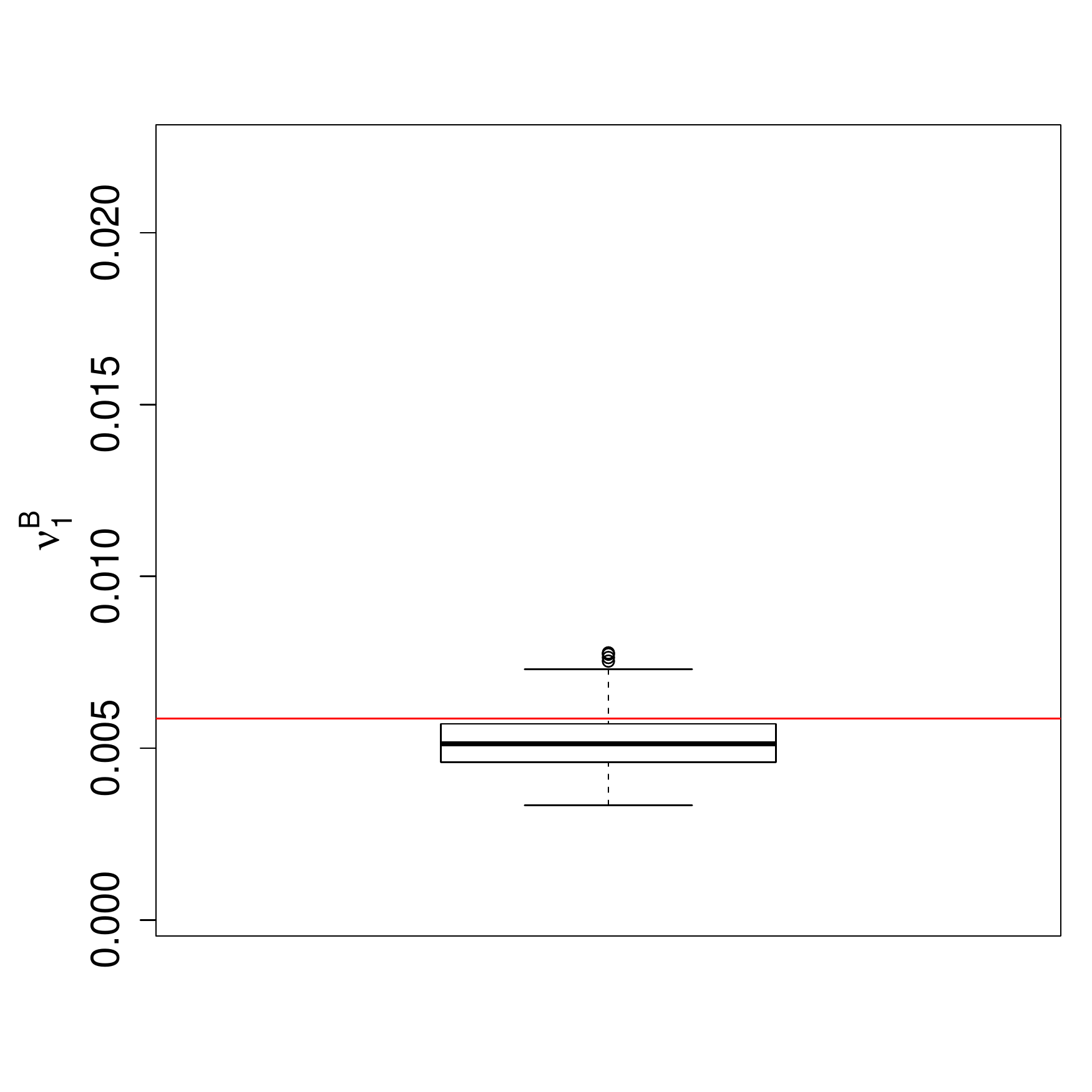}
\raisebox{0.15\textwidth}{\phantom{\textbf{B}}}
\includegraphics[width=0.25\textwidth,page=2]{figures/simulation/eigenvalues_normal_I9_J16_crossed_as_data_26_Dec_26_Dec.pdf}
\raisebox{0.15\textwidth}{\textbf{C}}
\includegraphics[width=0.25\textwidth,page=3]{figures/simulation/eigenvalues_normal_I9_J16_crossed_as_data_26_Dec_26_Dec.pdf}\\
\raisebox{0.15\textwidth}{\textbf{E}}
\includegraphics[width=0.25\textwidth,page=4]{figures/simulation/eigenvalues_normal_I9_J16_crossed_as_data_26_Dec_26_Dec.pdf}
\raisebox{0.15\textwidth}{\phantom{\textbf{B}}}
\includegraphics[width=0.25\textwidth,page=5]{figures/simulation/eigenvalues_normal_I9_J16_crossed_as_data_26_Dec_26_Dec.pdf}
\raisebox{0.15\textwidth}{\phantom{\textbf{B}}}
\includegraphics[width=0.25\textwidth,page=6]{figures/simulation/eigenvalues_normal_I9_J16_crossed_as_data_26_Dec_26_Dec.pdf}
\end{center}
\end{minipage}
\caption{Boxplots of the estimated eigenvalues of the auto-covariances of the crossed fRIs (top row), as well as the eigenvalues of the auto-covariance of the smooth error (bottom row) for all 200 simulations runs.}
\label{fig: boxplot eigenvalues crossed}
\end{center}
\end{figure}

\begin{figure}[h!]
\centering
\includegraphics[width=0.25\textwidth,page=1]{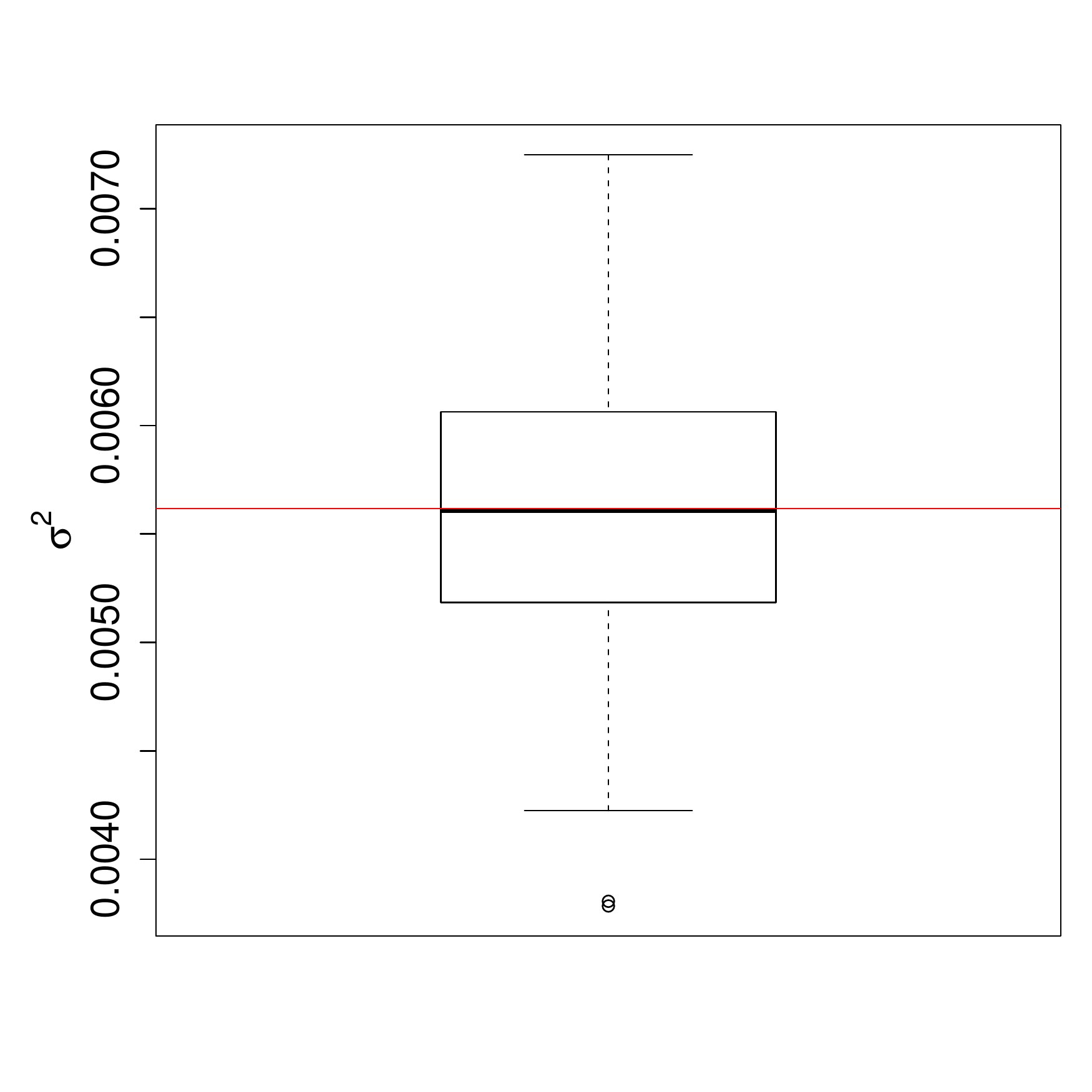}
\caption{Boxplots of the estimated error variances $\sigma^2$ for all 200 simulation runs.}
\label{fig: boxplot sigmasq crossed}
\end{figure}

\underline{Simulation results for the crossed-fRIs scenario}.
In the following, additional simulation results for the application-based crossed-fRIs scenario with centred and decorrelated basis weights are shown. Figure \ref{fig: mean crossed} shows the true and estimated mean functions. 
In Figure \ref{fig: boxplot eigenvalues crossed}, the boxplots of the estimated eigenvalues for each $B_i(t)$, $C_j(t)$, and $E_{ijh}(t)$ are depicted.
In Figure \ref{fig: boxplot sigmasq crossed}, we show the boxplot of the estimated error variances. 

\clearpage

\begin{figure}[h!]
\begin{center}
\includegraphics[width=0.2\textwidth,page=1]{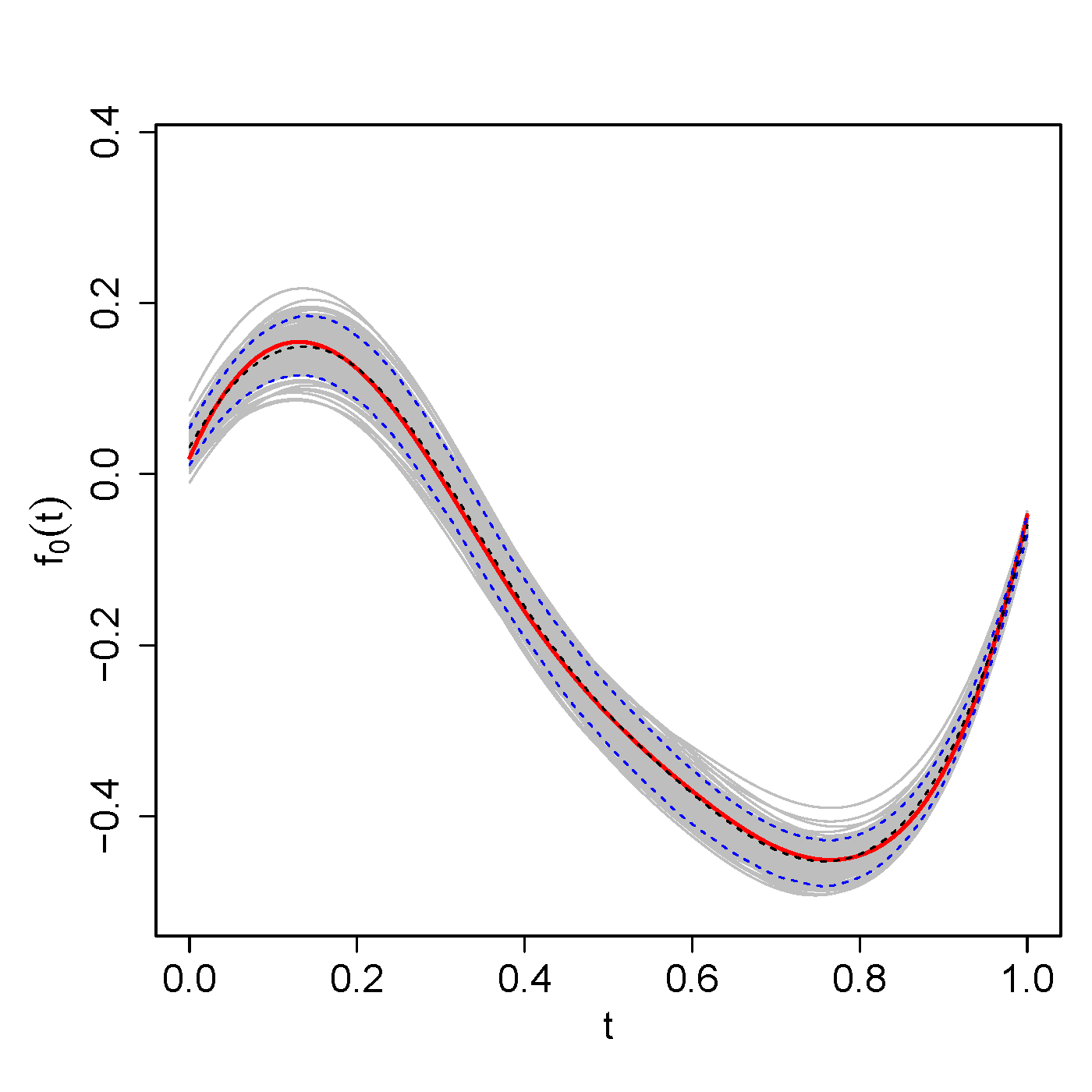}
\includegraphics[width=0.2\textwidth,page=3]{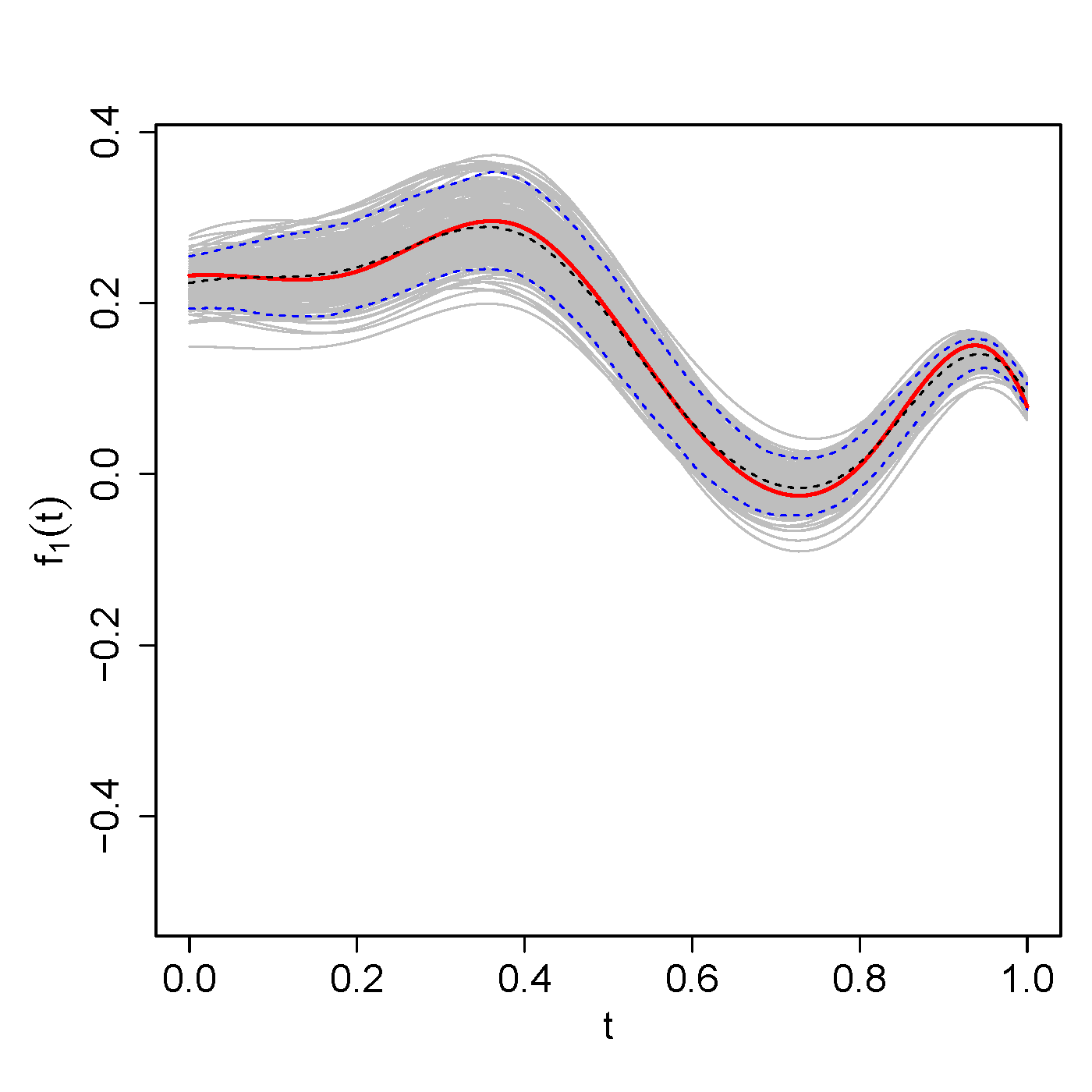}
\includegraphics[width=0.2\textwidth,page=5]{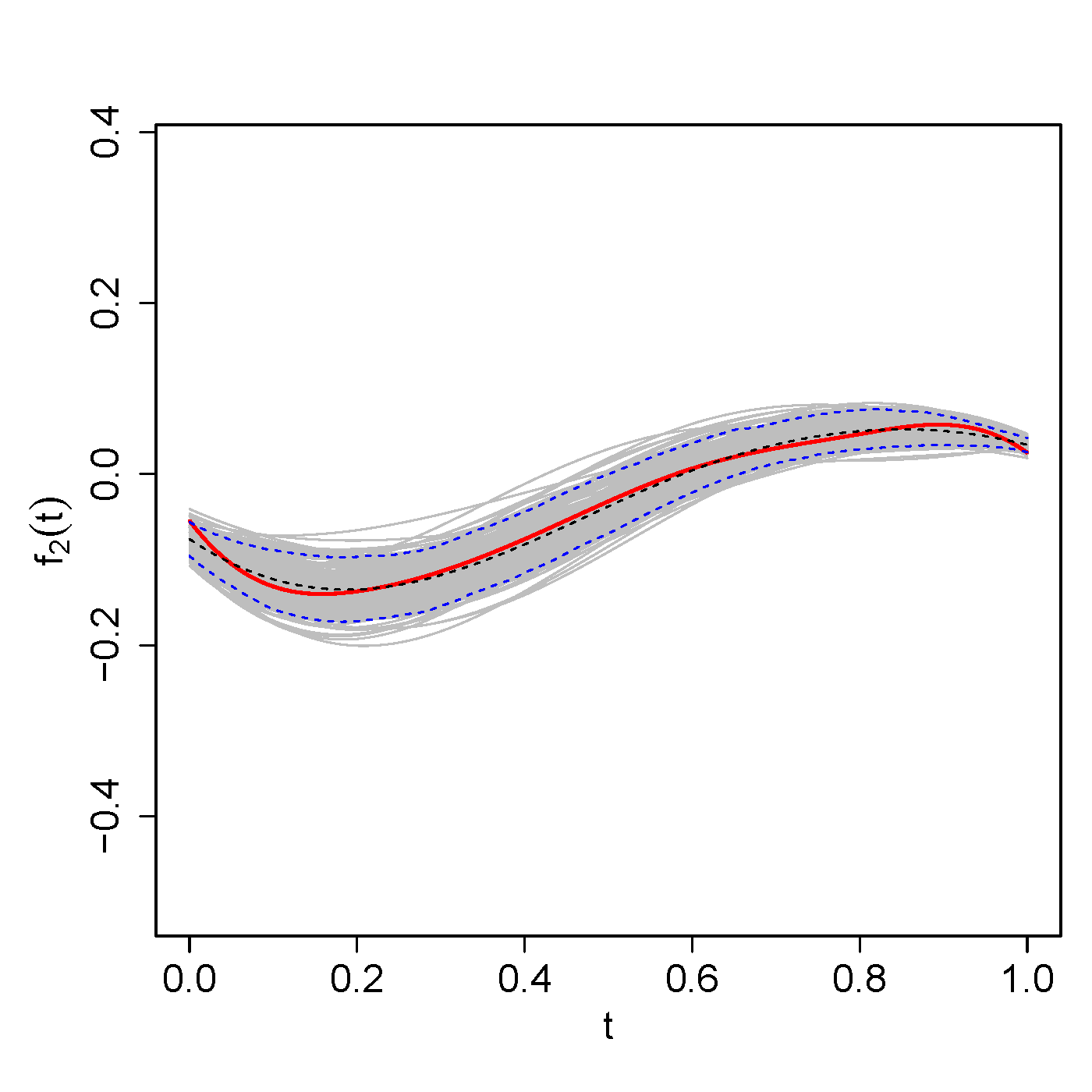}
\includegraphics[width=0.2\textwidth,page=7]{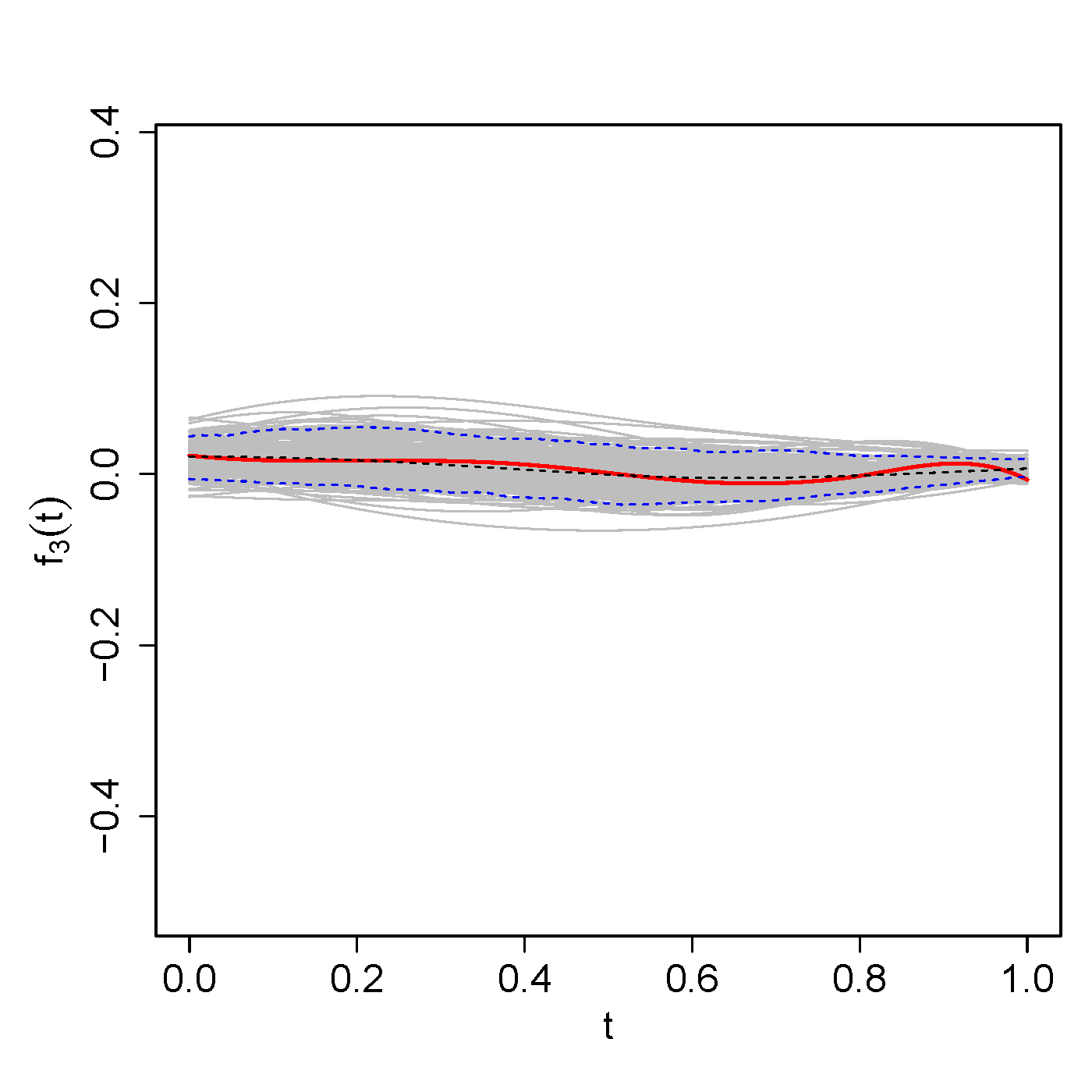}\\
\includegraphics[width=0.2\textwidth,page=9]{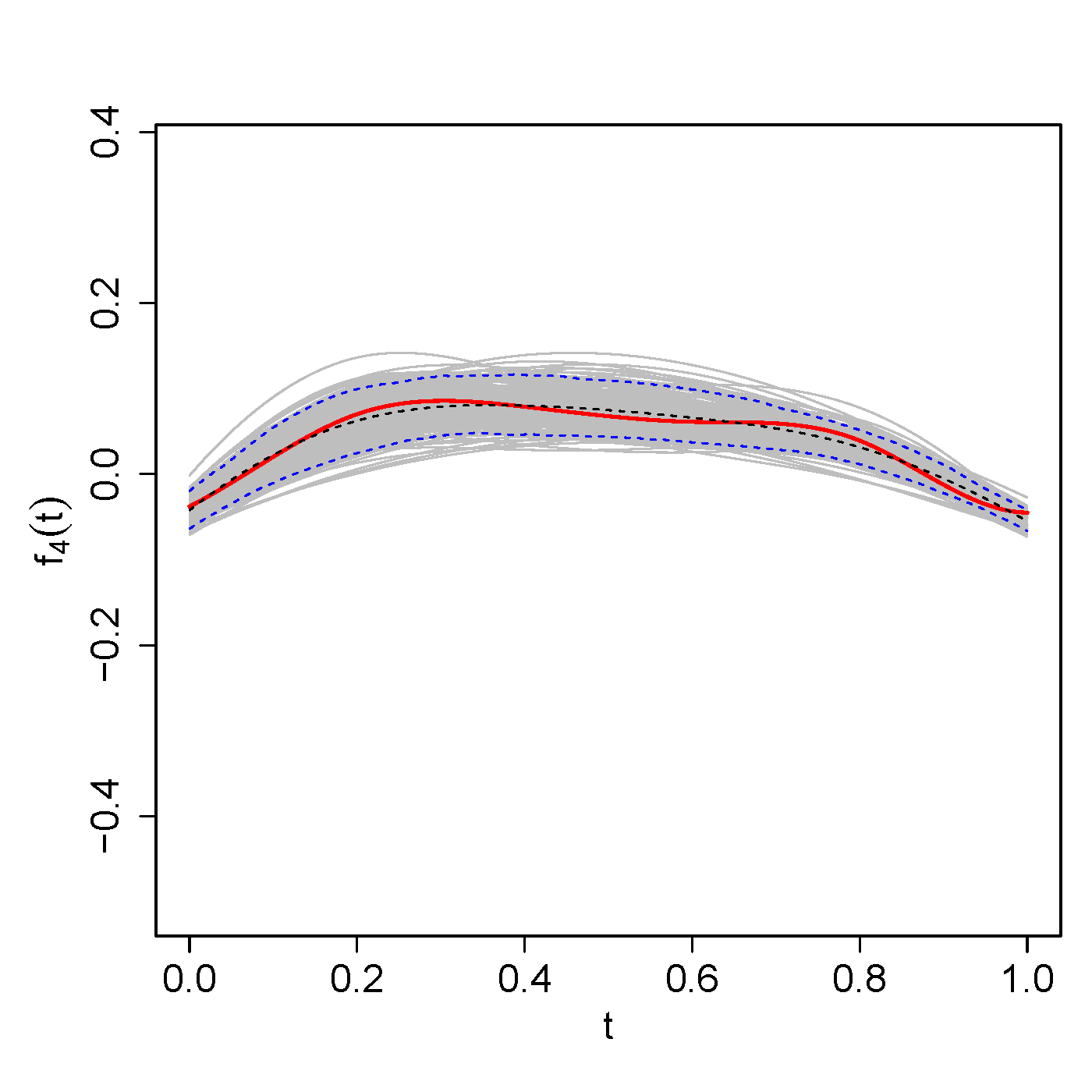}
\includegraphics[width=0.2\textwidth,page=11]{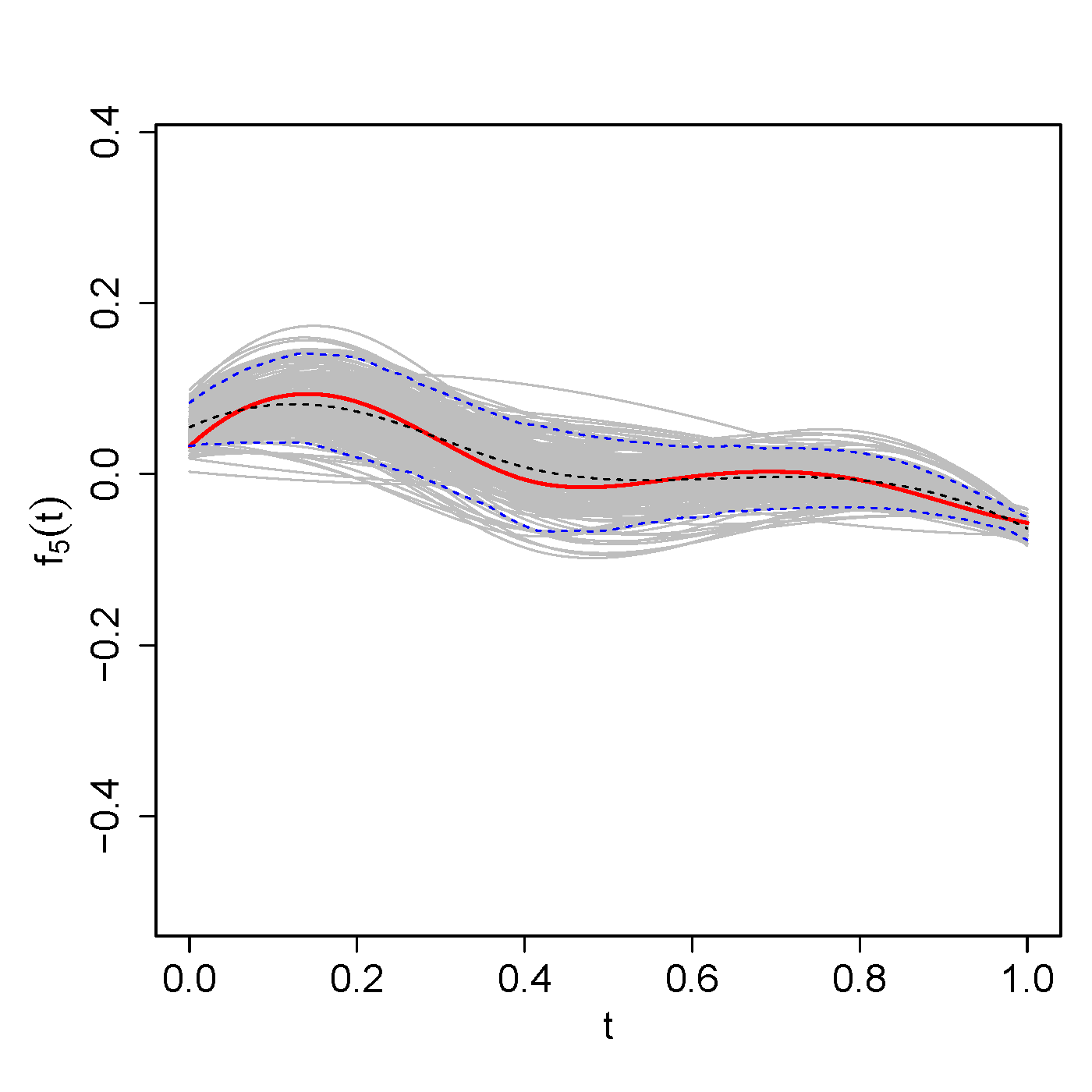}
\includegraphics[width=0.2\textwidth,page=13]{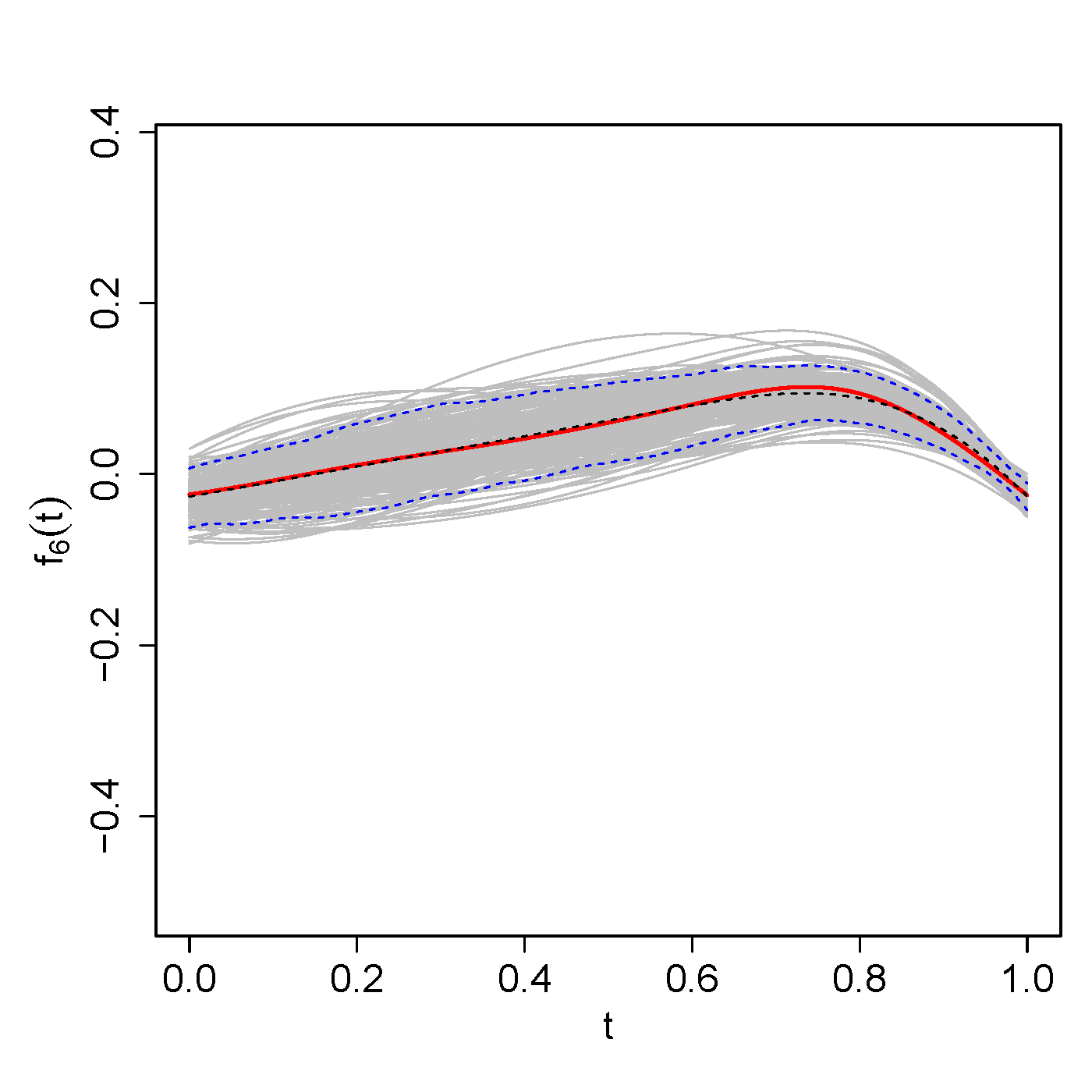}
\includegraphics[width=0.2\textwidth,page=15]{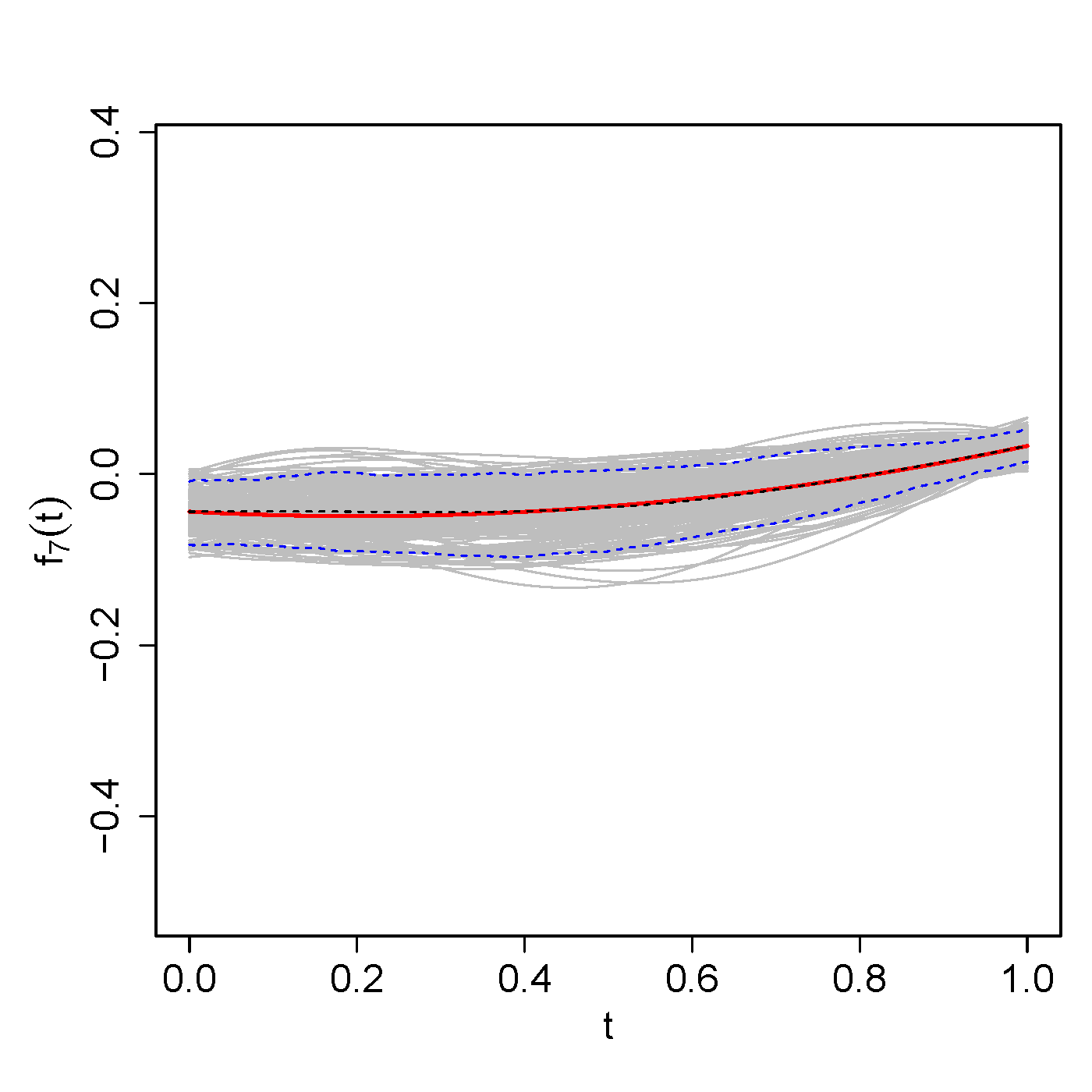}
\caption{True and estimated covariate and interaction effects estimated using the independence assumption. Shown are the true function (red), the mean of the estimated functions over 200 simulation runs (black dashed line), the point-wise 5th and 95th percentiles of the estimated functions (blue dashed lines), and the estimated functions of all 200 simulation runs (grey).}
\label{fig: mean RI}
\end{center}
\end{figure}

\begin{figure}[h]
\begin{center}
\includegraphics[width=0.2\textwidth,page=1]{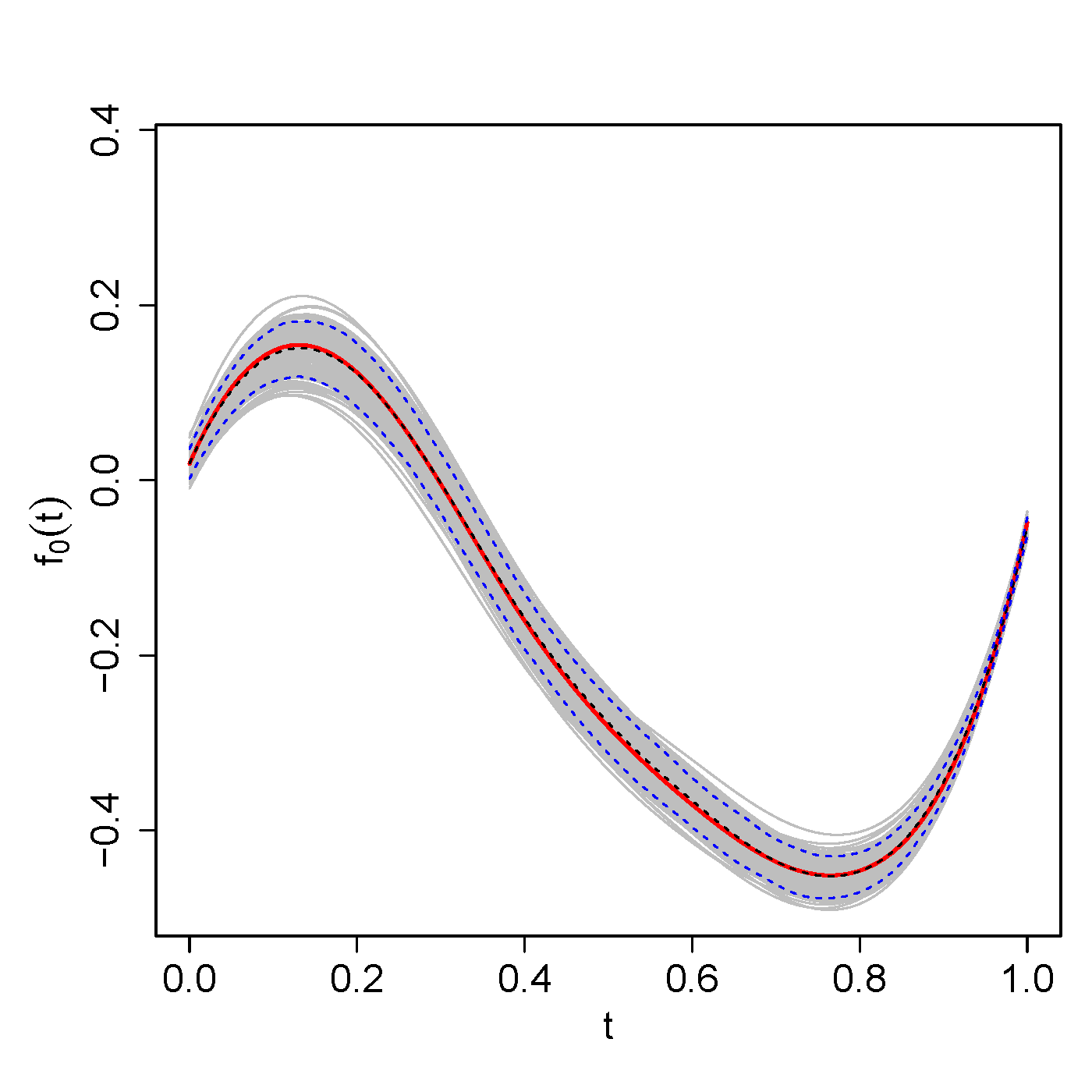}
\includegraphics[width=0.2\textwidth,page=3]{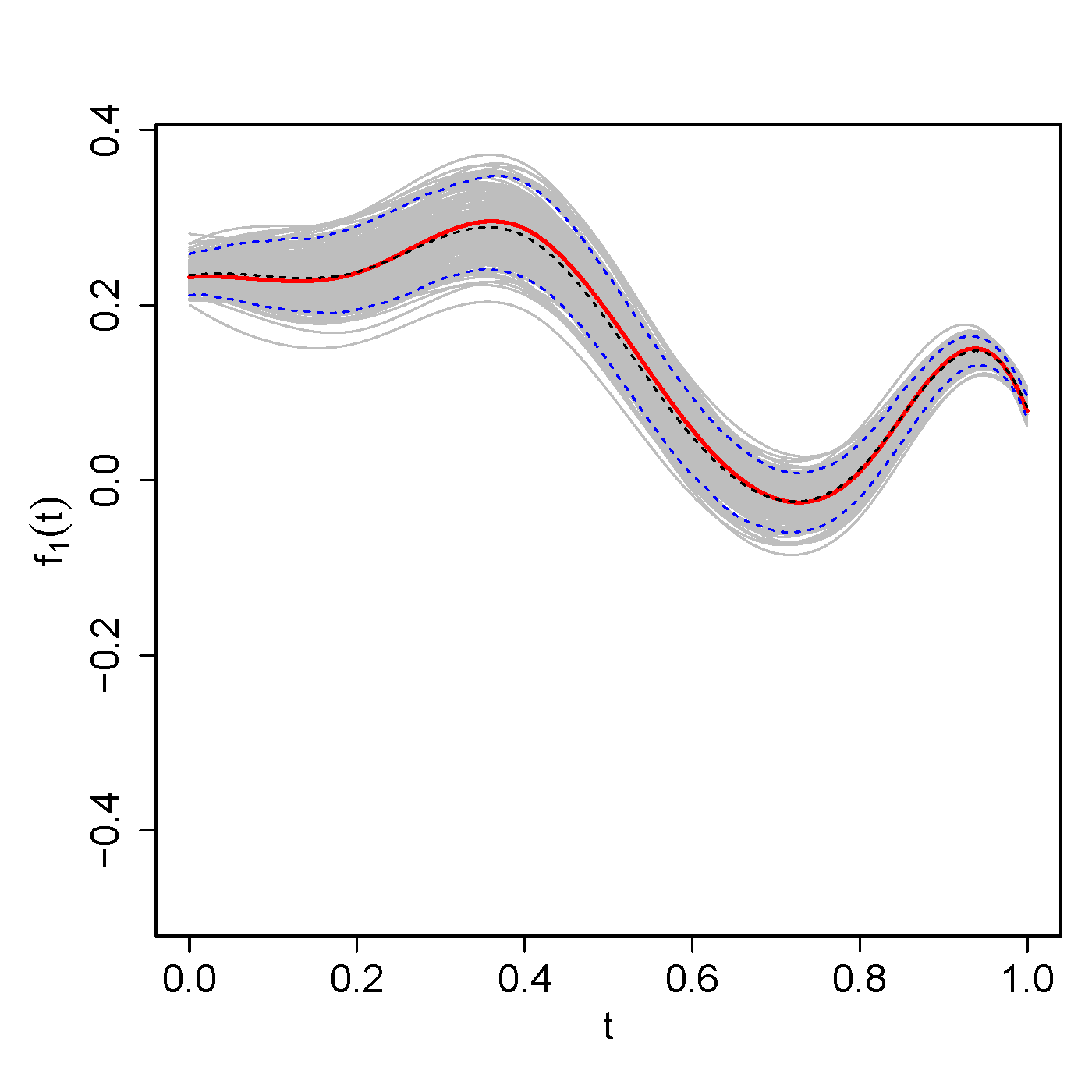}
\includegraphics[width=0.2\textwidth,page=5]{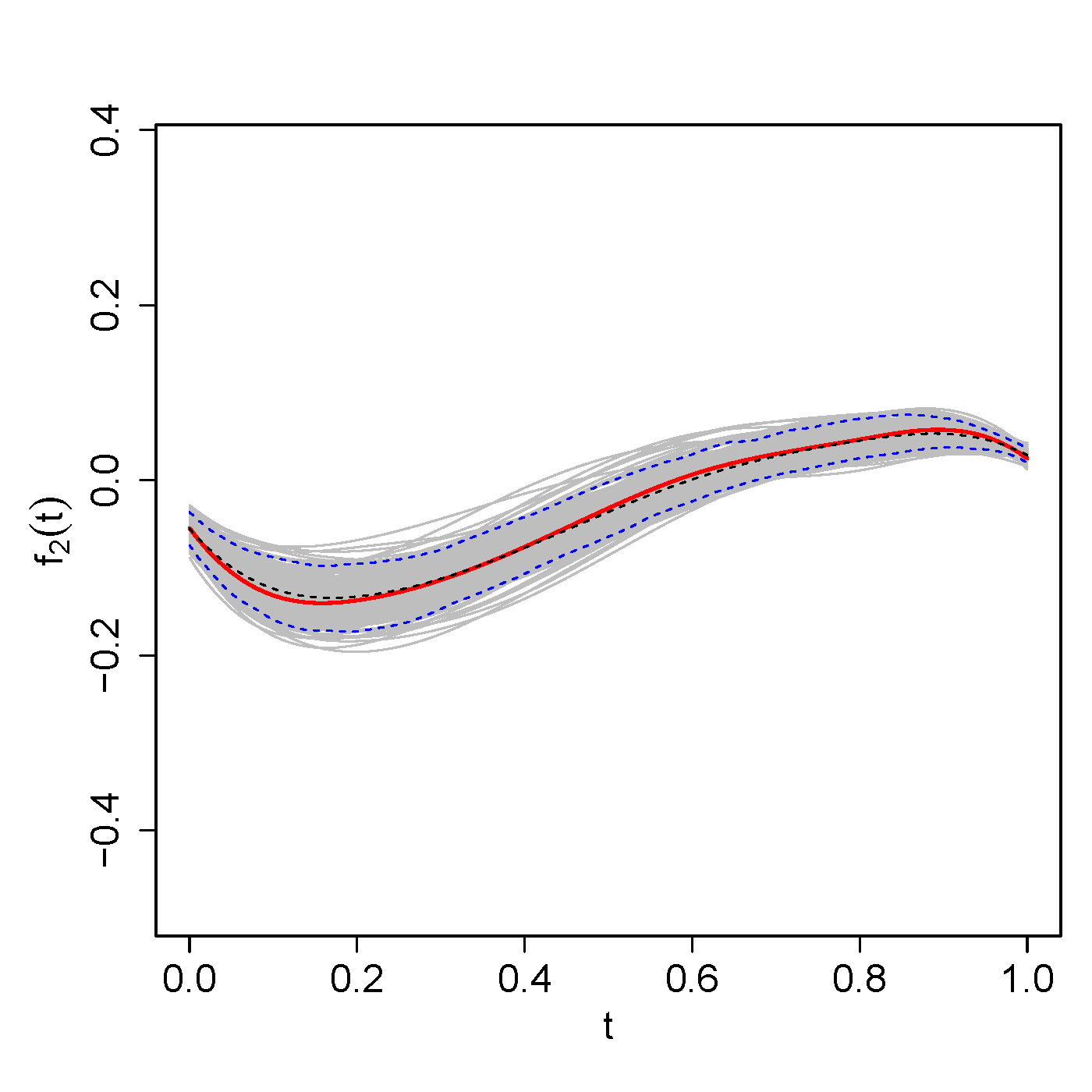}
\includegraphics[width=0.2\textwidth,page=7]{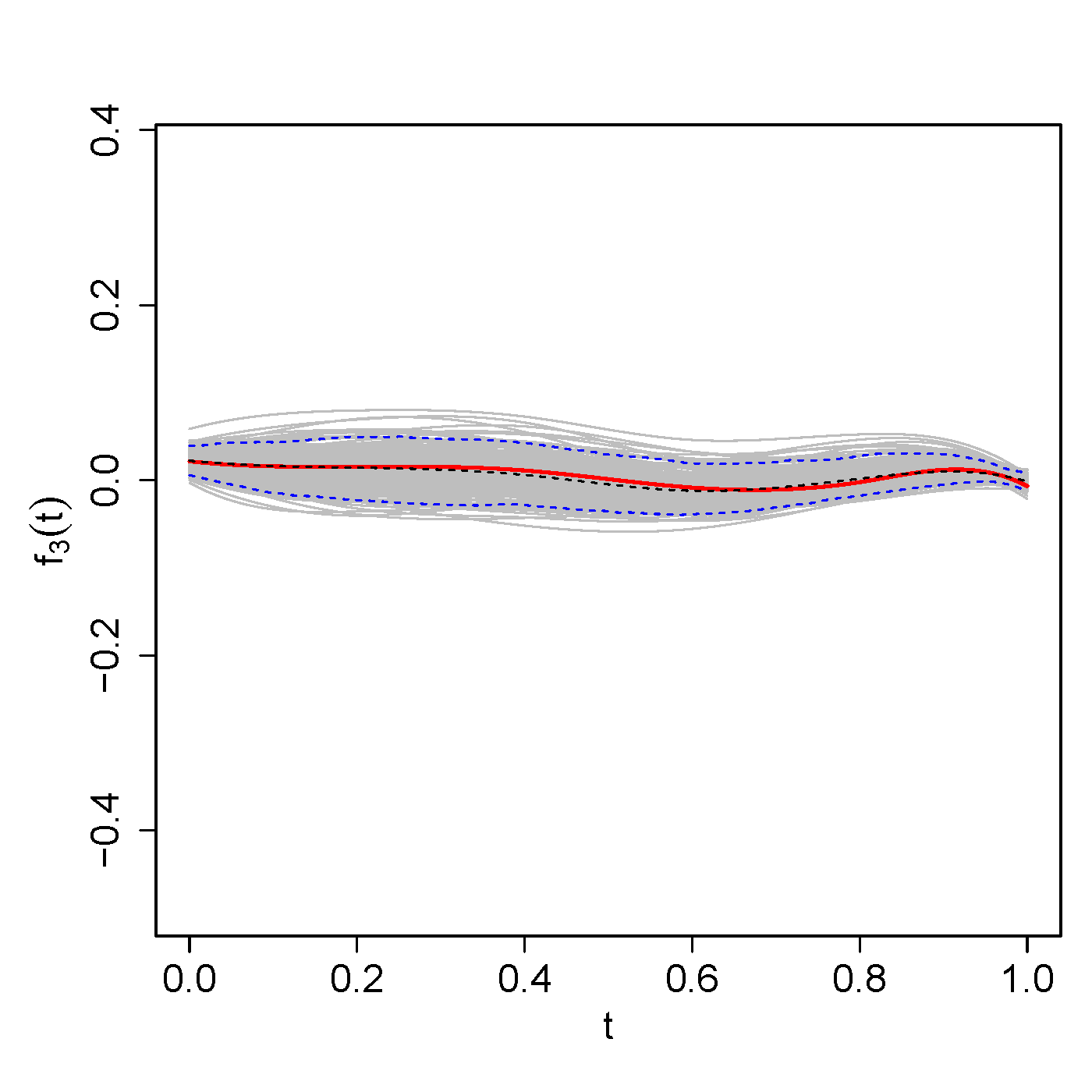}\\
\includegraphics[width=0.2\textwidth,page=9]{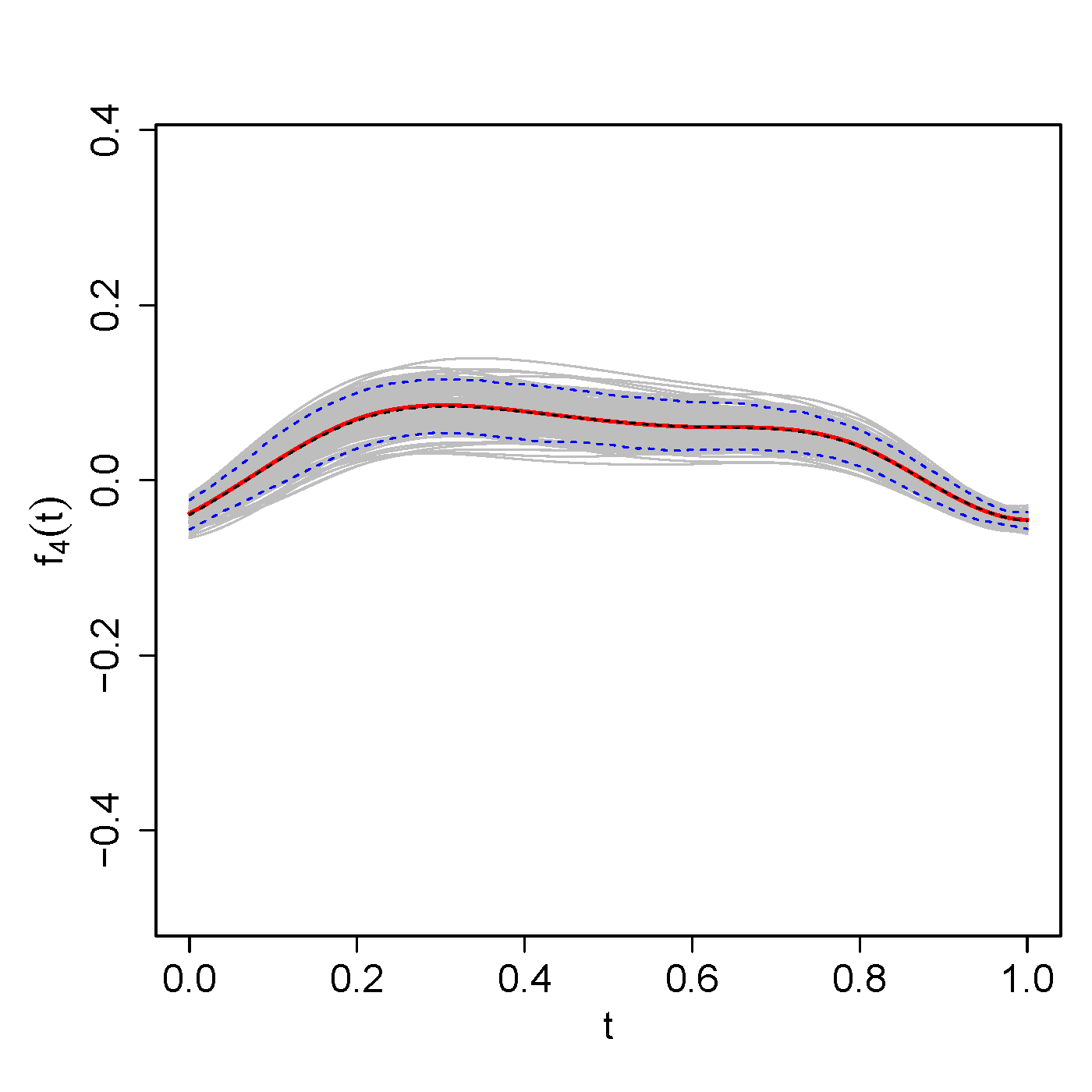}
\includegraphics[width=0.2\textwidth,page=11]{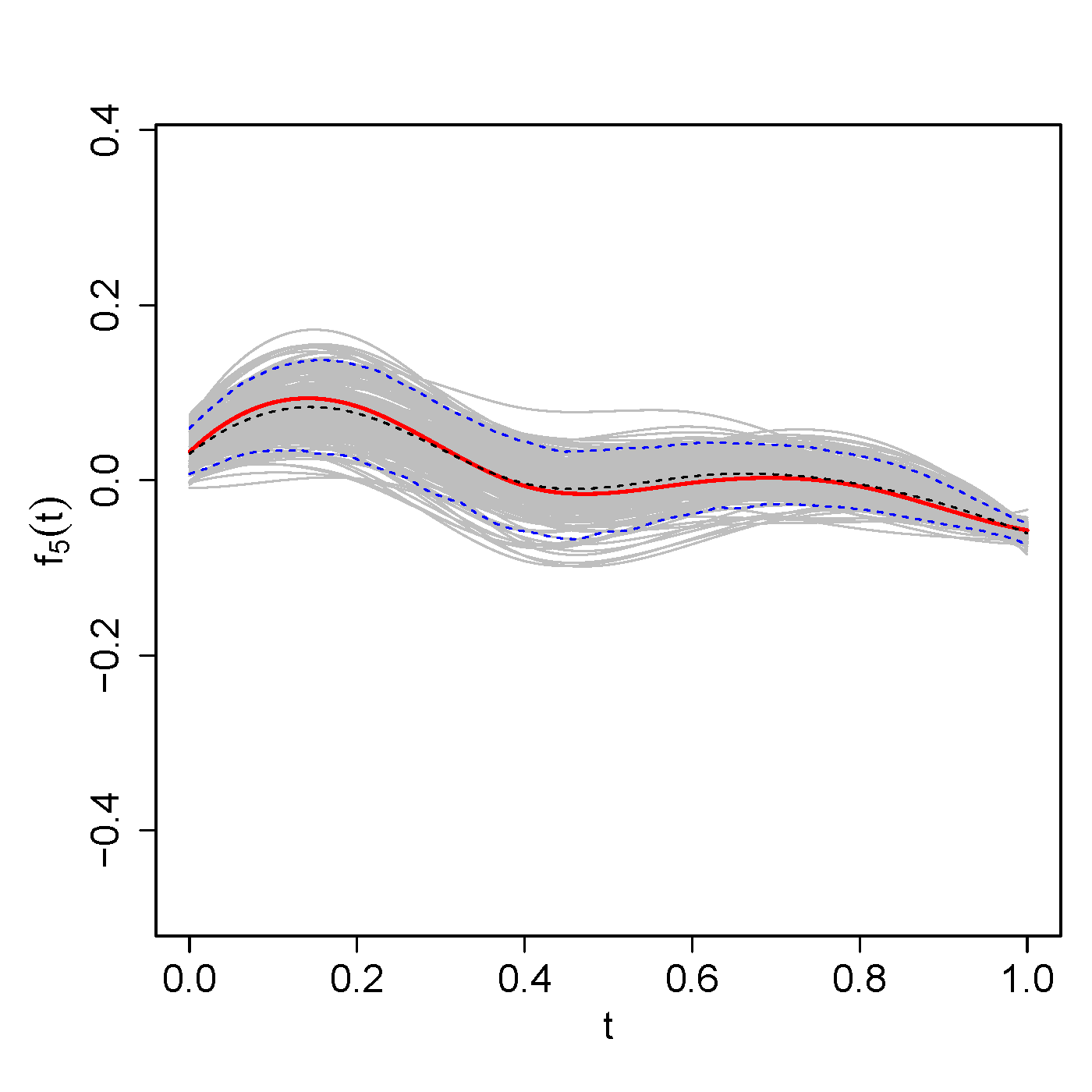}
\includegraphics[width=0.2\textwidth,page=13]{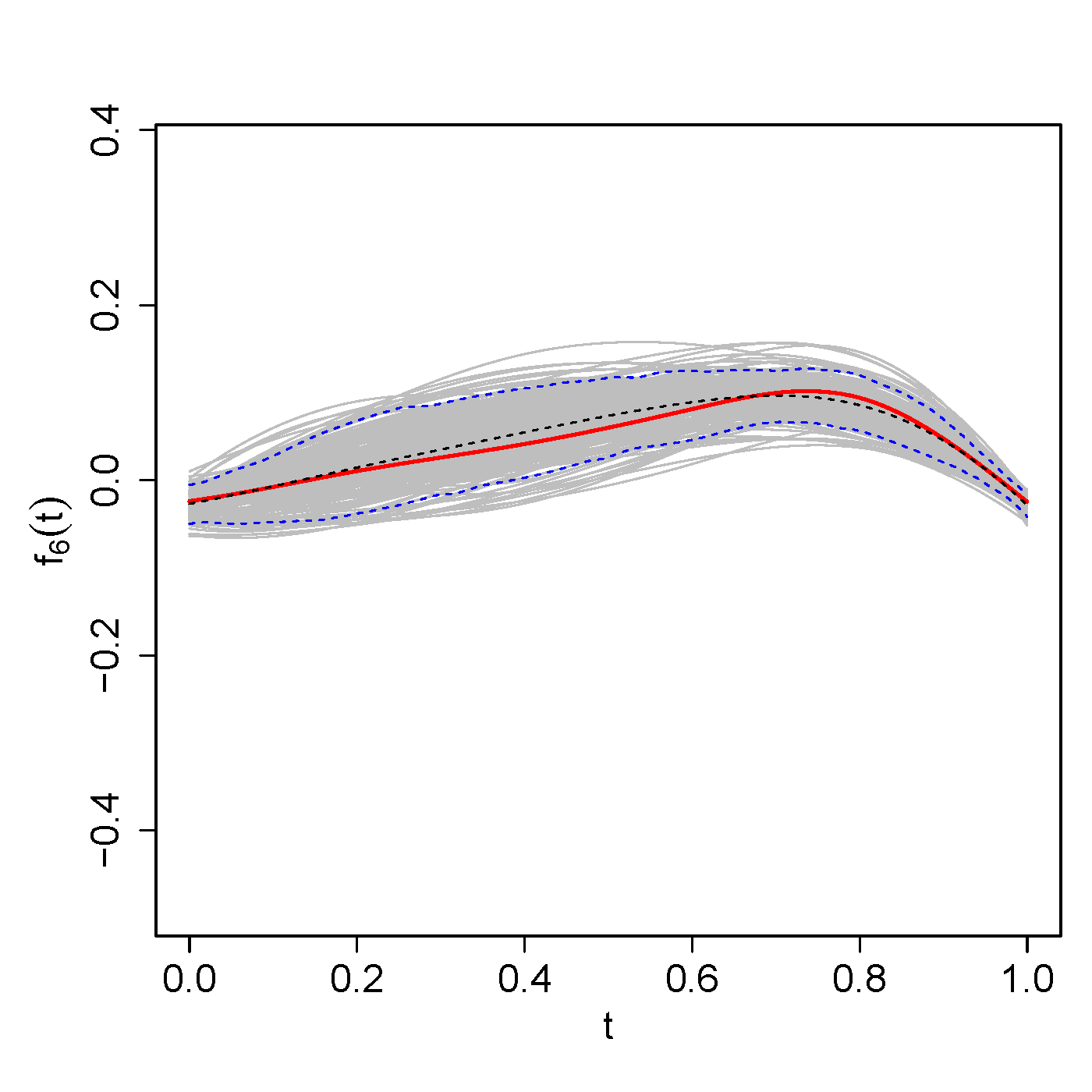}
\includegraphics[width=0.2\textwidth,page=15]{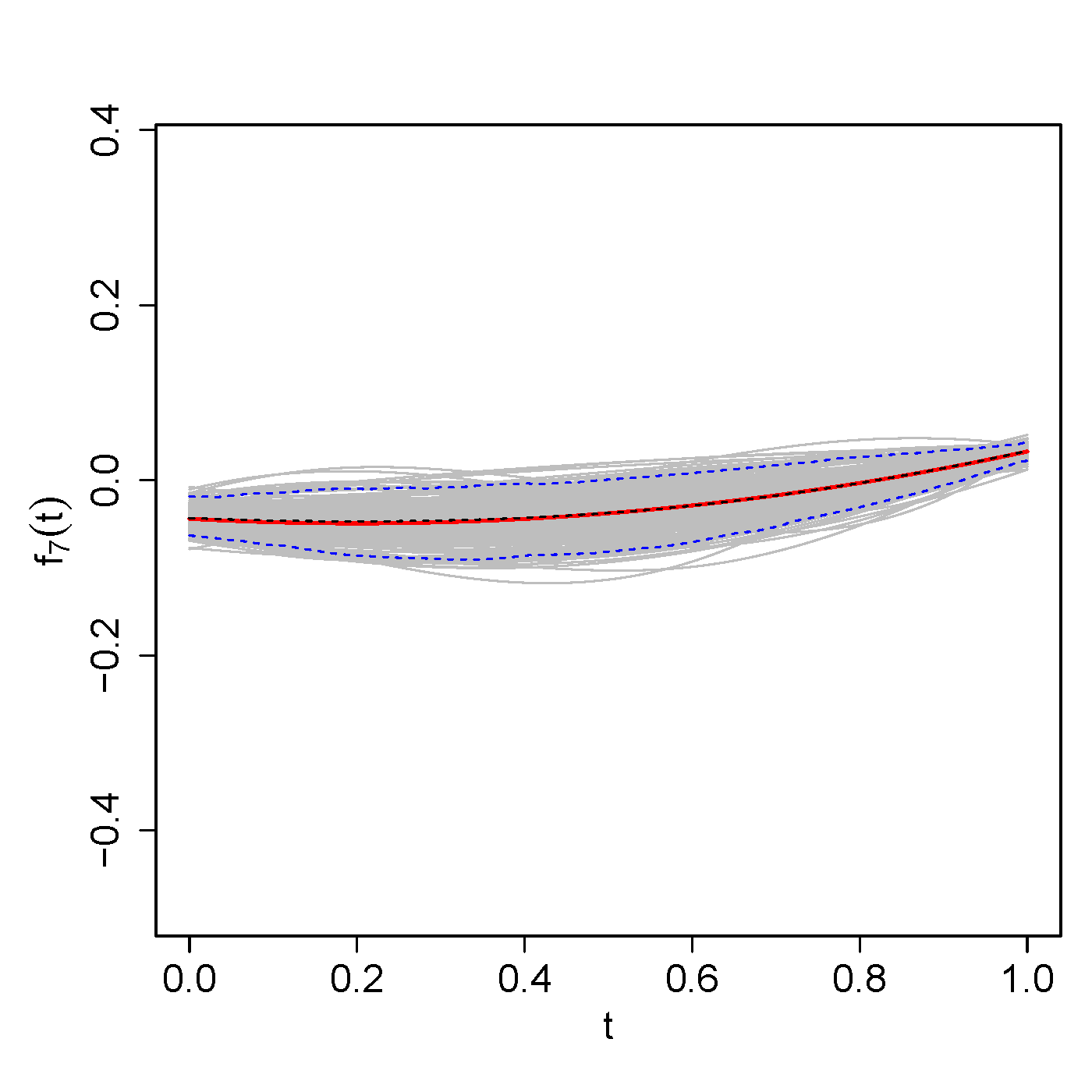}
\caption{True and estimated covariate and interaction effects estimated using FPC-FAMM. Shown are the true function (red), the mean of the estimated functions over 200 simulation runs (black dashed line), the point-wise 5th and 95th percentiles of the estimated functions (blue dashed lines), and the estimated functions of all 200 simulation runs (grey).}
\label{fig: mean RI pffr}
\end{center}
\end{figure}

\begin{figure}[h!]
\begin{center}
\includegraphics[width=0.2\textwidth,page=1]{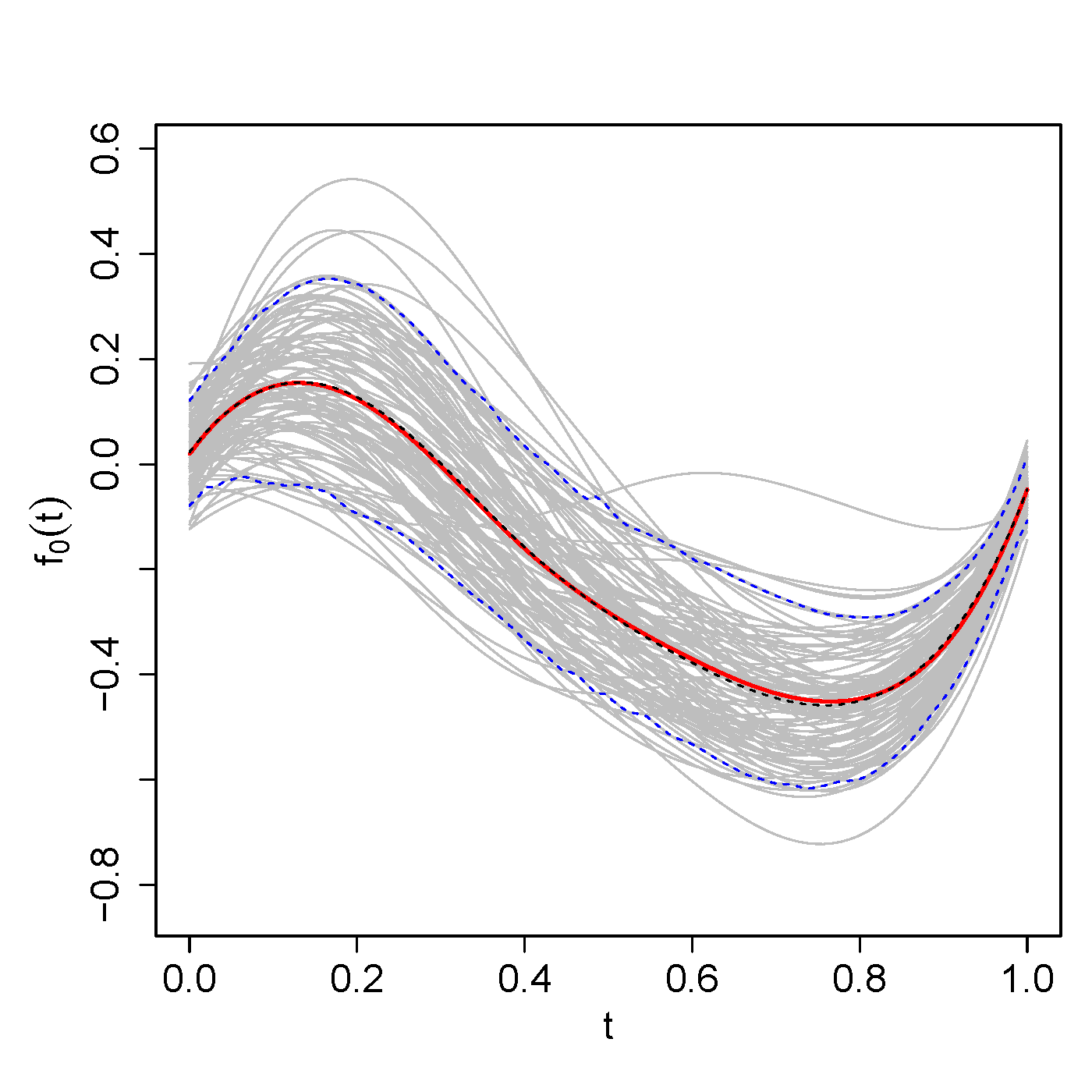}
\includegraphics[width=0.2\textwidth,page=3]{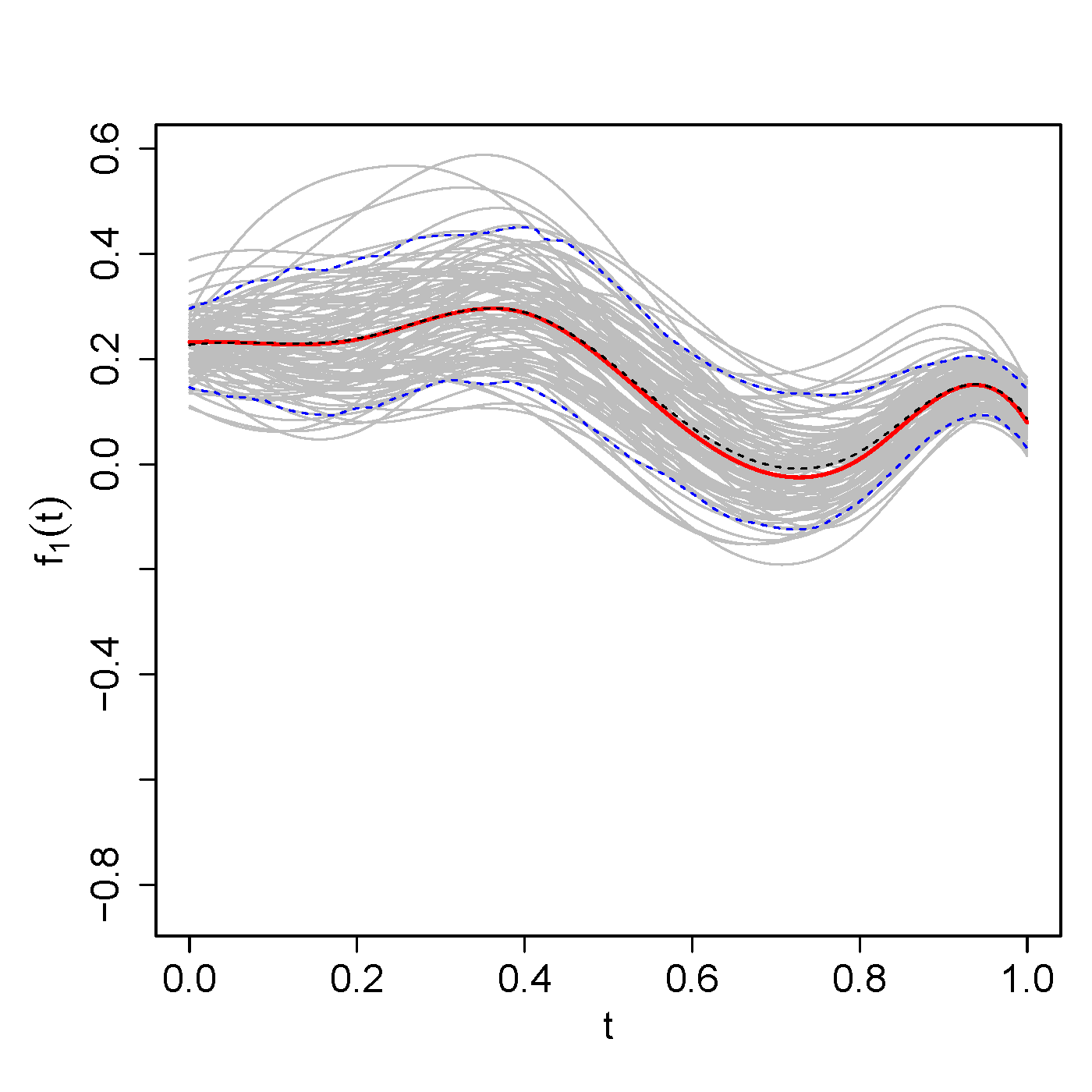}
\includegraphics[width=0.2\textwidth,page=5]{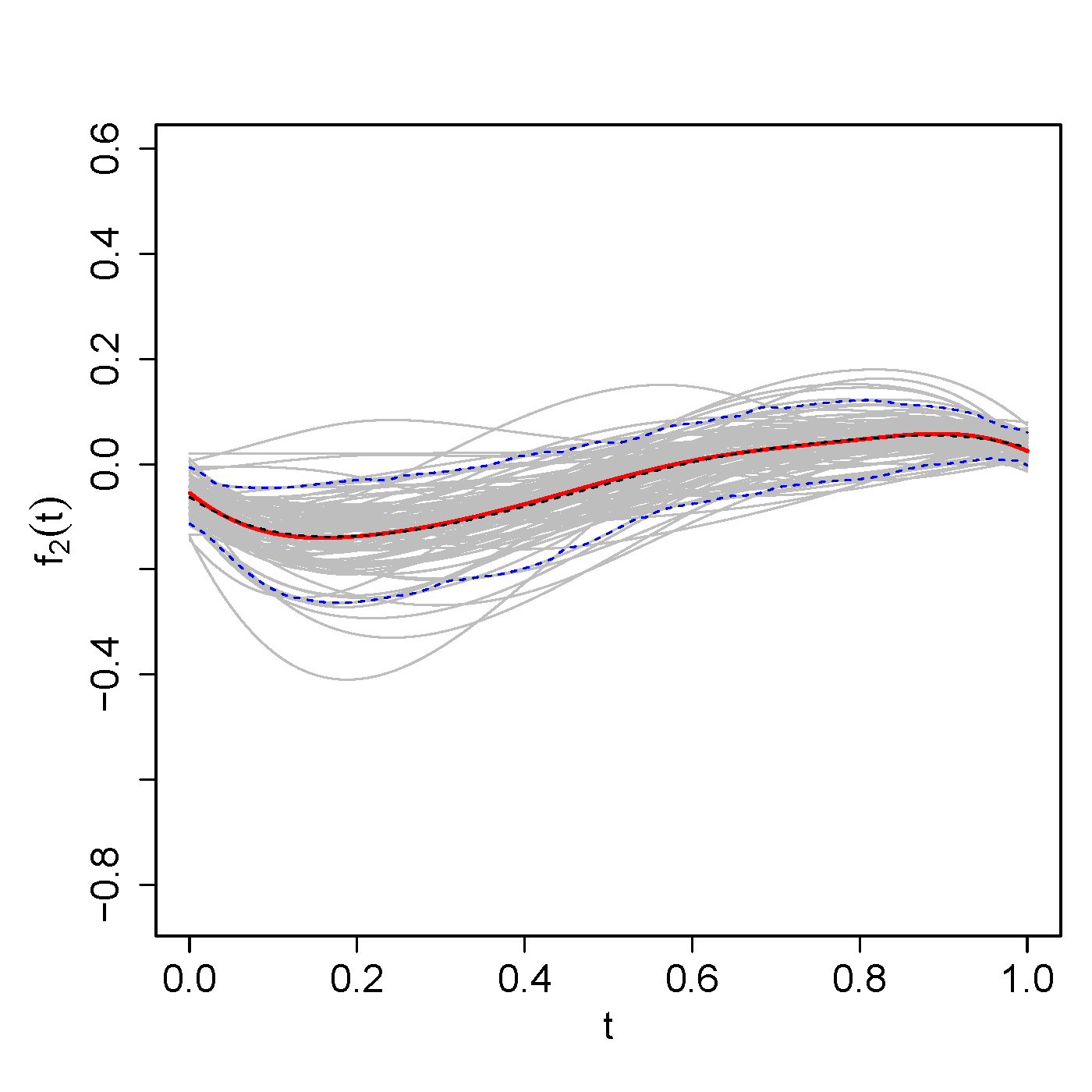}
\includegraphics[width=0.2\textwidth,page=7]{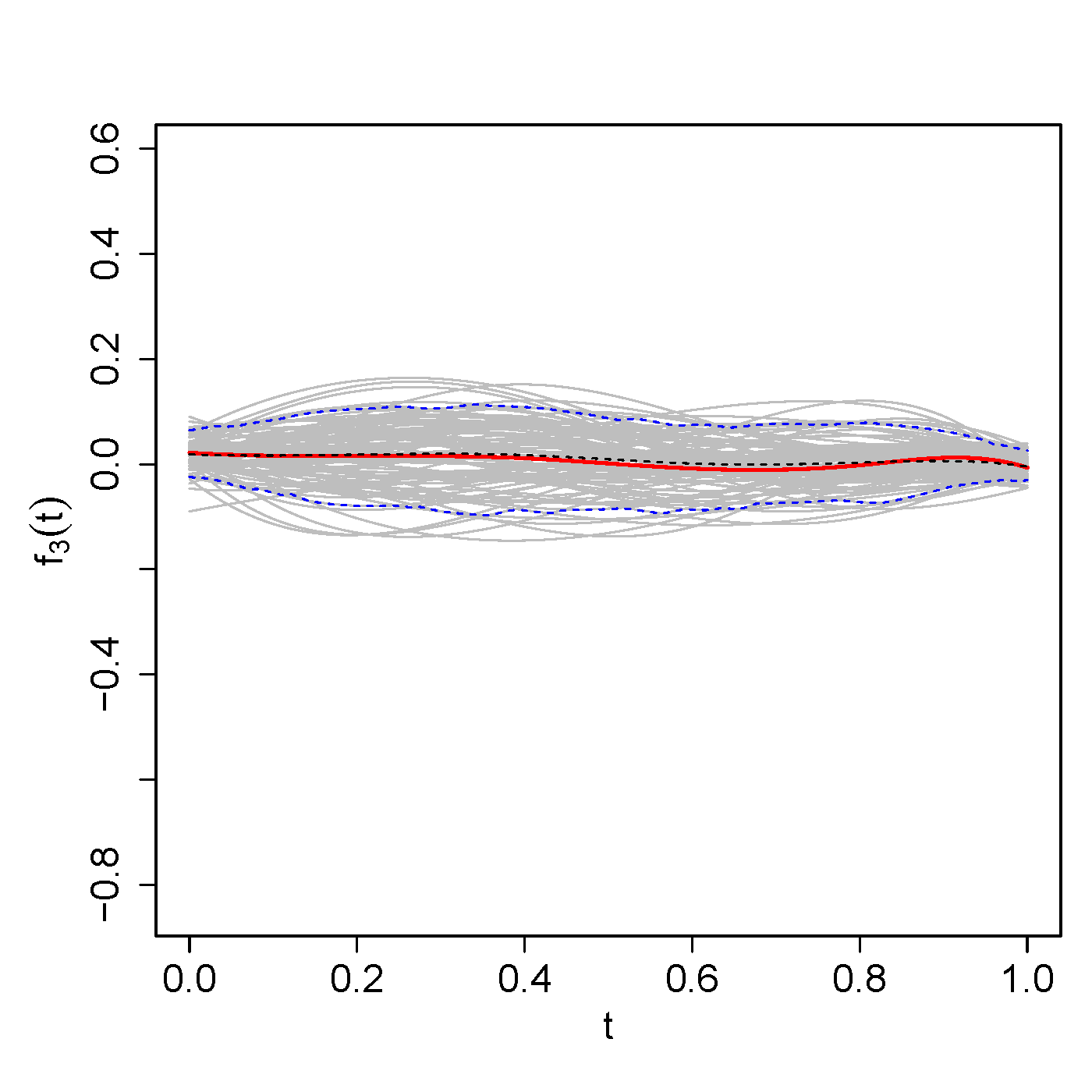}\\
\includegraphics[width=0.2\textwidth,page=9]{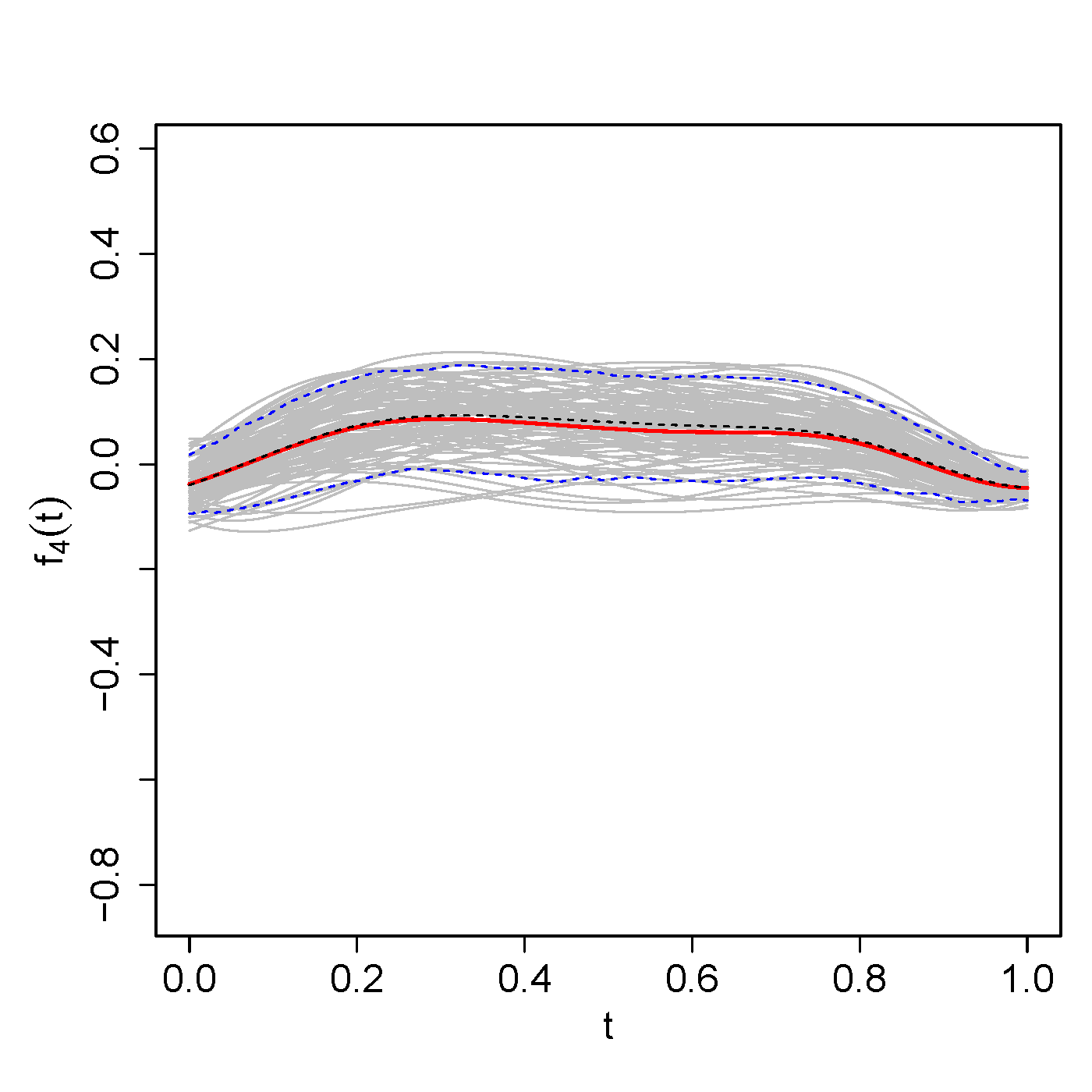}
\includegraphics[width=0.2\textwidth,page=11]{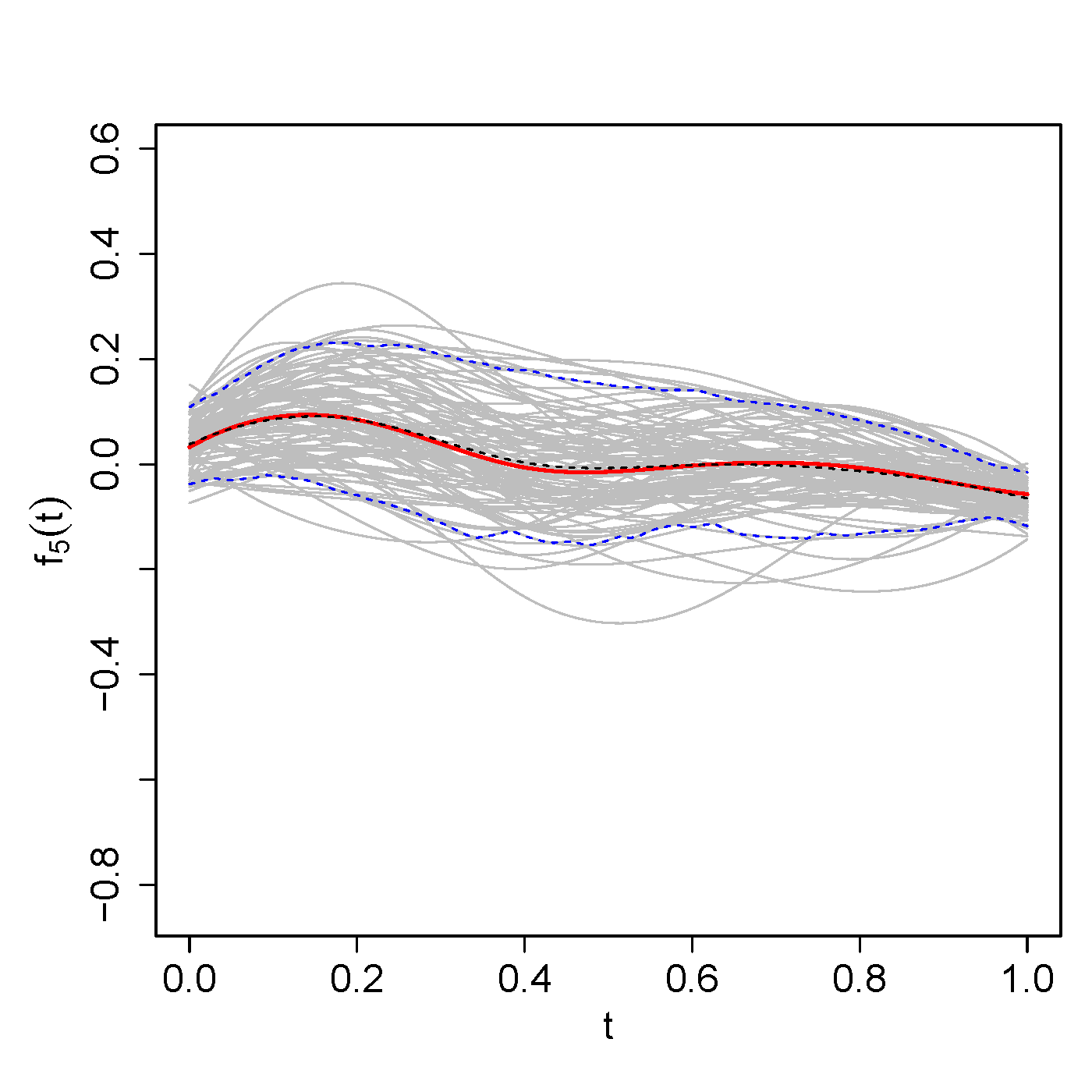}
\includegraphics[width=0.2\textwidth,page=13]{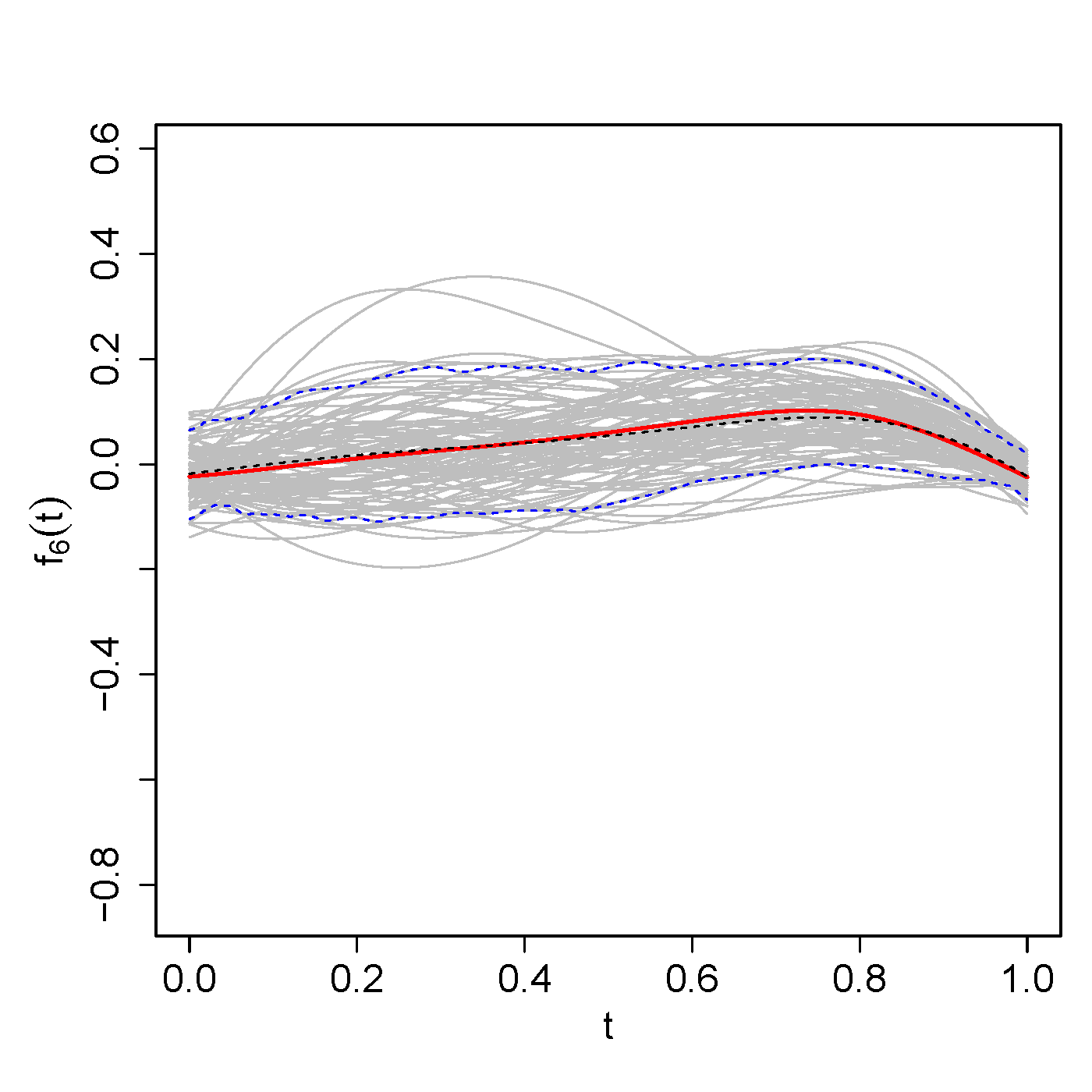}
\includegraphics[width=0.2\textwidth,page=15]{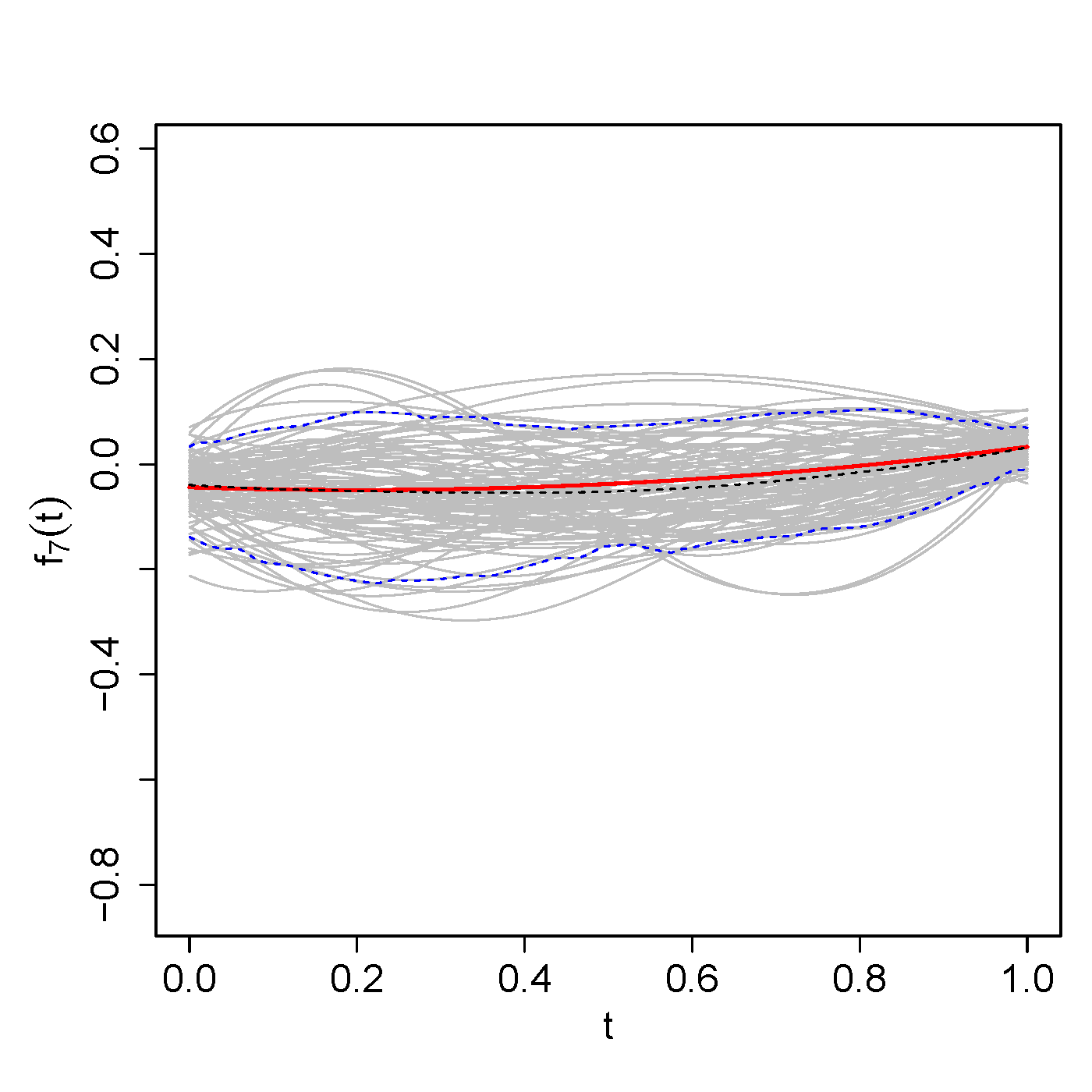}
\caption{True and estimated covariate and interaction effects estimated using spline-FAMM. Shown are the true function (red), the mean of the estimated functions over 100 simulation runs (black dashed line), the point-wise 5th and 95th percentiles of the estimated functions (blue dashed lines), and the estimated functions of all 100 simulation runs (grey).}
\label{fig: mean RI FAMM}
\end{center}
\end{figure}

\begin{figure}[h!]
\begin{center}
\includegraphics[width=0.2\textwidth,page=1]{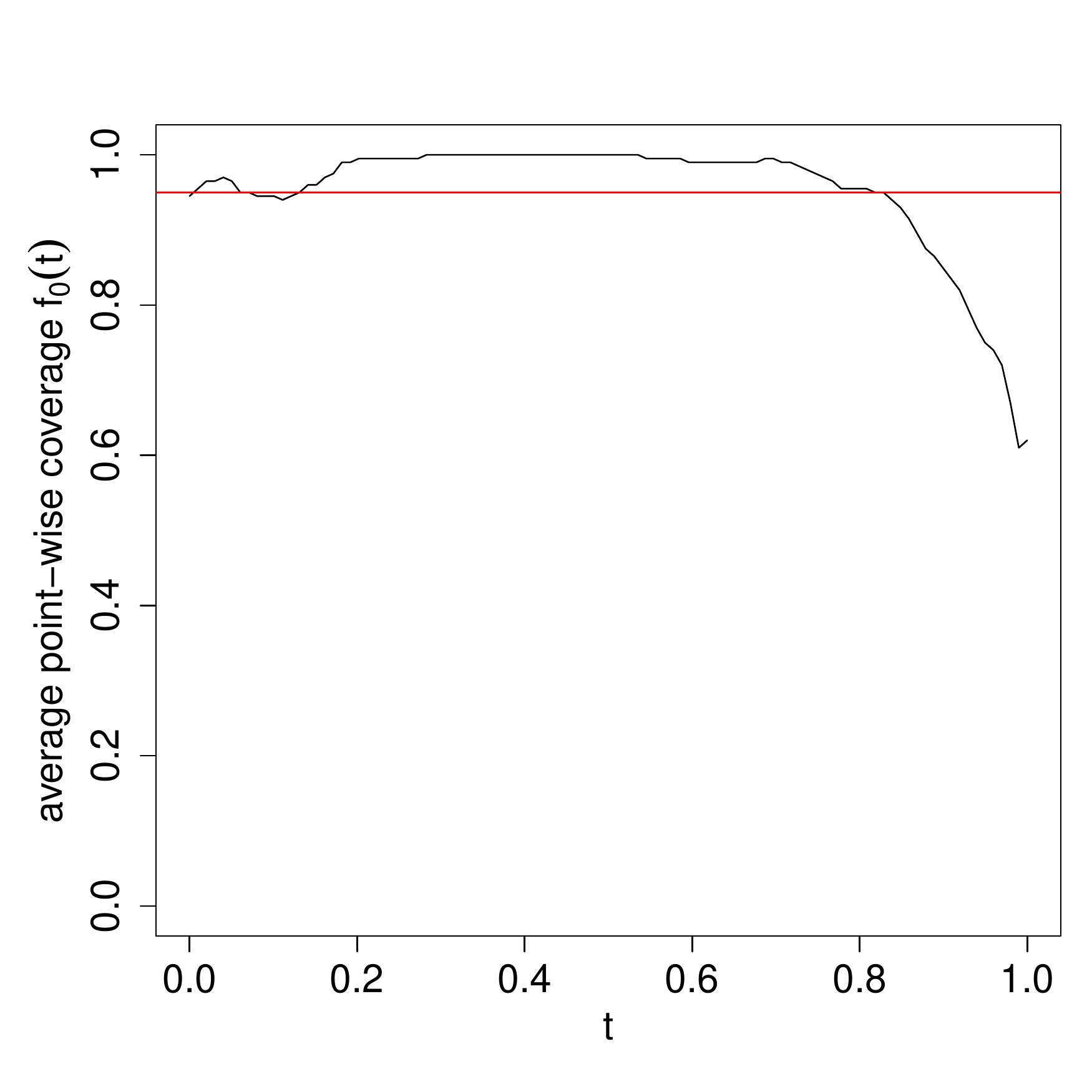}
\includegraphics[width=0.2\textwidth,page=2]{figures/simulation/covarage_mean_and_covariates_pffr_normal_I9_JNA_RI_as_data_24_Mar_24_Mar.pdf}
\includegraphics[width=0.2\textwidth,page=3]{figures/simulation/covarage_mean_and_covariates_pffr_normal_I9_JNA_RI_as_data_24_Mar_24_Mar.pdf}
\includegraphics[width=0.2\textwidth,page=4]{figures/simulation/covarage_mean_and_covariates_pffr_normal_I9_JNA_RI_as_data_24_Mar_24_Mar.pdf}\\
\includegraphics[width=0.2\textwidth,page=5]{figures/simulation/covarage_mean_and_covariates_pffr_normal_I9_JNA_RI_as_data_24_Mar_24_Mar.pdf}
\includegraphics[width=0.2\textwidth,page=6]{figures/simulation/covarage_mean_and_covariates_pffr_normal_I9_JNA_RI_as_data_24_Mar_24_Mar.pdf}
\includegraphics[width=0.2\textwidth,page=7]{figures/simulation/covarage_mean_and_covariates_pffr_normal_I9_JNA_RI_as_data_24_Mar_24_Mar.pdf}
\includegraphics[width=0.2\textwidth,page=8]{figures/simulation/covarage_mean_and_covariates_pffr_normal_I9_JNA_RI_as_data_24_Mar_24_Mar.pdf}
\caption{Average point-wise coverage of the point-wise CBs obtained by FPC-FAMM for all covariate and interaction effects. For each effect, the point-wise coverage averaged over 200 simulation runs (black line) is shown. The red line indicates the nominal value of 0.95.}
\label{fig: coverage RI pffr}
\end{center}
\end{figure}

\begin{figure}[h!]
\begin{center}
\includegraphics[width=0.2\textwidth,page=1]{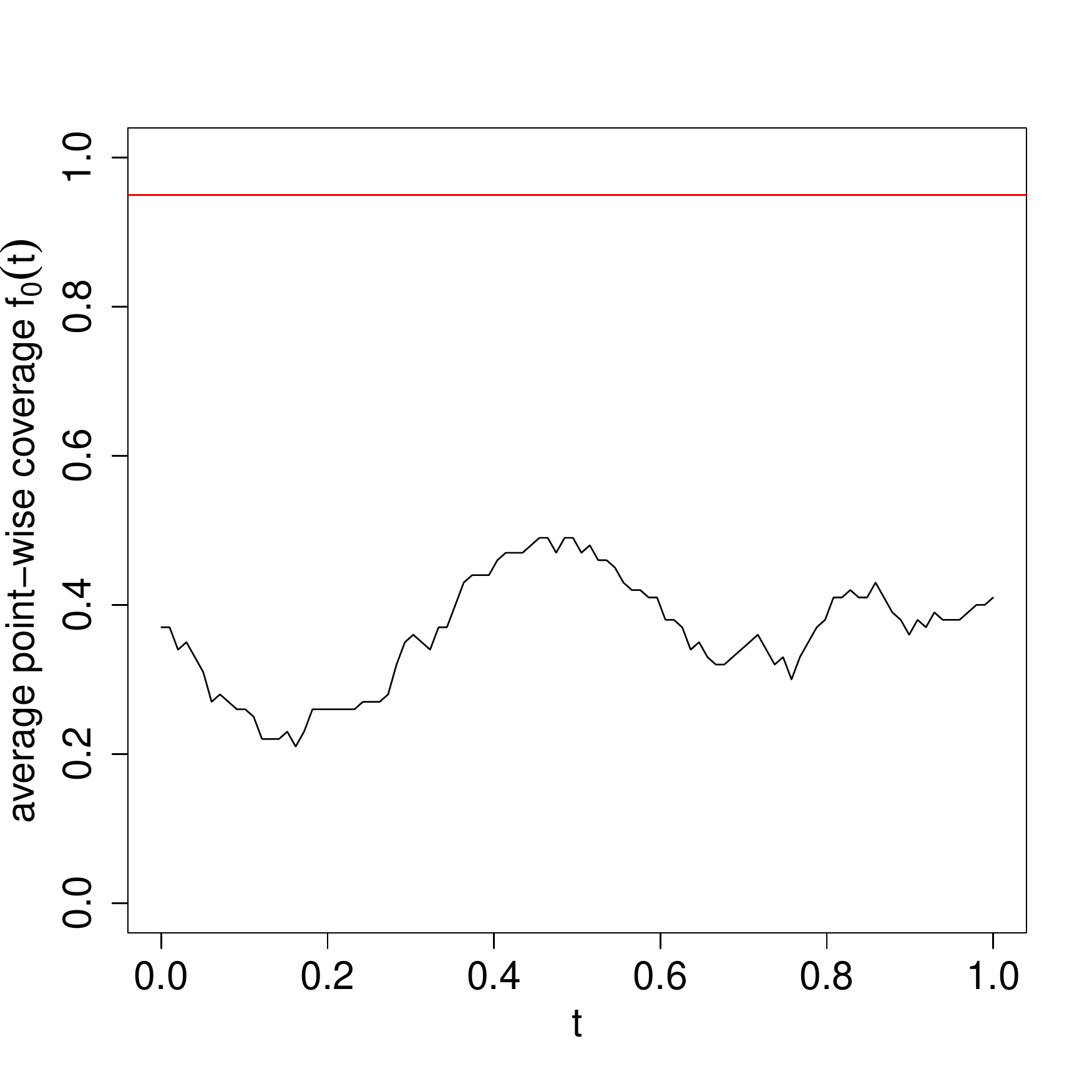}
\includegraphics[width=0.2\textwidth,page=2]{figures/simulation/covarage_mean_and_covariates_splines_normal_I9_JNA_RI_as_data_24_Mar_24_Mar_splines.pdf}
\includegraphics[width=0.2\textwidth,page=3]{figures/simulation/covarage_mean_and_covariates_splines_normal_I9_JNA_RI_as_data_24_Mar_24_Mar_splines.pdf}
\includegraphics[width=0.2\textwidth,page=4]{figures/simulation/covarage_mean_and_covariates_splines_normal_I9_JNA_RI_as_data_24_Mar_24_Mar_splines.pdf}\\
\includegraphics[width=0.2\textwidth,page=5]{figures/simulation/covarage_mean_and_covariates_splines_normal_I9_JNA_RI_as_data_24_Mar_24_Mar_splines.pdf}
\includegraphics[width=0.2\textwidth,page=6]{figures/simulation/covarage_mean_and_covariates_splines_normal_I9_JNA_RI_as_data_24_Mar_24_Mar_splines.pdf}
\includegraphics[width=0.2\textwidth,page=7]{figures/simulation/covarage_mean_and_covariates_splines_normal_I9_JNA_RI_as_data_24_Mar_24_Mar_splines.pdf}
\includegraphics[width=0.2\textwidth,page=8]{figures/simulation/covarage_mean_and_covariates_splines_normal_I9_JNA_RI_as_data_24_Mar_24_Mar_splines.pdf}
\caption{Average point-wise coverage of the point-wise CBs obtained by spline-FAMM for all covariate and interaction effects. For each effect, the point-wise coverage averaged over 100 simulation runs (black line) is shown. The red line indicates the nominal value of 0.95.}
\label{fig: coverage RI FAMM}
\end{center}
\end{figure}

\begin{figure}[h!]
\begin{center}
\begin{minipage}{1\textwidth}
\begin{center}
\raisebox{0.15\textwidth}{\textbf{B}}
\includegraphics[width=0.25\textwidth]{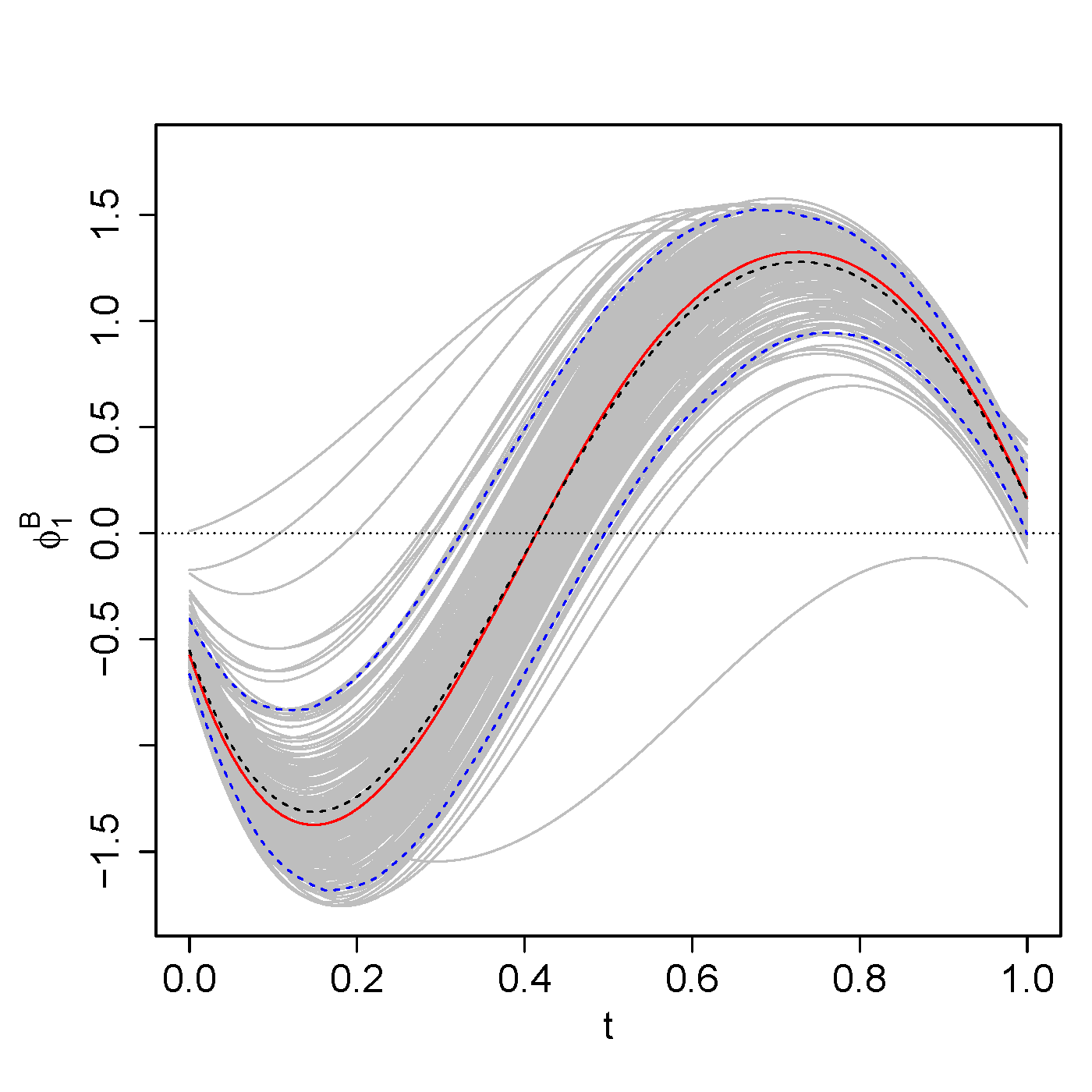}
\raisebox{0.15\textwidth}{\phantom{\textbf{B}}}
\includegraphics[width=0.25\textwidth,page=3]{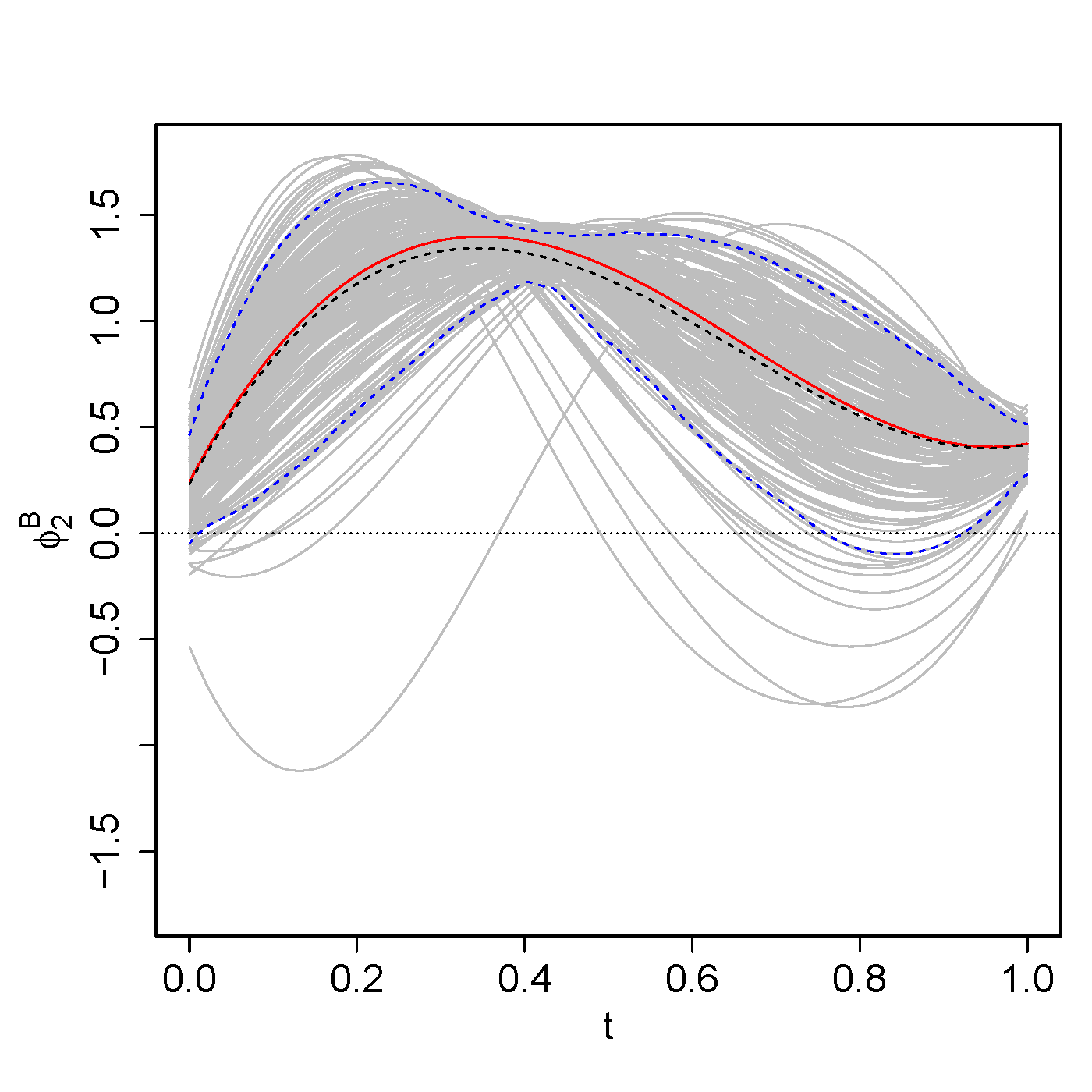}
\raisebox{0.15\textwidth}{\phantom{\textbf{B}}}
\includegraphics[width=0.25\textwidth,page=1]{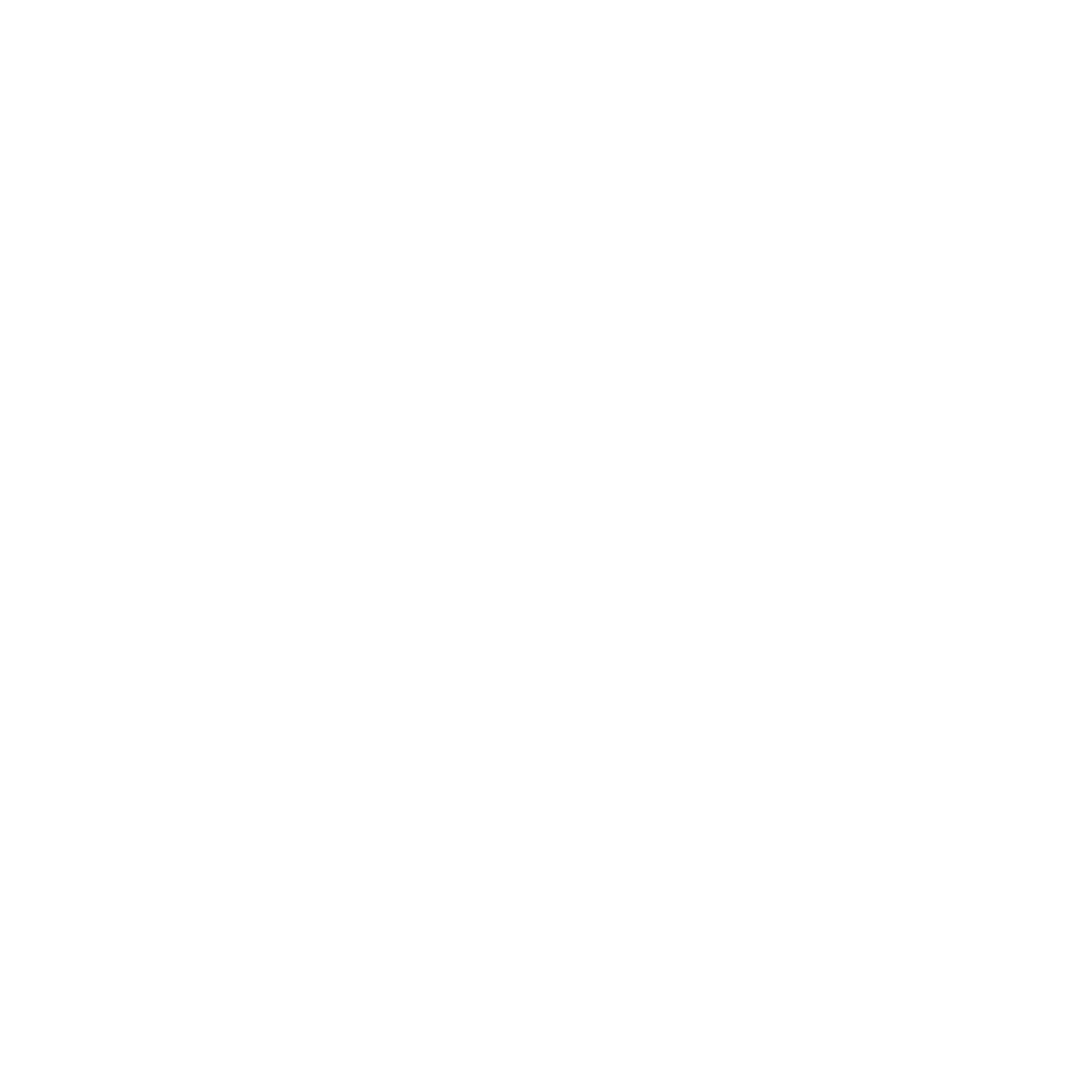}\\
\raisebox{0.15\textwidth}{\textbf{E}}
\includegraphics[width=0.25\textwidth,page=5]{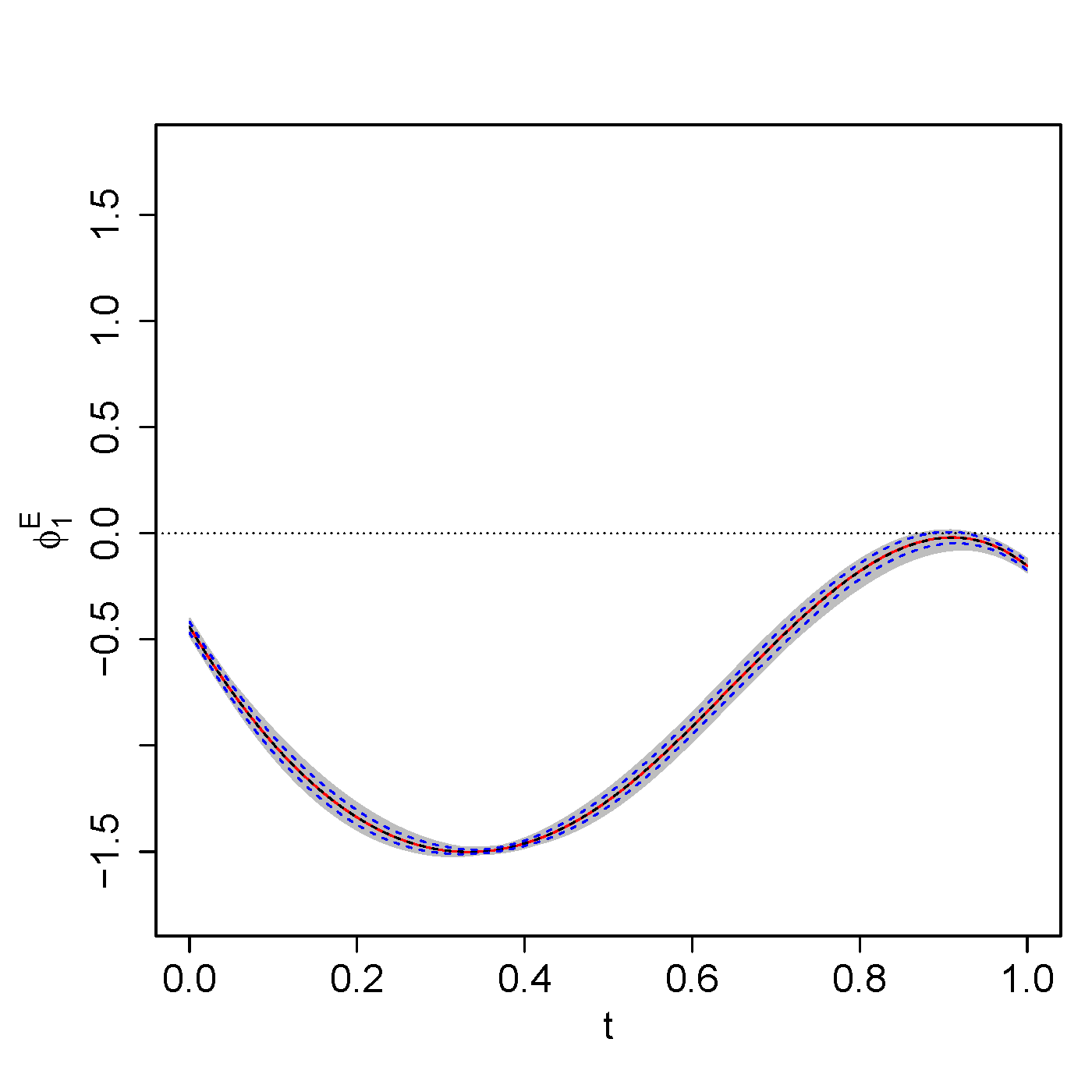}
\raisebox{0.15\textwidth}{\phantom{\textbf{B}}}
\includegraphics[width=0.25\textwidth,page=7]{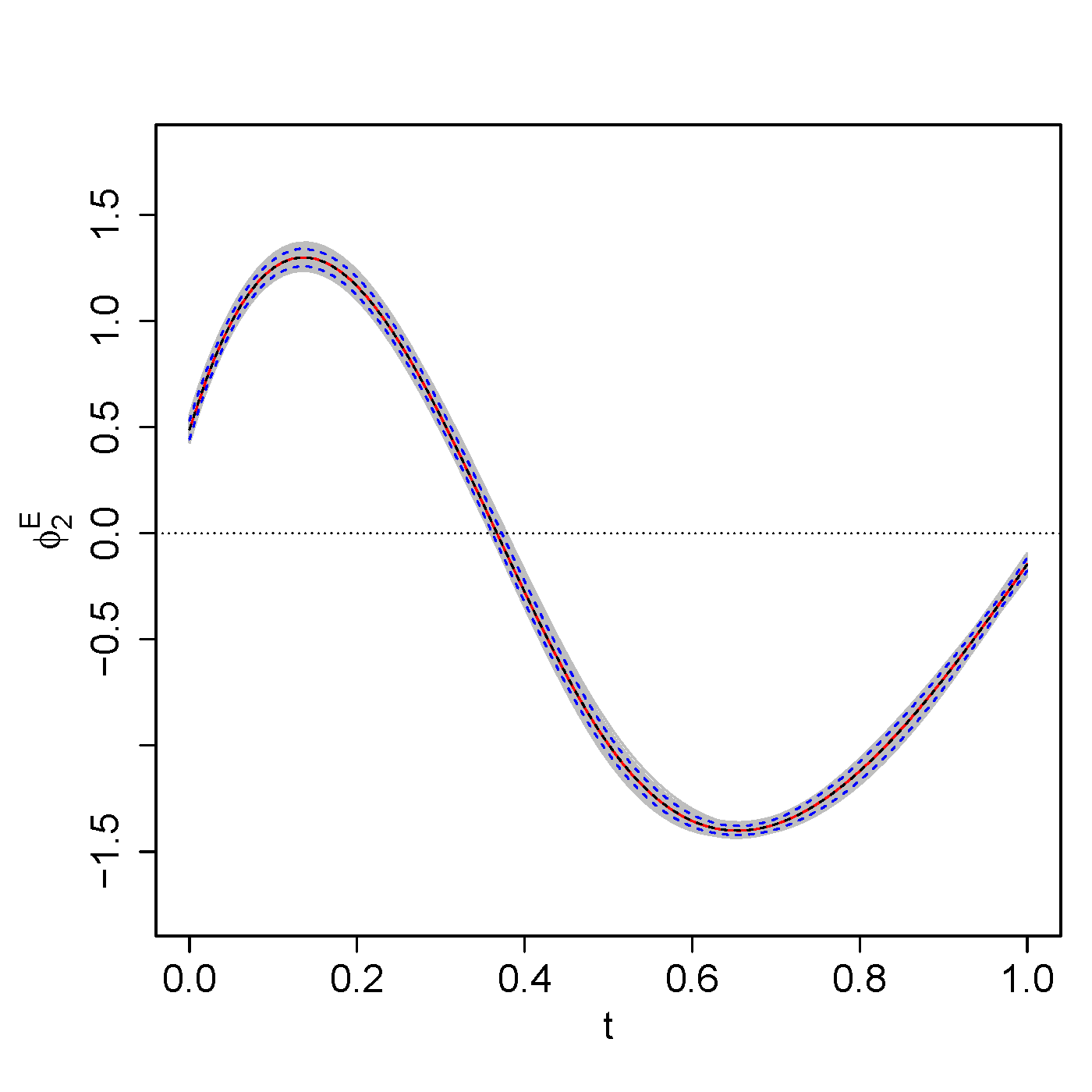}
\raisebox{0.15\textwidth}{\phantom{\textbf{B}}}
\includegraphics[width=0.25\textwidth,page=9]{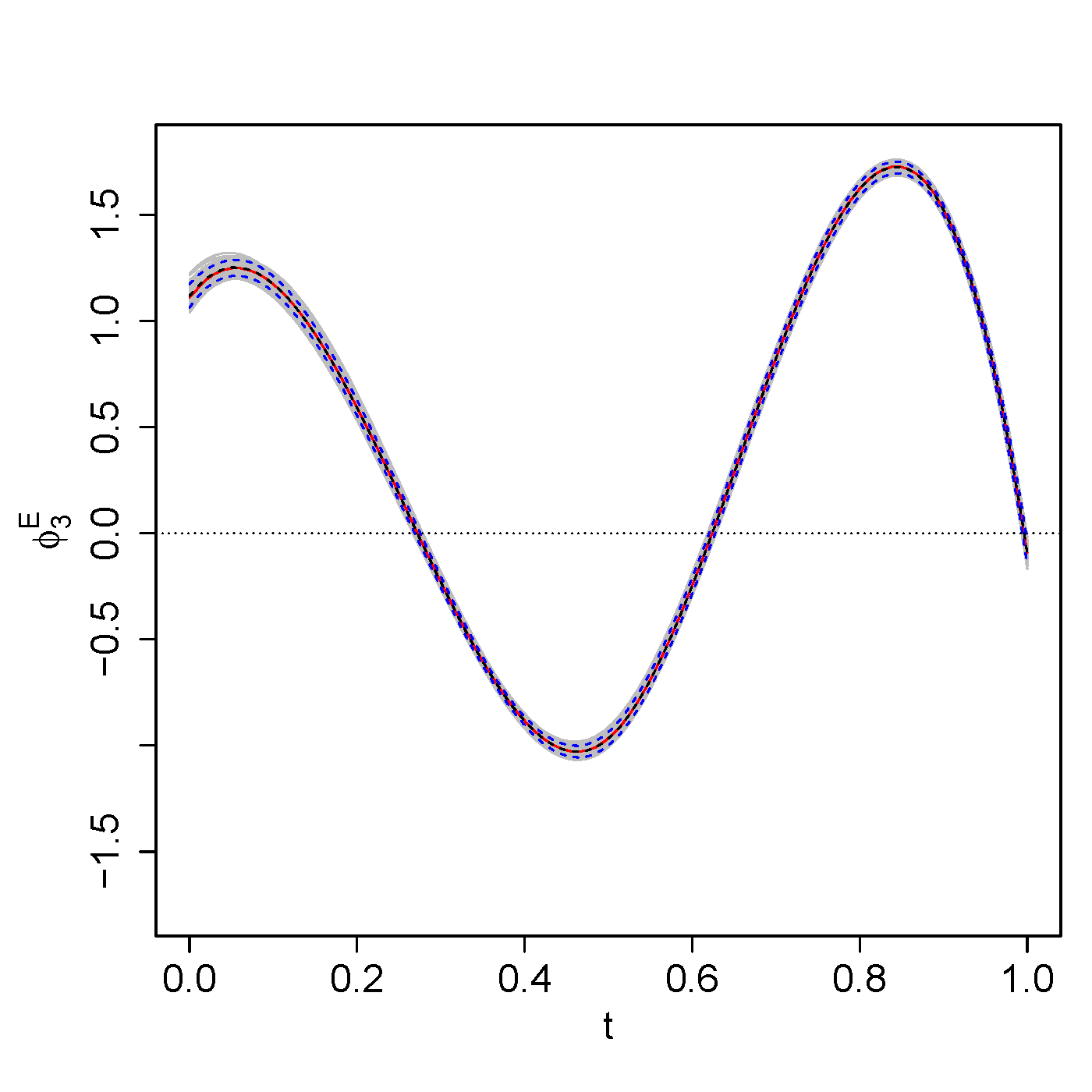}
\end{center}
\end{minipage}
\caption{True and estimated FPCs of the fRI (top row) and of the smooth error (bottom row). Shown are the true functions (red), the mean of the estimated functions over 200 simulation runs (black dashed line), the point-wise 5th and 95th percentiles of the estimated functions (blue dashed lines), and the estimated functions of all 200 simulation runs (grey).}
\label{fig: eigenfunctions RI}
\end{center}
\end{figure}

\begin{figure}[h!]
\begin{minipage}{1\textwidth}
\centering
\raisebox{0.15\textwidth}{\textbf{B}}
\includegraphics[width=0.25\textwidth,page=1]{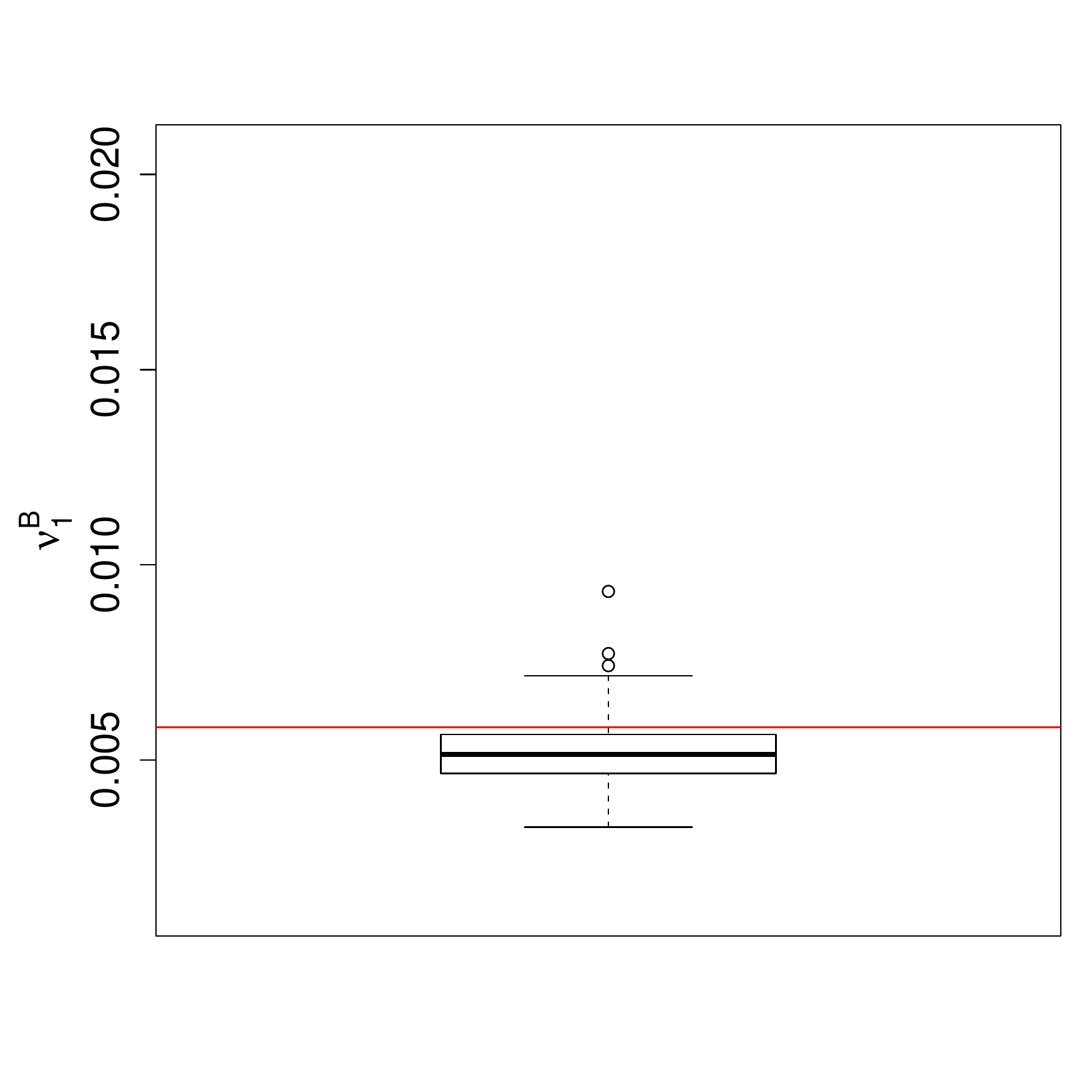}
\raisebox{0.15\textwidth}{\phantom{\textbf{B}}}
\includegraphics[width=0.25\textwidth,page=2]{figures/simulation/eigenvalues_normal_I9_JNA_RI_as_data_24_Mar_24_Mar.pdf}
\raisebox{0.15\textwidth}{\phantom{\textbf{B}}}
\includegraphics[width=0.25\textwidth,page=1]{figures/simulation/blank.pdf}\\
\raisebox{0.15\textwidth}{\textbf{E}}
\includegraphics[width=0.25\textwidth,page=3]{figures/simulation/eigenvalues_normal_I9_JNA_RI_as_data_24_Mar_24_Mar.pdf}
\raisebox{0.15\textwidth}{\phantom{\textbf{B}}}
\includegraphics[width=0.25\textwidth,page=4]{figures/simulation/eigenvalues_normal_I9_JNA_RI_as_data_24_Mar_24_Mar.pdf}
\raisebox{0.15\textwidth}{\phantom{\textbf{B}}}
\includegraphics[width=0.25\textwidth,page=5]{figures/simulation/eigenvalues_normal_I9_JNA_RI_as_data_24_Mar_24_Mar.pdf}
\end{minipage}
\caption{Boxplots of the estimated eigenvalues of the auto-covariances of the fRI (top row), and of the smooth error (bottom row) for all 200 simulations runs.}
\label{fig: boxplot eigenvalues RI}
\end{figure}

\begin{figure}[h!]
\centering
\includegraphics[width=0.25\textwidth,page=1]{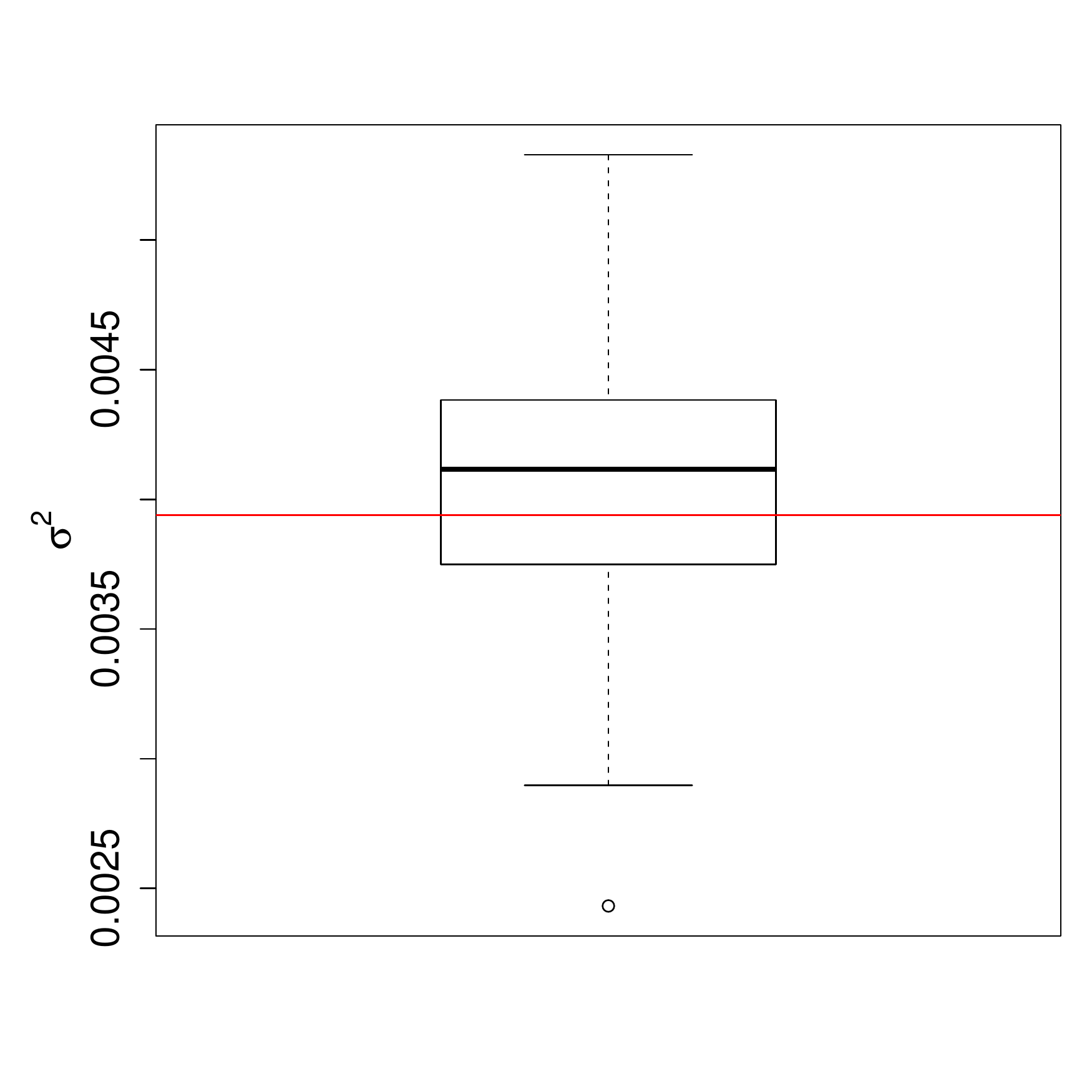}
\caption{Boxplot of the estimated error variances $\sigma^2$ for all 200 simulation runs.}
\label{fig: boxplot sigmasq RI}
\end{figure}

\underline{Simulation results for the fRI scenario}.
In the following, we show additional simulation results for the application-based fRI scenario with centred and decorrelated basis weights. Figure \ref{fig: mean RI}, Figure \ref{fig: mean RI pffr}, and Figure \ref{fig: mean RI FAMM} show the true and estimated covariate and interaction effects based on the independence assumption (Figure \ref{fig: mean RI}), on FPC-FAMM estimation (Figure \ref{fig: mean RI pffr}), and on the spline-FAMM alternative (Figure \ref{fig: mean RI FAMM}), respectively. We evaluate the performance of the point-wise CBs obtained by FPC-FAMM and spline-FAMM, by looking at the point-wise coverage shown in Figures \ref{fig: coverage RI pffr} and \ref{fig: coverage RI FAMM}. We additionally compare the simultaneous coverage of the point-wise CBs in terms of percentage of completely covered curves in Table \ref{tab: prop covered curves RI}.\\
Figure \ref{fig: eigenfunctions RI} depicts the true and estimated FPCs of the fRI and of the smooth error and Figure \ref{fig: boxplot eigenvalues RI} shows the boxplot of the estimated eigenvalues. The boxplot of the estimated error variances is shown in Figure \ref{fig: boxplot sigmasq RI}. Table \ref{tab: mean riMSEs RI} lists the average rrMSEs for all model components except for the covariate effects and Table \ref{tab: mean riMSEs RI covariates} lists the rrMSEs for the covariate effects.


\clearpage
\setlength{\tabcolsep}{1mm}
\begin{table}[h!]
\centering
\small
\caption{Simultaneous coverage of the point-wise CBs for FPC-FAMM and for spline-FAMM. Shown is the proportion of completely covered curves for all covariate and interaction effects. For FPC-FAMM, the coverage refers to 200 simulation runs, whereas for splines-FAMM, 100 simulation runs are taken into account.}
\begin{tabular}{l|rrrrrrrr}
&  $f_0(t)$ & $f_1(t)$ & $f_2(t)$ & $f_3(t)$ & $f_4(t)$ & $f_5(t)$ & $f_6(t)$ & $f_7(t)$ \\ 
    \hline
$\mu(t,\mx_{ijh})_{\FPCFAMM}$ & 43.5\%& 71.5\%& 70.5\%& 64.5\%&  76.0\%&  70.0\%&  66.0\%& 78.5\%\\ 
$\mu(t,\mx_{ijh})_{\FAMM}$ & 1.0\% & 5.0\% & 1.0\% & 1.0\% & 2.0\%&  2.0\% & 2.0\% &  5.0\% 
\end{tabular}
\label{tab: prop covered curves RI}
\end{table}


\renewcommand{\arraystretch}{0.8} 
\begin{table}[h!]
 \setlength{\tabcolsep}{1mm}
\small
\centering 
\caption{rrMSEs averaged over 200 simulation runs for all model components by random process. Rows 1-3: Number of grouping levels $L^X$ and average rrMSEs for the fRI and the smooth error. Last row: Average rrMSEs for the functional response, the mean, and the error variance.}
\begin{tabular}{c|c|rrrrrrrrrrrrr|rr}
$X$& $L^X$ & $K^X$ & $\phi^X_1$ & $\phi^X_2$ & $\phi^X_3$ & $\nu^X_1$ & $\nu^X_2$& $\nu^X_3$& $\xi^X_1$ & $\xi^X_2$& $\xi^X_3$ & $X$ & $X_{\FPCFAMM}$&$X_{\FAMM}$ &$\sigma^2$ \\ 
\hline
$B$ & 9& 0.23 & 0.21 & 0.22 & & 0.15 & 0.21 & & 0.22 & 0.30 & & 0.17 & 0.17 & 1.17 &\\ 
$E$ & 707 & 0.03 & 0.02 & 0.02 & 0.02 & 0.02 & 0.03 & 0.03 & 0.15 & 0.19 & 0.23 & 0.17 & 0.17 & 0.50 & \\ 
$Y$  &  & & &  &  & &  & &  &  &  & 0.09 &0.09 &0.15 &0.10
\end{tabular}
\label{tab: mean riMSEs RI}
\end{table}
\clearpage

 \setlength{\tabcolsep}{1mm}
\begin{table}[h!]
\small
\centering
\caption{Average rrMSEs for the estimated mean and covariate effects. Rows 1-2: rrMSEs for the estimation based on the independence assumption and for the estimation using FPC-FAMM averaged over 200 simulation runs. Last row: rrMSEs for the spline-based alternative averaged over 100 simulation runs.}
\begin{tabular}{l|rrrrrrrr}
& $f_0(t)$ & $f_1(t)$ & $f_2(t)$ & $f_3(t)$ & $f_4(t)$ & $f_5(t)$ & $f_6(t)$ & $f_7(t)$ \\ 
  \hline
$\mu(t,\mx_{ijh})$ & 0.06 & 0.13 & 0.22 & 1.43 & 0.30 & 0.58 & 0.39 & 0.60 \\  
$\mu(t,\mx_{ijh})_{\FPCFAMM}$ &0.06 & 0.12 & 0.20 & 1.34 & 0.25 & 0.53 & 0.39 & 0.47  \\ 
$\mu(t,\mx_{ijh})_{\FAMM}$ &0.36 & 0.38 & 0.58 & 3.83 & 0.84 & 1.54 & 1.10 & 1.85\\
\end{tabular}
\label{tab: mean riMSEs RI covariates}
\end{table}

The average rrISEs of the covariate and interaction effects lie between 0.06 ($f_0(t)$) for both estimation options and 1.43 or 1.34 ($f_3(t)$) for the estimation using the independence assumption or FPC-FAMM, respectively. Note that the high value for $f_3(t)$ is the result of the fact that the true covariate effect is very close to zero along the whole time interval, and to avoid dividing by values near zero it is more meaningful to look at the root mean squared error (rMSE) instead which is similar to the rMSEs of other covariates. 
\clearpage

\begin{figure}[h!]
\begin{center}
\includegraphics[width=0.25\textwidth]{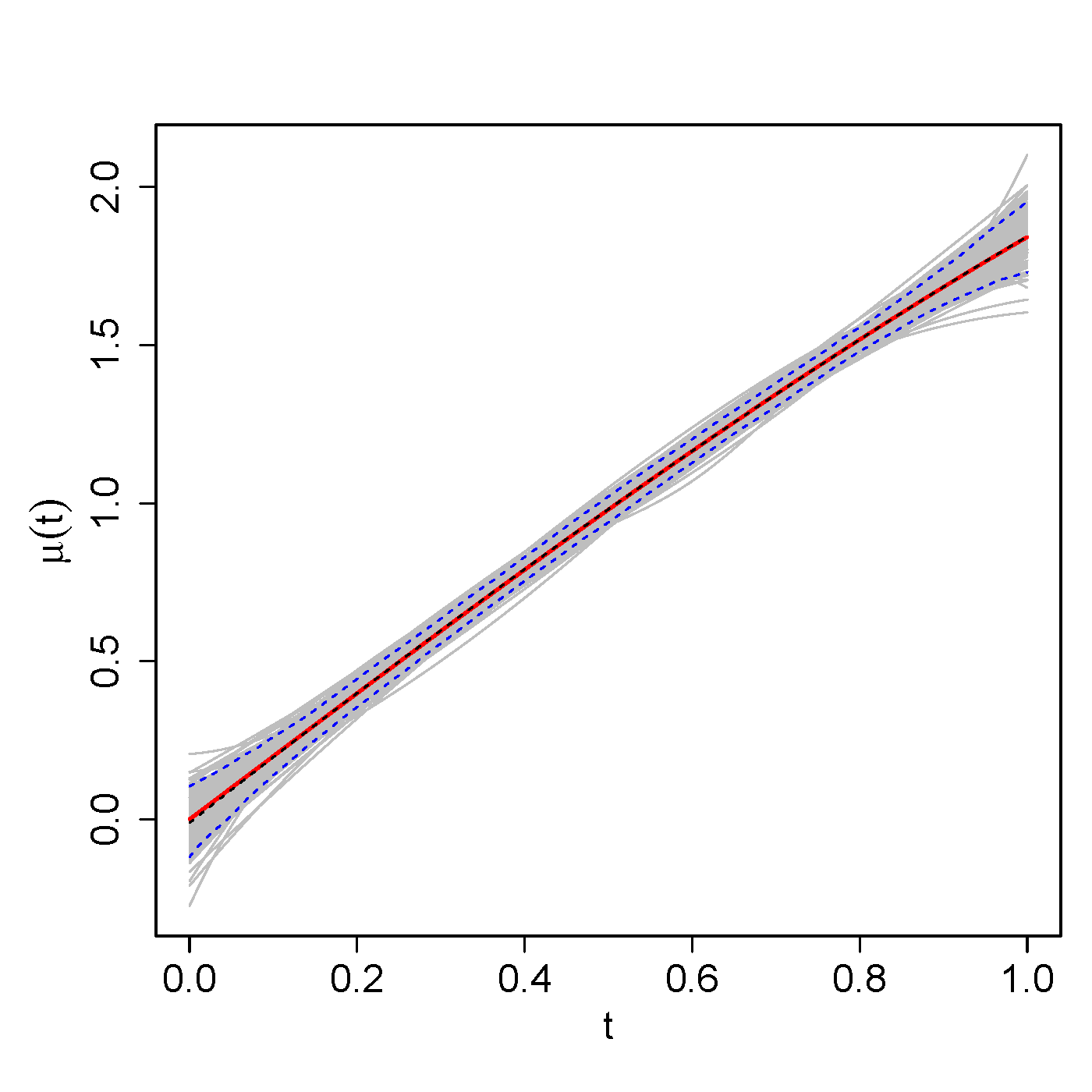}
\caption{True and estimated mean function $\mu(t,\mx_{ijh})$. Shown are the true function (red), the mean of the estimated functions over 200 simulation runs (black dashed line), the point-wise 5th and 95th percentiles of the estimated functions (blue dashed lines), and the estimated functions of all 200 simulation runs (grey).}
\label{fig: mean sparse}
\end{center}
\end{figure}

\begin{figure}[h!]
\begin{center}
\begin{minipage}{1\textwidth}
\begin{center}
\raisebox{0.15\textwidth}{\textbf{B}}
\includegraphics[width=0.25\textwidth,page=1]{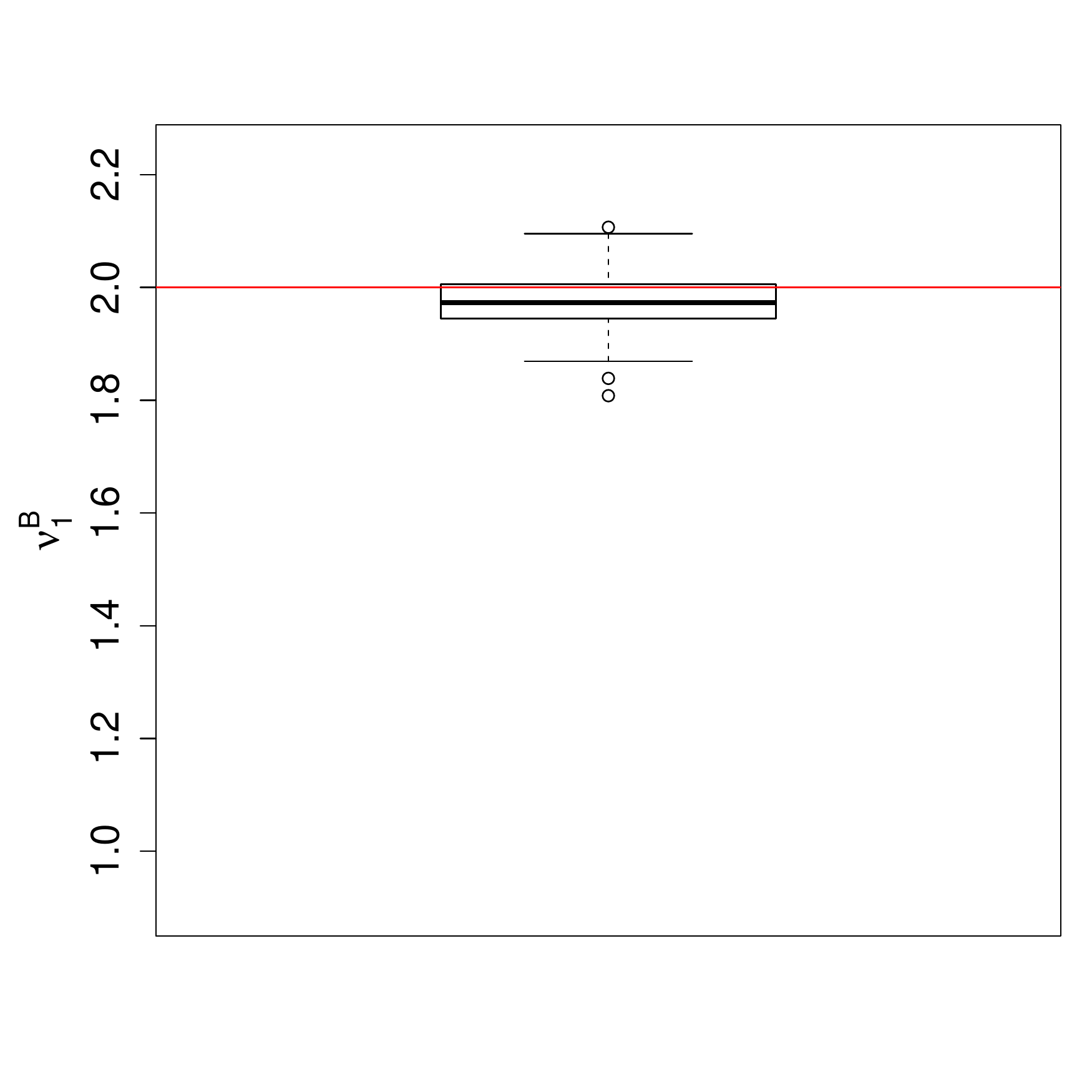}
\raisebox{0.15\textwidth}{\phantom{\textbf{B}}}
\includegraphics[width=0.25\textwidth,page=2]{figures/simulation/eigenvalues_normal_I40_J40_crossed_23_Dec_sparse_24_Dec.pdf}\\
\raisebox{0.15\textwidth}{\textbf{C}}
\includegraphics[width=0.25\textwidth,page=3]{figures/simulation/eigenvalues_normal_I40_J40_crossed_23_Dec_sparse_24_Dec.pdf}
\raisebox{0.15\textwidth}{\phantom{\textbf{B}}}
\includegraphics[width=0.25\textwidth,page=4]{figures/simulation/eigenvalues_normal_I40_J40_crossed_23_Dec_sparse_24_Dec.pdf}\\
\raisebox{0.15\textwidth}{\textbf{E}}
\includegraphics[width=0.25\textwidth,page=5]{figures/simulation/eigenvalues_normal_I40_J40_crossed_23_Dec_sparse_24_Dec.pdf}
\raisebox{0.15\textwidth}{\phantom{\textbf{B}}}
\includegraphics[width=0.25\textwidth,page=6]{figures/simulation/eigenvalues_normal_I40_J40_crossed_23_Dec_sparse_24_Dec.pdf}
\end{center}
\end{minipage}
\caption{Boxplots of the estimated eigenvalues of the auto-covariances of the crossed fRIs $B_i(t)$ (top row), $C_j(t)$ (middle row), as well as the eigenvalues of the auto-covariance of the smooth error $E_{ijh}(t)$ (bottom row) for all 200 simulations runs.}
\label{fig: boxplot eigenvalues sparse}
\end{center}
\end{figure}

\begin{figure}[h!]
\centering
\includegraphics[width=0.25\textwidth,page=1]{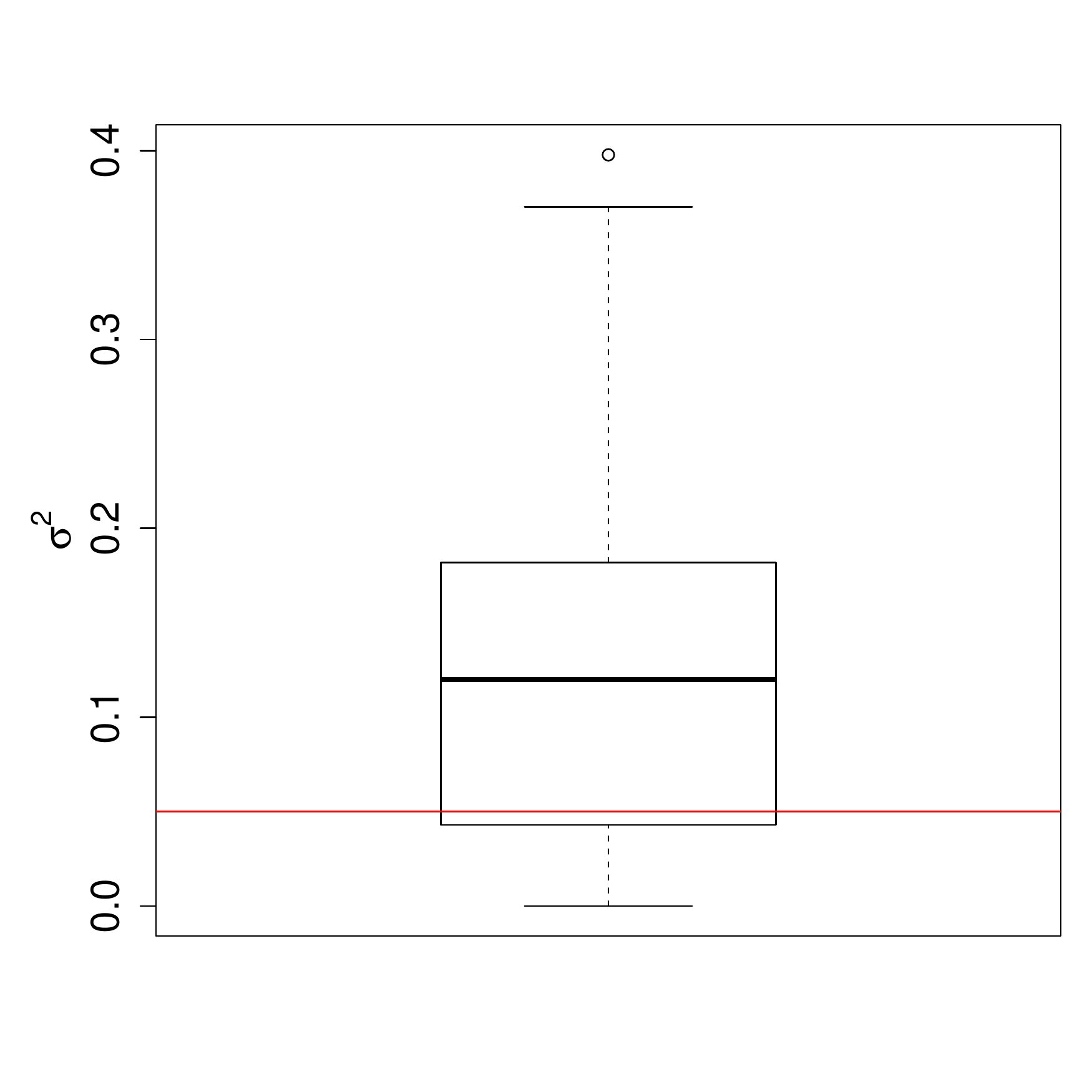}
\caption{Boxplots of the estimated error variances $\sigma^2$ for all 200 simulation runs.}
\label{fig: boxplot sigmasq sparse}
\end{figure}

\underline{Simulation results for the sparse scenario}.
In the following, additional results for the sparse scenario with centred and decorrelated basis weights are shown. Figure \ref{fig: mean sparse} shows the true and estimated mean functions and Figure \ref{fig: boxplot eigenvalues sparse} depicts the boxplot of the estimated eigenvalues for the two fRIs and for the smooth error. In Figure \ref{fig: boxplot sigmasq sparse}, we show the boxplot of the estimated error variances. In Table \ref{tab: mean riMSEs sparse}, the average relative errors for all model components are given.

\begin{figure}[h!]
\begin{center}
\raisebox{0.15\textwidth}{\textbf{B}}
\includegraphics[width=0.25\textwidth,page=1]{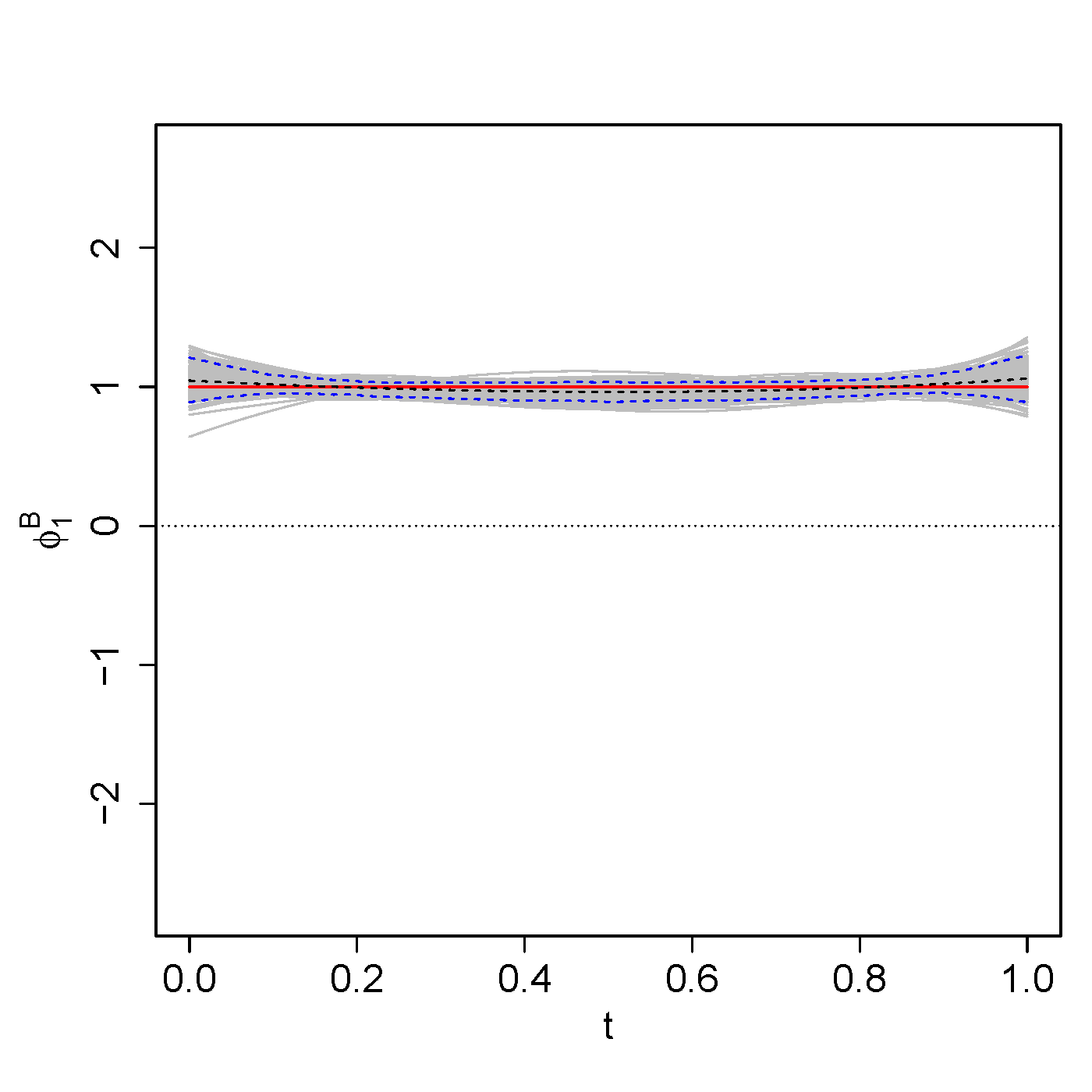}
\raisebox{0.15\textwidth}{\phantom{\textbf{B}}}
\includegraphics[width=0.25\textwidth,page=3]{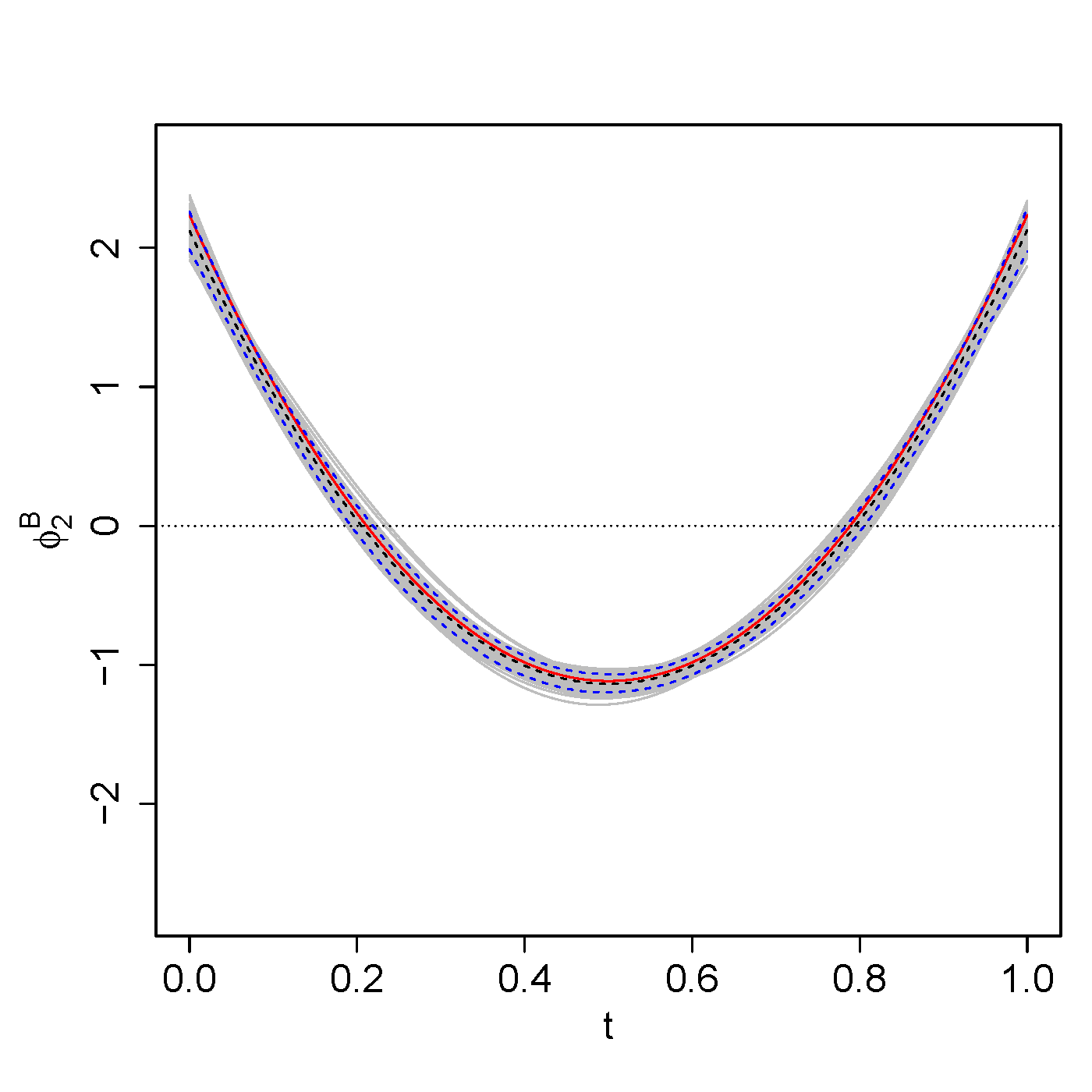}\\
\raisebox{0.15\textwidth}{\textbf{C}}
\includegraphics[width=0.25\textwidth,page=5]{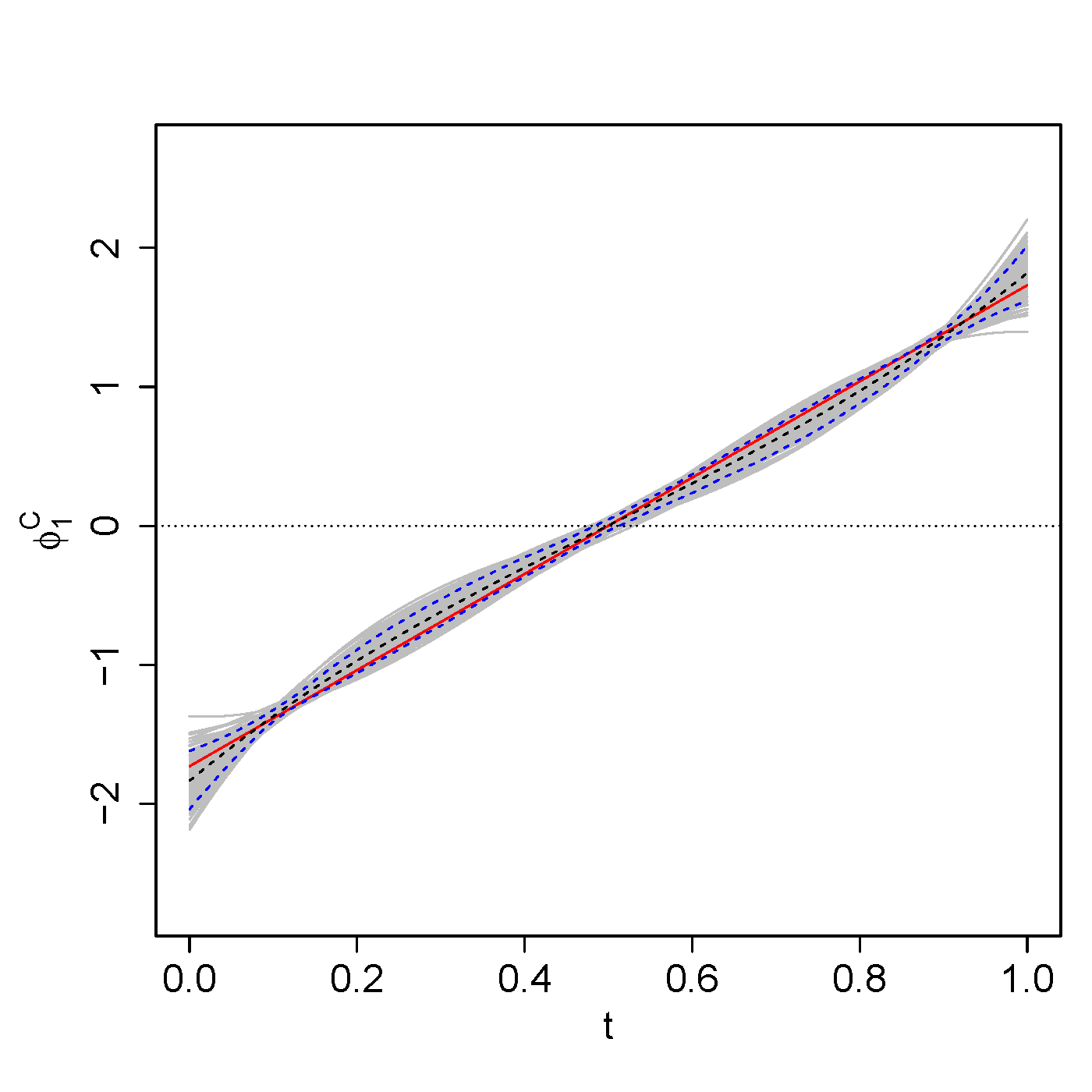}
\raisebox{0.15\textwidth}{\phantom{\textbf{B}}}
\includegraphics[width=0.25\textwidth,page=7]{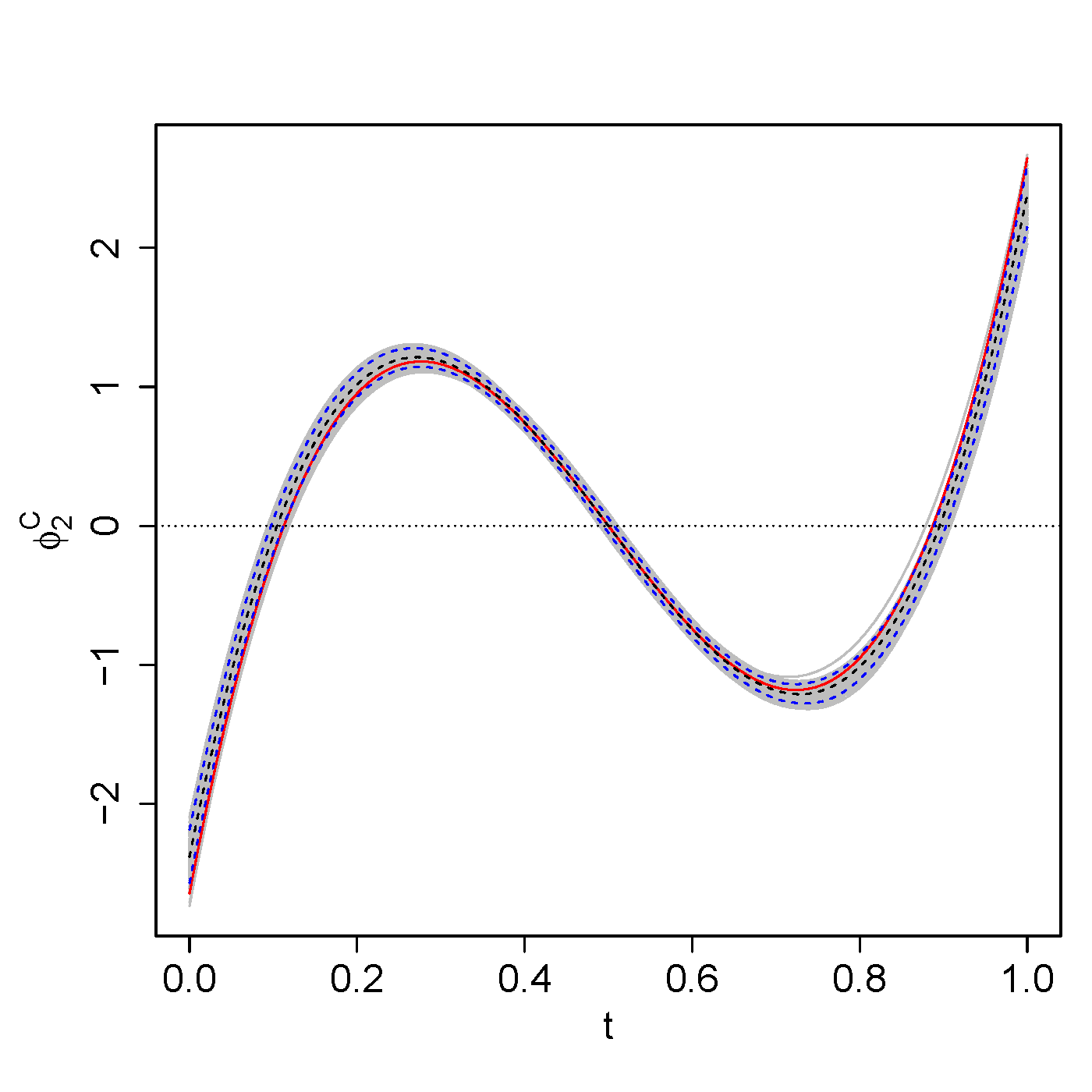}\\
\raisebox{0.15\textwidth}{\textbf{E}}
\includegraphics[width=0.25\textwidth,page=9]{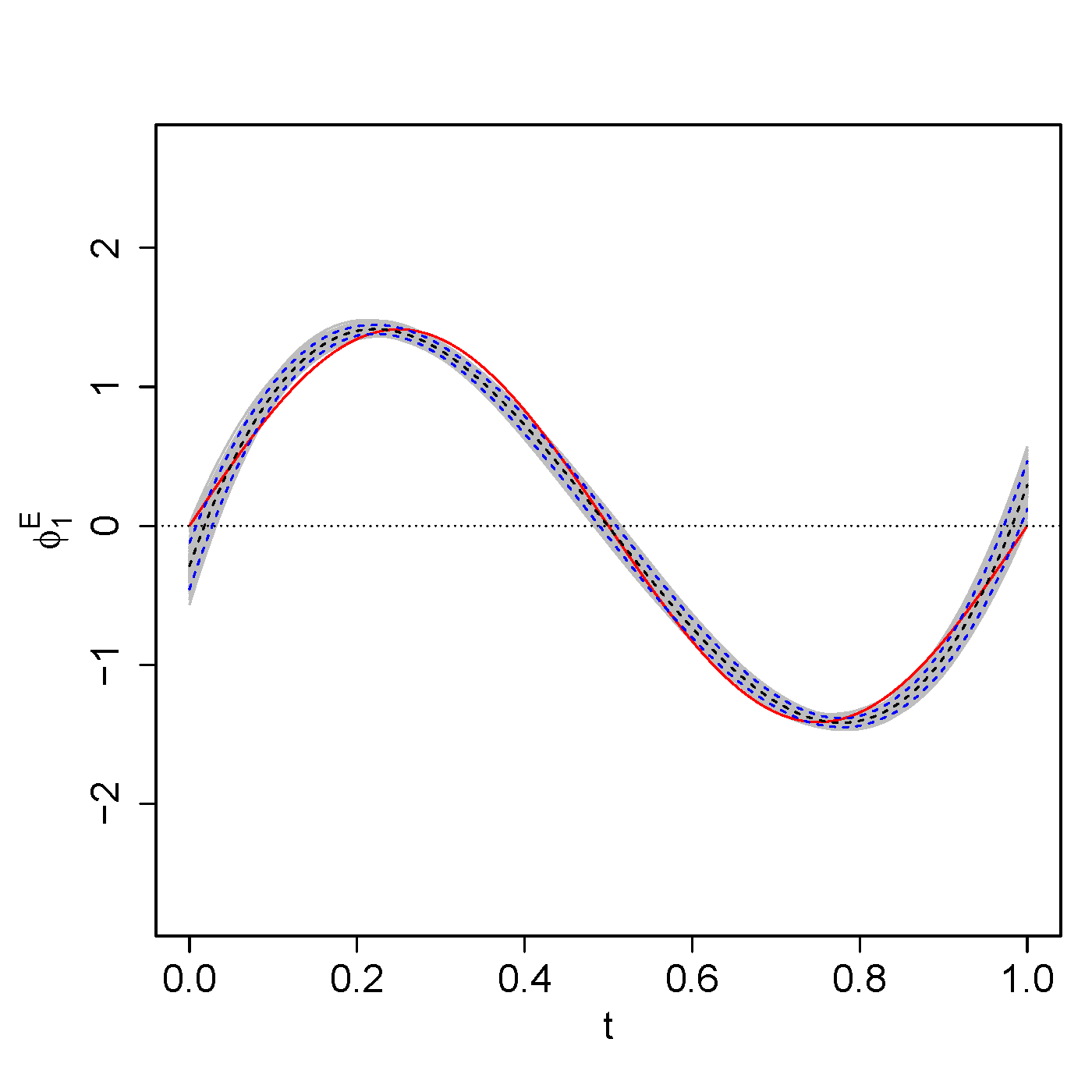}
\raisebox{0.15\textwidth}{\phantom{\textbf{B}}}
\includegraphics[width=0.25\textwidth,page=11]{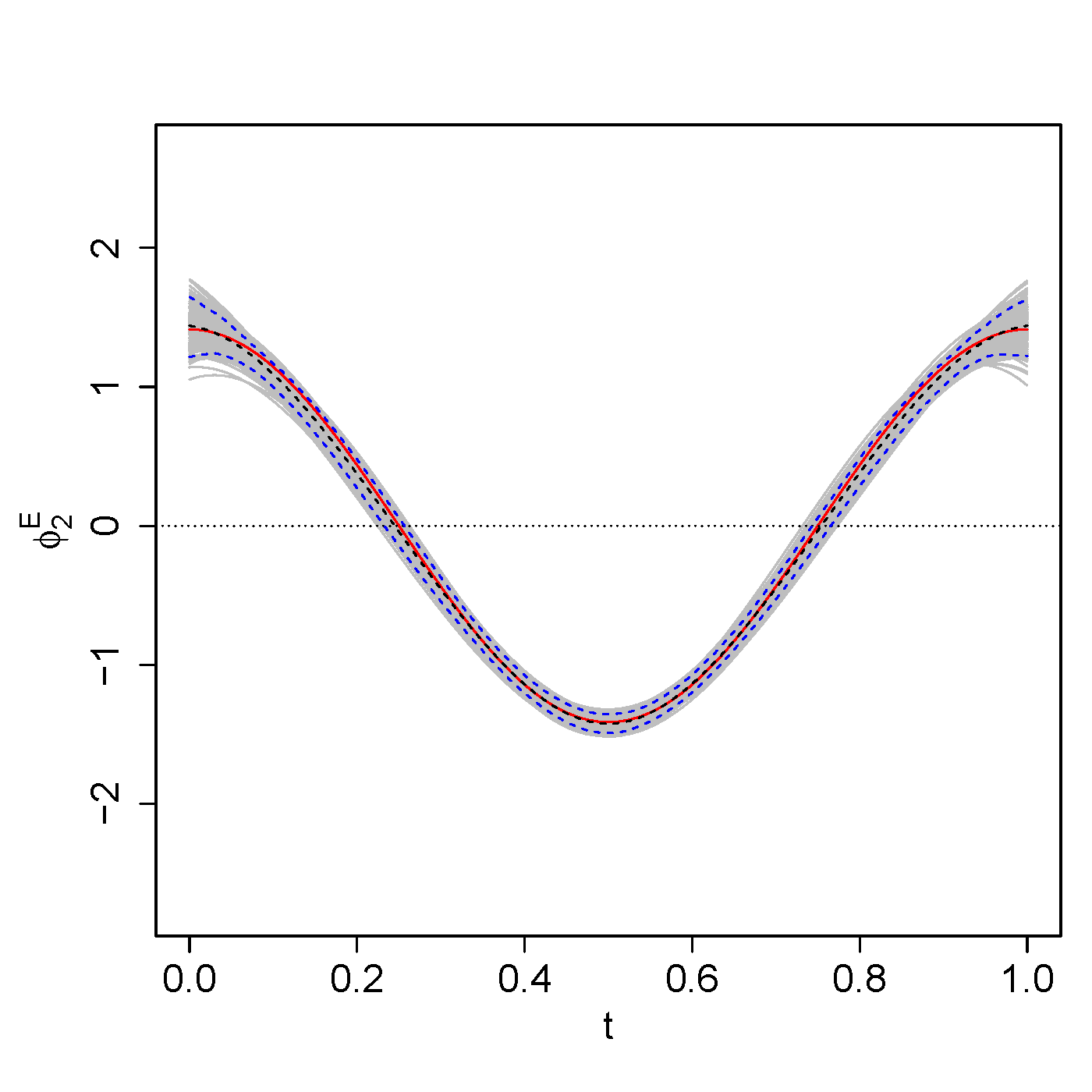}
\caption{True and estimated FPCs of the crossed fRIs $B_i(t)$ (top row) and $C_j(t)$ (middle row), as well as the FPCs of the smooth error $E_{ijh}(t)$ (bottom row). Shown are the true functions (red), the mean of the estimated functions over 200 simulation runs (black dashed line), the point-wise 5th and 95th percentiles of the estimated functions (blue dashed lines), and the estimated functions of all 200 simulation runs (grey).}
\label{fig: eigenfuncionts sparse}
\end{center}
\end{figure}

\renewcommand{\arraystretch}{0.8} 
\begin{table}[h!]
 \setlength{\tabcolsep}{1mm}
\centering
\small
\caption{rrMSEs averaged over 200 simulation runs for all model components by random process. Rows 1-3: Number of grouping levels $L^X$ and average relative errors for the functional random effects and their covariance decompositions. Last row: average rrMSE of the functional response, the mean, and the error variance.}
\begin{tabular}{c|c|rrrrrrrr|rr}
$X$&$L^X$ & $K^X$ & $\phi^X_1$ & $\phi^X_2$ & $\nu^X_1$ & $\nu^X_2$ & $\xi^X_1$ & $\xi^X_2$ & $X$&$\mu$&$\sigma^2$ \\ 
  \hline
$B$ & 40& 0.06 & 0.05 & 0.07 & 0.02 & 0.04 & 0.04 & 0.11 & 0.06 && \\ 
$C$ & 40&0.06 & 0.07 & 0.11 & 0.03 & 0.05 & 0.23 & 0.25 & 0.21 &&\\ 
$E$ & 4800 & 0.14 & 0.11 & 0.07 & 0.02 & 0.05 & 0.30 & 0.19 & 0.29 & & \\ 
$Y$ & & & & & & & & &0.09 & 0.03&1.81  
\end{tabular}
\label{tab: mean riMSEs sparse}
\end{table}

\clearpage
\subsection{Results for simulations with original basis weights}
Before we show the results for the simulation with the original (non-centred and non-decorrelated) basis weights in detail, we give a short summary and compare with the results for the simulations with centred and decorrelated basis weights.\\
We observe that with the original basis weights, the estimated FPCs are often permuted within one grouping variable, e.g.~the first and second FPC of the speakers are interchanged or are linear combinations of them, due to the correlation in the basis weights. As expected, this effect increases the smaller the corresponding number of independent levels, as the empirical FPC weights then have an empirical distribution far from the theoretical one. Moreover, we obtain higher average rrMSEs for the eigenvalues when using the original basis weights. Correspondingly, we also obtain worse results for the auto-covariances. For all three simulation settings, the rrMSEs for the basis weights tend to be higher with the original basis weights. Using the original basis weights results in higher rrMSEs for the functional random effects, especially for functional random effects with a small number of grouping levels. The estimation of the mean function for non-centred basis weights is much worse due to a shift of the mean by the respective FPC multiplied by the empirical mean of the basis weights. Correspondingly, also the coverage of the point-wise CBs decreases for most points. Yet, the functional response for the original basis weights is again estimated very well as shifts between mean and non-centred random effects cancel out. For the covariate effects and the coverage of the CBs, we hardly observe any differences to the simulations with centred and decorrelated basis weights. Also for the error variance, we do not observe a considerable change in the rrMSEs.

\begin{figure}[h!]
\begin{center}
\includegraphics[width=0.25\textwidth]{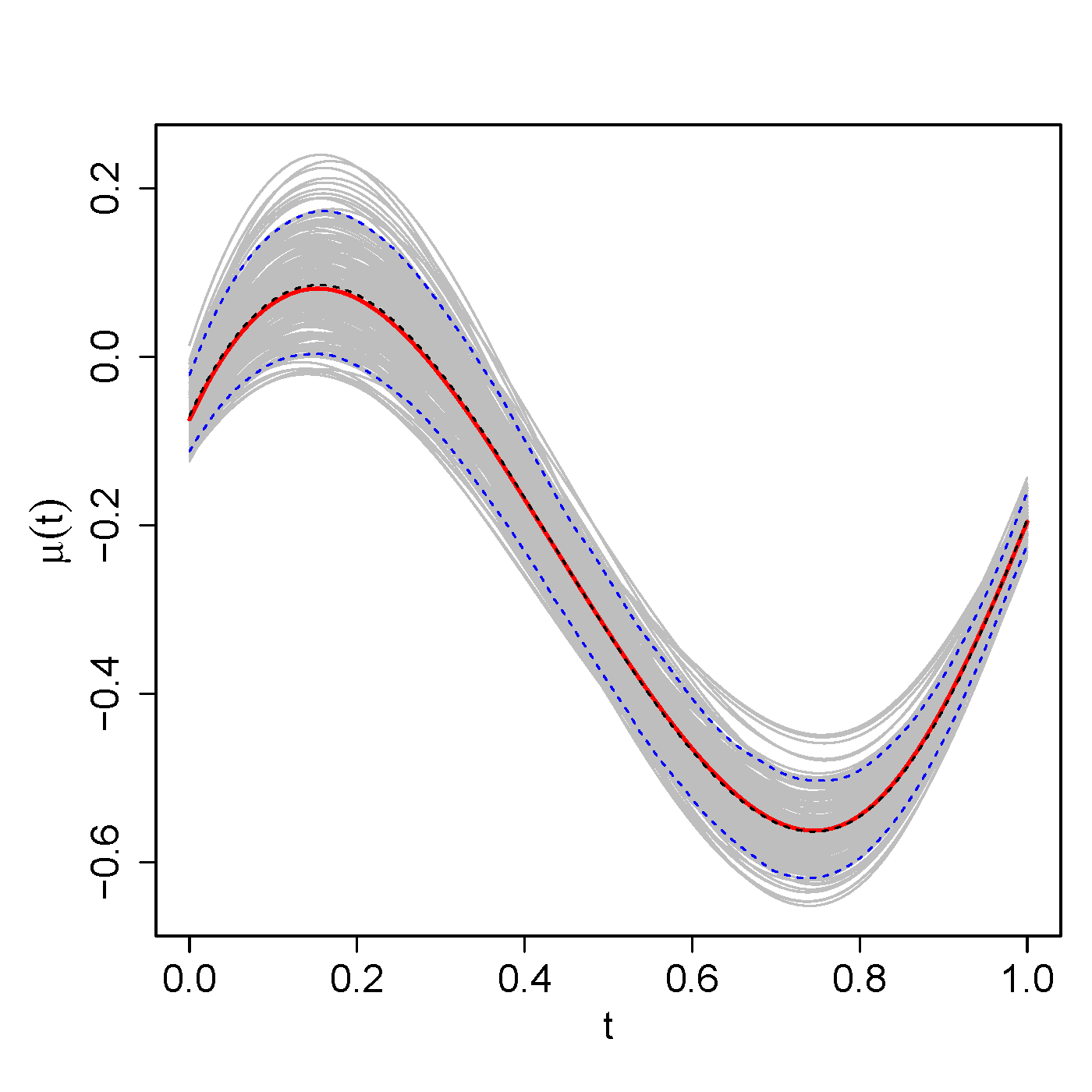}
\caption{True and estimated mean function $\mu(t,\mx_{ijh})$. Shown is the true function (red), the mean of the estimated functions over 200 simulation runs (black dashed line), the point-wise 5th and 95th percentiles of the estimated functions (blue dashed lines), and the estimated functions of all 200 simulation runs (grey).}
\label{fig: mean crossed non-centred}
\end{center}
\end{figure}

\begin{figure}[h!]
\begin{center}
\begin{minipage}{1\textwidth}
\begin{center}
\raisebox{0.15\textwidth}{\textbf{B}}
\includegraphics[width=0.25\textwidth,page=1]{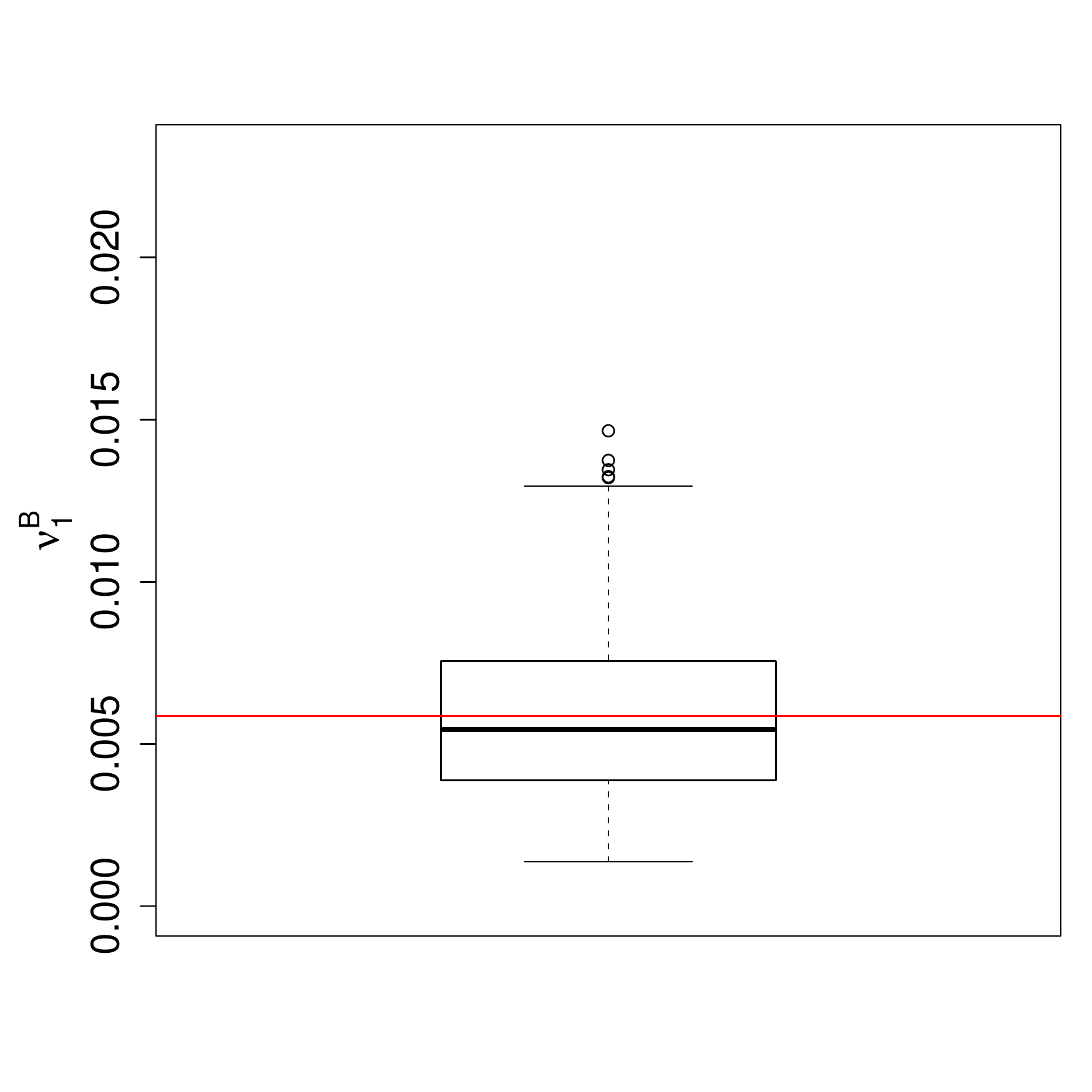}
\raisebox{0.15\textwidth}{\phantom{\textbf{B}}}
\includegraphics[width=0.25\textwidth,page=2]{figures/simulation/eigenvalues_normal_I9_J16_crossed_as_data_19_Jan_22_Jan.pdf}
\raisebox{0.15\textwidth}{\textbf{C}}
\includegraphics[width=0.25\textwidth,page=3]{figures/simulation/eigenvalues_normal_I9_J16_crossed_as_data_19_Jan_22_Jan.pdf}\\
\raisebox{0.15\textwidth}{\textbf{E}}
\includegraphics[width=0.25\textwidth,page=4]{figures/simulation/eigenvalues_normal_I9_J16_crossed_as_data_19_Jan_22_Jan.pdf}
\raisebox{0.15\textwidth}{\phantom{\textbf{B}}}
\includegraphics[width=0.25\textwidth,page=5]{figures/simulation/eigenvalues_normal_I9_J16_crossed_as_data_19_Jan_22_Jan.pdf}
\raisebox{0.15\textwidth}{\phantom{\textbf{B}}}
\includegraphics[width=0.25\textwidth,page=6]{figures/simulation/eigenvalues_normal_I9_J16_crossed_as_data_19_Jan_22_Jan.pdf}
\end{center}
\end{minipage}
\caption{Boxplots of the estimated eigenvalues of the auto-covariances of the crossed fRIs $B_i(t)$ and $C_j(t)$ (top row), as well as the eigenvalues of the auto-covariance of the smooth error $E_{ijh}(t)$ (bottom row) for all 200 simulations runs.}
\label{fig: boxplot eigenvalues crossed non-centred}
\end{center}
\end{figure}

\begin{figure}[h!]
\centering
\includegraphics[width=0.25\textwidth,page=1]{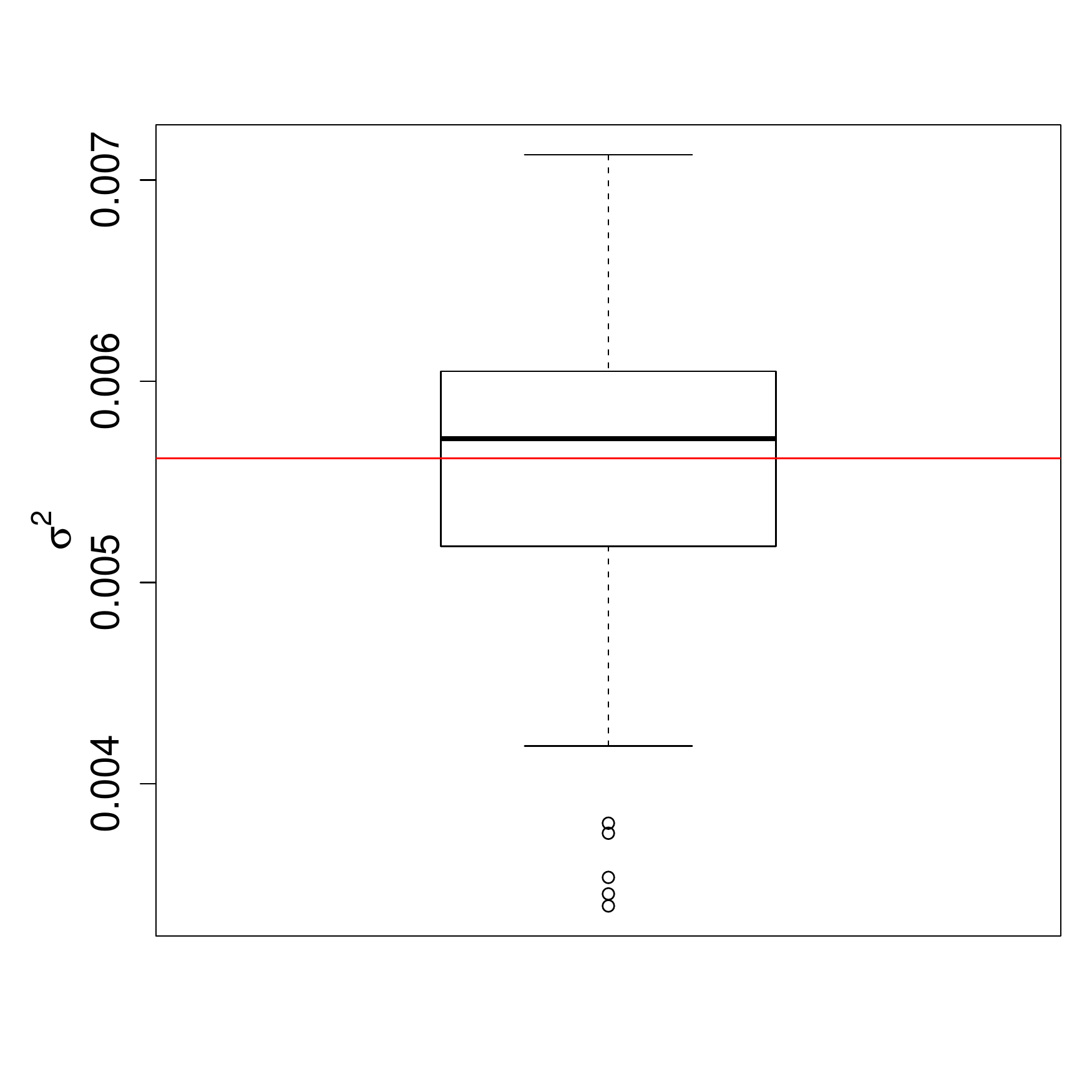}
\caption{Boxplots of the estimated error variances $\sigma^2$ for all 200 simulation runs.}
\label{fig: boxplot sigmasq crossed non-centred}
\end{figure}

\underline{Simulation results for crossed-fRIs scenario}.
In the following, we show the results for the simulations of the application-based crossed-fRIs scenario with non-centred and non-decorrelated basis weights. Figure \ref{fig: mean crossed non-centred} shows the true and estimated mean functions and Figure \ref{fig: boxplot eigenvalues crossed non-centred} depicts the boxplot of the estimated eigenvalues for the two fRIs and for the smooth error. In Figure \ref{fig: boxplot sigmasq crossed non-centred}, we show the boxplot of the estimated error variances. In Table \ref{tab: mean riMSEs crossed non-centred}, the average rrMSEs for all model components are given.

\begin{figure}[h!]
\begin{center}
\begin{minipage}{1\textwidth}
\begin{center}
\raisebox{0.15\textwidth}{\textbf{B}}
\includegraphics[width=0.25\textwidth,page=1]{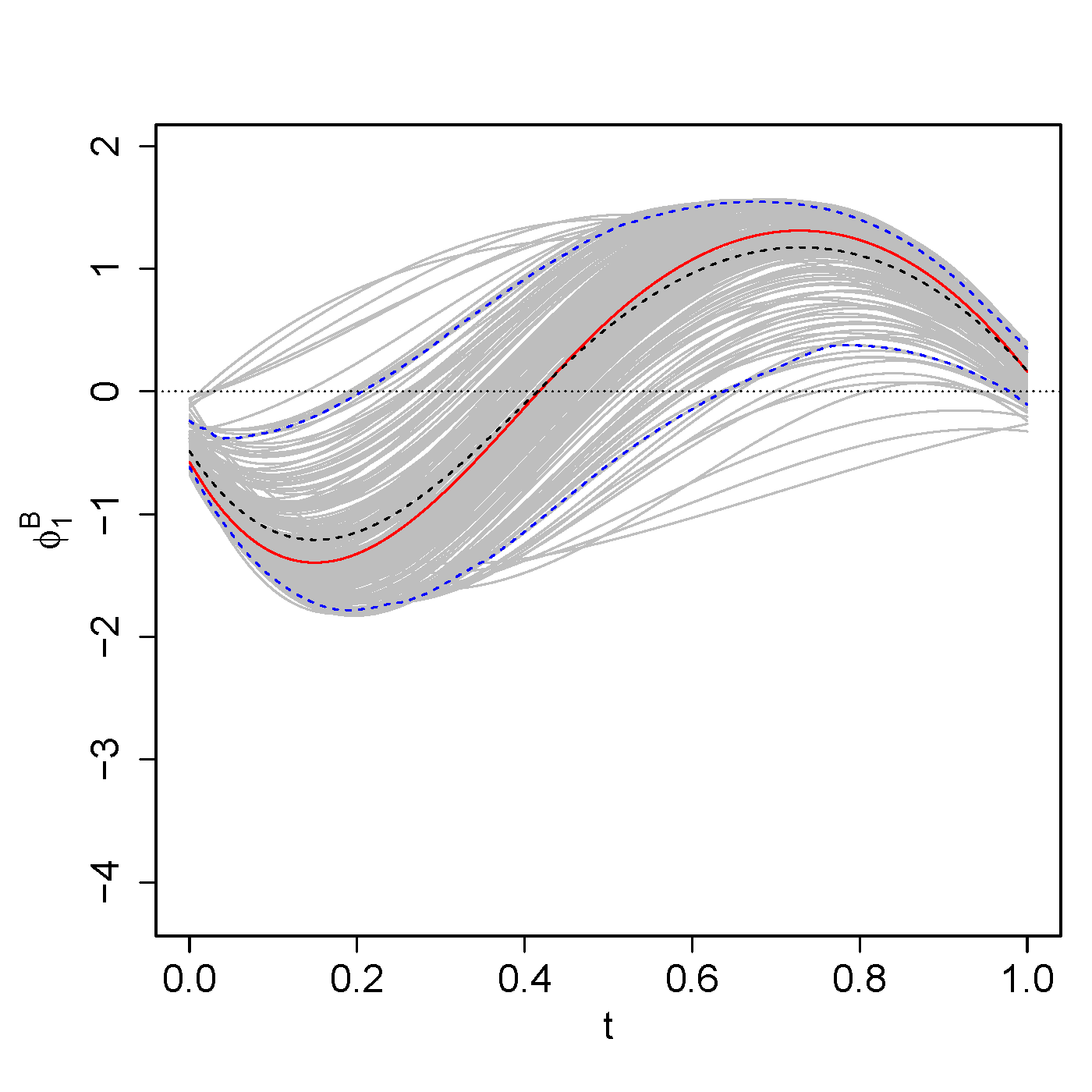}
\raisebox{0.15\textwidth}{\phantom{\textbf{B}}}
\includegraphics[width=0.25\textwidth,page=3]{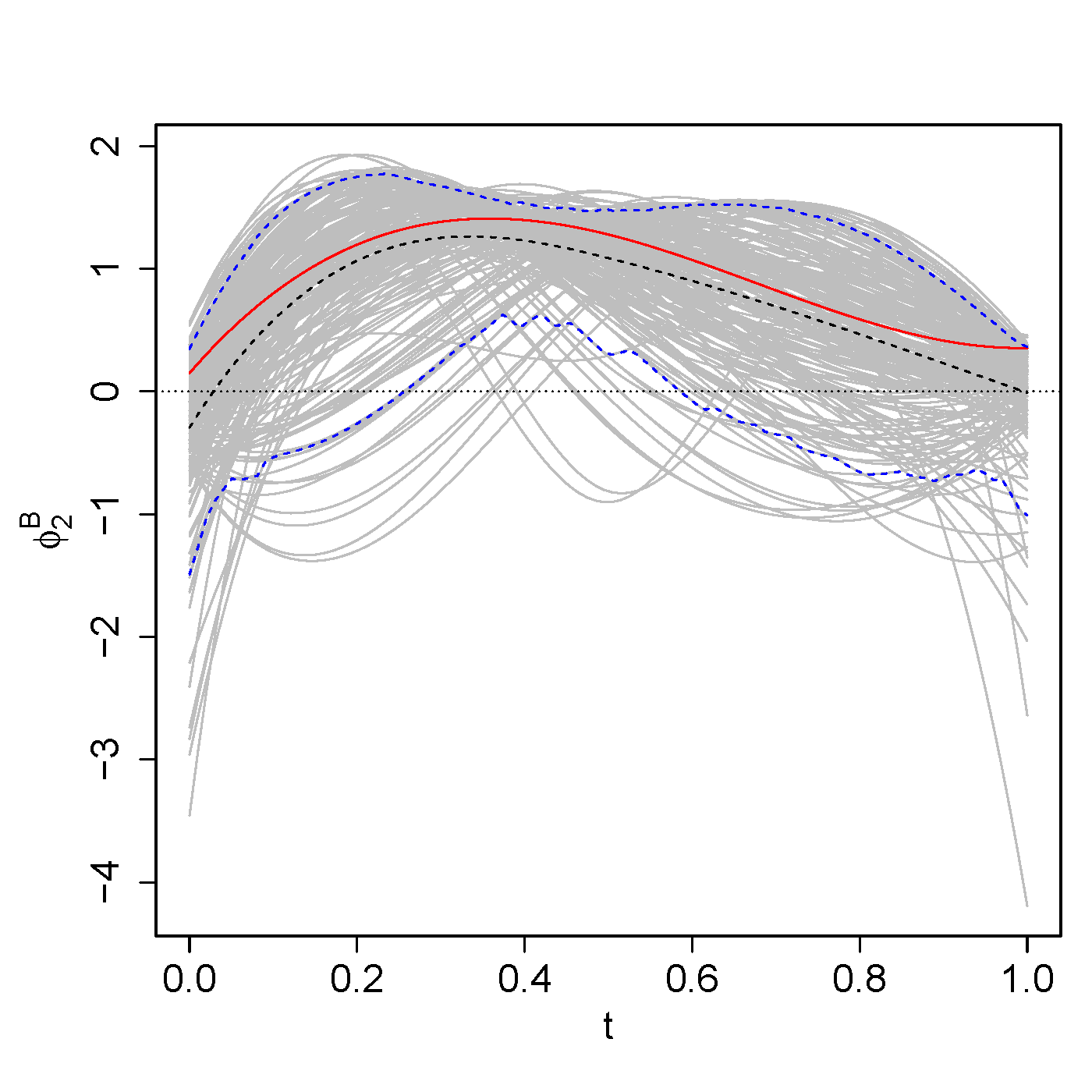}
\raisebox{0.15\textwidth}{\textbf{C}}
\includegraphics[width=0.25\textwidth,page=5]{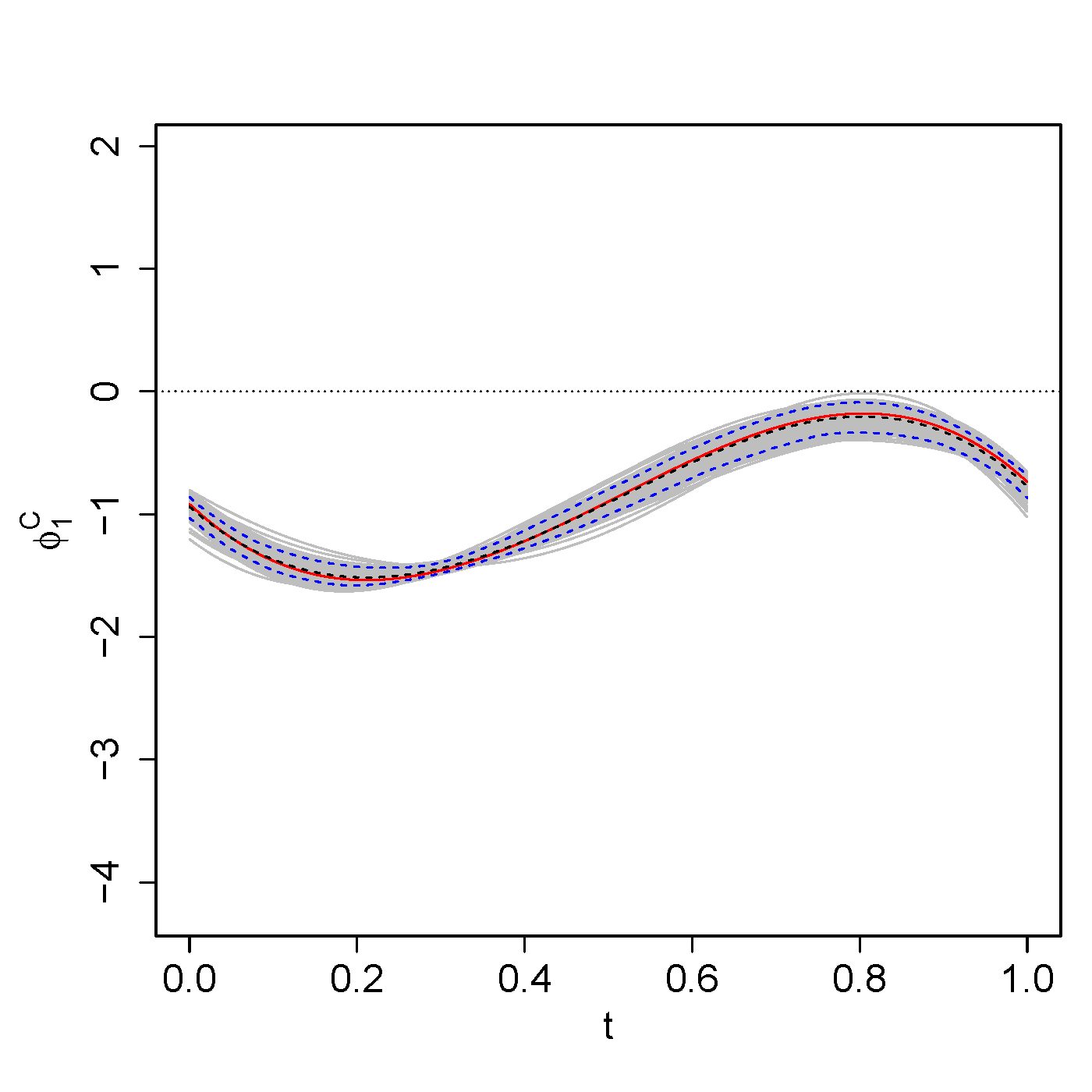}\\
\raisebox{0.15\textwidth}{\textbf{E}}
\includegraphics[width=0.25\textwidth,page=7]{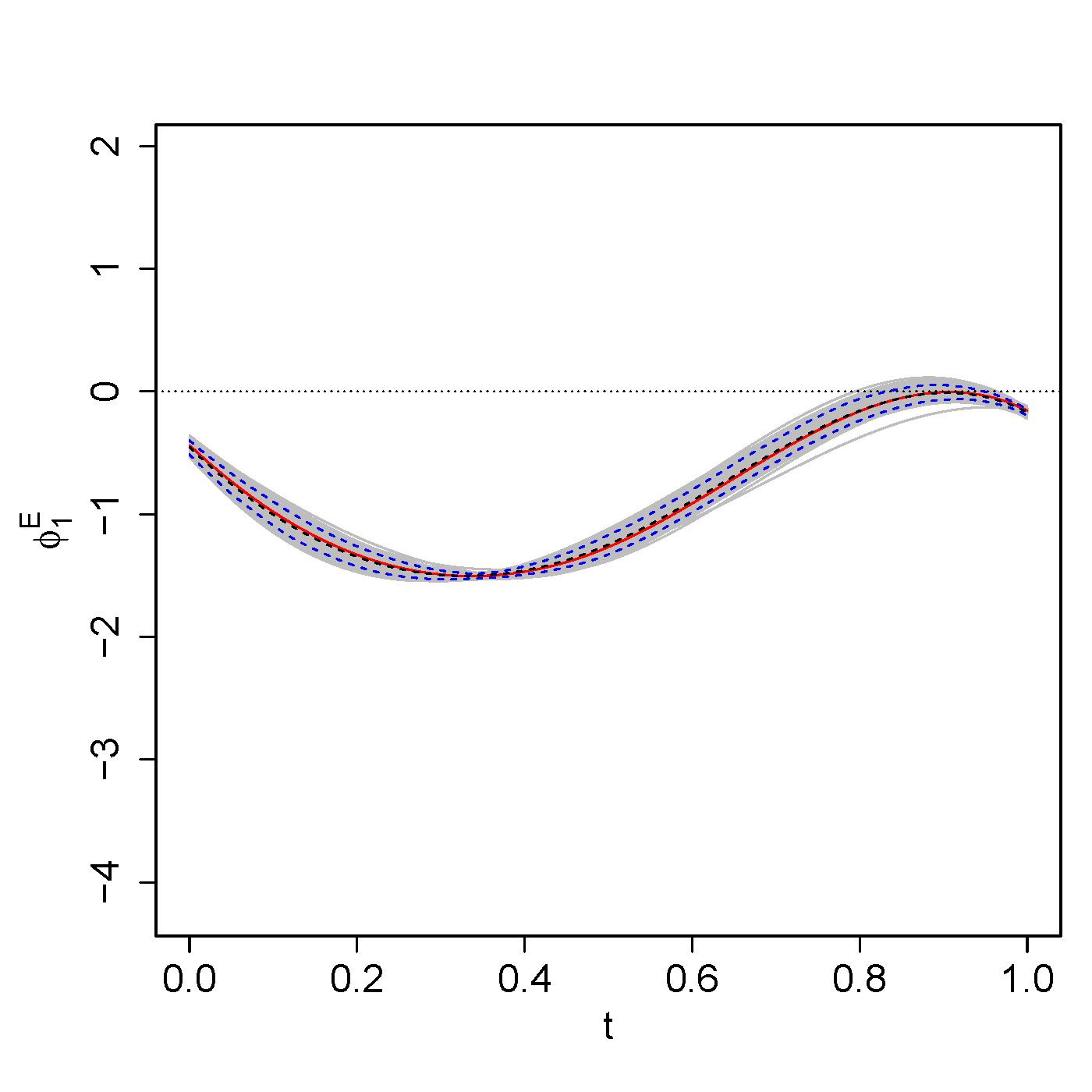}
\raisebox{0.15\textwidth}{\phantom{\textbf{B}}}
\includegraphics[width=0.25\textwidth,page=9]{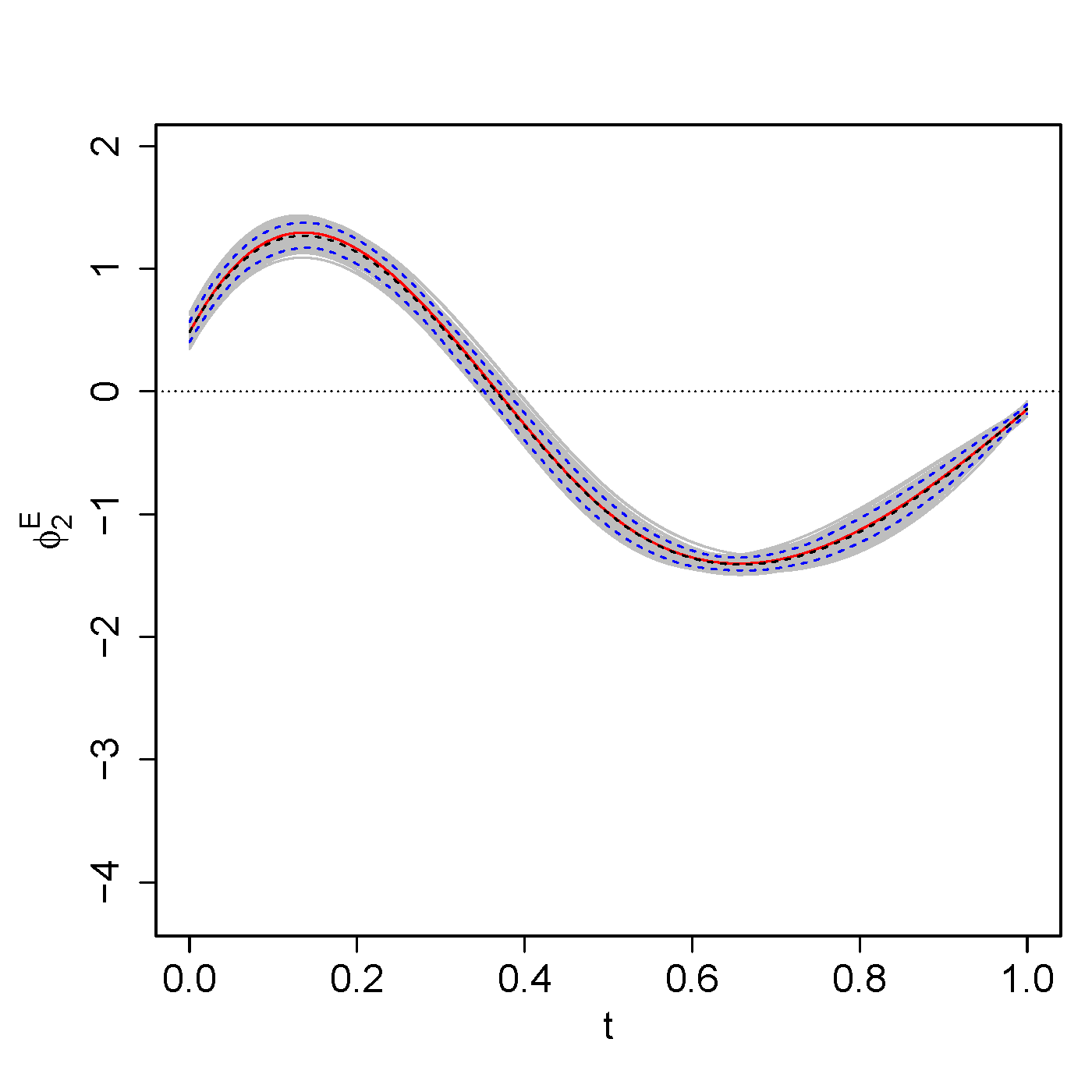}
\raisebox{0.15\textwidth}{\phantom{\textbf{B}}}
\includegraphics[width=0.25\textwidth,page=11]{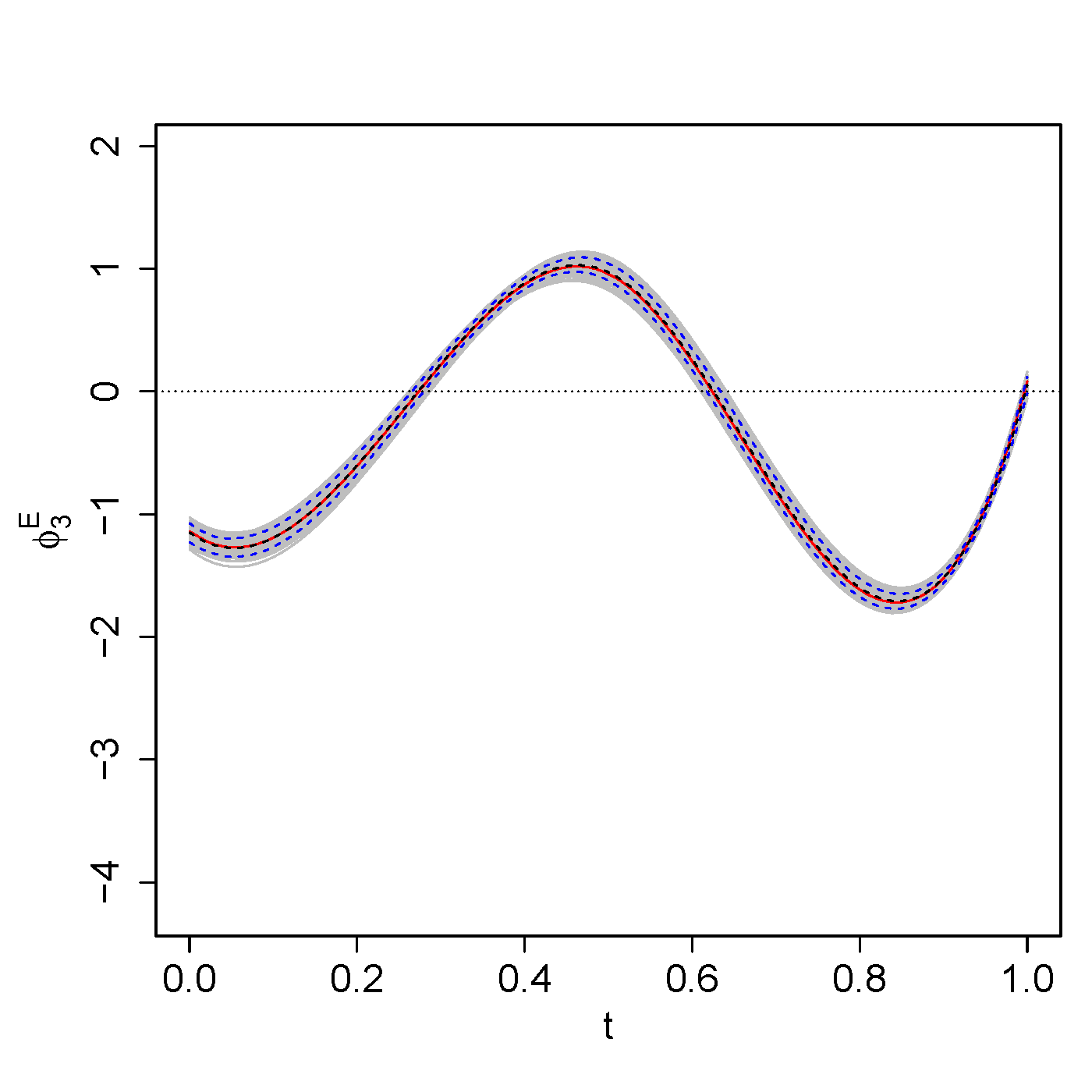}
\end{center}
\end{minipage}
\caption{True and estimated FPCs of the crossed fRIs $B_i(t)$ and $C_j(t)$ (top row), as well as the FPCs of the smooth error $E_{ijh}(t)$ (bottom row). Shown are the true functions (red), the mean of the estimated functions over 200 simulation runs (black dashed line), the point-wise 5th and 95th percentiles of the estimated functions (blue dashed lines), and the estimated functions of all 200 simulation runs (grey).}
\label{fig: eigenfuncionts crossed non-centred}
\end{center}
\end{figure}

\renewcommand{\arraystretch}{0.8} 
 \setlength{\tabcolsep}{1mm}
\begin{table}[h!]
\centering
\small
\caption{rrMSEs averaged over 200 simulation runs for all model components by random process. Rows 1-3: Number of grouping levels $L^X$ and average rrMSE for the functional random effects. Last row: Average rrMSE for the functional response, the mean, and the error variance.}
\begin{tabular}{c|c|rrrrrrrrrrr|rr}
$X$&$L^X$& $K^X$ & $\phi^X_1$ & $\phi^X_2$&  $\phi^X_3$& $\nu^X_1$ & $\nu^X_2$&$\nu^X_3$ & $\xi^X_1$ & $\xi^X_2$ &$\xi^X_3$ & $X$ &$\mu$&$\sigma^2$ \\   \hline
$B$ & 9& 0.55 & 0.37 & 0.46& & 0.38 & 0.53 && 0.41 & 0.70& & 0.38&& \\ 
$C$ & 16 & 0.32 & 0.05&& & 0.31 &&& 0.24 &&& 0.25&& \\ 
$E$ & 707&0.09 & 0.04 & 0.05 & 0.04 & 0.06 & 0.09 & 0.05 & 0.19 & 0.21 & 0.27 & 0.20 &&\\ 
$Y$ & & & & & & & & & & & & 0.10 &0.10 &0.10
\end{tabular}
\label{tab: mean riMSEs crossed non-centred}
\end{table}

\begin{figure}[h!]
\begin{center}
\includegraphics[width=0.2\textwidth,page=1]{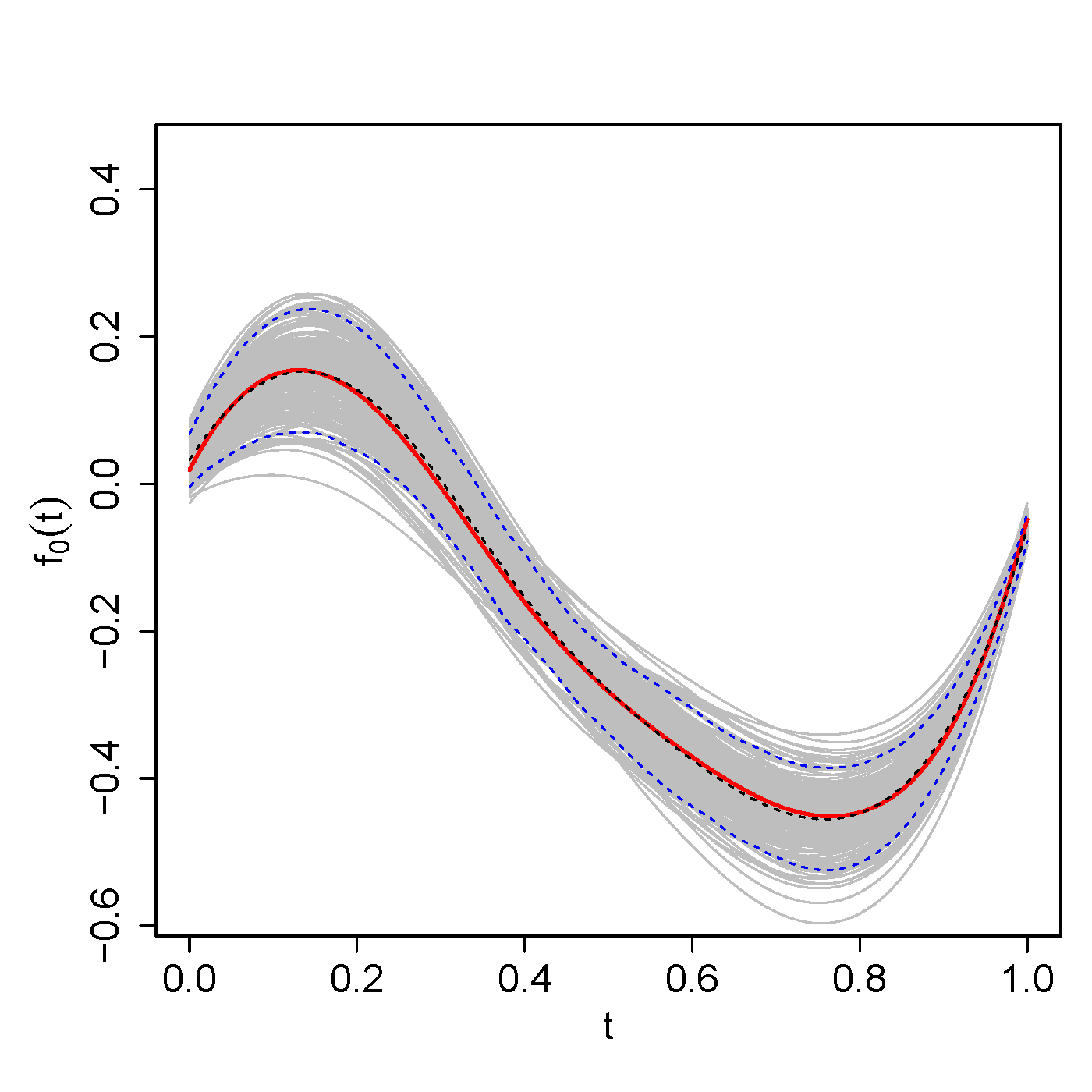}
\includegraphics[width=0.2\textwidth,page=3]{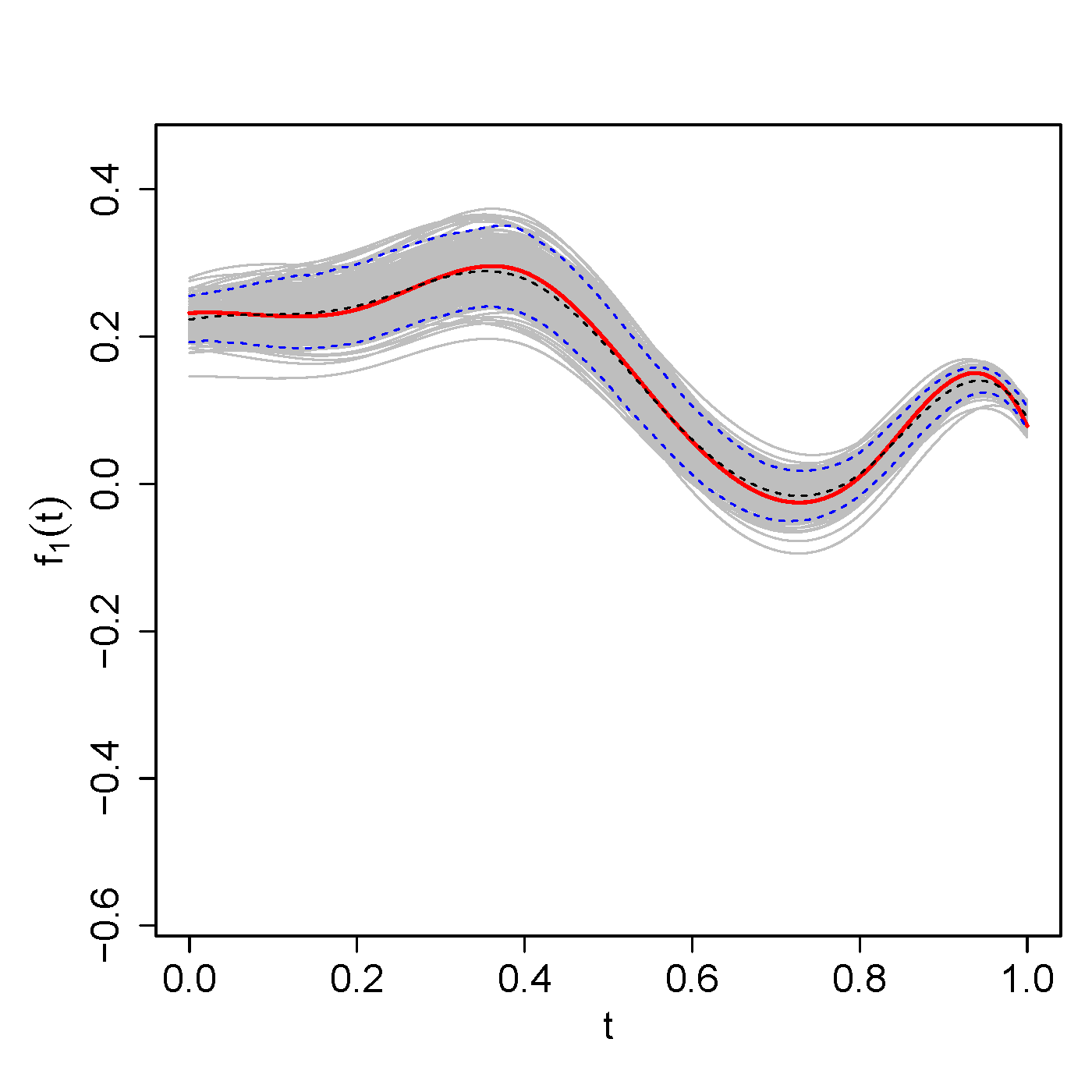}
\includegraphics[width=0.2\textwidth,page=5]{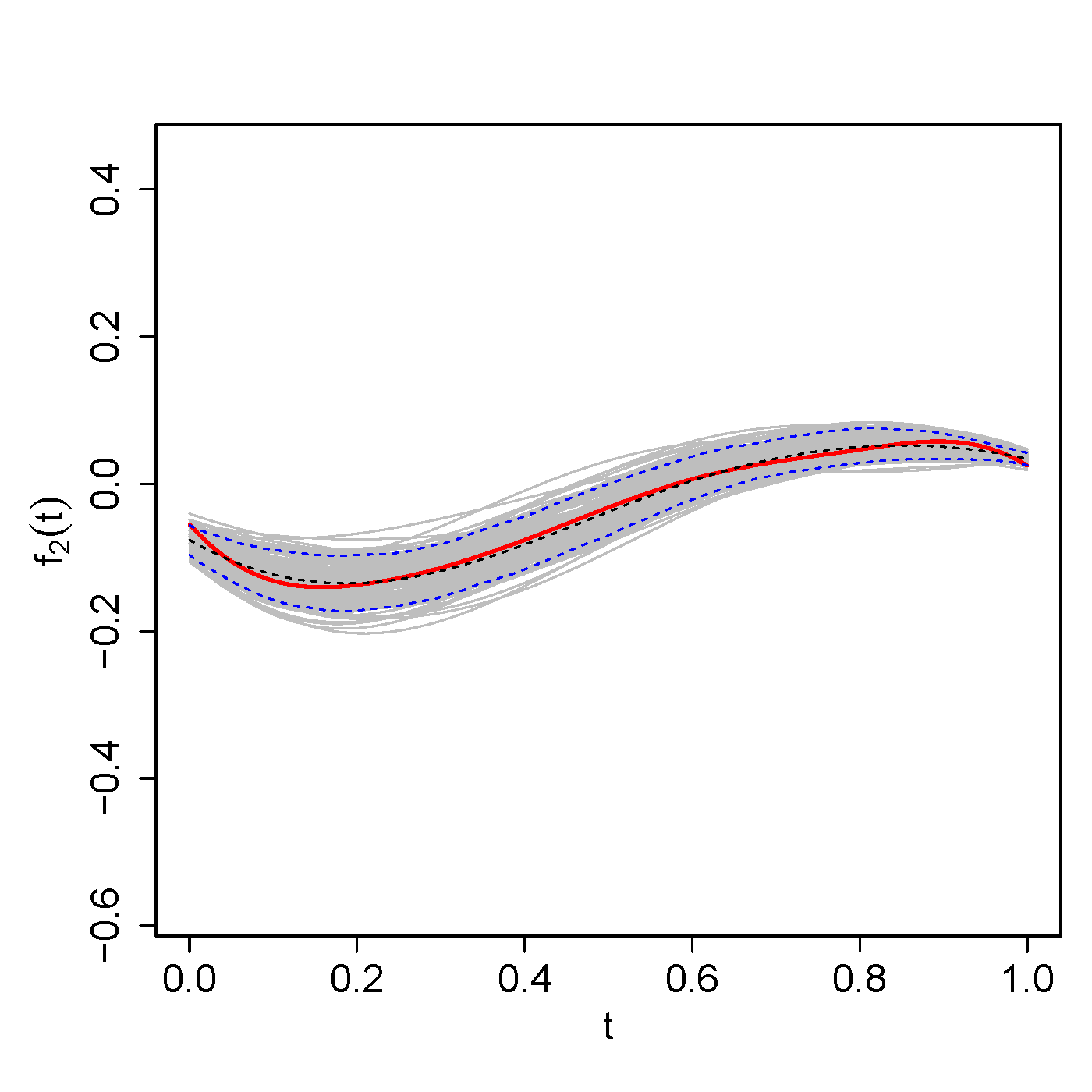}
\includegraphics[width=0.2\textwidth,page=7]{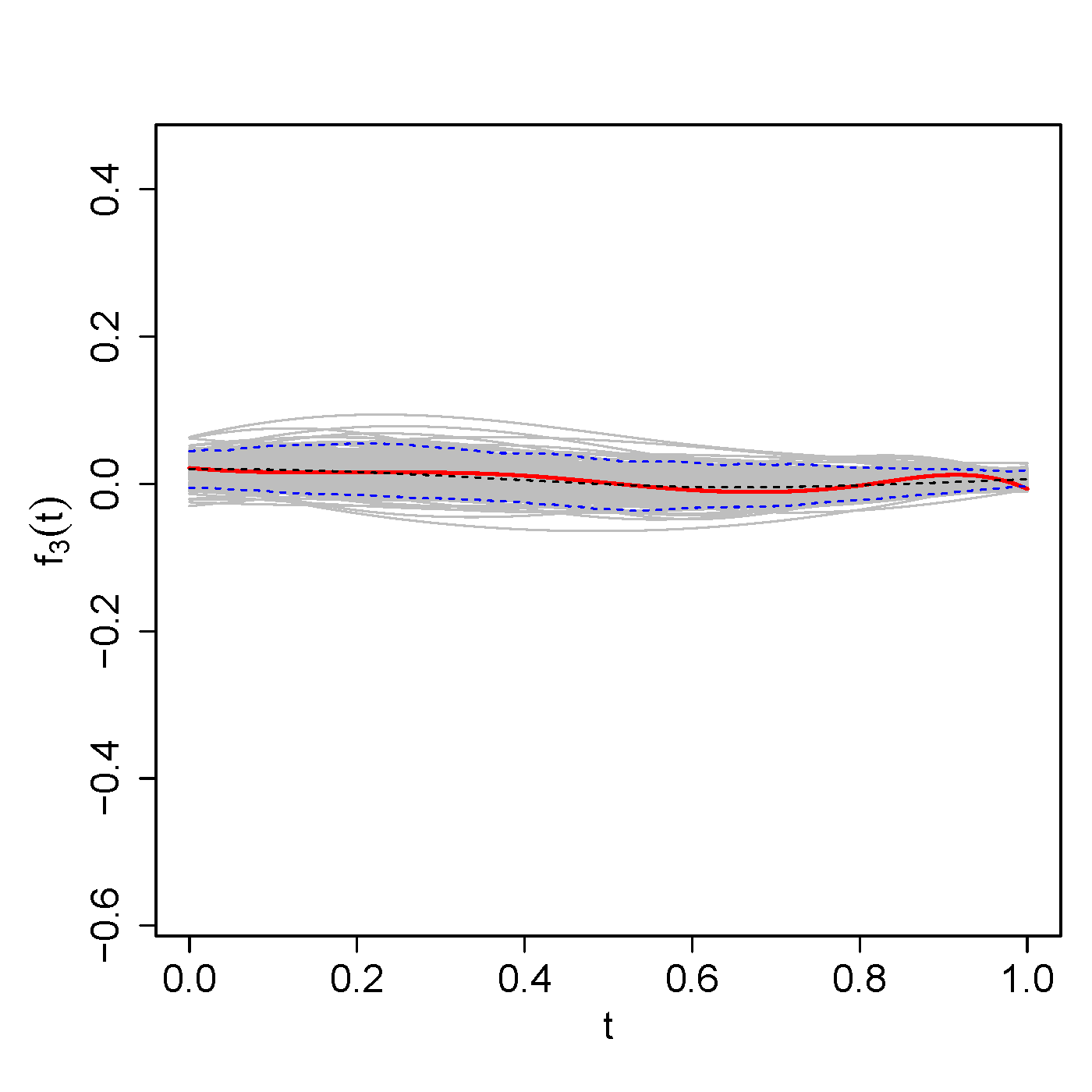}\\
\includegraphics[width=0.2\textwidth,page=9]{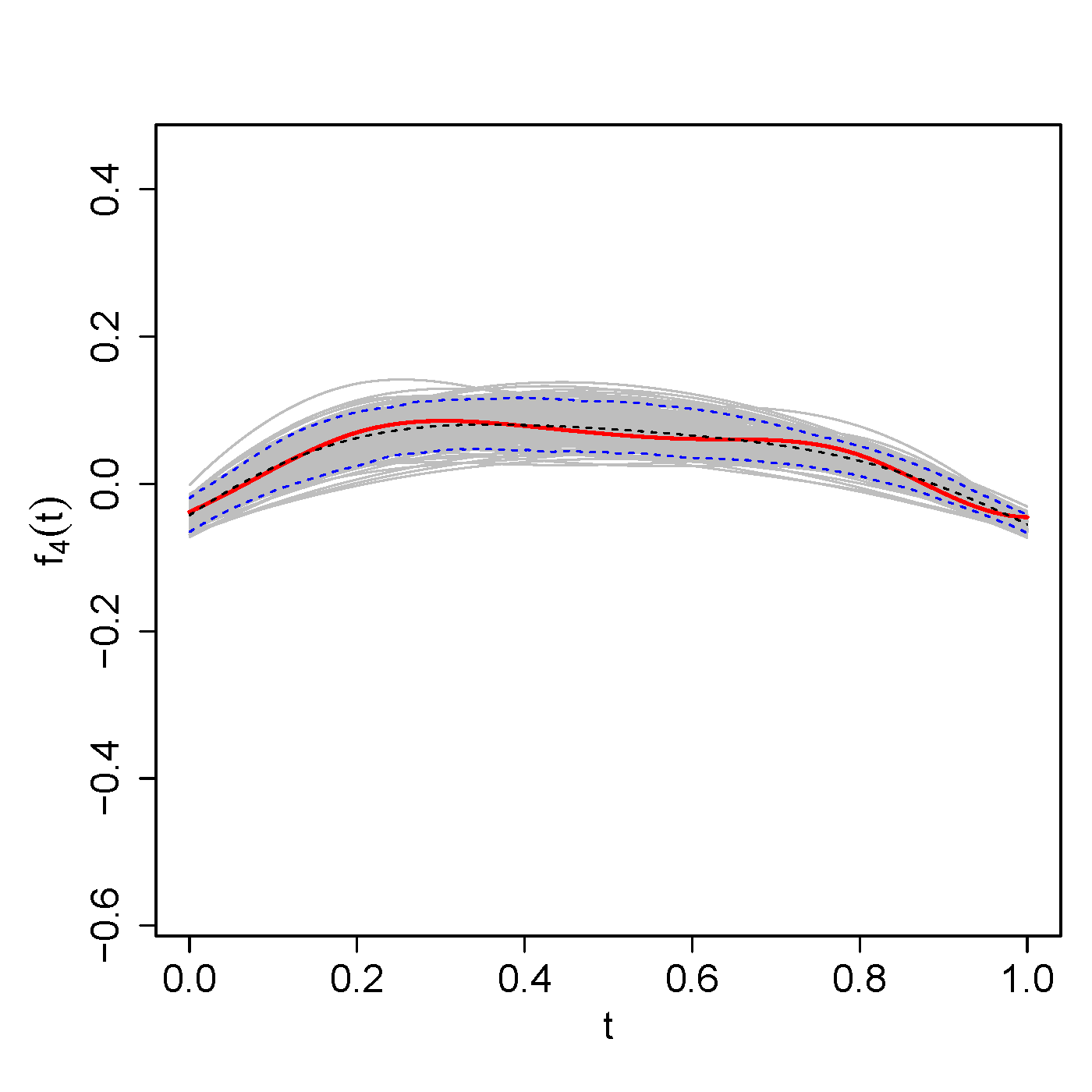}
\includegraphics[width=0.2\textwidth,page=11]{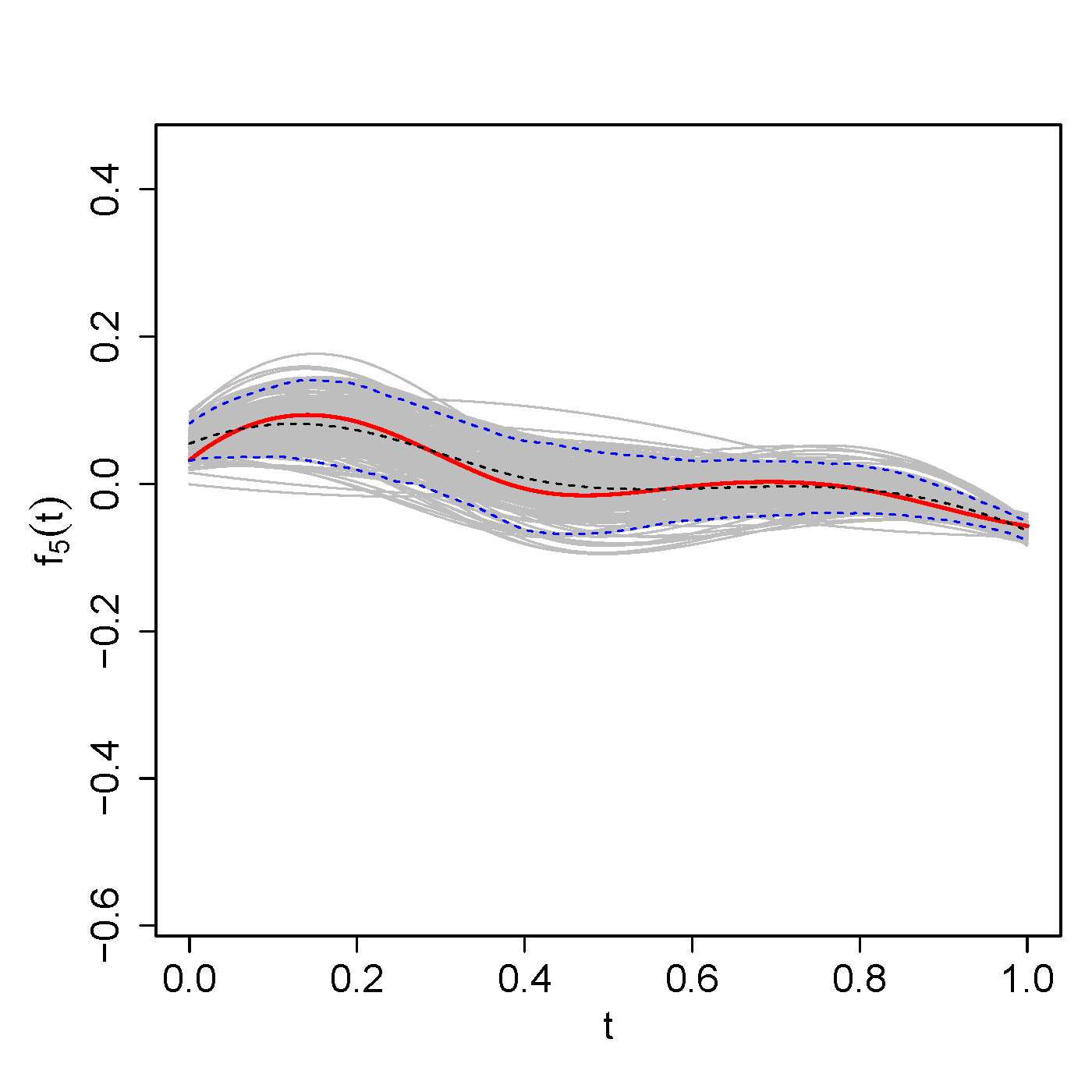}
\includegraphics[width=0.2\textwidth,page=13]{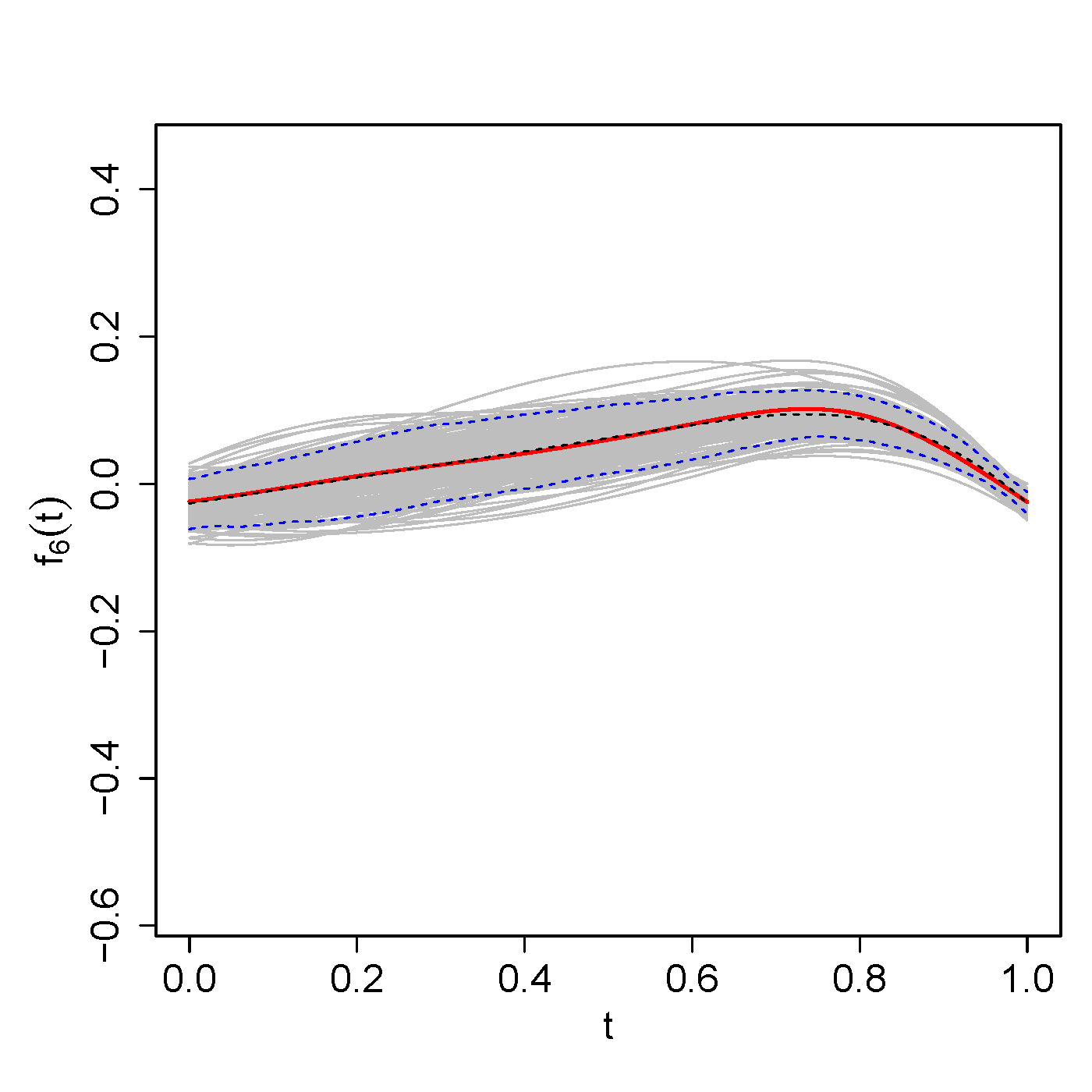}
\includegraphics[width=0.2\textwidth,page=15]{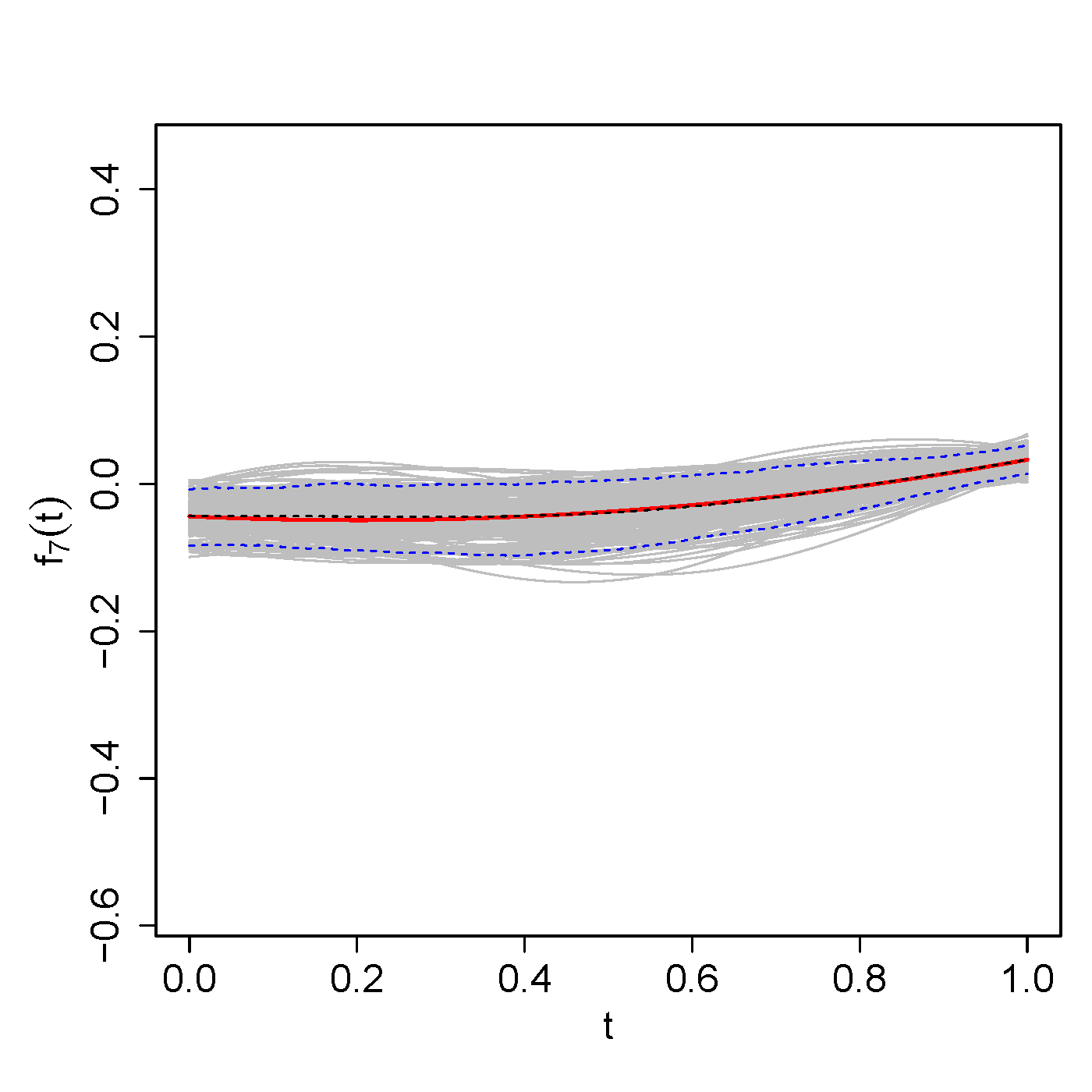}
\caption{True and estimated covariate and interaction effects estimated based on the independence assumption. Shown are the true function (red), the mean of the estimated functions over 200 simulation runs (black dashed line), the point-wise 5th and 95th percentiles of the estimated functions (blue dashed lines), and the estimated functions of all 200 simulation runs (grey).}
\label{fig: mean RI non-centred}
\end{center}
\end{figure}

\begin{figure}[h!]
\begin{center}
\includegraphics[width=0.2\textwidth,page=1]{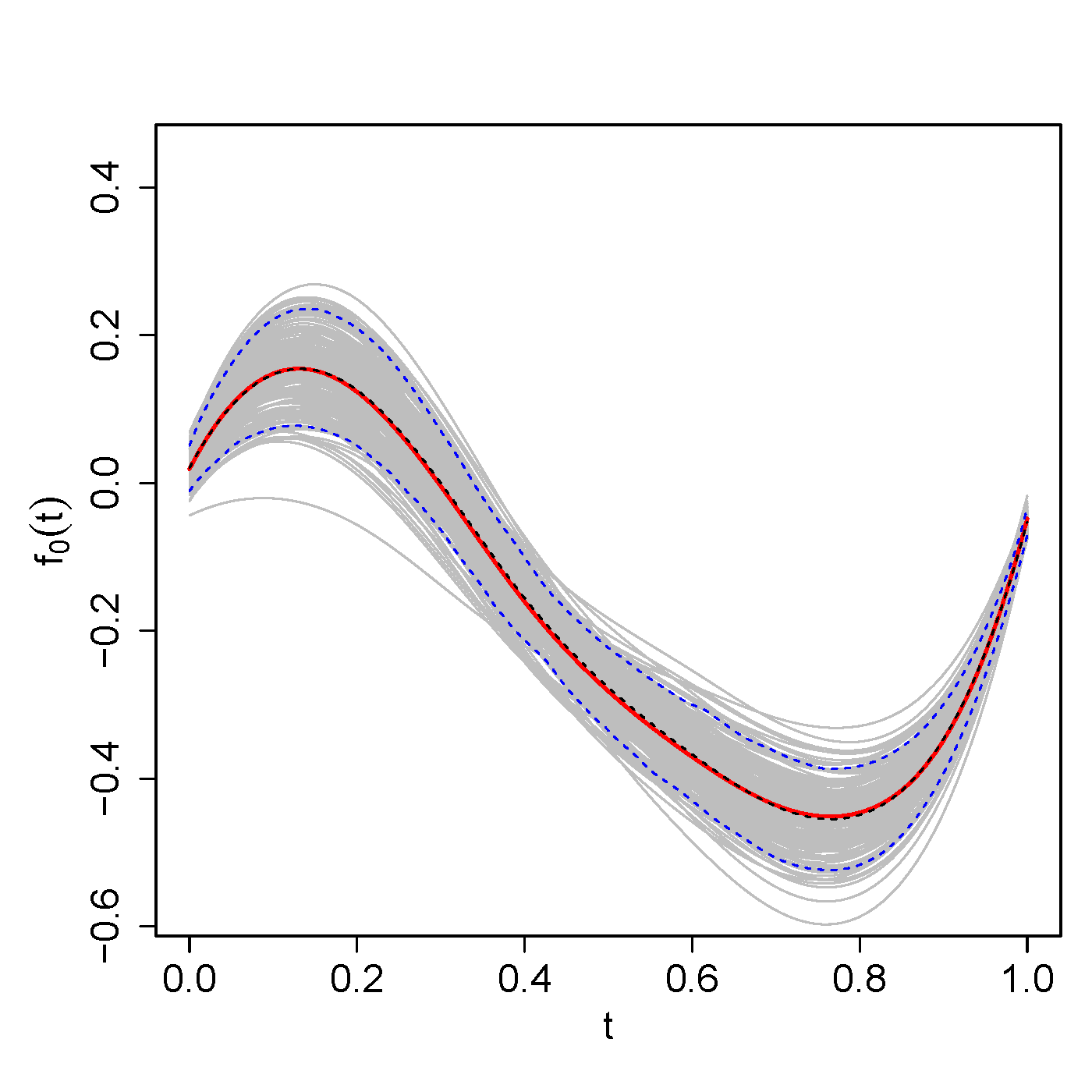}
\includegraphics[width=0.2\textwidth,page=3]{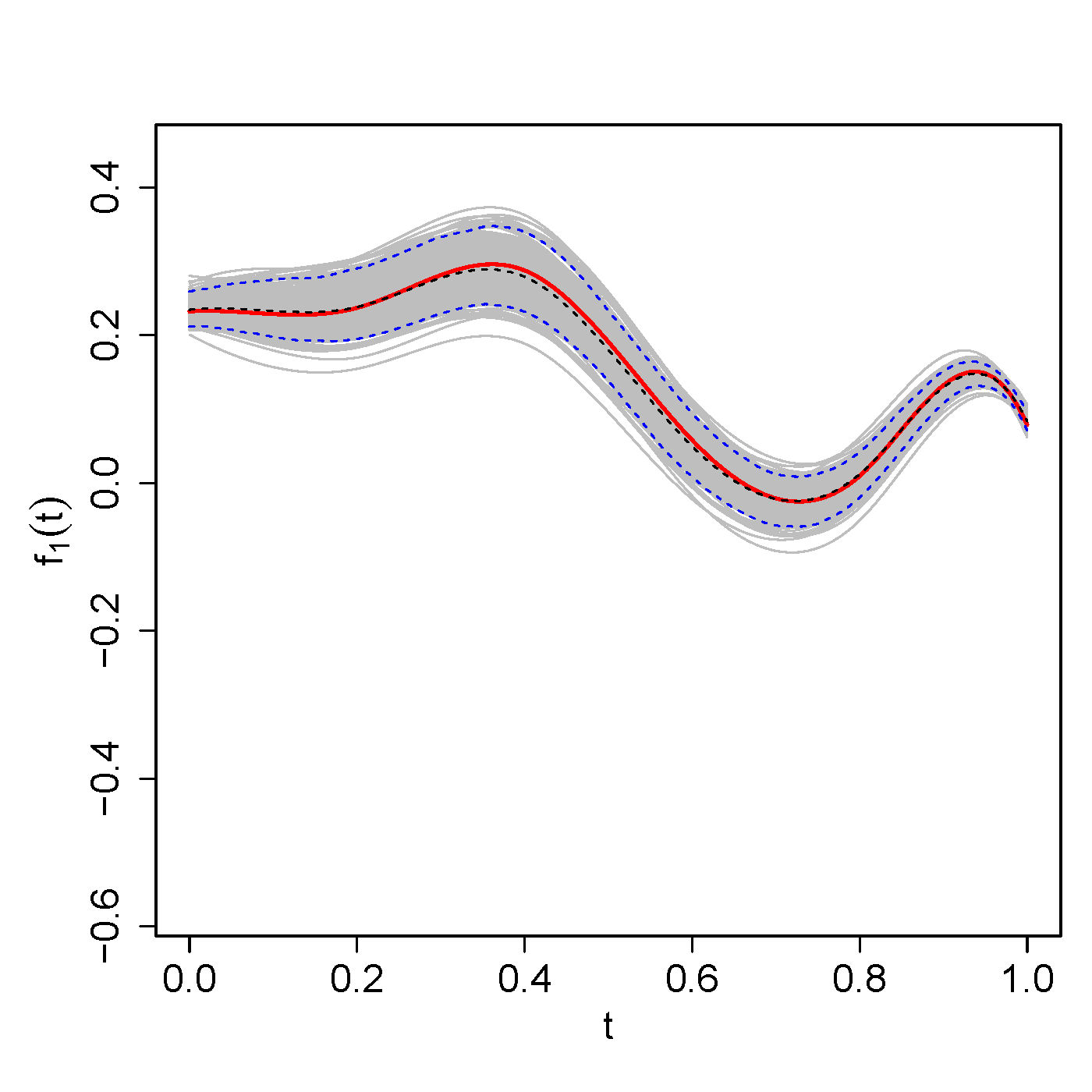}
\includegraphics[width=0.2\textwidth,page=5]{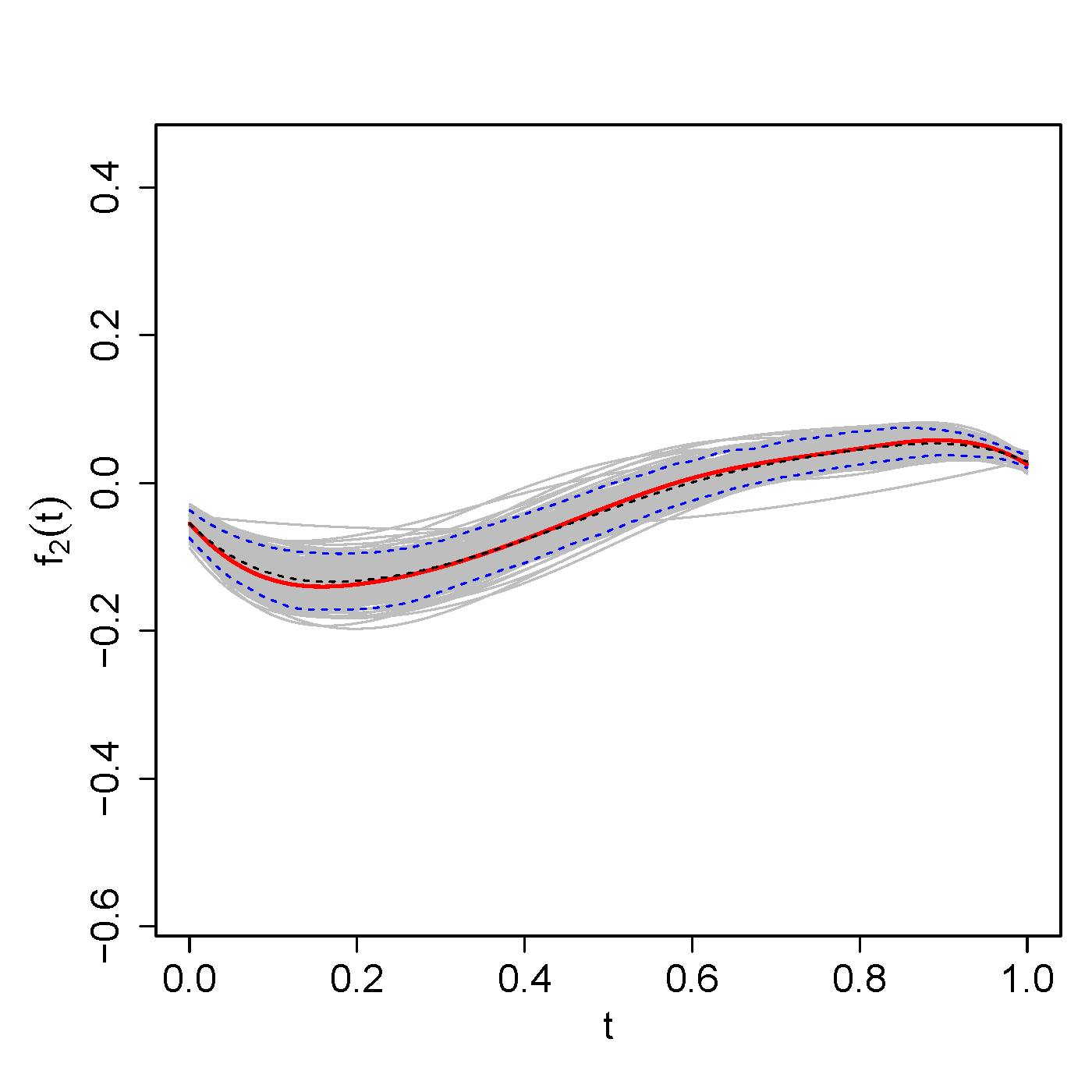}
\includegraphics[width=0.2\textwidth,page=7]{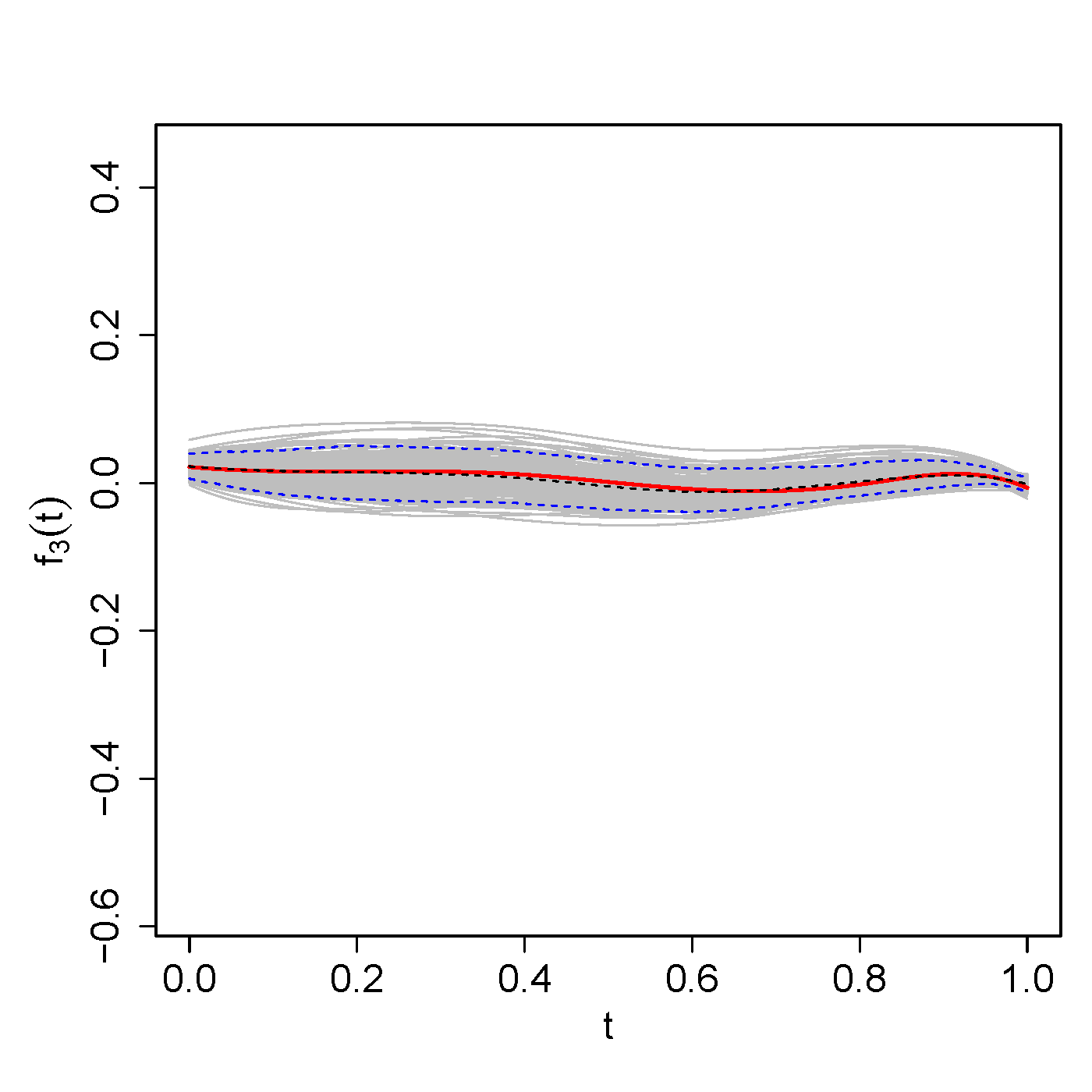}\\
\includegraphics[width=0.2\textwidth,page=9]{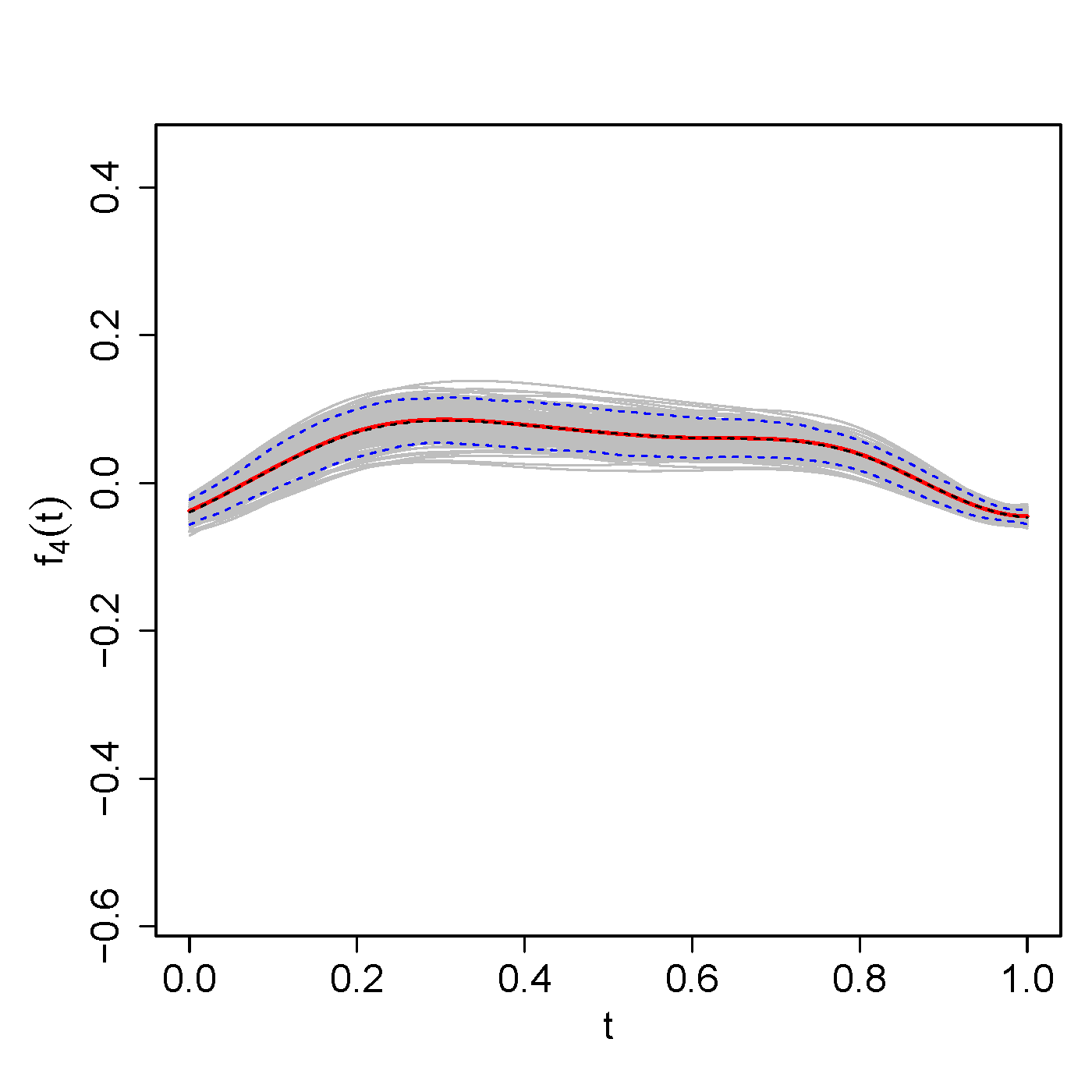}
\includegraphics[width=0.2\textwidth,page=11]{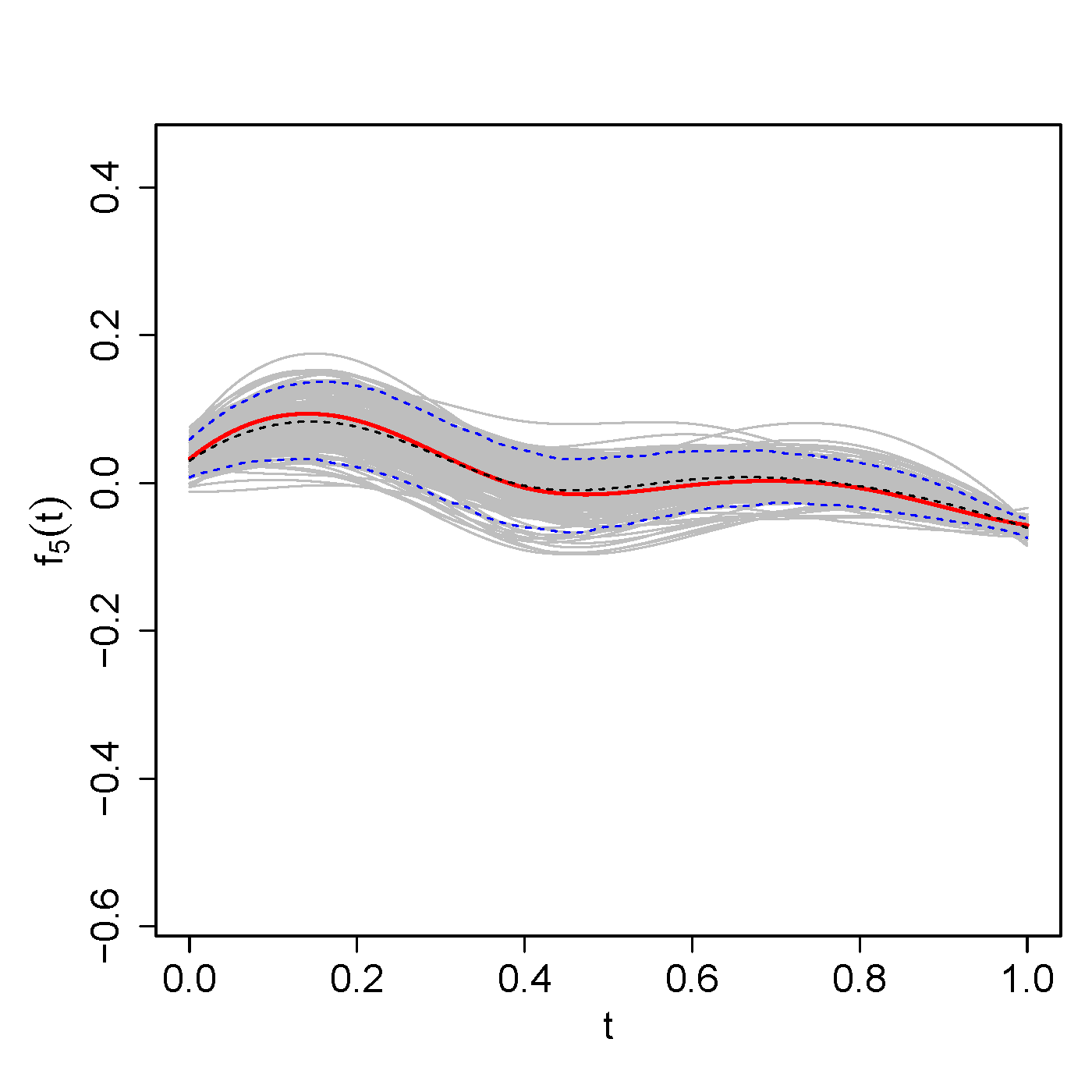}
\includegraphics[width=0.2\textwidth,page=13]{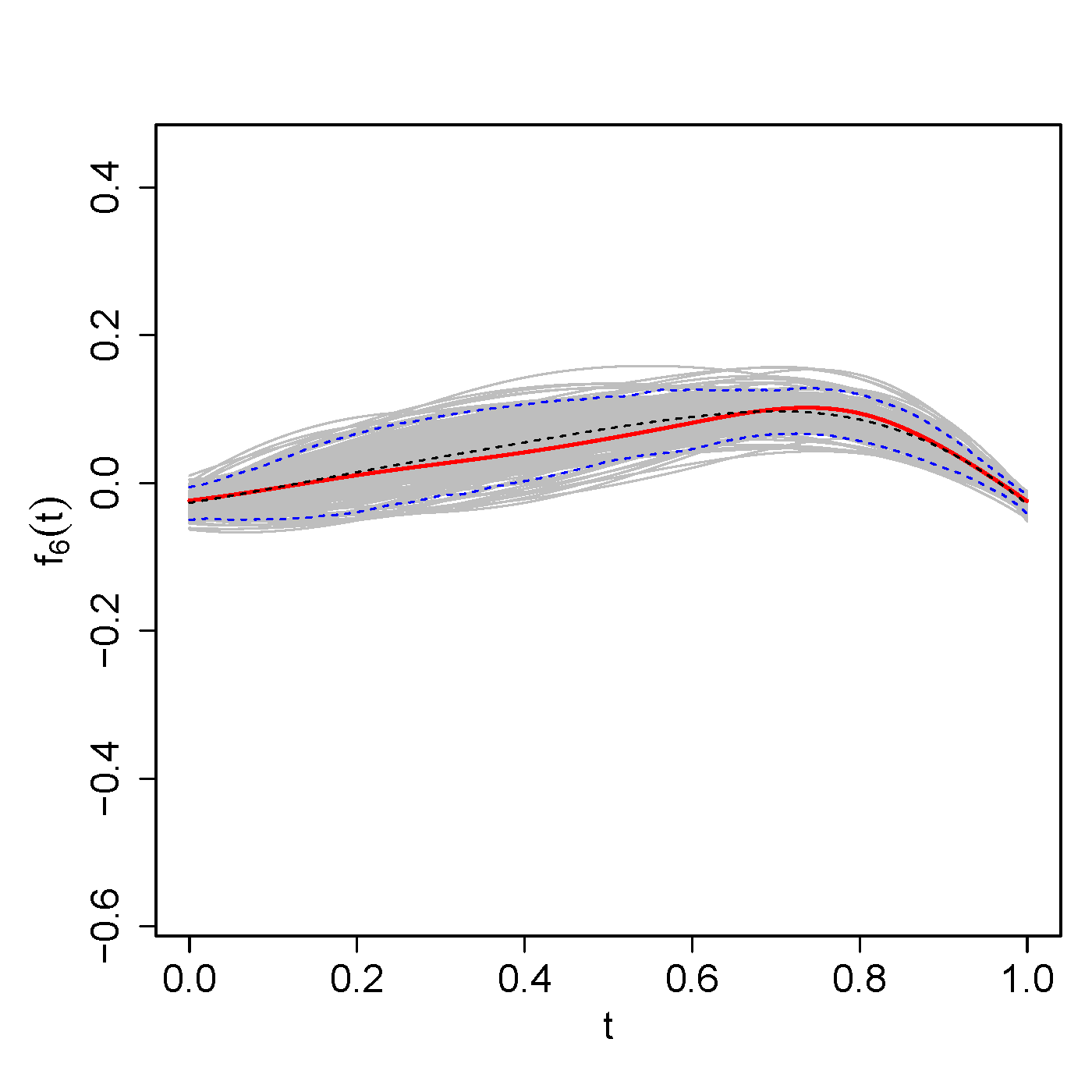}
\includegraphics[width=0.2\textwidth,page=15]{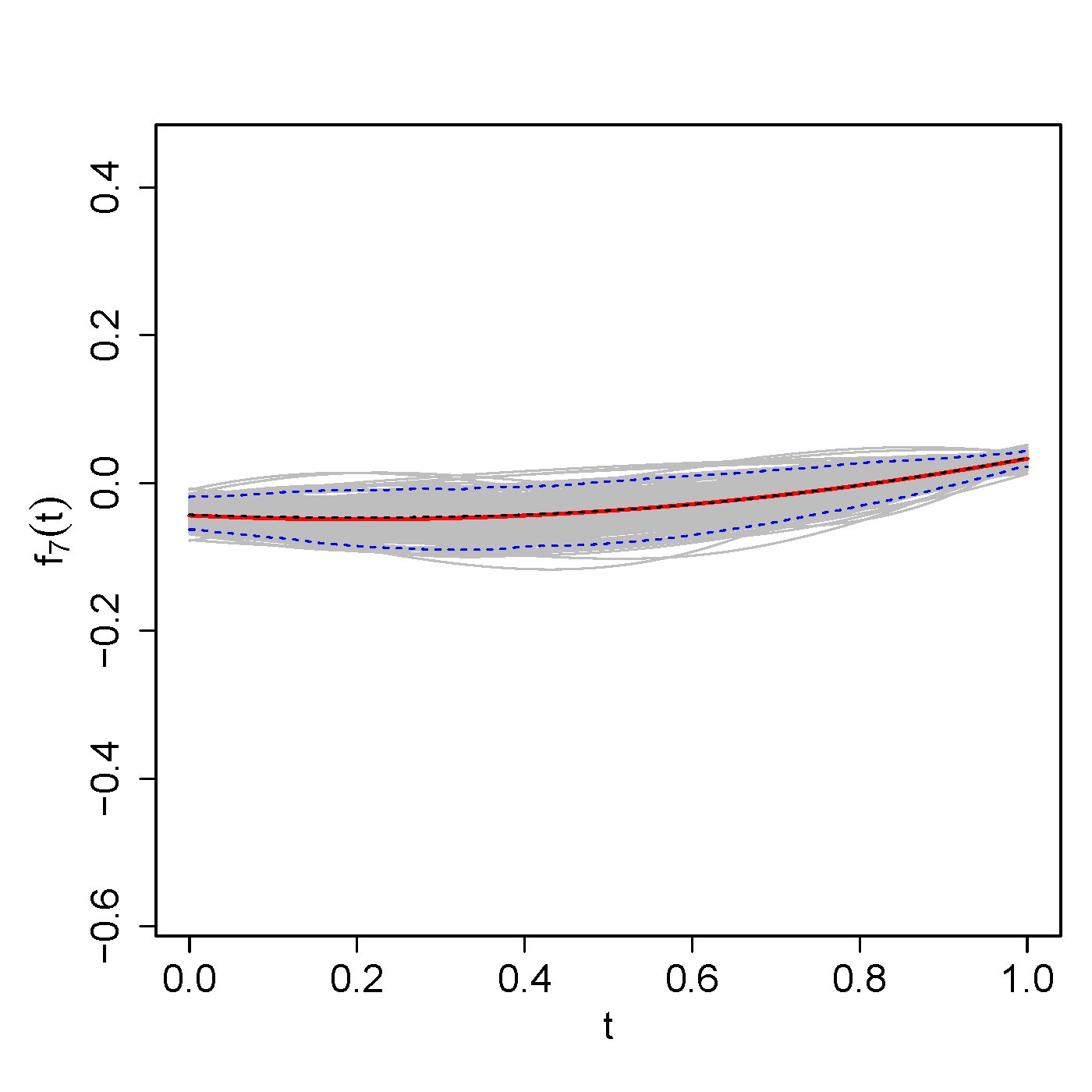}
\caption{True and estimated covariate and interaction effects using FPC-FAMM. Shown are the true function (red), the mean of the estimated functions over 200 simulation runs (black dashed line), the point-wise 5th and 95th percentiles of the estimated functions (blue dashed lines), and the estimated functions of all 200 simulation runs (grey).}
\label{fig: mean RI pffr non-centred}
\end{center}
\end{figure}

\begin{figure}[h!]
\begin{center}
\includegraphics[width=0.2\textwidth,page=1]{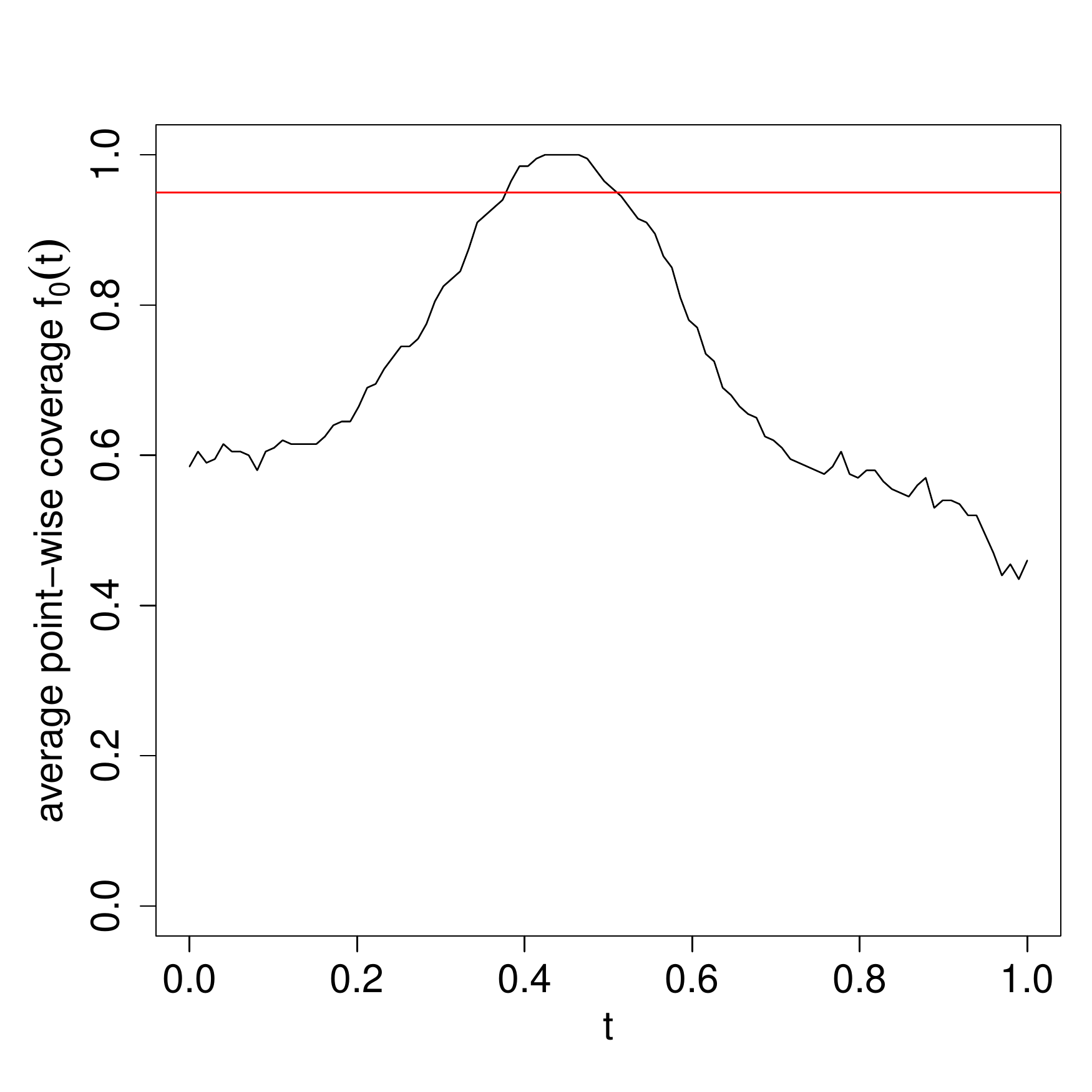}
\includegraphics[width=0.2\textwidth,page=2]{figures/simulation/covarage_mean_and_covariates_pffr_normal_I9_JNA_RI_as_data_30_Mar_30_Mar.pdf}
\includegraphics[width=0.2\textwidth,page=3]{figures/simulation/covarage_mean_and_covariates_pffr_normal_I9_JNA_RI_as_data_30_Mar_30_Mar.pdf}
\includegraphics[width=0.2\textwidth,page=4]{figures/simulation/covarage_mean_and_covariates_pffr_normal_I9_JNA_RI_as_data_30_Mar_30_Mar.pdf}\\
\includegraphics[width=0.2\textwidth,page=5]{figures/simulation/covarage_mean_and_covariates_pffr_normal_I9_JNA_RI_as_data_30_Mar_30_Mar.pdf}
\includegraphics[width=0.2\textwidth,page=6]{figures/simulation/covarage_mean_and_covariates_pffr_normal_I9_JNA_RI_as_data_30_Mar_30_Mar.pdf}
\includegraphics[width=0.2\textwidth,page=7]{figures/simulation/covarage_mean_and_covariates_pffr_normal_I9_JNA_RI_as_data_30_Mar_30_Mar.pdf}
\includegraphics[width=0.2\textwidth,page=8]{figures/simulation/covarage_mean_and_covariates_pffr_normal_I9_JNA_RI_as_data_30_Mar_30_Mar.pdf}
\caption{Average point-wise coverage of the point-wise CBs obtained by FPC-FAMM for all covariate and interaction effects. For each effect, the point-wise coverage averaged over 200 simulation runs (black line) is shown. The red line indicates the nominal value of 0.95.}
\label{fig: coverage RI pffr non-centred}
\end{center}
\end{figure}

\clearpage
\underline{Simulation results for the fRI scenario}.
In the following, we show additional simulation results for the simulations of the application-based fRI scenario with non-centred and non-decorrelated basis weights. Figure \ref{fig: mean RI non-centred} and Figure \ref{fig: mean RI pffr non-centred} show the true and estimated covariate and interaction effects estimated based on the independence assumption, and on FPC-FAMM, respectively. We evaluate the performance of the point-wise CBs by looking at the point-wise coverage shown in Figure \ref{fig: coverage RI pffr non-centred}. We additionally look at the simultaneous coverage of the point-wise CBs in terms of percentage of completely covered curves in Table \ref{tab: prop covered curves RI pffr non-centred}.

 \setlength{\tabcolsep}{1mm}
\begin{table}[h!]
\centering
\small
\caption{Simultaneous coverage of the point-wise CBs obtained by FPC-FAMM. Shown is the proportion of completely covered curves for all covariate and interaction effects. The coverage refers to 200 simulation runs.}
\begin{tabular}{l|rrrrrrrr}
 & $f_0(t)$ & $f_1(t)$ & $f_2(t)$ & $f_3(t)$ &$f_4(t)$ & $f_5(t)$ & $f_6(t)$ & $f_7(t)$ \\ 
  \hline 
$\mu(t,\mx_{ijh})_{\FPCFAMM}$&    11\% &  71\% & 71.5\% & 65.5\% & 74.5\% & 69\% &  67\% &  78\% \\ 
\end{tabular}
\label{tab: prop covered curves RI pffr non-centred}
\end{table}

\begin{figure}[h!]
\begin{center}
\begin{minipage}{1\textwidth}
\begin{center}
\raisebox{0.15\textwidth}{\textbf{B}}
\includegraphics[width=0.25\textwidth,page=1]{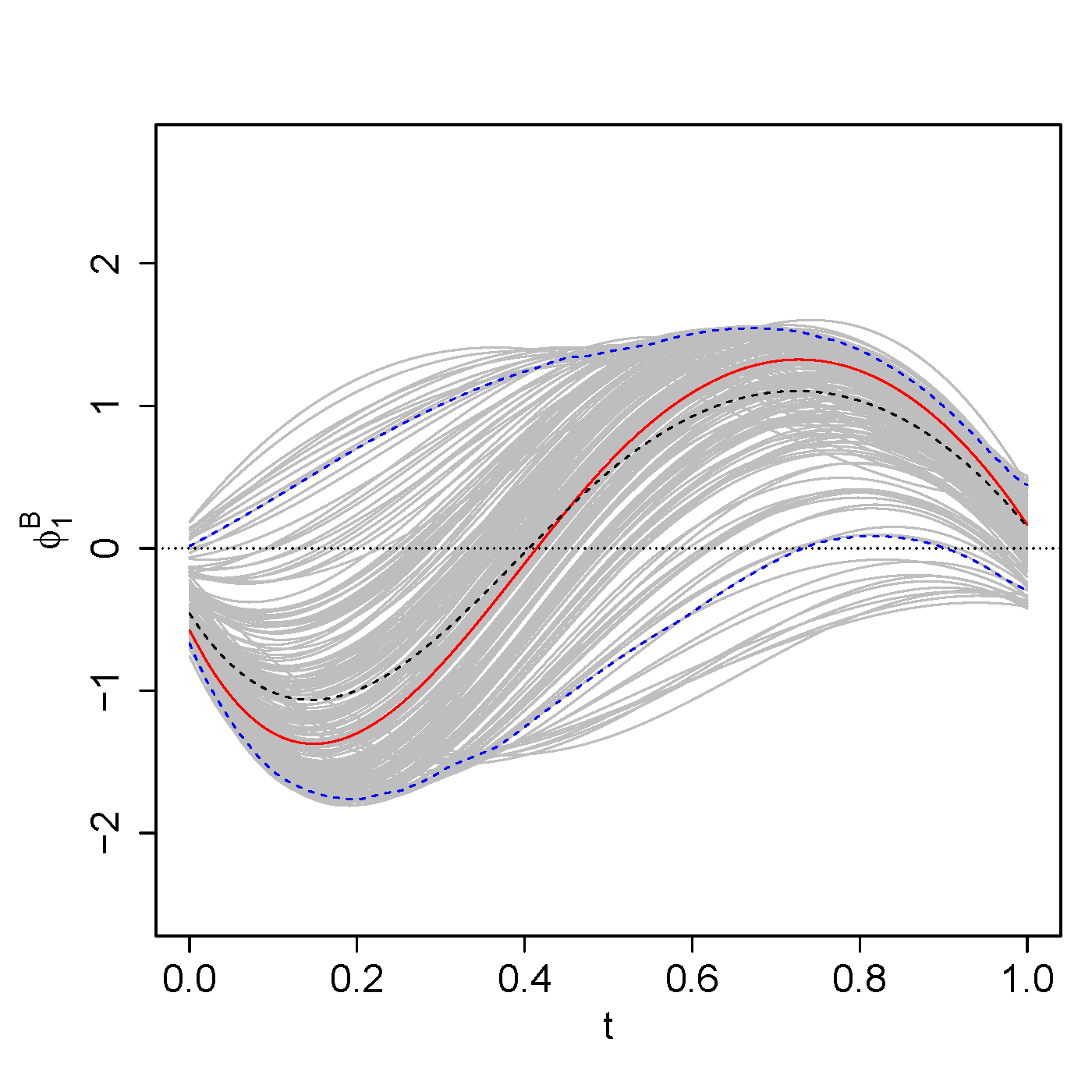}
\raisebox{0.15\textwidth}{\phantom{\textbf{B}}}
\includegraphics[width=0.25\textwidth,page=3]{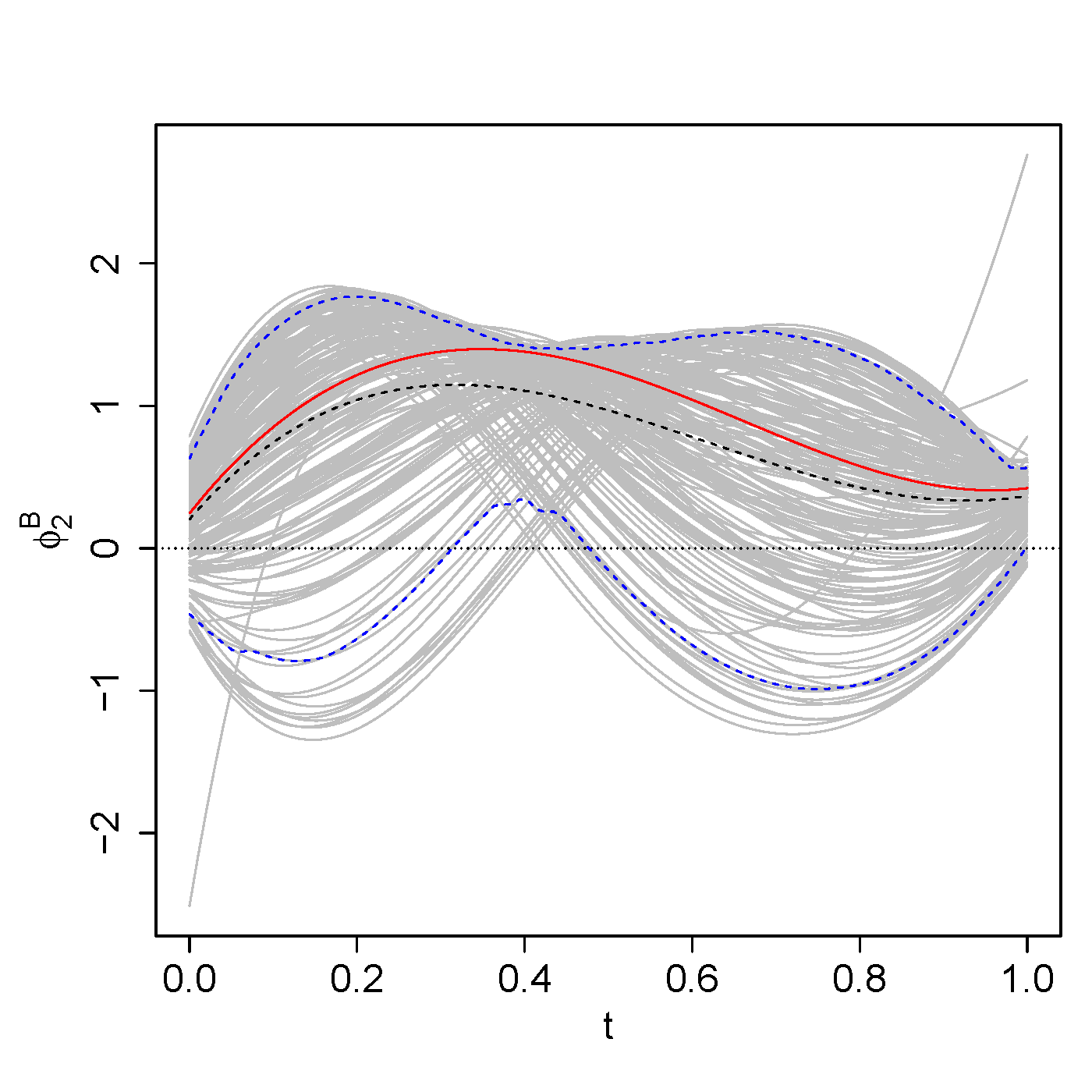}
\raisebox{0.15\textwidth}{\phantom{\textbf{B}}}
\includegraphics[width=0.25\textwidth,page=1]{figures/simulation/blank.pdf}\\
\raisebox{0.15\textwidth}{\textbf{E}}
\includegraphics[width=0.25\textwidth,page=5]{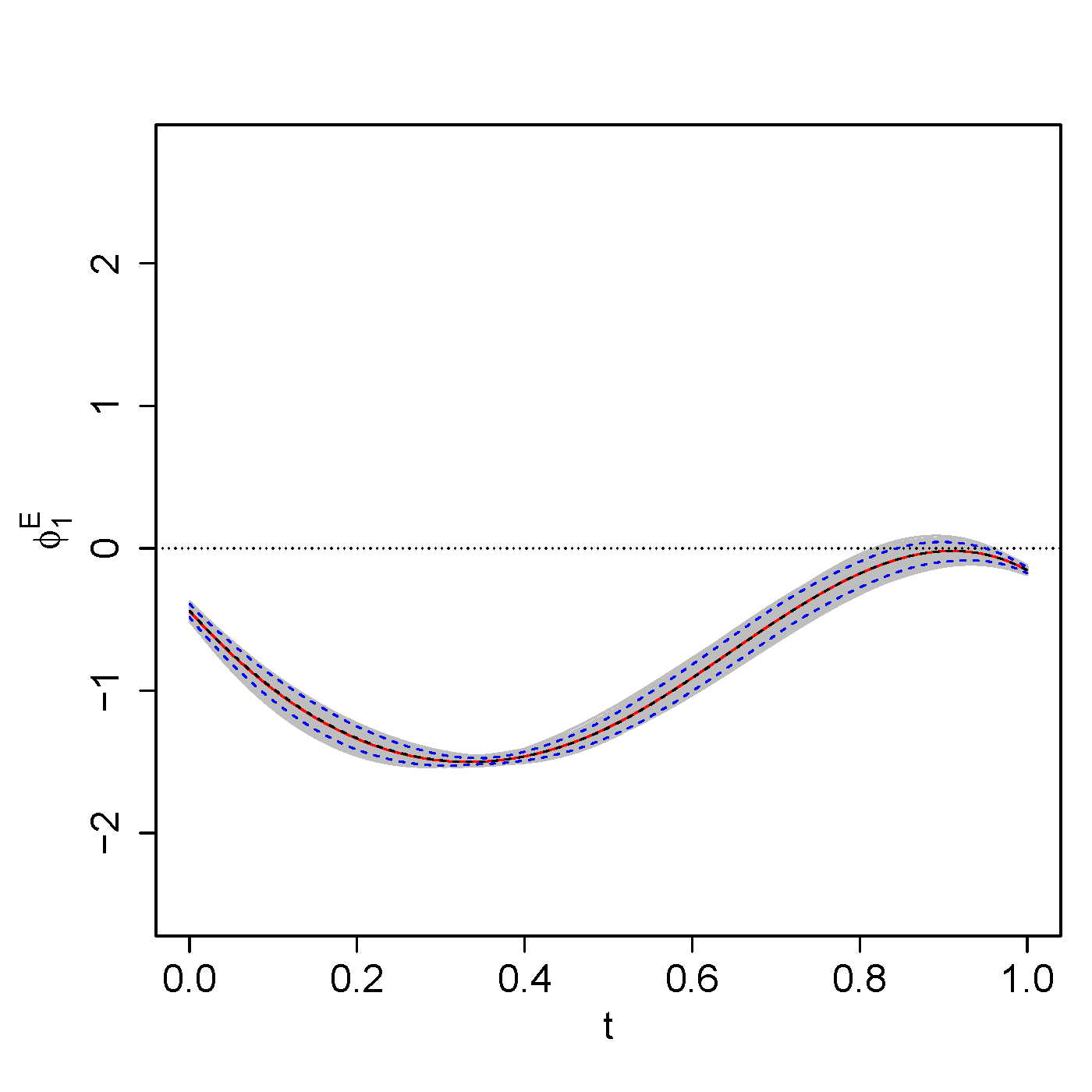}
\raisebox{0.15\textwidth}{\phantom{\textbf{B}}}
\includegraphics[width=0.25\textwidth,page=7]{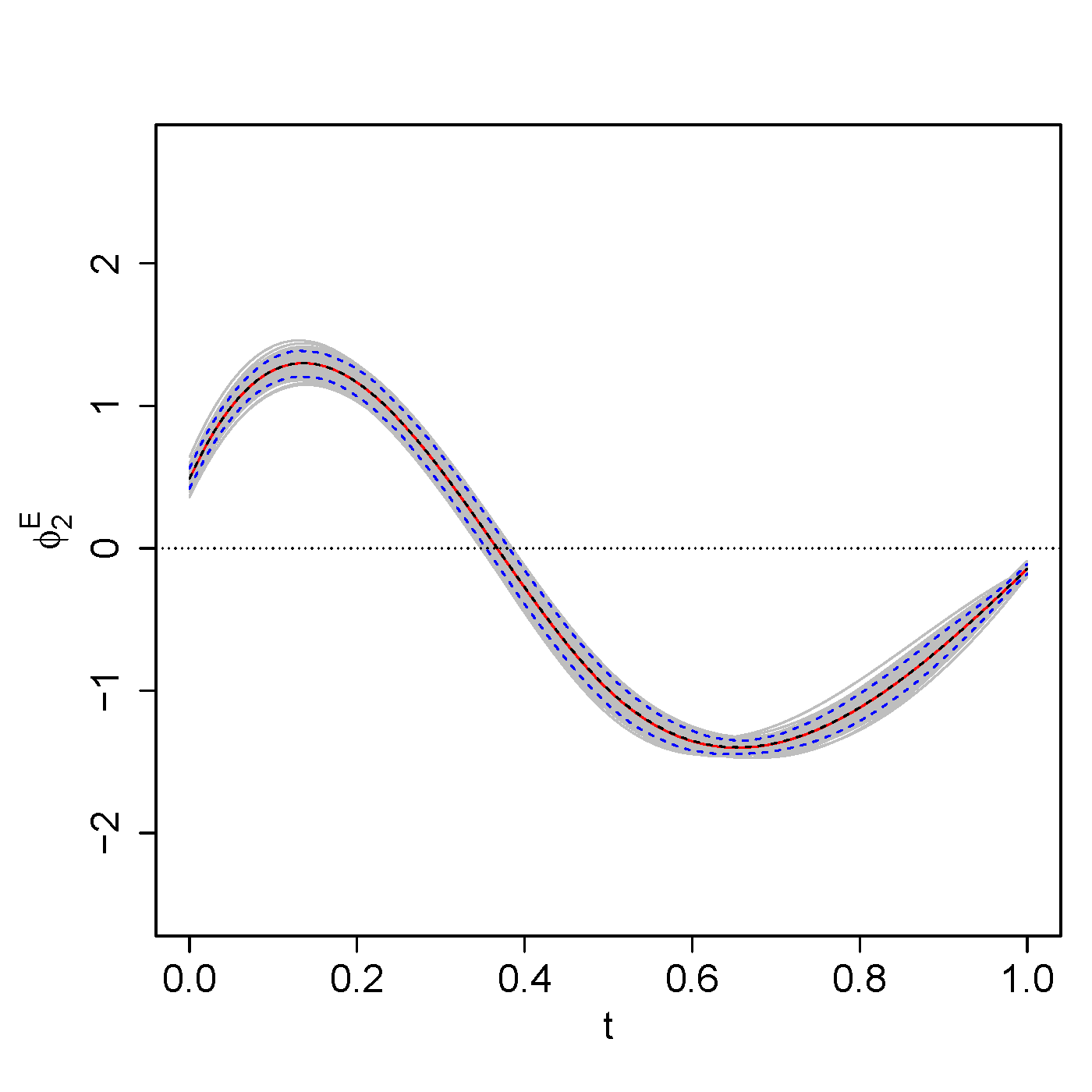}
\raisebox{0.15\textwidth}{\phantom{\textbf{B}}}
\includegraphics[width=0.25\textwidth,page=9]{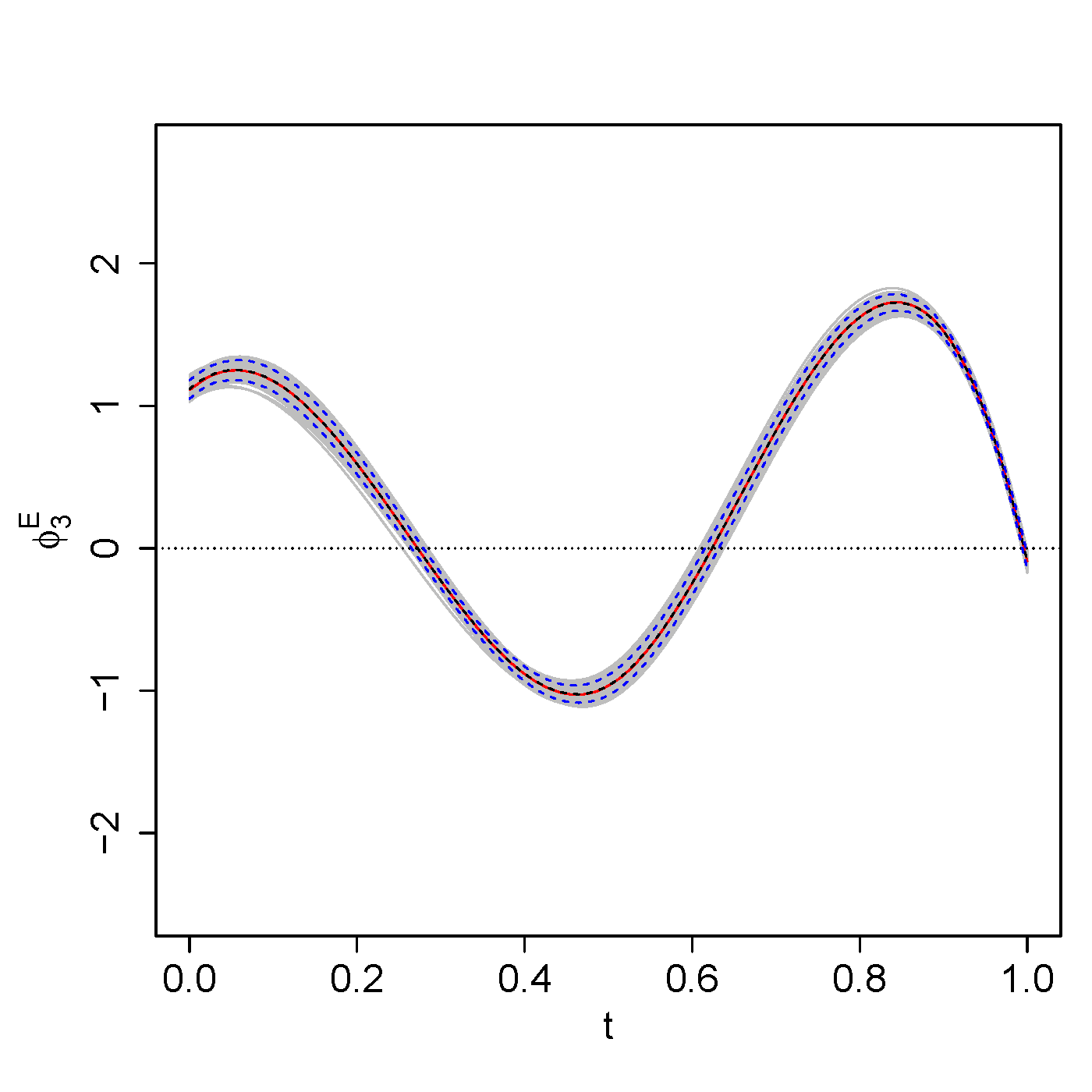}
\end{center}
\end{minipage}
\caption{True and estimated FPCs of the fRI $B_i(t)$ (top row) and of the smooth error (bottom row). Shown are the true functions (red), the mean of the estimated functions over 200 simulation runs (black dashed line), the point-wise 5th and 95th percentiles of the estimated functions (blue dashed lines), and the estimated functions of all 200 simulation runs (grey).}
\label{fig: eigenfuncionts RI non-centred}
\end{center}
\end{figure}

\begin{figure}[h!]
\begin{center}
\begin{minipage}{1\textwidth}
\begin{center}
\raisebox{0.15\textwidth}{\textbf{B}}
\includegraphics[width=0.25\textwidth,page=1]{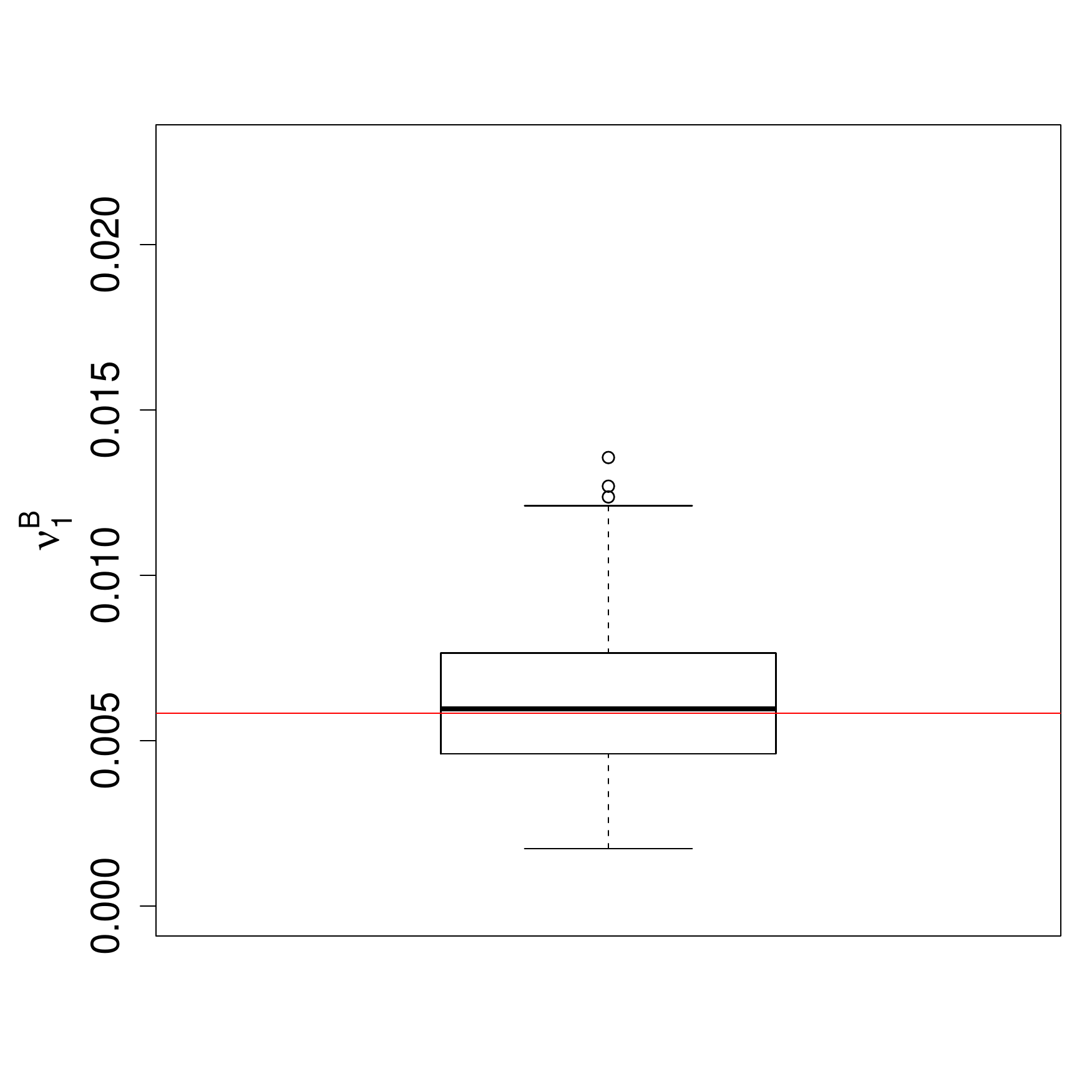}
\raisebox{0.15\textwidth}{\phantom{\textbf{B}}}
\includegraphics[width=0.25\textwidth,page=2]{figures/simulation/eigenvalues_normal_I9_JNA_RI_as_data_30_Mar_30_Mar.pdf}
\raisebox{0.15\textwidth}{\phantom{\textbf{B}}}
\includegraphics[width=0.25\textwidth,page=1]{figures/simulation/blank.pdf}\\
\raisebox{0.15\textwidth}{\textbf{E}}
\includegraphics[width=0.25\textwidth,page=3]{figures/simulation/eigenvalues_normal_I9_JNA_RI_as_data_30_Mar_30_Mar.pdf}
\raisebox{0.15\textwidth}{\phantom{\textbf{B}}}
\includegraphics[width=0.25\textwidth,page=4]{figures/simulation/eigenvalues_normal_I9_JNA_RI_as_data_30_Mar_30_Mar.pdf}
\raisebox{0.15\textwidth}{\phantom{\textbf{B}}}
\includegraphics[width=0.25\textwidth,page=5]{figures/simulation/eigenvalues_normal_I9_JNA_RI_as_data_30_Mar_30_Mar.pdf}
\end{center}
\end{minipage}
\caption{Boxplots of the estimated eigenvalues of the auto-covariances of the fRI $B_i(t)$ (top row), and of the smooth error (bottom row) for all 200 simulations runs.}
\label{fig: boxplot eigenvalues RI non-centred}
\end{center}
\end{figure}

\begin{figure}[h!]
\centering
\includegraphics[width=0.25\textwidth,page=1]{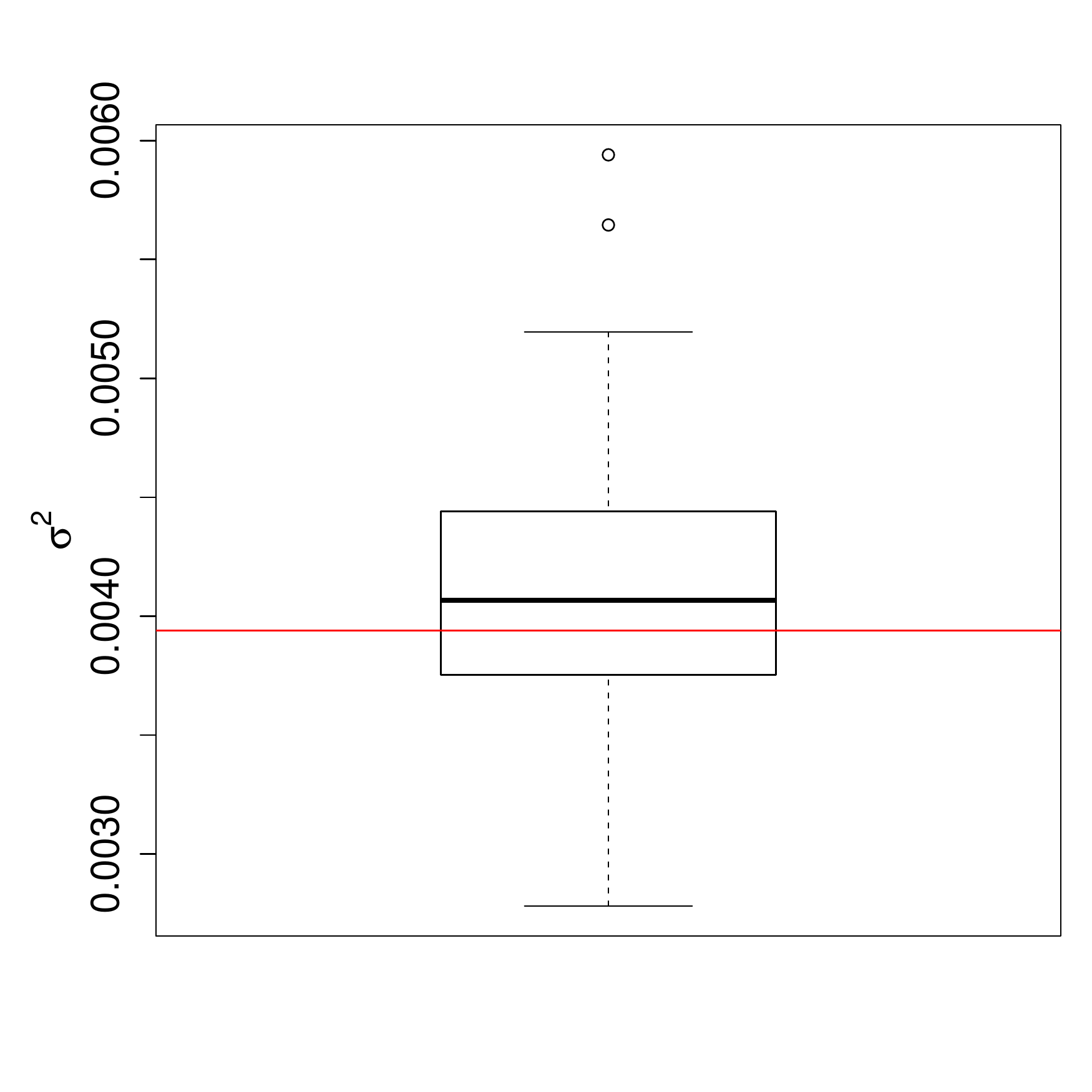}
\caption{Boxplots of the estimated error variances $\sigma^2$ for all 200 simulation runs.}
\label{fig: boxplot sigmasq RI non-centred}
\end{figure}

\renewcommand{\arraystretch}{0.8} 
 \setlength{\tabcolsep}{1mm}
\begin{table}[h!]
\centering
\small
\caption{rrMSEs averaged over 200 simulation runs for all model components by random process. Rows 1-3: Number of grouping levels $L^X$ and average rrMSEs for the fRI and the smooth error. Last row: Average rrMSEs for the functional response, the mean, and the error variance.}
\begin{tabular}{c|c|rrrrrrrrrrrr|rr}
$X$&$L^X$& $K^X$ & $\phi^X_1$ & $\phi^X_2$& $\phi^X_3$ & $\nu^X_1$ & $\nu^X_2$& $\nu^X_3$ & $\xi^X_1$ & $\xi^X_2$&$\xi^X_3$ & $X$ & $X_{\FPCFAMM}$ &$\sigma^2$ \\ 
\hline
$B$&9& 0.51 & 0.49 & 0.51& & 0.31 & 0.44 && 0.50 & 0.74 && 0.35 & 0.35 & \\ 
$E$ & 707& 0.07 & 0.04 & 0.05 & 0.04 & 0.05 & 0.05 & 0.05 & 0.16 & 0.20 & 0.25 & 0.18 & 0.17& \\ 
$Y$ & & & & & & & & & & & & 0.09 & 0.09 & 0.11 
\end{tabular}
\label{tab: mean riMSEs RI non-centred}
\end{table}

 \setlength{\tabcolsep}{1mm}
\begin{table}[h!]
\centering
\small
\caption{Average rrMSEs for the estimated mean and covariate effects for the estimation using the independence assumption (first row) and using FPC-FAMM (second row) averaged over 200 simulation runs.}
\begin{tabular}{l|rrrrrrrr}
& $f_0(t)$ & $f_1(t)$ & $f_2(t)$ & $f_3(t)$ & $f_4(t)$ & $f_5(t)$ & $f_6(t)$ & $f_7(t)$ \\ 
  \hline
$\mu(t\mx_{ijh})$& 0.13 & 0.13 & 0.22 & 1.43 & 0.30 & 0.58 & 0.39 & 0.59\\ 
$\mu(t\mx_{ijh})_{\FPCFAMM}$ & 0.12 & 0.12 & 0.21 & 1.34 & 0.25 & 0.53 & 0.38 & 0.47 \\ 
\end{tabular}
\label{tab: mean riMSEs RI non-centred covariates}
\end{table}

\begin{figure}[h!]
\begin{center}
\includegraphics[width=0.25\textwidth]{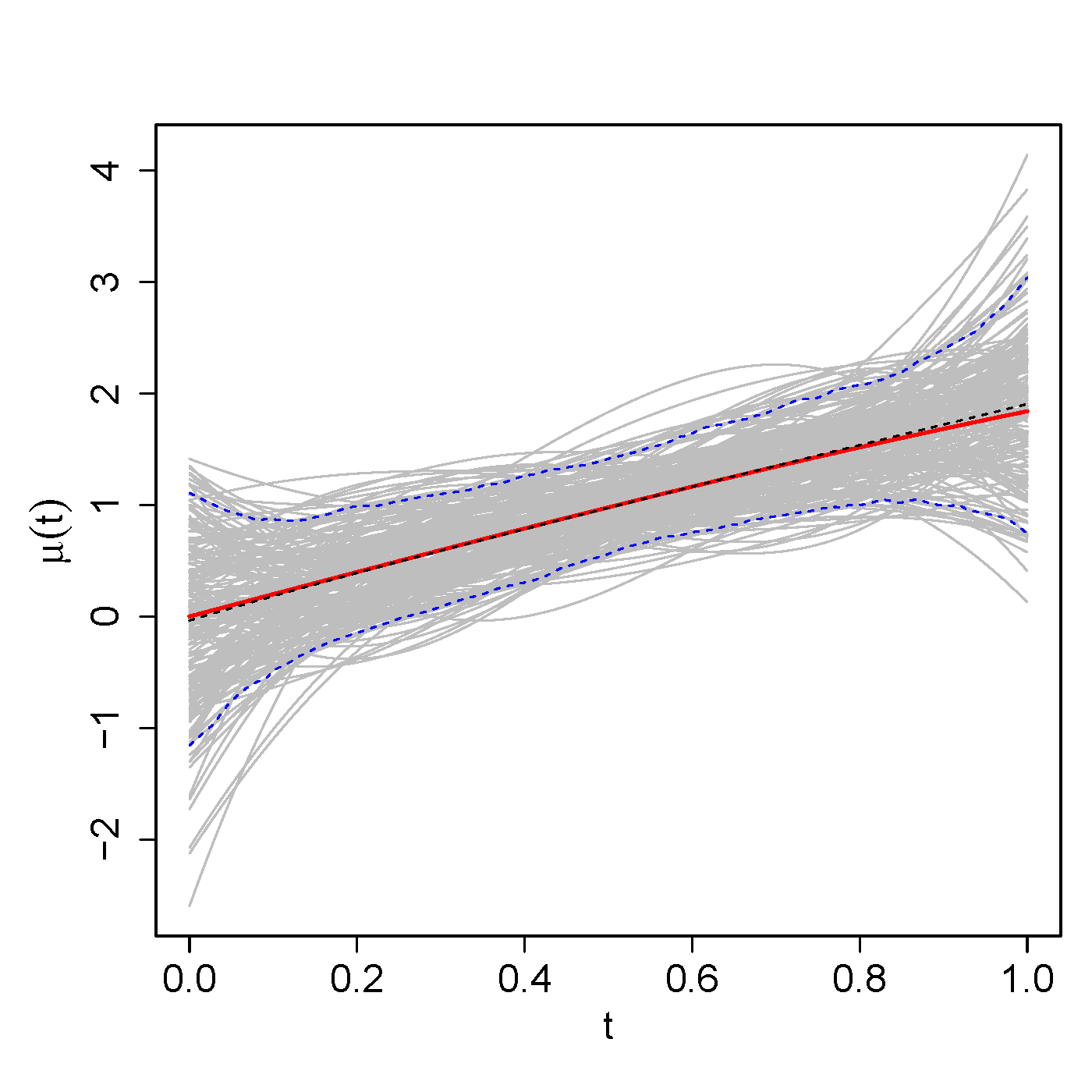}
\caption{True and estimated mean function $\mu(t,\mx_{ijh})$. Shown are the true function (red), the mean of the estimated functions over 200 simulation runs (black dashed line), the point-wise 5th and 95th percentiles of the estimated functions (blue dashed lines), and the estimated functions of all 200 simulation runs (grey).}
\label{fig: mean sparse non-centred}
\end{center}
\end{figure}

\begin{figure}[h!]
\begin{center}
\begin{minipage}{1\textwidth}
\begin{center}
\raisebox{0.15\textwidth}{\textbf{B}}
\includegraphics[width=0.25\textwidth,page=1]{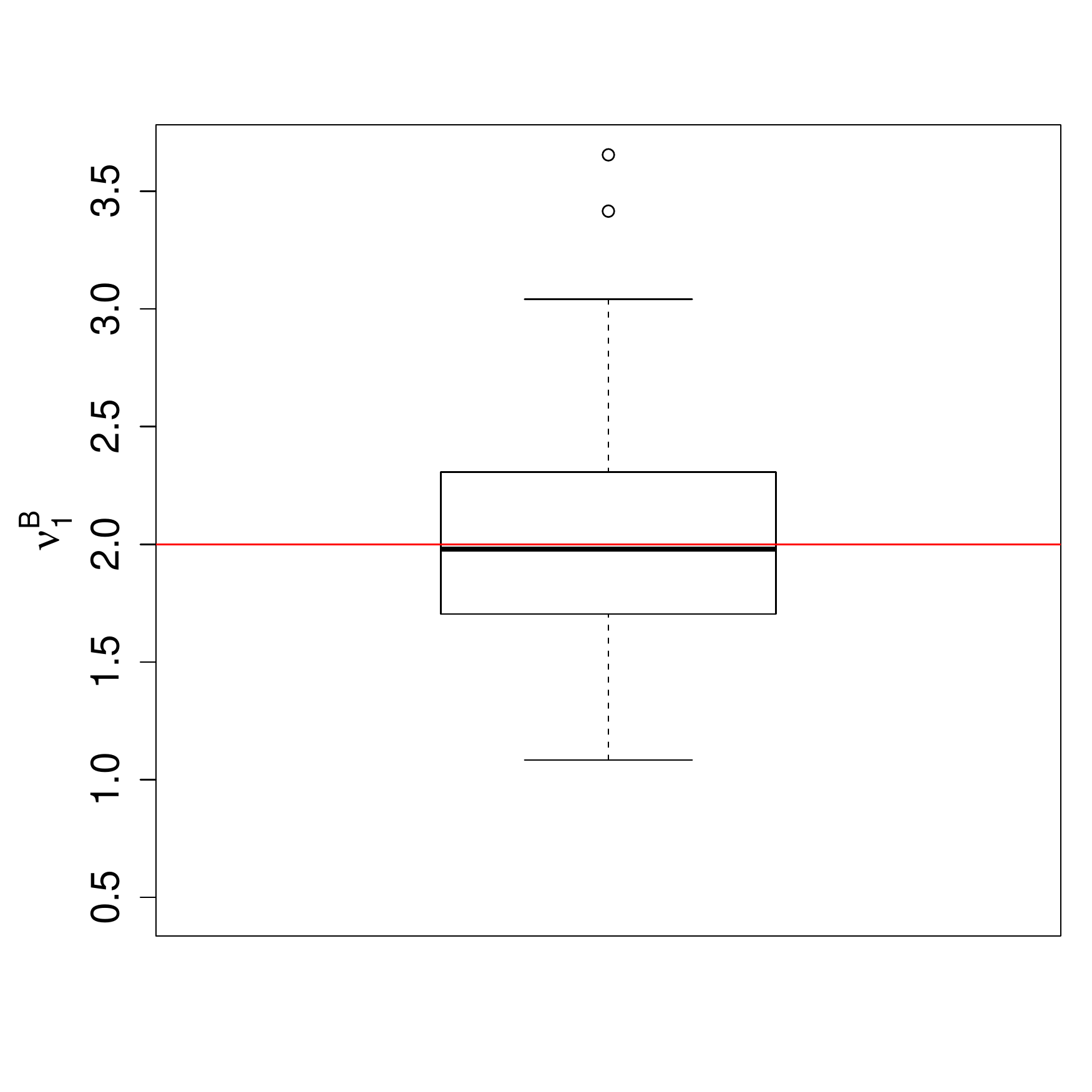}
\raisebox{0.15\textwidth}{\phantom{\textbf{B}}}
\includegraphics[width=0.25\textwidth,page=2]{figures/simulation/eigenvalues_normal_I40_J40_crossed_19_Jan_sparse_19_Jan.pdf}\\
\raisebox{0.15\textwidth}{\textbf{C}}
\includegraphics[width=0.25\textwidth,page=3]{figures/simulation/eigenvalues_normal_I40_J40_crossed_19_Jan_sparse_19_Jan.pdf}
\raisebox{0.15\textwidth}{\phantom{\textbf{B}}}
\includegraphics[width=0.25\textwidth,page=4]{figures/simulation/eigenvalues_normal_I40_J40_crossed_19_Jan_sparse_19_Jan.pdf}\\
\raisebox{0.15\textwidth}{\textbf{E}}
\includegraphics[width=0.25\textwidth,page=5]{figures/simulation/eigenvalues_normal_I40_J40_crossed_19_Jan_sparse_19_Jan.pdf}
\raisebox{0.15\textwidth}{\phantom{\textbf{B}}}
\includegraphics[width=0.25\textwidth,page=6]{figures/simulation/eigenvalues_normal_I40_J40_crossed_19_Jan_sparse_19_Jan.pdf}
\end{center}
\end{minipage}
\caption{Boxplots of the estimated eigenvalues of the auto-covariances of the crossed fRIs $B_i(t)$ (top row), $C_j(t)$ (middle row), as well as the eigenvalues of the auto-covariance of the smooth error $E_{ijh}(t)$ (bottom row) for all 200 simulations runs.}
\label{fig: boxplot eigenvalues sparse non-centred}
\end{center}
\end{figure}

\begin{figure}[h!]
\centering
\includegraphics[width=0.25\textwidth,page=1]{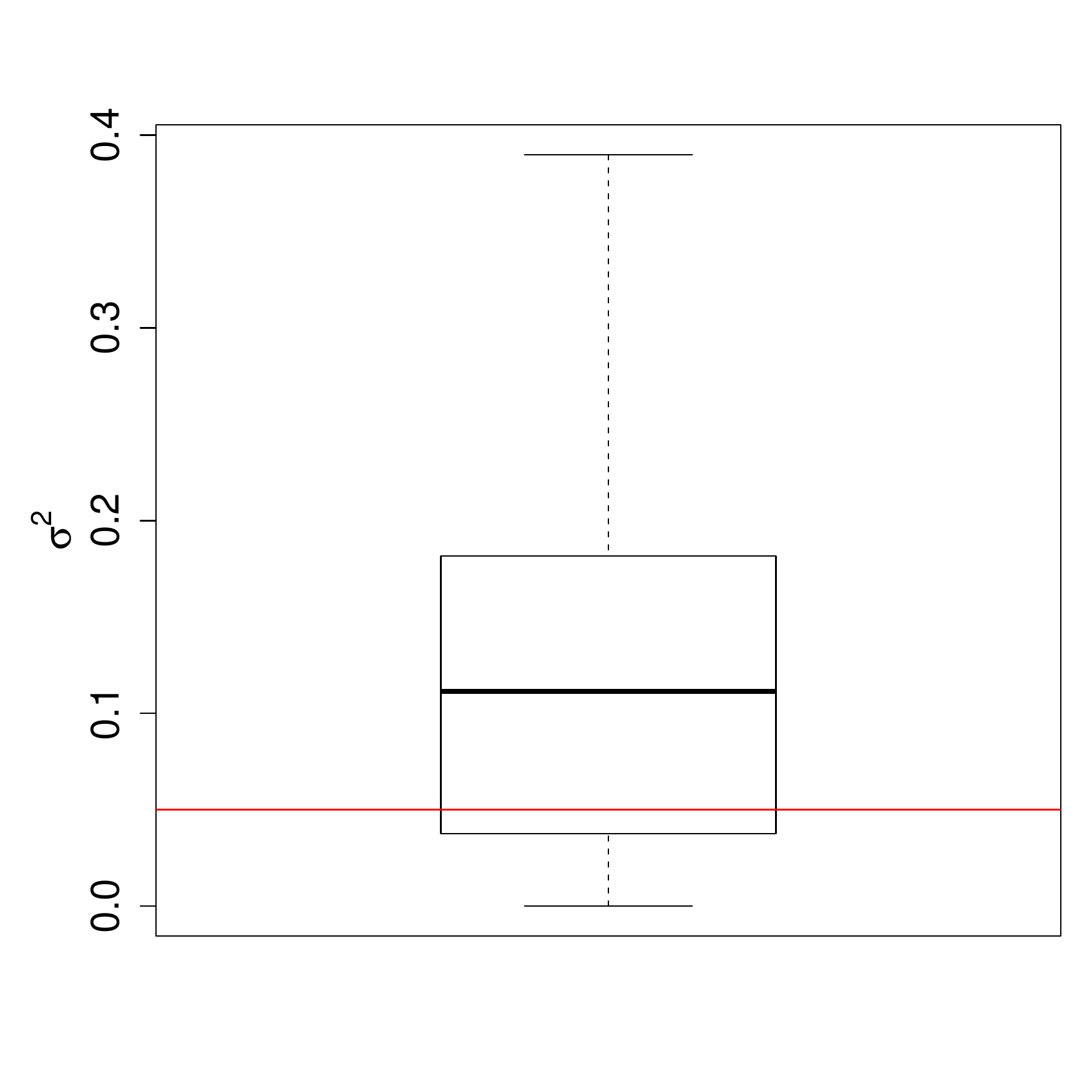}
\caption{Boxplots of the estimated error variances $\sigma^2$ for all 200 simulation runs.}
\label{fig: boxplot sigmasq sparse non-centred}
\end{figure}

\clearpage
\underline{Simulation results for the sparse scenario}.
In the following, additional results for the simulations of the sparse scenario with non-centred and non-decorrelated basis weights are shown. Figure \ref{fig: mean sparse non-centred} shows the true and estimated mean functions and Figure \ref{fig: boxplot eigenvalues sparse non-centred} depicts the boxplot of the estimated eigenvalues for the two fRIs and for the smooth error. In Figure \ref{fig: boxplot sigmasq sparse non-centred}, we show the boxplot of the estimated error variances. In Table \ref{tab: mean riMSEs sparse non-centred}, the average rrMSEs for all model components are given.

\begin{figure}[h!]
\begin{center}
\raisebox{0.15\textwidth}{\textbf{B}}
\includegraphics[width=0.25\textwidth,page=1]{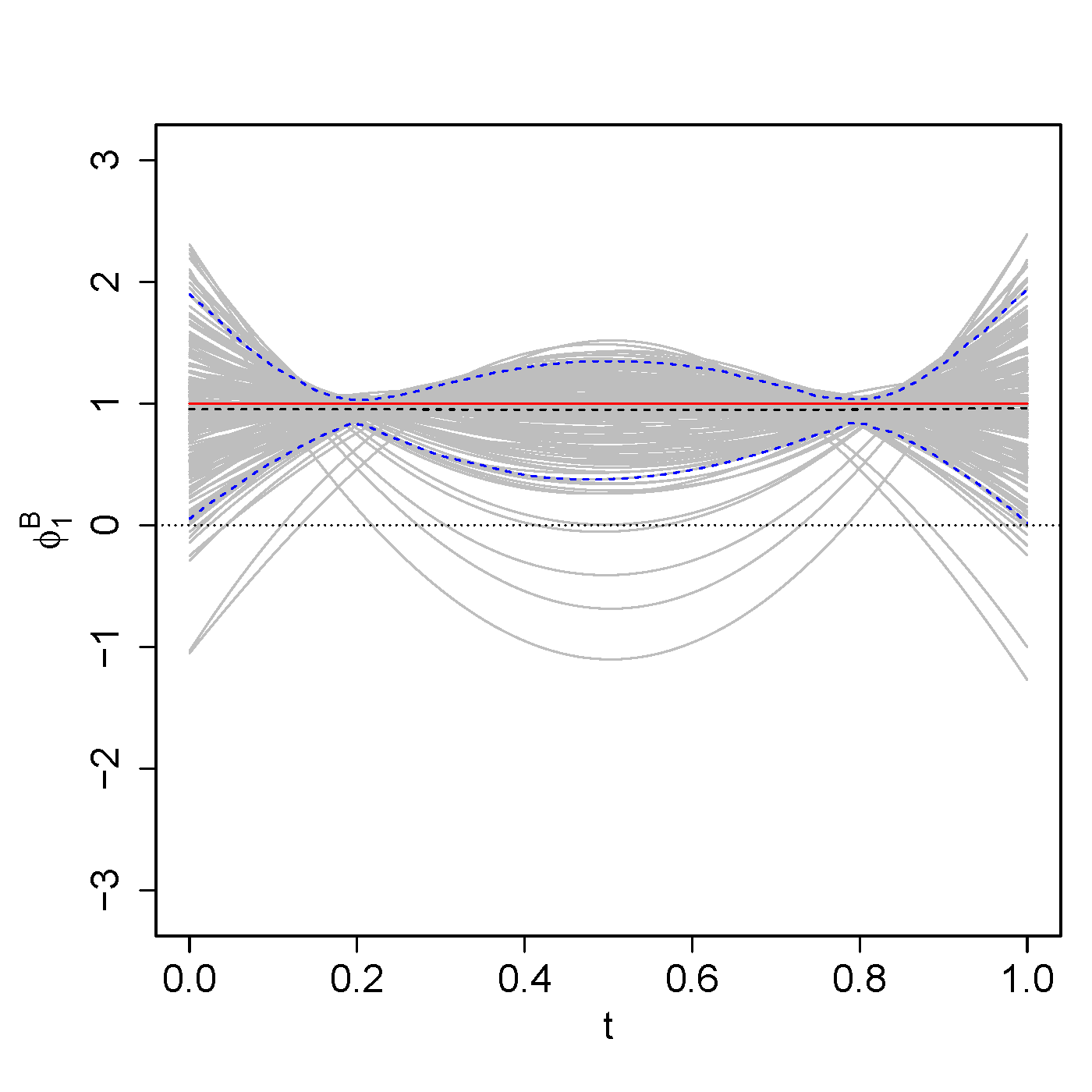}
\raisebox{0.15\textwidth}{\phantom{\textbf{B}}}
\includegraphics[width=0.25\textwidth,page=3]{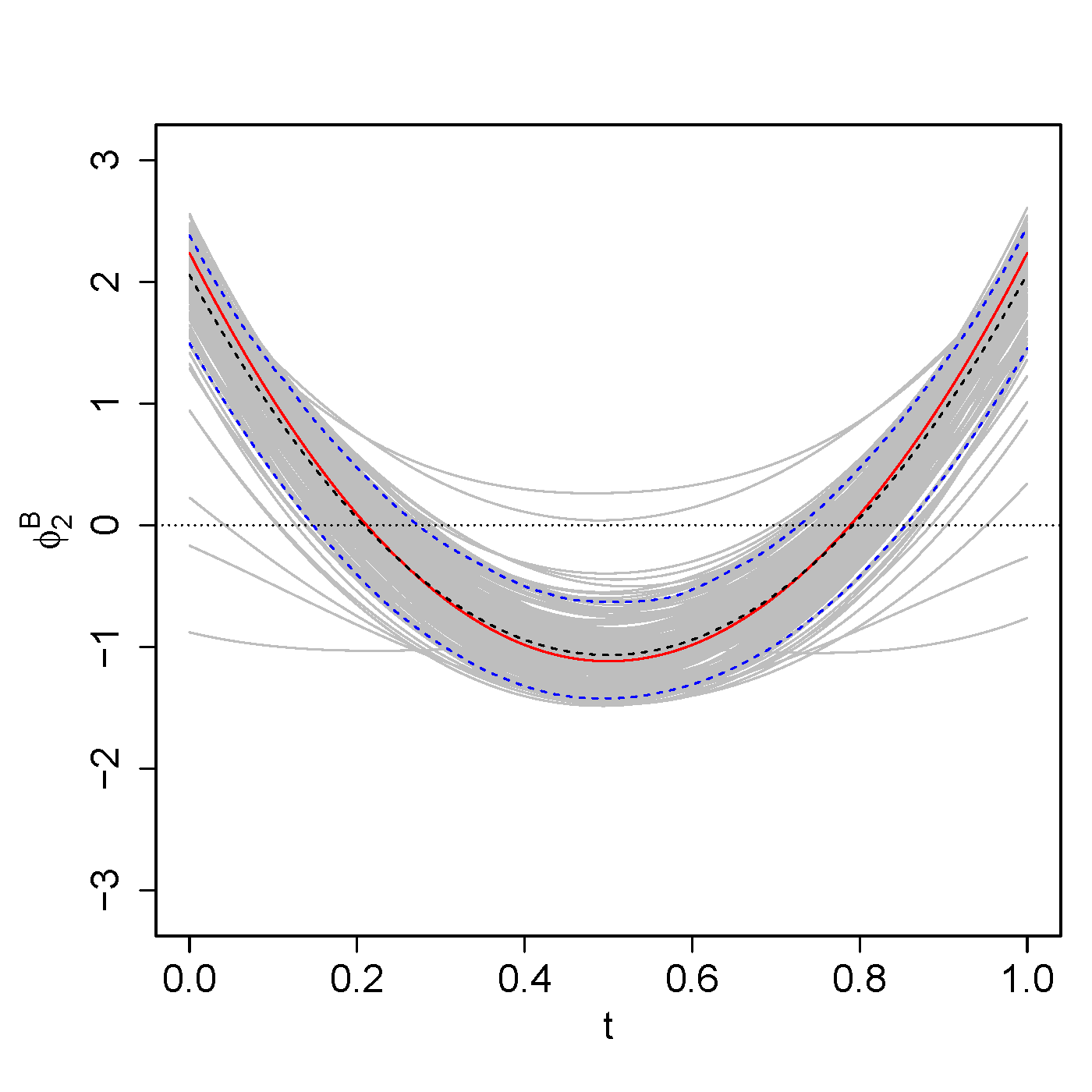}\\
\raisebox{0.15\textwidth}{\textbf{C}}
\includegraphics[width=0.25\textwidth,page=5]{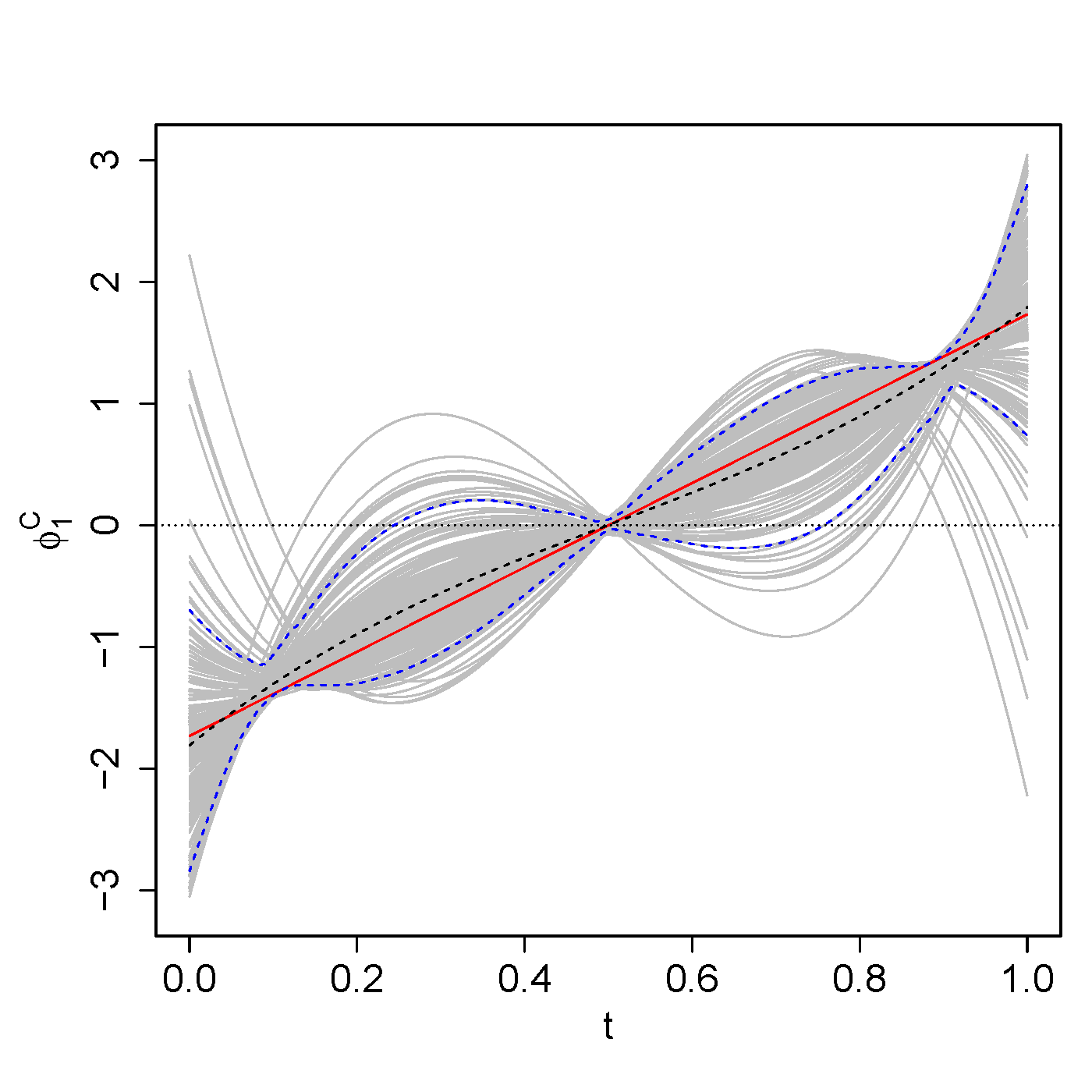}
\raisebox{0.15\textwidth}{\phantom{\textbf{B}}}
\includegraphics[width=0.25\textwidth,page=7]{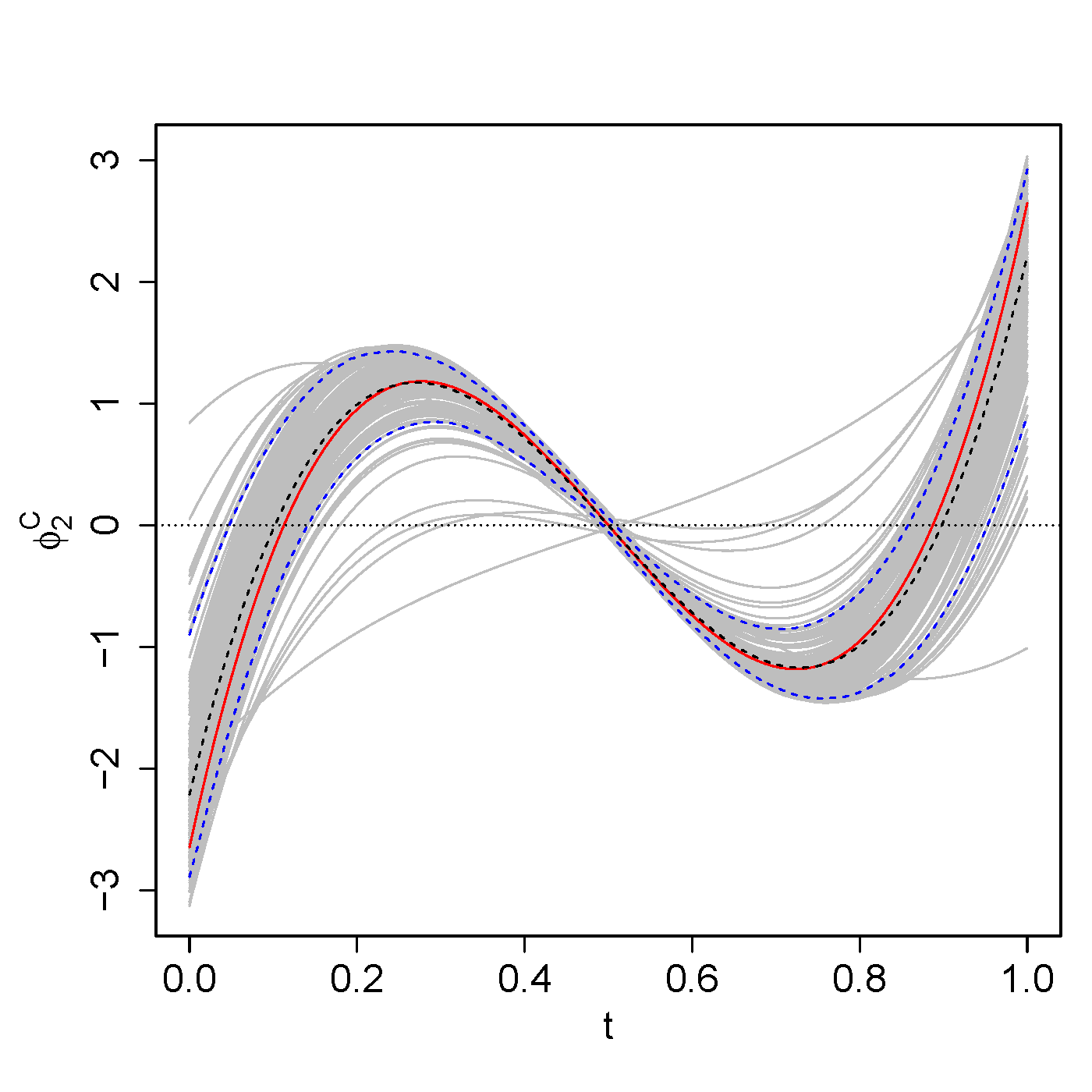}\\
\raisebox{0.15\textwidth}{\textbf{E}}
\includegraphics[width=0.25\textwidth,page=9]{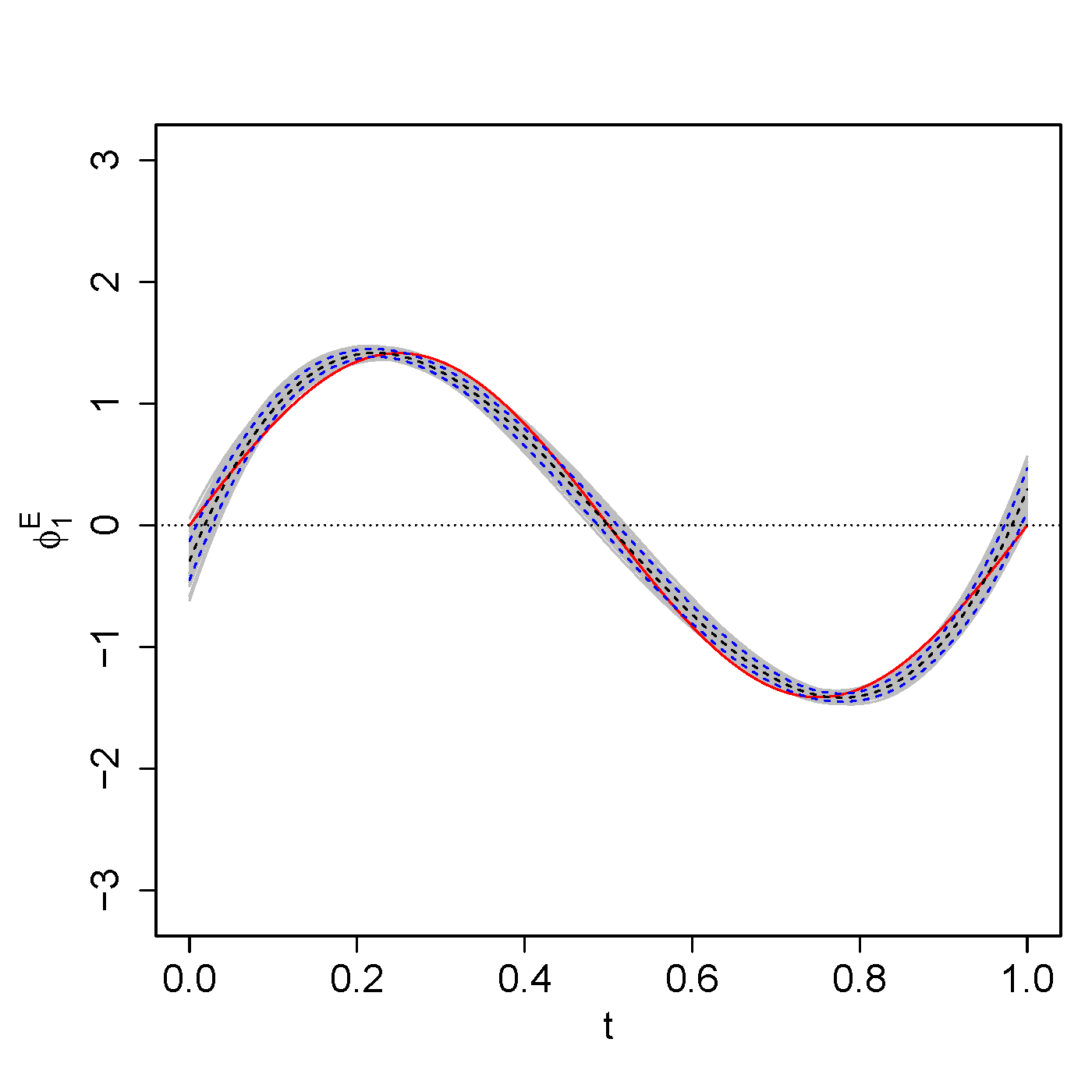}
\raisebox{0.15\textwidth}{\phantom{\textbf{B}}}
\includegraphics[width=0.25\textwidth,page=11]{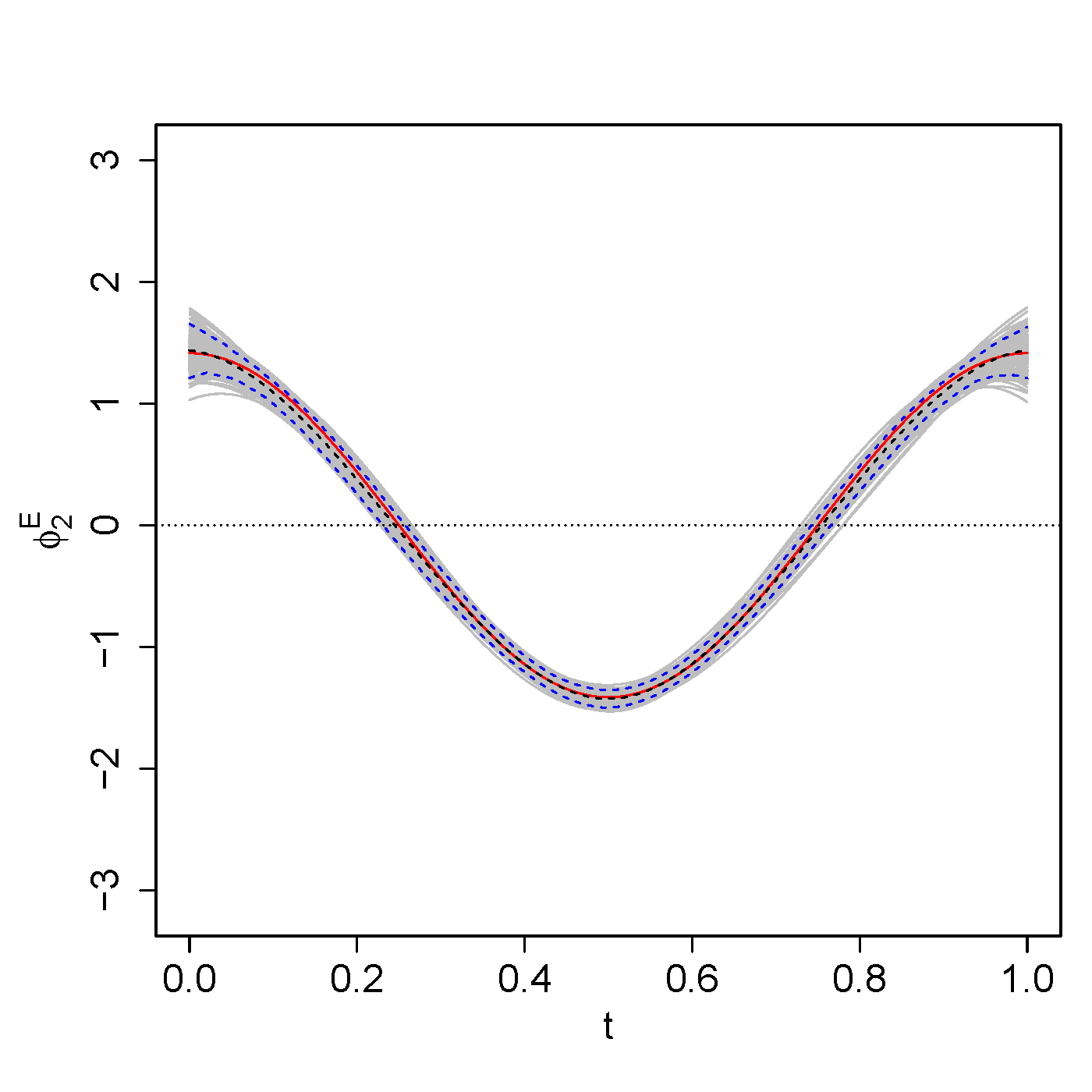}
\caption{True and estimated FPCs of the crossed fRIs $B_i(t)$ (top row) and $C_j(t)$ (middle row), as well as the FPCs of the smooth error $E_{ijh}(t)$ (bottom row). Shown are the true functions (red), the mean of the estimated functions over 200 simulation runs (black dashed line), the point-wise 5th and 95th percentiles of the estimated functions (blue dashed lines), and the estimated functions of all 200 simulation runs (grey).}
\label{fig: eigenfuncionts sparse non-centred}
\end{center}
\end{figure}

\renewcommand{\arraystretch}{0.8}
 \setlength{\tabcolsep}{1mm}
\begin{table}[h!]
\centering
\small
\caption{rrMSEs averaged over 200 simulation runs for all model components by random process. Rows 1-3: Number of grouping levels $L^X$ and average rrMSEs for the random processes. Last row: Average rrMSEs for the functional response, the mean, and the error variance.}
\begin{tabular}{c|c|rrrrrrrr|rr}
$X$&$L^X$& $K^X$ & $\phi^X_1$ & $\phi^X_2$ & $\nu^X_1$ & $\nu^X_2$ & $\xi^X_1$ & $\xi^X_2$ & $X$ & $\mu$&$\sigma^2$  \\ 
  \hline
$B$&40& 0.25 & 0.22 & 0.22 & 0.17 & 0.18 & 0.21 & 0.36 & 0.15 && \\ 
$C$ & 40& 0.24 & 0.24 & 0.26 & 0.17 & 0.20 & 0.36 & 0.48 & 0.28 &&\\ 
$E$ & 4800&0.14 & 0.11 & 0.07 & 0.03 & 0.05 & 0.30 & 0.20 & 0.30 &&\\ 
$Y$ & & &&&&&&& 0.09& 0.32 & 1.76
\end{tabular}
\label{tab: mean riMSEs sparse non-centred}
\end{table}

\end{document}

%% file: header_Sparse_FLMM.tex
\newcommand{\pkg}[1]{\textsf{#1}}
\newcommand{\bvec}{\left[\begin{array}{c}}
\newcommand{\evec}{\end{array}\right]}
\newcommand{\bmat}[1]{\left[\begin{array}{*{#1}{c}}}
\newcommand{\emat}{\end{array}\right]}

\newcommand{\tocite}[1]{\emph{\textcolor{red}{(CITE)}}}
\definecolor{mygray}{gray}{0.5}

\newcommand{\bi}{\begin{itemize}}
\newcommand{\ei}{\end{itemize}}
\newcommand{\ben}{\begin{enumerate}}
\newcommand{\een}{\end{enumerate}}
\newcommand{\beq}{\begin{equation}}
\newcommand{\eeq}{\end{equation}}
\newcommand{\bea}{\begin{eqnarray}}
\newcommand{\eea}{\end{eqnarray}}
\newcommand{\bc}{\begin{center}}
\newcommand{\ec}{\end{center}}
\newcommand{\beastern}{\begin{eqnarray*}}
\newcommand{\eeastern}{\end{eqnarray*}}

\newcommand{\tr}{^\top}
\newcommand{\pr}{\prime}
\newcommand{\eps}{\varepsilon}

\newcommand{\Var}{\operatorname{Var}}
\newcommand{\EV}{\mathds{E}}
\newcommand{\Cov}{\operatorname{Cov}}

\newcommand{\diag} {\operatorname{diag}}

\newcommand{\rrMSE}{\operatornamewithlimits{rrMSE}}

\newcommand{\FAMM}{\operatornamewithlimits{spline-FAMM}}
\newcommand{\FPCFAMM}{\operatornamewithlimits{FPC-FAMM}}
\newcommand{\sS}{/s\#sh/}
\newcommand{\Ss}{/sh\#s/}
\newcommand{\dint}{\operatornamewithlimits{d}}

\newcommand{\mG}{\bm{G}}

\newcommand{\mI}{\bm{I}}
\newcommand{\mK}{\bm{K}}

\newcommand{\mM}{\bm{M}}
\newcommand{\mP}{\bm{P}}

\newcommand{\mS}{\bm{S}}

\newcommand{\mU}{\bm{U}}

\newcommand{\mx}{\bm{x}}

\newcommand{\mY}{\bm{Y}}

\newcommand{\mz}{\bm{z}}

\newcommand{\mPhi}{\bm{\Phi}}
\newcommand{\mphi}{\bm{\phi}}
\newcommand{\mxi}{\bm{\xi}}

\newcommand{\mtheta}{\bm{\theta}}

\newcommand{\meps}{\bm{\varepsilon}}
\newcommand{\mPsi}{\bm{\Psi}}
\newcommand{\mmu}{\bm{\mu}}

\numberwithin{equation}{section}

\definecolor{LmuGreen}{RGB}{0,148,64}